\pdfoutput=1
 \documentclass[twocolumn,trackchanges]{aastex63}
\usepackage[figuresright]{rotating}
\usepackage{tabularx}
\newcommand{\kms}{{\rm km\,s}^{-1}}

\newcommand{\HI}{H\,\textsc{i}}

\newcommand{\dotmin}{\rlap.{'}}
\newcommand{\dotdeg}{\rlap.{^\circ}}

\shorttitle{FASHI 21\,cm \HI\ absorption galaxies}
\shortauthors{Zhang et al.}

\begin{document}

\title{FASHI DR2: A Catalog of 132 Low-Redshift H\,\small{I} 21 cm Absorption Systems}
\correspondingauthor{Chuan-Peng Zhang}
\email{cpzhang@nao.cas.cn}

\author[0000-0002-4428-3183]{Chuan-Peng Zhang}
\affiliation{State Key Laboratory of Radio Astronomy and Technology, National Astronomical Observatories, Chinese Academy of Sciences, Beijing 100101, China}
\affiliation{Guizhou Radio Astronomical Observatory, Guizhou University, Guiyang 550000, China}

\author{Ming Zhu}
\author{Peng Jiang}
\affiliation{State Key Laboratory of Radio Astronomy and Technology, National Astronomical Observatories, Chinese Academy of Sciences, Beijing 100101, China}
\affiliation{Guizhou Radio Astronomical Observatory, Guizhou University, Guiyang 550000, China}

\author{Hong Guo}
\affiliation{Shanghai Astronomical Observatory, Chinese Academy of Sciences, Nandan Road 80, Shanghai 200030, China}

\author{Yizhou Gu}
\affiliation{Tsung-Dao Lee Institute and Key Laboratory for Particle Physics, Astrophysics and Cosmology, Ministry of Education, Shanghai Jiao Tong University, Shanghai 201210, China}
\affiliation{Department of Astronomy, School of Physics and Astronomy, and Shanghai Key Laboratory for Particle Physics and Cosmology, Shanghai Jiao Tong University, Shanghai 200240, China}

\author{Cheng Cheng}
\affiliation{Chinese Academy of Sciences South America Center for Astronomy, National Astronomical Observatories, CAS, Beijing 100101, China}

\author{Jin-Long Xu}
\author{Nai-Ping Yu}
\author{Xiao-Lan Liu}
\author{Bo Zhang}
\affiliation{State Key Laboratory of Radio Astronomy and Technology, National Astronomical Observatories, Chinese Academy of Sciences, Beijing 100101, China}
\affiliation{Guizhou Radio Astronomical Observatory, Guizhou University, Guiyang 550000, China}




\begin{abstract}
We present an untargeted survey of 21 cm \HI\ absorption systems based on the second data release of the \textbf{F}AST \textbf{A}ll \textbf{S}ky \textbf{H\,{\footnotesize{I}}} survey (FASHI DR2), covering approximately 19,500 deg$^{2}$ at $z\lesssim0.09$. A total of 132 \HI\ absorbers are identified, including approximately 60 new discoveries, forming one of the largest homogeneous samples of low-redshift \HI\ absorbers assembled to date. The sample extends to continuum flux densities as low as 2.6 mJy, substantially below the limits of previous flux-limited surveys. The absorber population is dominated by narrow systems ($W_{50}<100$ km\,s$^{-1}$), while broad absorbers ($W_{50}>200$ km\,s$^{-1}$) account for 13.6\% of the sample. Most absorbers are optically thin, with a median optical depth of $\tau_{\rm HI}\approx0.14$. The velocity-offset distribution is broadly symmetric about the systemic velocities of the host galaxies. The associated absorbers are preferentially found in massive, actively star-forming galaxies. We find tentative evidence for a weak anti-correlation between \HI\ column density and stellar mass, although the relation exhibits substantial scatter. These results provide the first statistical characterization of the low-redshift \HI\ absorber population based on the FASHI DR2 sample and establish a valuable benchmark for future \HI\ absorption surveys with next-generation radio facilities.
\end{abstract}

\section{Introduction}
\label{sec:intro}

Neutral atomic hydrogen (\HI) plays a fundamental role in galaxy evolution by providing the primary reservoir from which molecular gas and stars ultimately form (e.g., \citealt{Walter2008,Leroy2008,Zhang2026na}). Over the past two decades, large-area 21-cm emission surveys have greatly improved our understanding of the global \HI\ content of galaxies in the local Universe (e.g., \citealt{Zwaan2005,Haynes2018,Zhang2024fashi,Zhang2026fashi}). These surveys have established the statistical properties of \HI-selected galaxies and their connection to galaxy structure, environment, and star formation. However, 21-cm emission observations mainly trace the bulk neutral gas reservoir and are significantly less sensitive to cold, compact, or low-column-density gas components.

In contrast, \HI\ 21-cm absorption provides a uniquely sensitive probe of cold neutral gas along the line of sight to radio continuum sources. Because the detectability of absorption depends primarily on the background continuum brightness rather than distance, absorption measurements can reveal cold gas structures that are difficult to detect in emission, including circumnuclear gas, compact clouds, and low-column-density components of the interstellar medium (ISM) \citep{Morganti2001,Curran2010}. \HI\ absorption studies therefore provide important constraints on gas accretion, ISM structure, and feedback processes associated with active galactic nuclei (AGN).

At low redshift ($z \lesssim 0.1$), \HI\ absorption systems are particularly valuable because the absorbing gas can be directly connected to the physical properties of the host galaxies, including stellar mass, star formation activity, morphology, and nuclear activity. Previous studies have shown that \HI\ absorption is commonly associated with compact radio sources and gas-rich galaxies, and may arise from a combination of rotating galactic disks, circumnuclear structures, and jet--ISM interactions \citep{Chandola2011,Gereb2015,Morganti2018}. However, most existing low-redshift absorption surveys have relied on targeted observations of pre-selected radio galaxies or AGN samples \citep{Curran2008,Allison2012}. As a result, current absorber catalogs remain relatively small and are subject to complex selection effects related to radio luminosity, source morphology, and survey strategy. These limitations have hindered robust statistical studies of the low-redshift \HI\ absorber population, including their incidence, kinematic properties, and host-galaxy demographics.

The advent of new-generation wide-field \HI\ surveys now provides an opportunity to overcome many of these limitations \citep{Allison2022,Yoon2025}. The Five-hundred-meter Aperture Spherical radio Telescope (FAST) \HI\ Survey (FASHI) combines the exceptional sensitivity of FAST with large sky coverage, enabling both emission and absorption studies over a cosmologically representative volume \citep{Zhang2025ab}. The second FASHI data release (DR2) contains more than 156,000 extragalactic \HI\ emission detections at $z \lesssim 0.09$, with a median sensitivity of 0.57 mJy beam$^{-1}$ at a velocity resolution of 6.4 km\,s$^{-1}$ \citep{Zhang2026fashi}. Building upon our previous FASHI DR1 absorber catalog \citep{Zhang2025ab}, which was based on approximately 10,000 deg$^{2}$ of sky coverage and contained 51 \HI\ absorption systems, the DR2 dataset nearly doubles the survey area to about 19,500 deg$^{2}$ and increases the absorber sample to 132 systems. This substantial expansion enables more robust statistical investigations of absorber demographics, line-width distributions, velocity offsets, and host-galaxy properties, while providing improved constraints on the low-redshift \HI\ absorber population.

In this paper, we present an untargeted search for \HI\ 21-cm absorption systems using the FASHI DR2 dataset. We identify a total of 132 absorbers, including approximately 60 new discoveries, forming one of the largest homogeneous samples of low-redshift \HI\ absorbers assembled to date. The sample extends to significantly fainter radio continuum sources than previous flux-limited surveys, demonstrating the capability of FAST to detect weak absorption systems that were previously inaccessible. Using this sample, we investigate the statistical properties of low-redshift absorbers, including their optical depths, column densities, line widths, velocity offsets, and host-galaxy properties. We additionally examine the occurrence of high-velocity blueshifted absorbers that may be associated with AGN-driven neutral gas outflows. These results provide new constraints on the demographics and kinematics of cold neutral gas in the local Universe and establish an important benchmark for future absorption-line studies with next-generation radio facilities such as the Square Kilometre Array \citep{Braun2015,Mahony2026}.

The structure of this paper is as follows. In Section~\ref{sec:data_reduc}, we describe the FASHI survey and the procedures used to identify \HI\ absorption candidates. Section~\ref{sec:result} presents the absorber catalog and the statistical properties of the sample. In Section~\ref{sec:discu}, we discuss the implications of the results for the low-redshift \HI\ absorber population and their connection to galaxy evolution. Finally, our main conclusions are summarized in Section~\ref{sec:summary}.

 \begin{figure*}[htp]
 \centering
 \includegraphics[width=0.99\textwidth, angle=0]{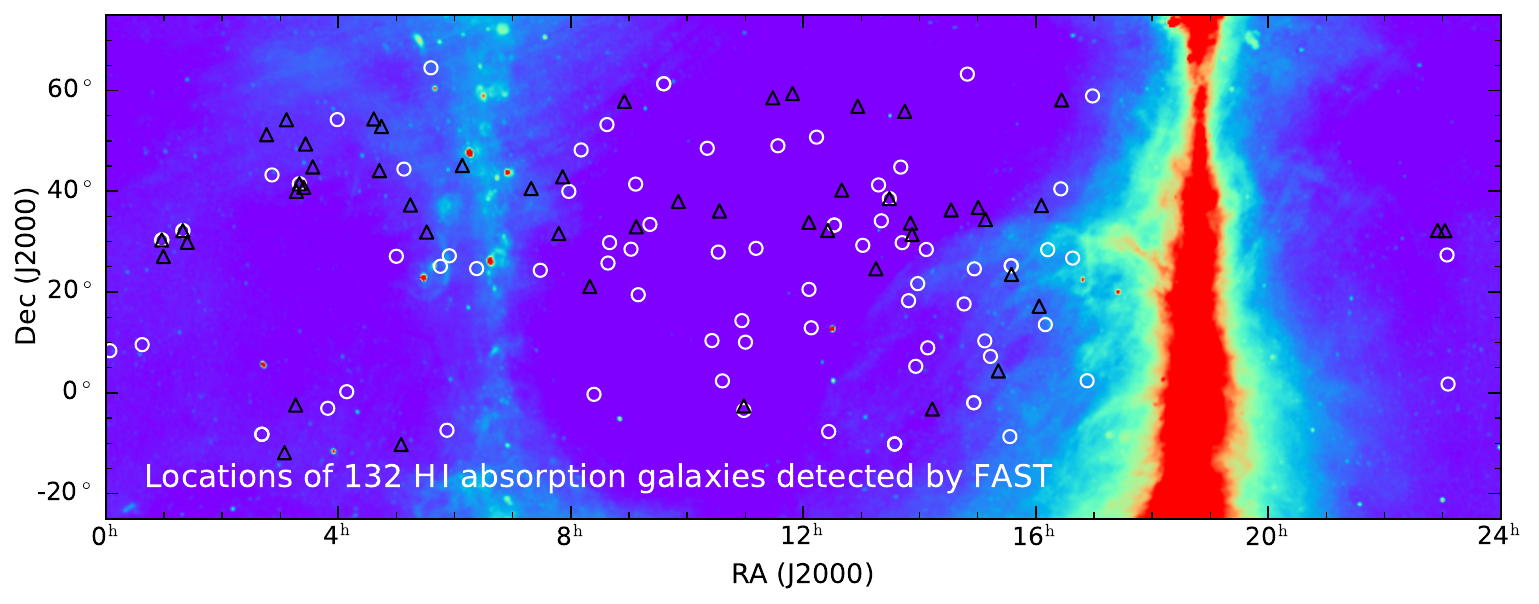}
 \caption{Locations of the 132 H\,{\scriptsize{I}} absorption galaxies detected by FASHI across about 19,500 deg$^{2}$. Circles indicate sources from this study, and triangles represent those from \citet{Zhang2025ab}. The background is the all-sky 21-cm radio continuum map from the Stockert 25-m \citep{Reich1982} and Villa Elisa 30-m \citep{Reich1986, Reich2001} surveys.}
 \label{Fig:location}
 \end{figure*}

\section{Observations and Data Processing}
\label{sec:data_reduc}

\subsection{Observations}

The \HI\ 21\,cm absorption spectra analyzed in this work are derived from the FASHI project \citep{Zhang2026fashi}, which utilizes the 19-beam receiver of FAST to achieve efficient sky coverage. FAST has a nominal aperture of 500\,m and an effective aperture of approximately 300\,m, yielding a beam size of $\sim$2.9$'$ at 1.4\,GHz, with a pointing accuracy better than $15''$. The spectral backend covers a total bandwidth of 500\,MHz (1000–1500\,MHz) with 65,536 channels, providing a raw frequency resolution of 7.63\,kHz, corresponding to a velocity resolution of 1.6\,$\kms$ at 1.4\,GHz. Intensity calibration is performed by injecting a $\sim$11\,K noise signal for 1 second every 32 or 64 seconds throughout the observations, with the conversion factor from flux density to antenna temperature (DPFU) ranging from 13 to 17\,K\,Jy$^{-1}$ at 1.4\,GHz for the 19 beams \citep{Nan2011,Jiang2019,Jiang2020}. The FASHI survey primarily adopts a drift-scan observing strategy: the azimuth arm is fixed along the meridian at a pre-assigned J2000 declination, with adjacent drift centers separated by $21\dotmin65$, and the feed array is rotated by $23\dotdeg4$ to achieve super-Nyquist sampling, ensuring that the sampling rate of the Earth-rotation drift-scan tracks is smaller than half of the beam FWHM for each individual beam.

\subsection{Data Reduction}

Due to severe radio frequency interference (RFI) and the relatively poor sensitivity at lower frequencies, our analysis is restricted to data above 1305.5\,MHz, corresponding to $z\lesssim0.09$ \citep[see][]{Zhang2022rfi}. The FASHI data are processed through \texttt{HiFAST}, the dedicated FAST spectral data reduction pipeline \citep{Jing2024}, which integrates modules for antenna temperature calibration, baseline fitting, RFI flagging, standing wave removal, gridding, flux calibration, and FITS cube generation. For baseline correction, we employ the Asymmetrically Reweighted Penalized Least Squares algorithm (\texttt{arPLS}; \citealp{Baek2015}). The FASHI spectral data exhibit standing waves with characteristic frequency periods of $\sim$1\,MHz, $\sim$2\,MHz, and $\sim$0.04\,MHz; correcting these standing waves significantly improves the spectral quality. Standing wave removal is primarily achieved through fitting and subtraction using the \texttt{SW} package within the \texttt{HiFAST} pipeline \citep{Jing2024,Xu2025}. The data are smoothed to a final spectral resolution of $\sim$6.4\,$\kms$ per channel, and the resulting data cubes are gridded with a pixel scale of 1\,arcmin. A heliocentric velocity correction is applied to account for the Doppler effect. Further details on the observational setup and reduction procedures can be found in the FASHI survey paper by \citet{Zhang2024fashi}.

\begin{figure*}[htp]
 \centering
 \includegraphics[height=0.24\textwidth, angle=0]{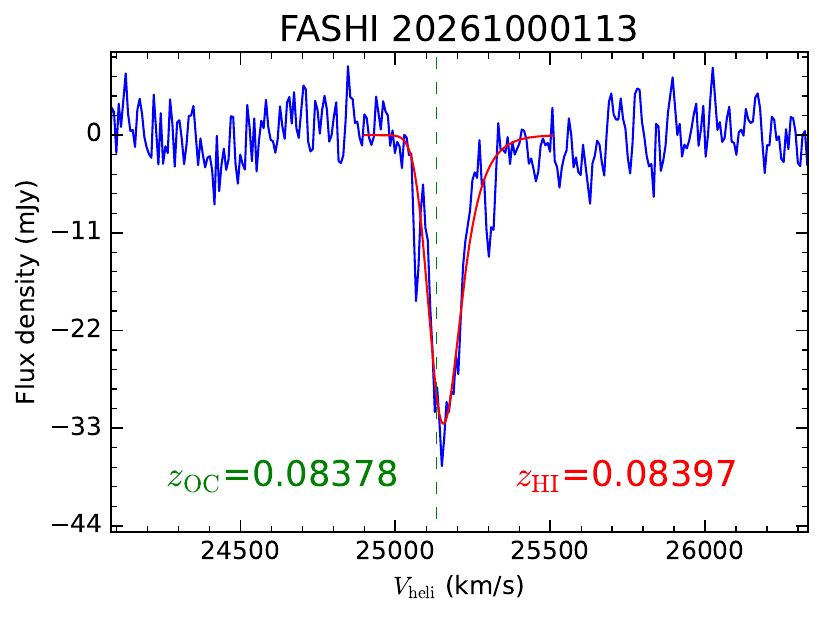}
 \includegraphics[height=0.27\textwidth, angle=0]{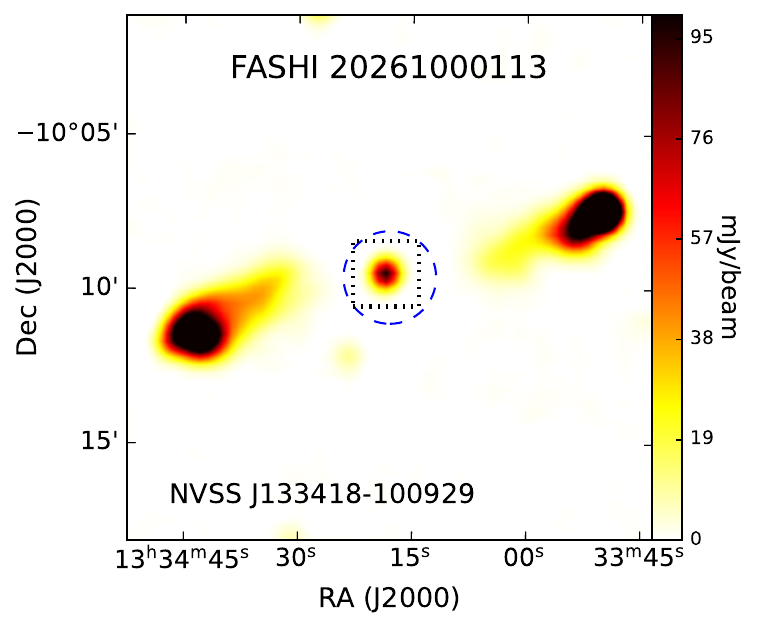}
 \includegraphics[height=0.27\textwidth, angle=0]{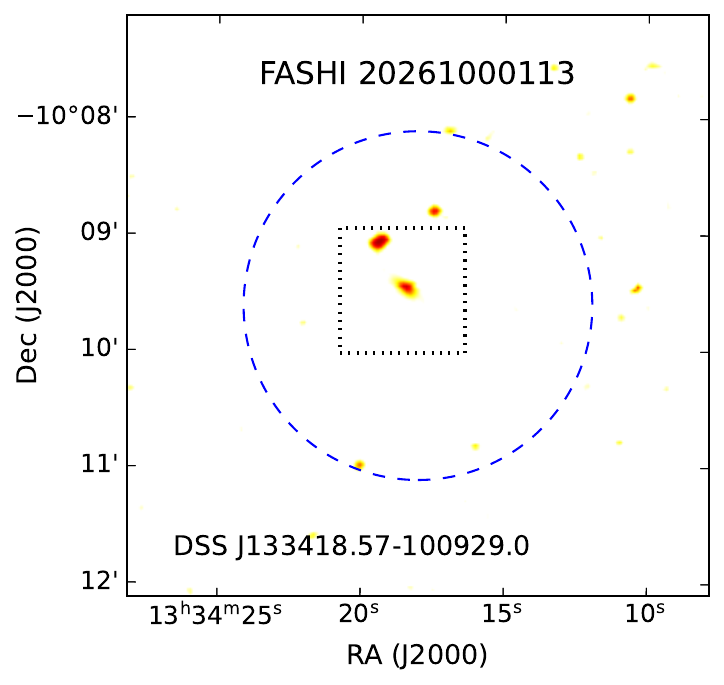}
 \caption{H\,{\scriptsize{I}} absorption galaxy ID\,20261000113. \textit{Left}: H\,{\scriptsize{I}} absorption spectrum. The \HI\ redshift measured by FAST (in red) and an optical spectroscopic redshift (in green) are presented in the panel. The red curve line is the fit to the spectrum, and the green dashed line is the optical spectroscopic redshift. \textit{Middle}: Blue circle indicates the position of \HI\ source with a beam size of $\sim$2.9$'$. The background shows 1.4\,GHz radio continuum distribution. The bright source within a square is the radio counterpart. \textit{Right}: The background shows DSS image. The bright source within a square is the DSS counterpart. The other 50 H\,{\scriptsize{I}} absorption galaxies are shown in the online version of Figure\,\ref{Fig:FASHI_hi}.}
 \label{Fig:FASHI_hi}
 \end{figure*}

\subsection{Source Finding and Candidate Identification}
\label{sec:extraction}

The identification of \HI\ absorption systems follows a two-step strategy that combines automated detection with interactive verification. For the initial automated step, we employ version 2 of the \HI\ Source Finding Application (\texttt{SoFiA})\footnote{\url{https://github.com/SoFiA-Admin/SoFiA-2}}, a widely used tool in large-scale \HI\ surveys \citep[e.g.,][]{Westmeier2022,Zhang2024fashi}. To search for absorption rather than emission, we set \texttt{input.invert = true}, which instructs \texttt{SoFiA} to detect negative signals. Apart from this parameter change, the absorption source finding procedure is identical to the emission source extraction described in \citet{Zhang2024fashi}.

Our \texttt{SoFiA} configuration adopts a detection threshold of $4.5\sigma$, where $\sigma$ denotes the local noise level within the source bounding box (see the \texttt{SoFiA} Cookbook for details). The smoothing kernels are set to \texttt{kernelsXY = 0, 3, 6} and \texttt{kernelsZ = 0, 3, 7, 15}. For source linking, the minimum source size is 5 pixels/channels in both spatial and spectral dimensions, while the maximum size is 50 pixels in the spatial domain (unlimited in the spectral domain). We disable \texttt{reliability.enable} to retain potential artifacts such as RFI and residual baseline fluctuations, which are subsequently removed during visual inspection.

Because automated detection inevitably produces false positives due to the RFI \citep[e.g.,][]{Zhang2022rfi}, we perform manual verification for all candidates. Each candidate is examined based on its moment maps (0th, 1st, and 2nd orders), integrated spectral profile, signal-to-noise ratio, spatial coordinates, flux density, optical counterpart, and radio continuum counterpart. Spurious detections are discarded during this interactive process. After this rigorous validation, we have identified and confirmed a total of 132 \HI\ absorbers (see Figures\,\ref{Fig:location} and \ref{Fig:FASHI_hi}).

\subsection{Spectral Line Fitting}
\label{sec:fit}

Deriving absorption line properties, such as profile width and the presence of multiple kinematic components, has traditionally been accomplished using Gaussian fitting \citep[e.g.,][]{Zhangbo2021,Hu2023}. However, this approach requires an a priori assumption about the number of Gaussian components, which is particularly problematic for complex, multi-peaked profiles such as those commonly found in our absorption sample. To overcome this limitation, we adopt the busy-function fitting method developed by \citet{Westmeier2014}, which has been successfully applied to \HI\ absorption studies by \citet{Gereb2015}. The busy function is a heuristic analytic function capable of fitting a wide range of line shapes, including symmetric, asymmetric, and multi-peaked profiles, without the need to pre-specify the number of components.

Our fitting procedure distinguishes three scenarios based on the presence and relative position of emission features. When an absorption line appears in isolation, the busy function is applied directly to the spectrum. When emission is present but does not blend with the absorption feature, we first mask the emission line and then fit the absorption line. When absorption and emission are blended such that the absorption line is embedded within the emission profile, we first mask the absorption feature and fit the emission component, then subtract the best-fit emission model from the original spectrum before fitting the residual absorption line. This approach has been successfully applied to absorption line fitting by \citet{Zhang2025ab}.

\section{Detection Results and Analysis}
\label{sec:result}

\subsection{Spatial Distribution of FASHI \HI\ Absorbers}

The FASHI survey has covered approximately 19,500 deg\(^2\) of the sky, spanning right ascensions from \(0^h\) to \(24^h\) and declinations from \(-14^\circ\) to \(+66^\circ\), with a typical spectral detection sensitivity of \(\sim\)1.04 mJy at a velocity resolution of \(\sim\)6.4 km s\(^{-1}\) at 1.4\,GHz. However, due to the schedule-filler nature of the project, the coverage is not fully uniform across the entire area. Figure~\ref{Fig:location} presents the spatial distribution of the 132 \HI\ absorption systems detected by FAST. The current detection rate of \HI\ absorbers is approximately 0.007 sources per square degree, compared to about 8.0 sources per square degree for \HI\ emission sources \citep{Zhang2026fashi}. Based on these numbers, the occurrence probability of \HI\ absorption among all \HI\ galaxies is estimated to be about 1/1180. Given that the survey sensitivity is not spatially uniform \citep{Zhang2026fashi}, this value should be regarded as an approximate estimate. Assuming the current detection sensitivity and extending the coverage to the full FASHI goal of 22,000 deg\(^2\) at \(z \lesssim 0.09\), the total number of \HI\ absorbers detectable by FASHI is expected to reach approximately 150.

\begin{table*}[htp]
\caption{Coordinates and counterparts of FASHI \HI\ absorption galaxies}
\label{tab:name}
\vskip 0pt
\centering \scriptsize
\renewcommand{\arraystretch}{0.80}
\setlength{\tabcolsep}{1.0mm}{
\begin{tabular}{ccccccccc}
\hline \hline
[1]  &  [2]  & [3]   & [4]  & [5] & [6]  & [7]   & [8]  & [9]   \\ 
FASHI ID & FASHI & RA & Dec & $l$ & $b$ & NVSS & OC & Galaxy \\
& J2000 & deg & deg & deg & deg & J2000 & J2000     \\
\hline
20261000001 & J135606.53+051545.8 & 209.0272 & 5.2627 &  340.9653 & 63.2540 & J135607+051517 & J135607.21+051517.20 & NGC5363 \\
20261000002 & J024105.46-081330.9 & 40.2728 & -8.2253 &  181.9784 & -57.9039 & J024104-081521 & J024104.80-084439.25 & NGC1052 \\
20261000005 & J055210.37-072613.4 & 88.0432 & -7.4371 &  212.9065 & -16.5406 & J055211-072722 & J055211.38-073237.49 & -- \\
20261000011 & J034906.63-030423.3 & 57.2776 & -3.0732 &  191.2754 & -41.2394 & J034904-030353 & J034905.49-035622.31 & PGC013905 \\
20261000013 & J151311.91+071345.2 & 228.2996 & 7.2292 &  8.8956 & 50.9603 & J151313+071331 & J151313.12+071331.88 & PGC054330 \\
20261000017 & J005748.80+302103.7 & 14.4534 & 30.3510 &  124.5628 & -32.5005 & J005748+302114 & J005748.88+302108.48 & NGC0315 \\
20261000021 & J083824.61+254533.8 & 129.6026 & 25.7594 &  198.8346 & 33.9767 & J083824+254516 & J083824.01+254516.35 & NGC2623 \\
20261000024 & J122611.82-074111.1 & 186.5493 & -7.6864 &  292.0797 & 54.6526 & J122615-074041 & J122616.20-071908.97 & NGC4404 \\
20261000026 & J165836.39+585605.2 & 254.6516 & 58.9348 &  88.0026 & 37.4204 & J165831+585615 & J165831.36+585610.11 & NGC6286 \\
20261000029 & J153315.49-084125.4 & 233.3145 & -8.6904 &  356.2682 & 36.9081 & J153320-084203 & J153320.70-081758.22 & -- \\
20261000031 & J140827.97+085556.1 & 212.1165 & 8.9323 &  351.5491 & 64.2636 & J140830+085556 & J140830.70+085554.87 & NGC5482 \\
20261000033 & J165258.74+022411.6 & 253.2448 & 2.4032 &  20.7309 & 27.2926 & J165258+022403 & J165258.90+022403.00 & NGC6240 \\
20261000034 & J230431.82+272146.3 & 346.1326 & 27.3629 &  95.5381 & -29.7609 & J230428+272127 & J230428.24+272126.50 & PGC070403 \\
20261000038 & J132852.60+383354.8 & 202.2192 & 38.5652 &  90.8730 & 76.1388 & J132852+383438 & J132852.13+383439.67 & -- \\
20261000040 & J090937.03+192803.4 & 137.4043 & 19.4676 &  209.0405 & 38.8868 & J090937+192807 & J090937.45+192808.28 & PGC025814 \\
20261000042 & J111112.86+284049.0 & 167.8036 & 28.6803 &  203.0491 & 67.8434 & J111113+284147 & J111113.20+284146.92 & NGC3561 \\
20261000043 & J040843.58+001303.3 & 62.1816 & 0.2176 &  191.3106 & -35.3356 & J040845+001306 & J040845.32+001307.22 & -- \\
20261000047 & J062249.36+243952.6 & 95.7057 & 24.6646 &  187.7997 & 5.1385 & J062249+243902 & J062249.86+243902.61 & -- \\
20261000050 & J153438.34+251324.0 & 233.6597 & 25.2233 &  39.3986 & 53.4940 & J153437+251311 & J153437.62+251311.38 & 2MASXJ15343758+2513114 \\
20261000054 & J110017.57+100302.0 & 165.0732 & 10.0505 &  240.7205 & 58.7443 & J110018+100256 & J110017.99+100256.54 & PGC200263 \\
20261000056 & J082357.32-001810.1 & 125.9888 & -0.3028 &  224.2355 & 20.3149 & J082357-001816 & J082357.69-004142.49 & PGC023563 \\
20261000059 & J160952.98+133206.2 & 242.4707 & 13.5351 &  26.8451 & 41.8658 & J160953+133147 & J160953.46+133147.86 & PGC057340 \\
20261000062 & J093552.16+612051.5 & 143.9673 & 61.3476 &  152.4827 & 42.9000 & J093551+612112 & J093551.59+612111.53 & UGC05101 \\
20261000065 & J003733.29+093343.1 & 9.3887 & 9.5620 &  117.2183 & -53.1557 & J003734+093324 & J003734.10+093324.44 & UGC00373 \\
20261000066 & J000410.85+082151.8 & 1.0452 & 8.3644 &  103.4011 & -52.7061 & J000410+082157 & J000410.60+082157.55 & PGC1345393 \\
20261000071 & J102543.20+102231.1 & 156.4300 & 10.3753 &  231.8853 & 51.8584 & J102544+102231 & J102544.21+102230.47 & PGC1378581 \\
20261000073 & J055437.98+271137.2 & 88.6582 & 27.1937 &  182.5158 & 0.8124 & J055437+271126 & J055437.16+271126.13 & -- \\
20261000075 & J103631.83+022202.7 & 159.1326 & 2.3674 &  244.5964 & 49.3235 & J103632+022144 & J103631.87+022144.00 & PGC031449 \\
20261000077 & J025134.46+431511.0 & 42.8936 & 43.2531 &  145.0409 & -14.4019 & J025134+431515 & J025135.45+431511.90 & -- \\
20261000085 & J011934.58+321048.7 & 19.8941 & 32.1802 &  129.8282 & -30.3145 & J011935+321050 & J011934.97+321049.90 & PGC004773 \\
20261000087 & J050745.52+442432.8 & 76.9397 & 44.4091 &  162.9513 & 2.3603 & J050746+442447 & J050746.57+442447.08 & -- \\
20261000088 & J083720.52+531452.9 & 129.3355 & 53.2480 &  165.0949 & 37.0680 & J083717+531517 & J083717.99+531516.85 & PGC2435890 \\
20261000089 & J045955.57+270617.0 & 74.9815 & 27.1047 &  175.8231 & -9.3631 & J045956+270602 & J045956.08+270602.07 & -- \\
20261000091 & J035857.00+541400.4 & 59.7375 & 54.2334 &  148.1877 & 0.8259 & J035853+541315 & J035853.90+541315.78 & -- \\
20261000093 & J230536.30+014435.4 & 346.4013 & 1.7432 &  77.2662 & -51.4929 & J230537+014441 & J230537.29+014441.89 & -- \\
20261000097 & J105840.15-033006.5 & 164.6673 & -3.5018 &  256.9401 & 49.0333 & J105839-033015 & J105839.33-032944.77 & PGC1070369 \\
20261000099 & J131740.72+411600.5 & 199.4197 & 41.2668 &  103.7259 & 74.8682 & J131739+411546 & J131739.20+411545.61 & PGC2178259 \\
20261000109 & J053537.73+643046.6 & 83.9072 & 64.5130 &  148.0752 & 16.6921 & J053538+643141 & J053538.97+643141.08 & -- \\
20261000110 & J105637.95+141952.6 & 164.1581 & 14.3313 &  232.6990 & 60.3676 & J105638+141929 & J105638.84+141930.50 & PGC3090734 \\
20261000111 & J072830.51+241925.2 & 112.1271 & 24.3237 &  194.5037 & 18.5675 & J072830+241911 & J072830.93+241911.04 & 3C182 \\
20261000112 & J145558.26-015744.8 & 223.9927 & -1.9624 &  353.7904 & 48.2226 & J145558-015745 & J145558.73-010214.66 & PGC1105739 \\
20261000113 & J133418.03-100936.7 & 203.5751 & -10.1602 &  319.9487 & 51.2898 & J133418-100929 & J133418.57-105030.96 & PGC047815 \\
20261000114 & J120813.53+125413.3 & 182.0564 & 12.9037 &  265.4022 & 72.5474 & J120814+125412 & J120814.45+125412.52 & -- \\
20261000115 & J054526.82+250640.6 & 86.3617 & 25.1113 &  183.2456 & -2.0253 & J054527+250634 & J054527.11+250634.41 & -- \\
20261000117 & J031948.16+413042.1 & 49.9507 & 41.5117 &  150.5758 & -13.2612 & J031948+413042 & J031948.16+413042.10 & -- \\
20261000124 & J132035.42+340841.5 & 200.1476 & 34.1449 &  82.9506 & 80.5957 & J132035+340822 & J132035.41+340821.58 & IC0883 \\
20261000126 & J130124.47+291652.1 & 195.3519 & 29.2811 &  77.9400 & 86.9246 & J130125+291850 & J130125.27+291849.46 & NGC4922 \\
20261000127 & J092143.97+332557.5 & 140.4332 & 33.4326 &  191.6002 & 44.7591 & J092151+332407 & J092151.49+332406.70 & -- \\
20261000128 & J120552.47+203153.5 & 181.4686 & 20.5315 &  242.8587 & 77.6777 & J120551+203118 & J120551.47+203119.07 & NGC4093 \\
20261000133 & J090657.88+412601.2 & 136.7412 & 41.4337 &  180.3109 & 42.3551 & J090652+412429 & J090652.80+412429.64 & PGC025574 \\
20261000136 & J162549.98+402919.4 & 246.4583 & 40.4887 &  64.2239 & 44.2465 & J162549+402921 & J162549.97+402919.32 & NGC6150 \\
20261000139 & J121339.91+504522.9 & 183.4163 & 50.7563 &  137.3163 & 65.3039 & J121329+504430 & J121329.29+504429.41 & NGC4187 \\
20261000141 & J113348.56+490235.1 & 173.4523 & 49.0431 &  152.1011 & 63.4562 & J113353+490323 & J113359.23+490343.39 & IC0708 \\
20261000151 & J144921.58+631614.0 & 222.3399 & 63.2706 &  103.1773 & 49.0888 & J144921+631613 & J144921.59+631613.95 & IC1065 \\
20261000165 & J102053.71+483243.8 & 155.2238 & 48.5455 &  166.2996 & 53.9341 & J102051+483306 & J102053.68+483124.11 & PGC030258 \\
20261000166 & J161217.46+282457.9 & 243.0727 & 28.4161 &  46.8018 & 45.8815 & J161217+282546 & J161217.62+282546.46 & PGC1835500 \\
20261000184 & J134035.20+444817.0 & 205.1467 & 44.8047 &  97.0388 & 69.7716 & J134035+444817 & J134035.20+444817.40 & -- \\
20261000186 & J163803.35+264435.7 & 249.5139 & 26.7432 &  46.2539 & 39.9467 & J163803+264330 & J163803.66+264330.61 & PGC1785438 \\
20261000188 & J144559.83+173822.2 & 221.4993 & 17.6395 &  19.1917 & 61.9512 & J144557+173830 & J144558.04+173820.58 & -- \\
20261000191 & J075758.54+395917.1 & 119.4939 & 39.9881 &  180.1972 & 29.2242 & J075756+395936 & J075756.72+395936.10 & PGC2158936 \\
20261000193 & J135806.05+214021.1 & 209.5252 & 21.6725 &  16.7623 & 73.8894 & J135806+214021 & J135806.05+214021.30 & PGC097421 \\
20261000196 & J134840.55+181659.2 & 207.1689 & 18.2831 &  2.0540 & 74.1310 & J134840+181716 & J134840.10+181716.09 & -- \\
20261000201 & J081040.06+481243.6 & 122.6669 & 48.2121 &  171.1285 & 32.7487 & J081040+481230 & J081040.30+481233.16 & -- \\
20261000203 & J150721.87+101844.8 & 226.8411 & 10.3124 &  11.7874 & 53.8718 & J150721+101846 & J150721.88+101844.95 & PGC053985 \\
20261000205 & J123201.68+331819.7 & 188.0070 & 33.3055 &  155.9063 & 82.5365 & J123200+331745 & J123200.57+331747.21 & -- \\
20261000220 & J090209.87+283042.9 & 135.5411 & 28.5119 &  197.1903 & 39.7876 & J090209+283042 & J090209.88+283042.93 & PGC1837961 \\
20261000222 & J103214.01+275601.6 & 158.0584 & 27.9338 &  203.2403 & 59.2339 & J103213+275610 & J103214.02+275601.68 & PGC1821863 \\
20261000526 & J084002.57+294849.0 & 130.0107 & 29.8136 &  194.1906 & 35.3866 & J084001+294853 & J084002.37+294902.64 & PGC024369 \\
20261000549 & J140700.39+282714.7 & 211.7516 & 28.4541 &  41.8625 & 73.2517 & J140700+282714 & J140700.40+282714.66 & PGC050352 \\
20261001007 & J134204.50+294828.2 & 205.5188 & 29.8078 &  49.4433 & 78.5603 & J134159+294655 & J134159.72+294653.50 & PGC1877944 \\
20261001010 & J145633.30+243650.5 & 224.1387 & 24.6140 &  34.9205 & 61.8153 & J145631+243637 & J145631.33+243634.90 & PGC1711650 \\
20261100002 & J024105.88-081405.4 & 40.2745 & -8.2348 &  181.9944 & -57.9087 & J024104-081521 & J024104.80-084439.25 & NGC1052 \\
20261100017 & J005748.80+302103.7 & 14.4534 & 30.3510 &  124.5628 & -32.5005 & J005748+302114 & J005748.88+302108.48 & NGC0315 \\
20261100050 & J153438.88+251348.4 & 233.6620 & 25.2301 &  39.4105 & 53.4934 & J153437+251311 & J153437.62+251311.38 & 2MASXJ15343758+2513114 \\
20261100062 & J093557.67+612055.1 & 143.9903 & 61.3486 &  152.4754 & 42.9097 & J093551+612112 & J093551.59+612111.53 & UGC05101 \\
20261100085 & J011934.58+321048.7 & 19.8941 & 32.1802 &  129.8282 & -30.3145 & J011935+321050 & J011934.97+321049.90 & PGC004773 \\
20261100112 & J145558.56-015754.3 & 223.9940 & -1.9651 &  353.7890 & 48.2198 & J145558-015745 & J145558.73-010214.66 & PGC1105739 \\
20261100113 & J133415.71-100934.6 & 203.5655 & -10.1596 &  319.9343 & 51.2929 & J133418-100929 & J133418.57-105030.96 & PGC047815 \\
20261100205 & J123200.25+331812.5 & 188.0011 & 33.3035 &  155.9464 & 82.5353 & J123200+331745 & J123200.57+331747.21 & -- \\
20261200002 & J024109.55-081339.8 & 40.2898 & -8.2277 &  182.0022 & -57.8925 & J024104-081521 & J024104.80-084439.25 & NGC1052 \\
20261200113 & J133418.41-100933.5 & 203.5767 & -10.1593 &  319.9515 & 51.2903 & J133418-100929 & J133418.57-105030.96 & PGC047815 \\
\hline
\end{tabular}
}
\end{table*}

\begin{table*}[htp]
\caption{Physical parameters of FASHI \HI\ absorption galaxies}
\label{tab:Physics}
\vskip 0pt
\centering \scriptsize
\renewcommand{\arraystretch}{0.78}
\setlength{\tabcolsep}{1.2mm}{
\begin{tabular}{cccccccccc}
\hline \hline
[1]  &  [2]  & [3]   & [4]  & [5] & [6]  & [7]  & [8] & [9] & [10] \\ 
FASHI ID & $V_{\rm heli}$ & $z_{\rm HI}$ & $z_{\rm OC}$   & $W_{50}$ & $S_{\rm HI}$ & $S_{\rm 1.4GHz}$  & ${\rm log}L_{\rm 1.4GHz}$  & $\tau_{\rm HI}$ & $N_{\rm HI}$ \\
& $\kms$      &     &  &   $\kms$  & mJy & mJy & W\,Hz$^{-1}$ & & 10$^{21}$cm$^{-2}$    \\
\hline
20261000001 & 1184.0 & 0.00395 & 0.00380 & $\phantom{0} 96.7\pm  2.9$ & $ 11.94\pm  0.59$ & $ 160.3\pm   5.7$ & $ 21.65$ & $ 0.077\pm 0.005$ & $  1.36\pm  0.09$ \\
20261000002 & 1649.7 & 0.00550 & 0.00495 & $\phantom{0} 22.7\pm  1.5$ & $ 19.23\pm  0.96$ & $ 912.5\pm  27.4$ & $ 22.64$ & $ 0.021\pm 0.001$ & $  0.09\pm  0.01$ \\
20261000005 & 2637.8 & 0.00880 & 0.00750 & $\phantom{0} 76.3\pm  3.8$ & $  8.34\pm  0.76$ & $ 298.8\pm   9.0$ & $ 22.51$ & $ 0.028\pm 0.003$ & $  0.39\pm  0.04$ \\
20261000011 & 3829.4 & 0.01277 & 0.01340 & $\phantom{0} 22.6\pm  1.8$ & $ 18.74\pm  1.06$ & $ 169.0\pm   5.9$ & $ 22.77$ & $ 0.118\pm 0.008$ & $  0.49\pm  0.03$ \\
20261000013 & 3911.2 & 0.01305 & 0.01300 & $\phantom{0}120.8\pm  2.9$ & $ 14.23\pm  0.53$ & $  53.3\pm   1.6$ & $ 22.25$ & $ 0.311\pm 0.017$ & $  6.84\pm  0.38$ \\
20261000017 & 5020.3 & 0.01675 & 0.01659 & $\phantom{0} 41.3\pm  4.0$ & $  7.96\pm  0.61$ & $ 772.1\pm  25.3$ & $ 23.62$ & $ 0.010\pm 0.001$ & $  0.08\pm  0.01$ \\
20261000021 & 5524.6 & 0.01843 & 0.01820 & $\phantom{0}110.1\pm  2.2$ & $ 12.77\pm  0.78$ & $  95.7\pm   2.9$ & $ 22.80$ & $ 0.143\pm 0.011$ & $  2.87\pm  0.21$ \\
20261000024 & 5590.7 & 0.01865 & 0.01843 & $\phantom{0} 32.5\pm  4.1$ & $  6.94\pm  0.55$ & $  40.3\pm   1.6$ & $ 22.43$ & $ 0.189\pm 0.018$ & $  1.12\pm  0.11$ \\
20261000026 & 5636.2 & 0.01880 & 0.01876 & $\phantom{0} 89.9\pm  1.2$ & $ 21.83\pm  0.96$ & $ 156.7\pm   5.6$ & $ 23.04$ & $ 0.150\pm 0.009$ & $  2.46\pm  0.15$ \\
20261000029 & 6989.3 & 0.02331 & 0.02302 & $\phantom{0}192.2\pm  1.8$ & $ 29.32\pm  0.88$ & $ 213.0\pm   6.4$ & $ 23.35$ & $ 0.148\pm 0.007$ & $  5.19\pm  0.24$ \\
20261000031 & 7225.4 & 0.02410 & 0.02370 & $\phantom{0}207.8\pm  5.4$ & $  5.27\pm  0.37$ & $  12.6\pm   0.6$ & $ 22.15$ & $ 0.542\pm 0.061$ & $ 20.52\pm  2.31$ \\
20261000033 & 7350.5 & 0.02452 & 0.02433 & $\phantom{0}304.6\pm  1.7$ & $ 25.20\pm  2.51$ & $ 426.3\pm  15.0$ & $ 23.70$ & $ 0.061\pm 0.007$ & $  3.38\pm  0.37$ \\
20261000034 & 7702.1 & 0.02569 & 0.02556 & $\phantom{0} 95.2\pm  3.5$ & $  9.56\pm  1.55$ & $ 116.6\pm   3.5$ & $ 23.18$ & $ 0.086\pm 0.015$ & $  1.48\pm  0.25$ \\
20261000038 & 7626.5 & 0.02544 & 0.02651 & $\phantom{0}116.4\pm 11.1$ & $  1.66\pm  0.37$ & $  11.1\pm   1.3$ & $ 22.19$ & $ 0.162\pm 0.044$ & $  3.44\pm  0.94$ \\
20261000040 & 8370.9 & 0.02792 & 0.02780 & $\phantom{0} 92.6\pm  3.8$ & $  7.42\pm  0.55$ & $  68.9\pm   2.1$ & $ 23.03$ & $ 0.114\pm 0.010$ & $  1.92\pm  0.16$ \\
20261000042 & 8720.1 & 0.02909 & 0.02875 & $\phantom{0} 64.8\pm  2.7$ & $  3.04\pm  0.20$ & $  40.9\pm   1.3$ & $ 22.83$ & $ 0.077\pm 0.006$ & $  0.91\pm  0.07$ \\
20261000043 & 8936.0 & 0.02981 & 0.03030 & $\phantom{0} 43.9\pm  2.6$ & $ 21.37\pm  1.22$ & $  48.4\pm   1.5$ & $ 22.95$ & $ 0.582\pm 0.051$ & $  4.67\pm  0.41$ \\
20261000047 & 9255.5 & 0.03087 & 0.03088 & $\phantom{0} 46.8\pm  3.0$ & $  4.62\pm  0.70$ & $  43.3\pm   1.4$ & $ 22.92$ & $ 0.113\pm 0.018$ & $  0.96\pm  0.16$ \\
20261000050 & 10099.1 & 0.03369 & 0.03395 & $\phantom{0} 30.5\pm  1.6$ & $  7.76\pm  0.49$ & $  50.1\pm   1.6$ & $ 23.07$ & $ 0.168\pm 0.013$ & $  0.93\pm  0.07$ \\
20261000054 & 10787.1 & 0.03598 & 0.03600 & $\phantom{0} 82.2\pm  1.1$ & $ 72.62\pm  0.96$ & $ 184.1\pm   6.5$ & $ 23.68$ & $ 0.502\pm 0.025$ & $  7.52\pm  0.37$ \\
20261000056 & 10529.4 & 0.03512 & 0.03511 & $\phantom{0} 53.9\pm  8.4$ & $ 12.10\pm  3.01$ & $  13.6\pm   0.6$ & $ 22.53$ & $ 2.205\pm 2.037$ & $ 21.69\pm 20.03$ \\
20261000059 & 10632.0 & 0.03546 & 0.03570 & $\phantom{0} 67.6\pm  3.2$ & $ 11.34\pm  1.08$ & $  37.4\pm   1.2$ & $ 22.99$ & $ 0.361\pm 0.044$ & $  4.45\pm  0.54$ \\
20261000062 & 11656.7 & 0.03888 & 0.03940 & $\phantom{0}329.7\pm  3.3$ & $ 10.37\pm  0.38$ & $ 170.1\pm   5.8$ & $ 23.73$ & $ 0.063\pm 0.003$ & $  3.78\pm  0.20$ \\
20261000065 & 12021.3 & 0.04010 & 0.04013 & $\phantom{0} 85.8\pm  2.9$ & $ 22.83\pm  2.39$ & $  75.0\pm   2.3$ & $ 23.39$ & $ 0.363\pm 0.048$ & $  5.67\pm  0.75$ \\
20261000066 & 11904.9 & 0.03971 & 0.03946 & $\phantom{0}137.1\pm  3.6$ & $  5.32\pm  0.32$ & $  67.4\pm   2.1$ & $ 23.33$ & $ 0.082\pm 0.006$ & $  2.06\pm  0.14$ \\
20261000071 & 13684.8 & 0.04565 & 0.04568 & $\phantom{0} 42.3\pm  0.8$ & $ 44.14\pm  2.16$ & $  76.6\pm   2.3$ & $ 23.52$ & $ 0.859\pm 0.078$ & $  6.63\pm  0.60$ \\
20261000073 & 14327.5 & 0.04779 & 0.00000 & $\phantom{0}112.5\pm  3.2$ & $  8.64\pm  0.54$ & $  18.1\pm   0.7$ & $ 22.93$ & $ 0.649\pm 0.067$ & $ 13.31\pm  1.38$ \\
20261000075 & 15046.0 & 0.05019 & 0.05040 & $\phantom{0}295.0\pm  1.5$ & $ 20.25\pm  0.31$ & $ 262.1\pm   9.1$ & $ 24.14$ & $ 0.080\pm 0.003$ & $  4.32\pm  0.17$ \\
20261000077 & 15440.8 & 0.05151 & 0.05194 & $\phantom{0} 22.4\pm  2.9$ & $ 27.29\pm  2.61$ & $1224.9\pm  36.7$ & $ 24.84$ & $ 0.023\pm 0.002$ & $  0.09\pm  0.01$ \\
20261000085 & 17933.8 & 0.05982 & 0.05996 & $\phantom{0}135.0\pm  0.5$ & $129.19\pm  2.15$ & $2635.2\pm  79.1$ & $ 25.30$ & $ 0.050\pm 0.002$ & $  1.24\pm  0.04$ \\
20261000087 & 18309.6 & 0.06107 & 0.00000 & $\phantom{0} 22.5\pm  2.3$ & $ 19.50\pm  2.08$ & $ 251.8\pm   7.6$ & $ 24.30$ & $ 0.081\pm 0.009$ & $  0.33\pm  0.04$ \\
20261000088 & 18419.6 & 0.06144 & 0.06140 & $\phantom{0}218.9\pm  2.5$ & $  6.11\pm  0.75$ & $   6.3\pm   0.4$ & $ 22.70$ & $ 3.488\pm 4.375$ & $139.16\pm174.55$ \\
20261000089 & 18784.6 & 0.06266 & 0.06097 & $\phantom{0} 42.8\pm  2.9$ & $ 19.32\pm  2.20$ & $ 926.5\pm  27.8$ & $ 24.86$ & $ 0.021\pm 0.003$ & $  0.16\pm  0.02$ \\
20261000091 & 20019.8 & 0.06678 & 0.00000 & $\phantom{0}532.1\pm  5.2$ & $ 11.66\pm  0.46$ & $  10.0\pm   0.5$ & $ 22.98$ & $ 2.303\pm 1.000$ & $223.36\pm 97.00$ \\
20261000093 & 19706.1 & 0.06573 & 0.06567 & $\phantom{0} 21.4\pm  2.2$ & $ 41.01\pm  3.20$ & $  33.4\pm   1.1$ & $ 23.48$ & $ 2.303\pm 1.000$ & $  8.98\pm  3.90$ \\
20261000097 & 19852.8 & 0.06622 & 0.06595 & $\phantom{0} 89.8\pm  5.9$ & $ 11.99\pm  1.14$ & $  31.1\pm   1.0$ & $ 23.46$ & $ 0.487\pm 0.063$ & $  7.97\pm  1.03$ \\
20261000099 & 19952.6 & 0.06655 & 0.06610 & $\phantom{0}105.5\pm  4.7$ & $  8.62\pm  1.47$ & $ 266.0\pm   8.0$ & $ 24.39$ & $ 0.033\pm 0.006$ & $  0.63\pm  0.11$ \\
20261000109 & 24080.8 & 0.08032 & 0.00000 & $\phantom{0} 60.2\pm  2.9$ & $ 35.92\pm  2.12$ & $  61.2\pm   1.9$ & $ 23.93$ & $ 0.884\pm 0.095$ & $  9.71\pm  1.04$ \\
20261000110 & 24402.5 & 0.08140 & 0.08130 & $\phantom{0}221.0\pm  3.9$ & $  8.83\pm  0.53$ & $ 169.0\pm   5.1$ & $ 24.38$ & $ 0.054\pm 0.004$ & $  2.16\pm  0.15$ \\
20261000111 & 24673.0 & 0.08230 & 0.00000 & $\phantom{0} 93.6\pm  1.7$ & $ 27.01\pm  0.91$ & $ 920.8\pm  27.6$ & $ 25.13$ & $ 0.030\pm 0.001$ & $  0.51\pm  0.02$ \\
20261000112 & 24972.1 & 0.08330 & 0.08350 & $\phantom{0} 62.7\pm  3.0$ & $ 14.29\pm  0.91$ & $  48.1\pm   1.5$ & $ 23.86$ & $ 0.353\pm 0.030$ & $  4.03\pm  0.34$ \\
20261000113 & 25174.0 & 0.08397 & 0.08378 & $\phantom{0}105.5\pm  2.4$ & $ 32.50\pm  2.85$ & $ 100.4\pm   3.0$ & $ 24.18$ & $ 0.391\pm 0.044$ & $  7.52\pm  0.85$ \\
20261000114 & 25200.7 & 0.08406 & 0.08400 & $\phantom{0} 48.5\pm  3.3$ & $ 16.90\pm  2.06$ & $  21.1\pm   0.7$ & $ 23.51$ & $ 1.615\pm 0.508$ & $ 14.27\pm  4.49$ \\
20261000115 & 25264.7 & 0.08427 & 0.00000 & $\phantom{0}245.4\pm  8.4$ & $  7.49\pm  1.37$ & $  14.5\pm   0.6$ & $ 23.35$ & $ 0.727\pm 0.200$ & $ 32.52\pm  8.95$ \\
20261000117 & 5286.6 & 0.01763 & 0.01767 & $\phantom{0}466.7\pm  2.3$ & $ 63.06\pm  1.24$ & $22829.2\pm 684.9$ & $ 25.15$ & $ 0.003\pm 0.000$ & $  0.24\pm  0.01$ \\
20261000124 & 6921.9 & 0.02309 & 0.02310 & $\phantom{0}226.4\pm  4.3$ & $  6.49\pm  1.27$ & $ 104.4\pm   3.2$ & $ 23.04$ & $ 0.064\pm 0.013$ & $  2.65\pm  0.54$ \\
20261000126 & 7024.8 & 0.02343 & 0.02340 & $\phantom{0} 29.5\pm 13.9$ & $  4.72\pm  1.48$ & $  38.8\pm   1.2$ & $ 22.63$ & $ 0.130\pm 0.044$ & $  0.70\pm  0.23$ \\
20261000127 & 7651.0 & 0.02552 & 0.02480 & $\phantom{0} 48.5\pm  9.3$ & $  1.47\pm  0.49$ & $ 112.9\pm   3.9$ & $ 23.14$ & $ 0.013\pm 0.004$ & $  0.12\pm  0.04$ \\
20261000128 & 7200.9 & 0.02402 & 0.02380 & $\phantom{0}104.8\pm  5.3$ & $  3.16\pm  0.89$ & $  89.7\pm   2.7$ & $ 23.01$ & $ 0.036\pm 0.010$ & $  0.69\pm  0.20$ \\
20261000133 & 8143.5 & 0.02716 & 0.02740 & $\phantom{0}311.5\pm  2.9$ & $  2.52\pm  0.79$ & $  51.6\pm   1.6$ & $ 22.89$ & $ 0.050\pm 0.016$ & $  2.84\pm  0.92$ \\
20261000136 & 8776.8 & 0.02928 & 0.02920 & $\phantom{0} 12.6\pm  1.9$ & $ 19.95\pm  1.56$ & $  31.7\pm   1.3$ & $ 22.73$ & $ 0.993\pm 0.150$ & $  2.28\pm  0.34$ \\
20261000139 & 9259.4 & 0.03089 & 0.03076 & $\phantom{0}114.6\pm  7.8$ & $  2.06\pm  0.50$ & $  96.3\pm   2.9$ & $ 23.26$ & $ 0.022\pm 0.005$ & $  0.45\pm  0.11$ \\
20261000141 & 9384.1 & 0.03130 & 0.03160 & $\phantom{0} 50.0\pm  6.9$ & $  4.48\pm  1.34$ & $ 315.1\pm  10.9$ & $ 23.80$ & $ 0.014\pm 0.004$ & $  0.13\pm  0.04$ \\
20261000151 & 12657.8 & 0.04222 & 0.04170 & $\phantom{0}158.9\pm  5.1$ & $ 10.20\pm  2.24$ & $3006.0\pm  90.2$ & $ 25.03$ & $ 0.003\pm 0.001$ & $  0.10\pm  0.02$ \\
20261000165 & 15895.4 & 0.05302 & 0.05320 & $\phantom{0} 76.0\pm 10.3$ & $  8.06\pm  1.47$ & $ 294.0\pm  10.0$ & $ 24.24$ & $ 0.028\pm 0.005$ & $  0.39\pm  0.07$ \\
20261000166 & 15928.8 & 0.05313 & 0.05310 & $\phantom{0} 69.1\pm  5.8$ & $  3.29\pm  0.47$ & $  76.7\pm   2.3$ & $ 23.65$ & $ 0.044\pm 0.007$ & $  0.55\pm  0.08$ \\
20261000184 & 19613.8 & 0.06542 & 0.06540 & $\phantom{0} 15.5\pm  2.8$ & $ 17.83\pm  2.03$ & $  77.7\pm   2.4$ & $ 23.85$ & $ 0.261\pm 0.035$ & $  0.74\pm  0.10$ \\
20261000186 & 19675.5 & 0.06563 & 0.06593 & $\phantom{0} 97.9\pm  4.4$ & $  6.80\pm  0.84$ & $  35.1\pm   1.1$ & $ 23.51$ & $ 0.215\pm 0.031$ & $  3.84\pm  0.55$ \\
20261000188 & 19848.9 & 0.06621 & 0.06554 & $\phantom{0} 26.5\pm  5.2$ & $  8.07\pm  0.97$ & $ 826.9\pm  26.6$ & $ 24.88$ & $ 0.010\pm 0.001$ & $  0.05\pm  0.01$ \\
20261000191 & 19740.8 & 0.06585 & 0.06580 & $\phantom{0}133.3\pm  4.8$ & $  2.77\pm  0.83$ & $  99.5\pm   3.0$ & $ 23.96$ & $ 0.028\pm 0.009$ & $  0.69\pm  0.21$ \\
20261000193 & 19919.7 & 0.06644 & 0.06642 & $\phantom{0} 27.2\pm  4.6$ & $ 18.73\pm  2.35$ & $  91.7\pm   3.2$ & $ 23.93$ & $ 0.228\pm 0.033$ & $  1.13\pm  0.17$ \\
20261000196 & 21659.5 & 0.07225 & 0.07310 & $\phantom{0} 86.9\pm  4.6$ & $  2.74\pm  0.34$ & $  37.1\pm   1.2$ & $ 23.63$ & $ 0.077\pm 0.010$ & $  1.22\pm  0.16$ \\
20261000201 & 23273.7 & 0.07763 & 0.07749 & $\phantom{0} 72.2\pm  4.8$ & $  2.69\pm  0.16$ & $  38.8\pm   1.5$ & $ 23.70$ & $ 0.072\pm 0.005$ & $  0.95\pm  0.07$ \\
20261000203 & 23422.7 & 0.07813 & 0.07800 & $\phantom{0}292.6\pm 10.5$ & $  3.52\pm  0.44$ & $ 403.2\pm  12.1$ & $ 24.72$ & $ 0.009\pm 0.001$ & $  0.47\pm  0.06$ \\
20261000205 & 23555.8 & 0.07857 & 0.07882 & $\phantom{0} 30.0\pm  3.3$ & $  6.66\pm  1.18$ & $ 105.2\pm   3.2$ & $ 24.15$ & $ 0.065\pm 0.012$ & $  0.36\pm  0.07$ \\
20261000220 & 25455.9 & 0.08491 & 0.08490 & $\phantom{0}185.2\pm  6.3$ & $  7.91\pm  0.48$ & $  31.8\pm   1.0$ & $ 23.70$ & $ 0.286\pm 0.023$ & $  9.65\pm  0.77$ \\
20261000222 & 25528.9 & 0.08516 & 0.08520 & $\phantom{0}274.2\pm  6.8$ & $  6.02\pm  0.88$ & $  62.3\pm   2.3$ & $ 23.99$ & $ 0.102\pm 0.016$ & $  5.08\pm  0.80$ \\
20261000526 & 19487.0 & 0.06500 & 0.06480 & $\phantom{0} 85.0\pm  2.9$ & $  6.47\pm  0.62$ & $ 626.0\pm  21.9$ & $ 24.74$ & $ 0.010\pm 0.001$ & $  0.16\pm  0.02$ \\
20261000549 & 23052.1 & 0.07689 & 0.07700 & $\phantom{0} 52.3\pm  7.1$ & $  7.19\pm  1.45$ & $ 816.6\pm  24.5$ & $ 25.02$ & $ 0.009\pm 0.002$ & $  0.08\pm  0.02$ \\
20261001007 & 13550.7 & 0.04520 & 0.04490 & $\phantom{0} 16.7\pm  4.7$ & $  3.98\pm  0.65$ & $  10.3\pm   0.5$ & $ 22.63$ & $ 0.488\pm 0.107$ & $  1.49\pm  0.33$ \\
20261001010 & 9879.3 & 0.03295 & 0.03280 & $\phantom{0} 32.3\pm  3.2$ & $  6.80\pm  1.02$ & $  19.1\pm   0.7$ & $ 22.62$ & $ 0.440\pm 0.085$ & $  2.59\pm  0.50$ \\
20261100002 & 1520.6 & 0.00507 & 0.00495 & $\phantom{0} 12.0\pm  1.4$ & $  8.82\pm  0.80$ & $ 912.5\pm  27.4$ & $ 22.64$ & $ 0.010\pm 0.001$ & $  0.02\pm  0.00$ \\
20261100017 & 5412.7 & 0.01805 & 0.01659 & $\phantom{0}  9.8\pm  0.3$ & $100.78\pm 10.98$ & $ 772.1\pm  25.3$ & $ 23.62$ & $ 0.140\pm 0.017$ & $  0.25\pm  0.03$ \\
20261100050 & 10193.7 & 0.03400 & 0.03395 & $\phantom{0} 28.4\pm  3.5$ & $  5.16\pm  1.14$ & $  50.1\pm   1.6$ & $ 23.07$ & $ 0.109\pm 0.026$ & $  0.56\pm  0.13$ \\
20261100062 & 12012.4 & 0.04007 & 0.03940 & $\phantom{0}148.2\pm  4.4$ & $  4.56\pm  0.49$ & $ 170.1\pm   5.8$ & $ 23.73$ & $ 0.027\pm 0.003$ & $  0.73\pm  0.08$ \\
20261100085 & 18147.6 & 0.06053 & 0.05996 & $\phantom{0} 14.7\pm  0.8$ & $ 74.45\pm  2.27$ & $2635.2\pm  79.1$ & $ 25.30$ & $ 0.029\pm 0.001$ & $  0.08\pm  0.00$ \\
20261100112 & 25110.5 & 0.08376 & 0.08350 & $\phantom{0} 75.4\pm  4.9$ & $  7.42\pm  1.36$ & $  48.1\pm   1.5$ & $ 23.86$ & $ 0.168\pm 0.034$ & $  2.30\pm  0.47$ \\
20261100113 & 25303.9 & 0.08440 & 0.08378 & $\phantom{0} 26.8\pm  4.4$ & $  9.25\pm  1.36$ & $ 100.4\pm   3.0$ & $ 24.18$ & $ 0.097\pm 0.015$ & $  0.47\pm  0.07$ \\
20261100205 & 23672.0 & 0.07896 & 0.07882 & $\phantom{0} 27.4\pm  9.0$ & $  3.08\pm  0.96$ & $ 105.2\pm   3.2$ & $ 24.15$ & $ 0.030\pm 0.009$ & $  0.15\pm  0.05$ \\
20261200002 & 1547.1 & 0.00516 & 0.00495 & $\phantom{0} 12.5\pm  2.2$ & $  5.21\pm  0.80$ & $ 912.5\pm  27.4$ & $ 22.64$ & $ 0.006\pm 0.001$ & $  0.01\pm  0.00$ \\
20261200113 & 25068.1 & 0.08362 & 0.08378 & $\phantom{0} 25.3\pm  4.3$ & $ 14.04\pm  1.68$ & $ 100.4\pm   3.0$ & $ 24.18$ & $ 0.151\pm 0.020$ & $  0.69\pm  0.09$ \\
\hline
\end{tabular}
}
\end{table*}

\begin{table*}
\caption{\textbf{Cross-matched sources between the absorbers and GSWLC.}}
\label{tab:GSWLC}
\centering 
\setlength{\tabcolsep}{1.1mm}{
\begin{tabular}{cccccccccccccc}
\hline \hline
[1]  &  [2]  & [3]   & [4]  & [5] & [6]  & [7]  & [8] & [9] & [10] & [11] & [12]  \\
FASHI ID & OBJID & RA & Dec & $z_{\odot}$ & log($M_{\star})$ & log($\rm SFR_{SED})$ & $A_{\rm V}$  & $g$ & $r$ & $z$ & $Q$  \\
     &    & deg & deg &   & $\rm M_{\odot}$ & $\rm M_{\odot}\,yr^{-1}$ & mag & mag &  mag &  mag \\
\hline
20261000021 & 1237664668961538066 & 129.600 & 25.755 & 0.0182 & 10.55$\pm$0.02 & -0.09$\pm$0.07 & 0.12$\pm$0.06 & 14.04 & 13.34 & 12.76 & 0.78 \\
20261000024 & -- & -- & -- & -- & 10.97$\pm$2.19 & -0.80$\pm$0.16 & -- & 13.21 & 12.39 & 11.82 & 0.83 \\
20261000026 & -- & -- & -- & -- & 10.41$\pm$2.08 & 0.32$\pm$0.06 & -- & 14.13 & 13.23 & 13.16 & 0.44 \\
20261000031 & 1237662263778672703 & 212.128 & 8.932 & 0.0237 & 11.19$\pm$0.01 & 0.07$\pm$0.21 & 0.41$\pm$0.04 & 13.14 & 12.32 & 11.68 & 0.63 \\
20261000034 & -- & -- & -- & -- & 11.23$\pm$2.25 & -0.83$\pm$0.17 & -- & 13.98 & 13.04 & 12.29 & 0.60 \\
20261000038 & -- & -- & -- & -- & 10.20$\pm$2.04 & -0.72$\pm$0.14 & -- & -- & -- & -- & 0.43 \\
20261000040 & 1237667293188718632 & 137.406 & 19.469 & 0.0278 & 11.00$\pm$0.02 & 0.23$\pm$0.17 & 0.31$\pm$0.05 & 14.52 & 13.64 & 12.96 & 0.80 \\
20261000042 & 1237667254009266187 & 167.804 & 28.712 & 0.0293 & 10.66$\pm$0.04 & 0.55$\pm$0.06 & 0.38$\pm$0.06 & 13.82 & 13.06 & 12.52 & 0.87 \\
20261000059 & 1237665565545464010 & 242.473 & 13.530 & 0.0357 & 10.75$\pm$0.04 & 0.49$\pm$0.09 & 0.69$\pm$0.08 & 15.65 & 14.76 & 13.95 & 0.18 \\
20261000062 & 1237651272966275163 & 143.965 & 61.353 & 0.0394 & 11.25$\pm$0.03 & 1.16$\pm$0.01 & 0.75$\pm$0.00 & 14.63 & 13.98 & 13.74 & 0.59 \\
20261000065 & -- & -- & -- & -- & 11.14$\pm$2.23 & 0.87$\pm$0.17 & -- & 15.15 & 14.14 & 13.26 & 0.27 \\
20261000066 & -- & -- & -- & -- & 11.02$\pm$2.20 & -0.84$\pm$0.17 & -- & 15.01 & 14.09 & 13.43 & 0.65 \\
20261000075 & 1237651754001170530 & 159.133 & 2.362 & 0.0504 & 10.57$\pm$0.04 & 0.77$\pm$0.15 & 0.50$\pm$0.14 & 15.81 & 15.15 & 14.76 & 0.69 \\
20261000088 & 1237651272424489158 & 129.325 & 53.255 & 0.0614 & 11.17$\pm$0.02 & 0.22$\pm$0.17 & 0.65$\pm$0.07 & 16.78 & 15.79 & 15.18 & 0.50 \\
20261000093 & -- & -- & -- & -- & 10.98$\pm$2.20 & 0.08$\pm$0.02 & -- & -- & -- & -- & 0.57 \\
20261000097 & -- & -- & -- & -- & 11.23$\pm$2.25 & 0.81$\pm$0.16 & -- & 15.62 & 14.83 & 14.23 & 0.63 \\
20261000099 & 1237661966897971323 & 199.413 & 41.263 & 0.0661 & 11.46$\pm$0.03 & -0.35$\pm$0.95 & 0.47$\pm$0.07 & 15.49 & 14.53 & 14.13 & 0.48 \\
20261000112 & 1237655498673356983 & 223.995 & -1.963 & 0.0835 & 10.96$\pm$0.03 & 0.23$\pm$0.26 & 0.35$\pm$0.17 & 16.88 & 15.98 & 15.30 & 0.68 \\
20261000124 & 1237665128549384425 & 200.148 & 34.139 & 0.0231 & 10.51$\pm$0.05 & 0.33$\pm$0.21 & 0.37$\pm$0.17 & 14.34 & 13.73 & 13.52 & 0.50 \\
20261000126 & 1237665441515962401 & 195.355 & 29.314 & 0.0234 & 10.77$\pm$0.03 & 0.06$\pm$0.18 & 0.23$\pm$0.17 & 13.85 & 13.13 & 12.57 & 0.66 \\
20261000128 & 1237668298739548227 & 181.464 & 20.522 & 0.0238 & 10.79$\pm$0.03 & -0.20$\pm$0.26 & 0.52$\pm$0.11 & 14.40 & 13.57 & 12.94 & 0.86 \\
20261000133 & 1237657775543746601 & 136.720 & 41.408 & 0.0274 & 10.92$\pm$0.02 & -0.73$\pm$0.29 & 0.25$\pm$0.09 & 14.41 & 13.56 & 13.26 & 0.68 \\
20261000136 & 1237659330851700829 & 246.458 & 40.489 & 0.0292 & 11.27$\pm$0.01 & -0.73$\pm$0.33 & 0.26$\pm$0.04 & 13.64 & 12.77 & 12.35 & 0.48 \\
20261000139 & 1237658205584818218 & 183.372 & 50.741 & 0.0308 & 11.49$\pm$0.01 & 0.50$\pm$0.14 & 0.66$\pm$0.07 & 13.33 & 12.41 & 12.29 & 0.76 \\
20261000141 & 1237658612519075853 & 173.497 & 49.062 & 0.0316 & 11.22$\pm$0.01 & -0.24$\pm$0.15 & 0.33$\pm$0.02 & 13.59 & 12.73 & 12.58 & 0.67 \\
20261000151 & 1237651539792953375 & 222.340 & 63.271 & 0.0417 & 11.26$\pm$0.02 & 0.60$\pm$0.17 & 0.34$\pm$0.07 & 14.17 & 13.41 & 12.98 & -- \\
20261000165 & 1237657856602865751 & 155.224 & 48.523 & 0.0532 & 11.00$\pm$0.02 & -0.58$\pm$0.95 & 0.40$\pm$0.09 & 15.89 & 14.98 & 14.61 & 0.84 \\
20261000166 & 1237662335183290648 & 243.073 & 28.430 & 0.0531 & 11.34$\pm$0.03 & 0.09$\pm$0.30 & 0.44$\pm$0.17 & 15.01 & 14.11 & 13.46 & 0.82 \\
20261000184 & 1237661433240813676 & 205.147 & 44.805 & 0.0654 & 10.67$\pm$0.06 & 0.68$\pm$0.06 & 0.31$\pm$0.05 & -- & -- & -- & 0.42 \\
20261000186 & 1237661465992626579 & 249.522 & 26.697 & 0.0659 & 10.88$\pm$0.04 & 0.65$\pm$0.08 & 0.67$\pm$0.07 & 16.84 & 15.91 & 15.19 & 0.22 \\
20261000188 & 1237667781776113805 & 221.440 & 17.623 & 0.0657 & 9.93$\pm$0.06 & -0.06$\pm$0.14 & 0.25$\pm$0.10 & -- & -- & -- & 0.13 \\
20261000191 & 1237653589018018166 & 119.486 & 39.993 & 0.0658 & 10.82$\pm$0.05 & 0.72$\pm$0.20 & 0.45$\pm$0.10 & 16.28 & 15.45 & 15.05 & 0.91 \\
20261000201 & -- & -- & -- & -- & 10.89$\pm$2.18 & 0.59$\pm$0.12 & -- & -- & -- & -- & 0.41 \\
20261000205 & -- & -- & -- & -- & 9.47$\pm$1.89 & 0.56$\pm$0.11 & -- & -- & -- & -- & 0.59 \\
20261000220 & 1237664668427223246 & 135.541 & 28.512 & 0.0849 & 11.31$\pm$0.02 & 0.14$\pm$0.41 & 0.41$\pm$0.06 & 16.52 & 15.47 & 14.72 & 0.64 \\
20261001007 & -- & -- & -- & -- & 11.04$\pm$2.21 & -0.90$\pm$0.18 & -- & 15.05 & 14.17 & 13.53 & 0.52 \\
20261001010 & -- & -- & -- & -- & 10.50$\pm$2.10 & -0.14$\pm$0.03 & -- & 16.08 & 15.19 & 14.44 & 0.33 \\
20261100062 & 1237651272966275163 & 143.965 & 61.353 & 0.0394 & 11.25$\pm$0.03 & 1.16$\pm$0.01 & 0.75$\pm$0.00 & 14.63 & 13.98 & 13.74 & 0.59 \\
20261100112 & 1237655498673356983 & 223.995 & -1.963 & 0.0835 & 10.96$\pm$0.03 & 0.23$\pm$0.26 & 0.35$\pm$0.17 & 16.88 & 15.98 & 15.30 & 0.68 \\
20261100205 & -- & -- & -- & -- & 9.47$\pm$1.89 & 0.56$\pm$0.11 & -- & -- & -- & -- & 0.59 \\
\hline
\end{tabular}}
\end{table*}

\subsection{Source Coordinates and Counterparts}
\label{sec:coord_cata}

Table~\ref{tab:name} lists the coordinates and multi-wavelength counterparts of the FASHI \HI\ absorption systems. The columns are described as follows:

\begin{itemize}
\item Column 1: Unique index number for each FASHI \HI\ source.
\item Column 2: J2000 centroid coordinate in sexagesimal format (\texttt{Jhhmmss.ss$\pm$ddmmss.s}).
\item Columns 3–4: Right ascension and declination in decimal degrees (J2000).
\item Columns 5–6: Galactic longitude and latitude ($l$, $b$) in decimal degrees.
\item Column 7: J2000 coordinate or source name of the NVSS radio counterpart in the format \texttt{Jhhmmss$\pm$ddmmss} \citep{Condon1998}.
\item Column 8: J2000 coordinate of the optical counterpart in sexagesimal format.
\item Column 9: Name or identifier of the corresponding host galaxy.
\end{itemize}

\subsection{Physical Parameters of the Absorbers}
\label{sec:Physics}

Table~\ref{tab:Physics} summarizes the physical properties derived for each FASHI \HI\ absorption system. The columns are defined as follows:

\begin{itemize}
\item Column 1: Unique index number for each FASHI \HI\ source.
\item Column 2: Heliocentric velocity of the \HI\ absorption line (optical definition) in km s$^{-1}$.
\item Column 3: Redshift $z_{\rm HI}$ corresponding to the heliocentric velocity in Column 2.
\item Column 4: Redshift $z_{\rm OC}$ of the optical counterpart obtained from the literature.
\item Column 5: Line width $W_{50}$ (in km s$^{-1}$) measured at 50\% of the peak flux using the busy-function fitting method \citep{Westmeier2014}.
\item Column 6: Peak depth of the \HI\ absorption line $S_{\rm HI}$ (negative value, in mJy).
\item Column 7: Integrated 1.4 GHz continuum flux density $S_{\rm 1.4GHz}$ (in mJy) from the NVSS catalog \citep{Condon1998}.
\item Column 8: Radio luminosity (power) at 1.4 GHz, assuming isotropic emission, calculated as \citep{Yun2001}:
\begin{equation}
\log \frac{L_{\rm 1.4GHz}}{\rm W\,Hz^{-1}} = 17.08 + 2\log \frac{D_{\rm L}}{\rm Mpc} + \log \frac{S_{\rm 1.4GHz}}{\rm mJy},
\label{eq:lumeq}
\end{equation}
where $D_{\rm L}$ is the luminosity distance in Mpc computed using the cosmology calculator of \citet{Wright2006}.
\item Column 9: Peak optical depth $\tau_{\rm HI}$, estimated from \citep[e.g.,][]{Wolfe1975}:
\begin{equation}
\tau_{\rm HI} \approx -\ln\left(1 + \frac{S_{\rm HI}}{c_{\rm f} S_{\rm 1.4GHz}}\right),
\end{equation}
where we adopt a covering factor $c_{\rm f}=1$ \citep{Morganti2018}.
\item Column 10: \HI\ column density $N_{\rm HI}$, derived from \citep[e.g.,][]{Wolfe1975}:
\begin{equation}
N_{\rm HI} = 1.823 \times 10^{18} \, T_{\rm s} \int \tau_{\rm HI} \, dV,
\end{equation}
assuming a spin temperature $T_{\rm s} = 100$ K for the \HI\ gas \citep{Morganti2018}.
\end{itemize}

\subsection{Cross-matching with GSWLC}
\label{sec:catalog_GSWLC}

Table~\ref{tab:GSWLC} presents the results of cross-matching the FASHI \HI\ absorbers with the GALEX–SDSS–WISE Legacy Catalog (GSWLC) \citep{Salim2016}, adopting matching tolerances of $\delta_{\rm RA}\leq3'$, $\delta_{\rm Dec}\leq3'$, and $\delta_{\rm v}\leq300$ km s$^{-1}$. 
The columns in Table~\ref{tab:GSWLC} are described as follows:

\begin{itemize}
\item Column 1: Unique index number for each FASHI \HI\ source.
\item Column 2: OBJID, the SDSS photometric identifier from GSWLC.
\item Columns 3–4: Right ascension and declination (J2000, decimal degrees) from GSWLC.
\item Column 5: Redshift $z_{\odot}$ from GSWLC.
\item Column 6: Stellar mass $\log(M_{\star})$ (with uncertainty) from GSWLC.
\item Column 7: UV/optical spectral energy distribution (SED) star formation rate $\log(\mathrm{SFR}_{\mathrm{SED}})$ (with uncertainty) from GSWLC.
\item Column 8: Dust attenuation $A_{\rm V}$ in the rest-frame $V$ band from GSWLC.
\item Columns 9–11: $g$, $r$, and $z$-band photometry (AB magnitudes) from the Siena Galaxy Atlas (SGA).
\item Column 12: $Q = b/a$, the minor-to-major axis ratio, where $Q = 1$ corresponds to face-on ($i = 0^\circ$) and $Q = 0.2$ to edge-on ($i \gtrsim 78^\circ$). The axis ratio is derived from the ellipticity modulus $|\varepsilon|$ using the relation $b/a = (1-|\varepsilon|)/(1+|\varepsilon|)$, following the gravitational-lensing definition of ellipticity adopted in the Legacy Survey DR9 Tractor catalog \citep{2019AJ....157..168D}. Further details are available in the online documentation at \url{https://www.legacysurvey.org/dr9/catalogs/}.
\end{itemize}

 \begin{figure}[htp]
 \centering
 \includegraphics[width=0.48\textwidth, angle=0]{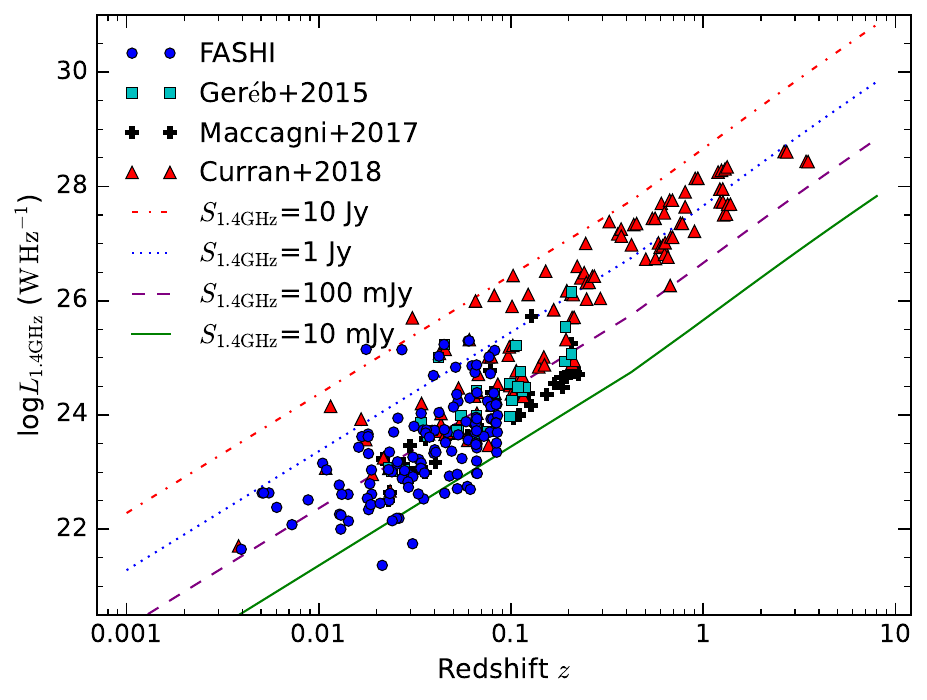}
 \includegraphics[width=0.48\textwidth, angle=0]{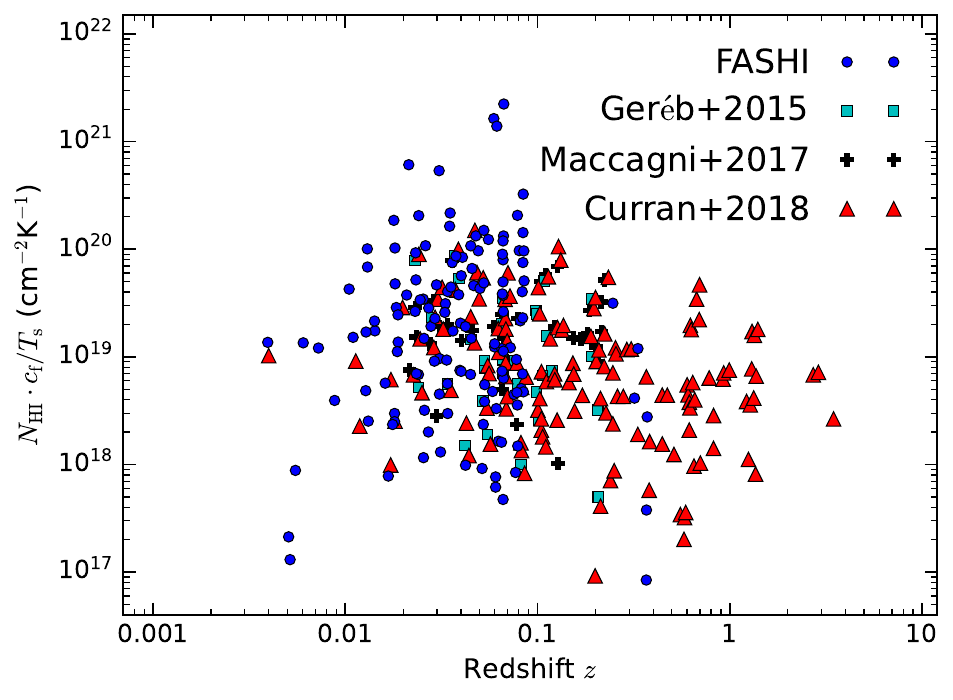}
 \includegraphics[width=0.48\textwidth, angle=0]{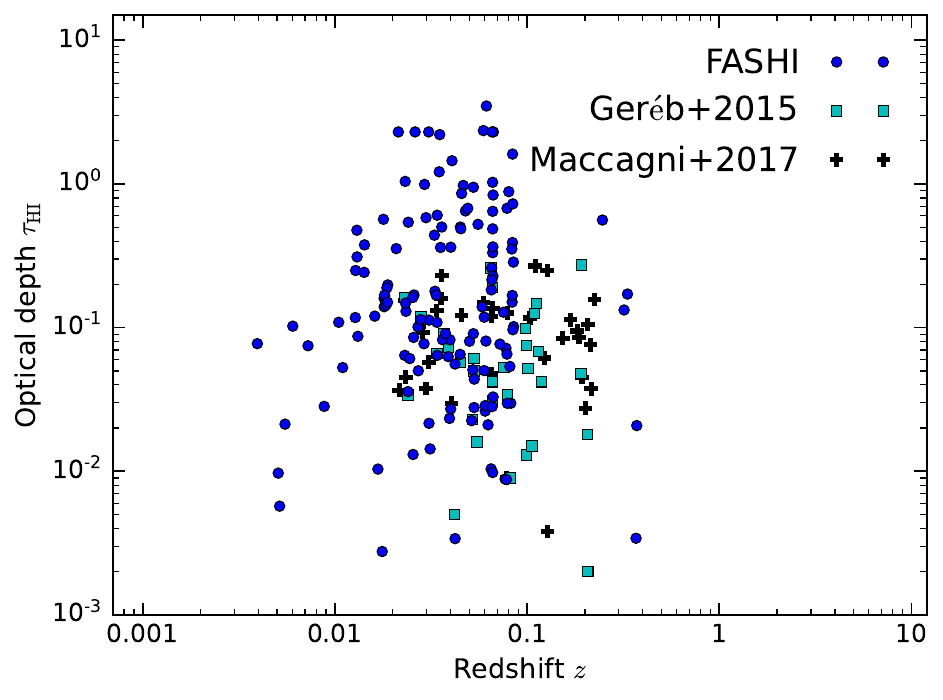}
 \caption{The rest-frame 1.4\,GHz continuum luminosity ($upper$) of the background sources, the \HI\ column density ($middle$), and the \HI\ peak optical depth versus redshift ($lower$) for the 21-cm \HI\ absorbers, including data points from \citet{Gereb2015}, \citet{Maccagni2017}, \citet{Curran2018}, and FASHI. In the upper panel, the different lines indicate different 1.4\,GHz continuum flux density.}
 \label{Fig:para-z}
 \end{figure}


\section{Discussion}
\label{sec:discu}

\subsection{Comparison with Previous Flux-limited Surveys}

One of the main advantages of the blind FASHI survey is its ability to probe substantially fainter radio continuum sources than previous flux-limited \HI\ absorption surveys. Figure~\ref{Fig:para-z} (upper panel) compares the 1.4 GHz radio luminosity distribution of the FASHI absorbers with those reported by \citet{Gereb2015}, \citet{Maccagni2017}, and \citet{Curran2018}. Previous samples are largely concentrated at $\log L_{\rm 1.4GHz} \gtrsim 24.5$, corresponding to continuum flux densities of $S_{\rm 1.4GHz} \gtrsim 100$ mJy. In contrast, the FASHI sample extends smoothly down to $\log L_{\rm 1.4GHz} \sim 22.5$, with continuum flux densities as low as $2.6\pm0.4$ mJy.

This result demonstrates that detectable \HI\ absorption is not confined to powerful radio galaxies, but can also be identified in significantly fainter radio systems. The expanded luminosity coverage provided by FASHI therefore reduces the strong continuum-selection bias present in many previous targeted surveys.

The lower panels of Figure~\ref{Fig:para-z} show that neither the \HI\ column density nor the optical depth exhibits a strong dependence on redshift, consistent with the results of \citet{Gupta2006}. The weak negative trends visible when combining different samples are more likely attributable to variations in survey sensitivity and selection effects than to genuine cosmic evolution.

\subsection{Line Width and Optical Depth Distributions}

 \begin{figure}[htp]
 \centering
 \includegraphics[width=0.48\textwidth, angle=0]{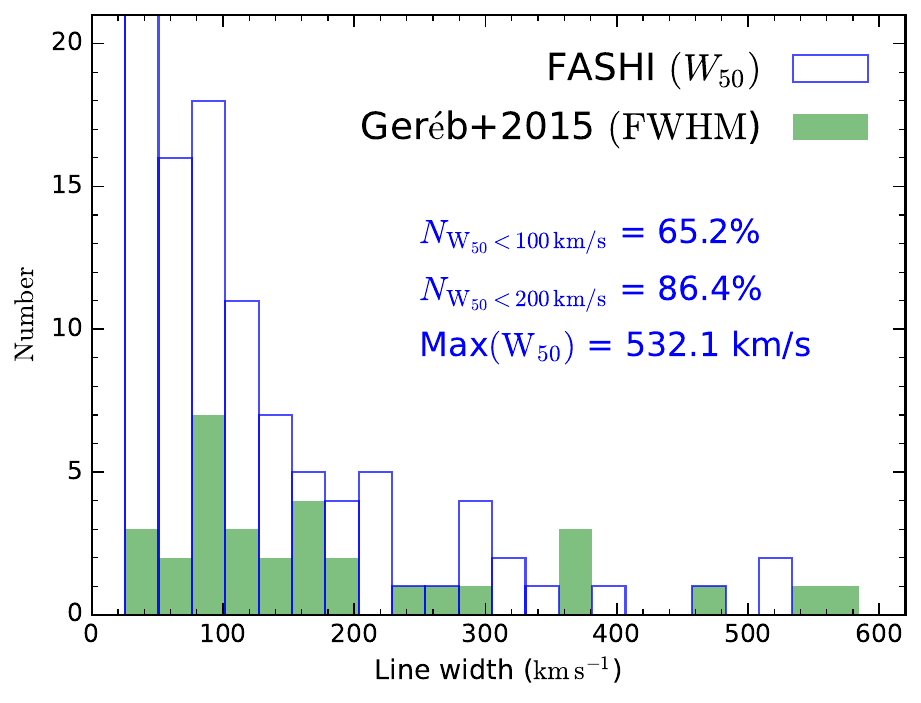}
  \includegraphics[width=0.48\textwidth, angle=0]{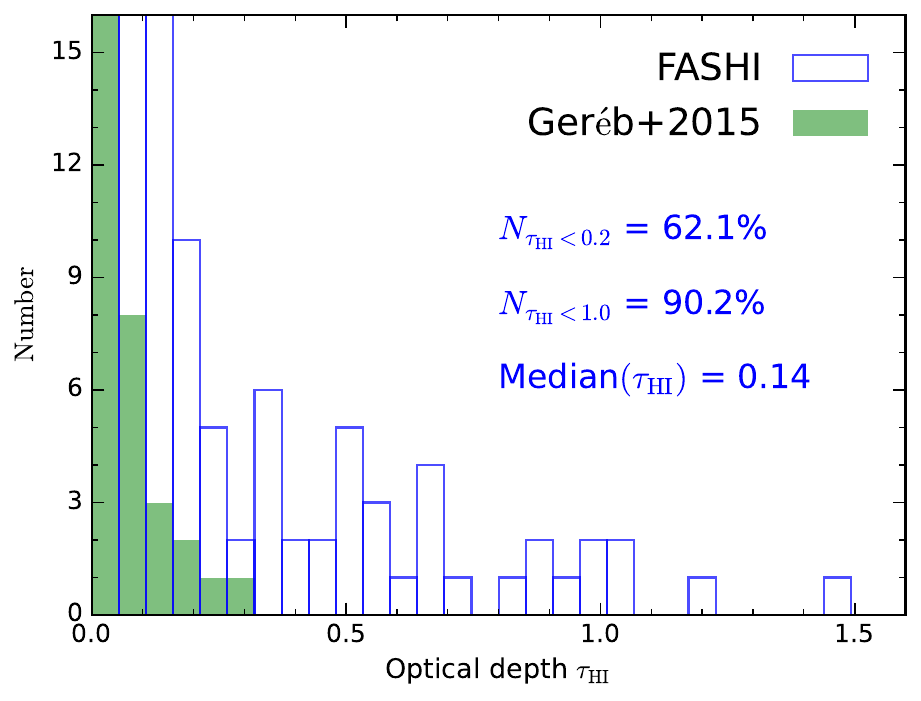}
\caption{The histograms of line width ($upper$) and peak optical depth ($lower$) for the 21\,cm \HI\ absorbers in FASHI and \citet{Gereb2015}.}
 \label{Fig:hist_para}
 \end{figure}

 \begin{figure}[htp]
 \centering
 \includegraphics[width=0.47\textwidth, angle=0]{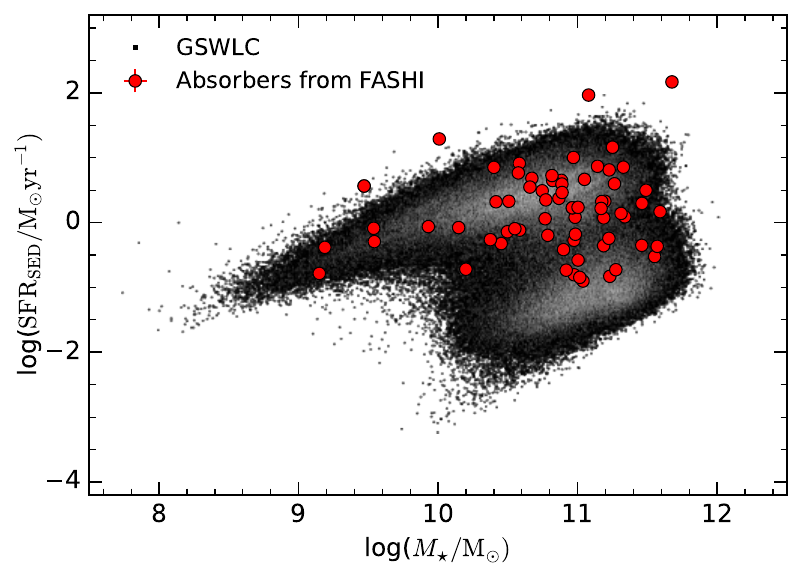}
 \includegraphics[width=0.47\textwidth, angle=0]{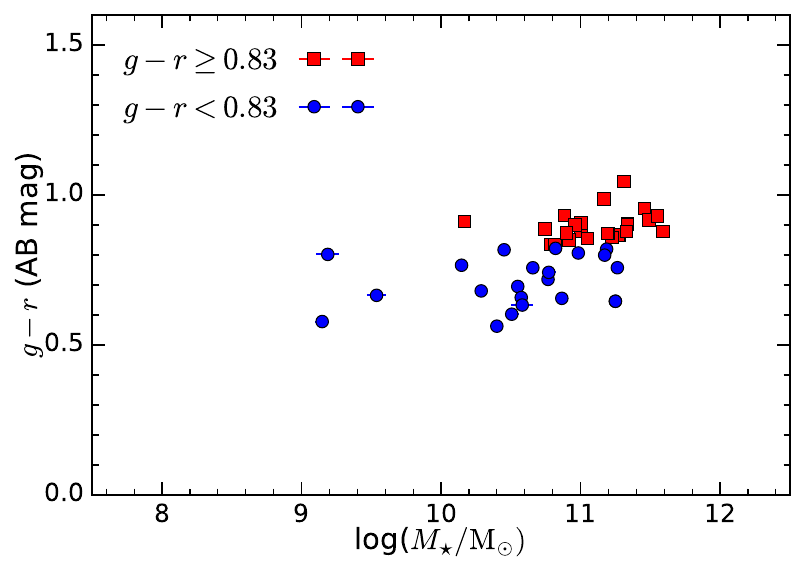}
 \caption{$Upper$: Two-dimensional distributions of star formation rate log($\rm SFR_{SED}$) and stellar mass log($M_{\star}$). The background shows all the data in the GSWLC. The red dots indicate cross-matched and associated \HI\ absorbers detected by FASHI. $Lower$: Color-stellar mass diagram of the FASHI absorbers. The blue and red galaxies are separated by $g-r=0.83$ \citep{Yang2006,Weinmann2006}.}
 \label{Fig:GSWLC}
 \end{figure}

Figure~\ref{Fig:hist_para} compares the distributions of $W_{50}$ and $\tau_{\rm HI}$ for the FASHI absorbers with those reported by \citet{Gereb2015}. The overall distributions are broadly similar between the two samples. Narrow absorption systems ($W_{50}<100$ km\,s$^{-1}$) dominate the FASHI sample, accounting for approximately 65.2\% of all absorbers, whereas broad systems ($W_{50}>200$ km\,s$^{-1}$) represent about 13.6\%.

Most absorbers are optically thin. Approximately 63.0\% of the FASHI absorbers have $\tau_{\rm HI}<0.2$, and nearly 90.2\% have $\tau_{\rm HI}<1.0$, with a median optical depth of $\tau_{\rm HI}\approx0.14$.

The larger size of the FASHI sample additionally reveals two noteworthy features. First, the fraction of very broad absorbers ($W_{50}>200$ km\,s$^{-1}$) is slightly higher than that found in the comparison sample. This difference may be related to the broader continuum-flux range probed by FASHI, although observational selection effects may also contribute. Second, while the median optical depth is similar to previous surveys, the FASHI distribution extends to significantly higher values ($\tau_{\rm HI}>2$ in several cases), indicating that a subset of the absorbers probe particularly dense or cold neutral gas along the line of sight. The difference between the FASHI and Ger\'eb et al. (2015) optical depth distributions is likely driven by the different survey strategies. As an untargeted wide-field survey, FASHI probes a large number of sightlines over a broad sky area, enabling the detection of rare but high-optical-depth systems ($\tau_{\rm HI} > 0.3$) that are statistically less likely to be captured in typical targeted surveys. This rarity of high-$\tau$ absorbers was previously quantified by \citet{Braun2012} (see their Figure 6), who showed that the number of sightlines with large optical depths drops rapidly with increasing $\tau$. Supporting this interpretation, the FLASH survey \citep{Yoon2025}, another untargeted wide-field absorption survey with ASKAP, finds an optical depth distribution and a median value ($\tau_{\rm HI} \approx 0.14$) that are strikingly similar to those of the FASHI sample, reinforcing the view that untargeted surveys provide a more representative census of the intrinsic optical depth distribution, including its rare high-$\tau$ tail.

\subsection{Host Galaxy Properties of the FASHI Absorbers}

 \begin{figure}[htp]
 \centering
 \includegraphics[width=0.47\textwidth, angle=0]{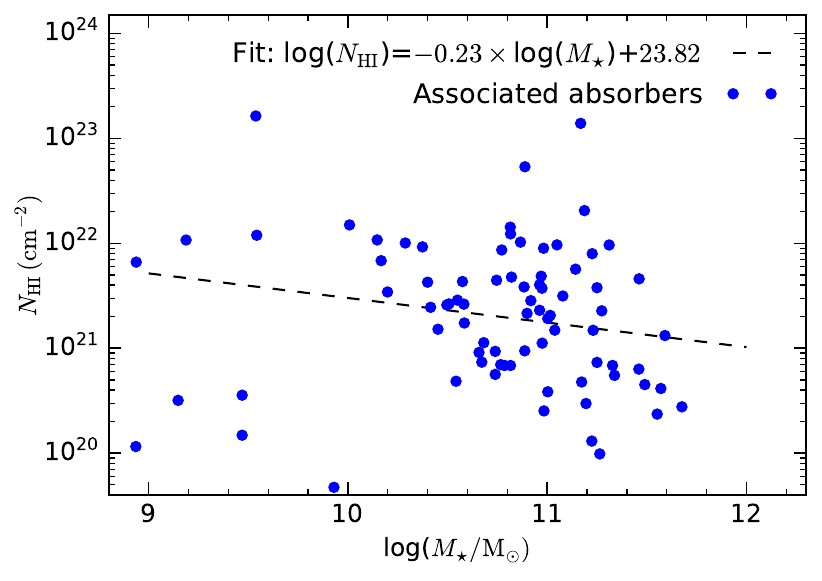}
 \caption{The relationship between the \HI\ column density and the stellar mass in the host galaxy. The blue dots indicate cross-matched and associated \HI\ absorbers detected by FASHI.}
 \label{Fig:N_Mstar}
 \end{figure}

The multi-wavelength properties of the host galaxies can be investigated for the subset of FASHI absorbers with reliable counterparts in the GSWLC catalog. These systems are classified as associated absorbers because their absorption velocities are consistent with originating within the same galaxies that host the background radio continuum sources.

As shown in the upper panel of Figure~\ref{Fig:GSWLC}, the absorbers span a broad range in both stellar mass ($10^{9.15}<M_{\star}<10^{11.68}\,M_{\odot}$) and star formation rate ($-0.90<\log({\rm SFR}/M_{\odot}{\rm yr}^{-1})<2.17$), with median values of $\log(M_{\star}/M_{\odot})=10.89$ and $\log({\rm SFR}/M_{\odot}{\rm yr}^{-1})=0.24$. Relative to the full GSWLC parent population, the absorbers are preferentially associated with galaxies of higher stellar mass and moderately elevated star formation activity, suggesting that \HI\ absorption is more frequently detected in massive, gas-rich systems.

This trend is further supported by the lower panel of Figure~\ref{Fig:GSWLC}, which presents the $g-r$ color as a function of stellar mass. Using the empirical division between blue and red galaxies at $g-r=0.83$ \citep{Yang2006,Weinmann2006}, most FASHI absorbers lie within the blue-cloud region, indicating that they are predominantly hosted by star-forming galaxies rather than passive ellipticals.

Figure~\ref{Fig:N_Mstar} shows that the associated absorbers exhibit an overall anti-correlation between $N_{\rm HI}$ and stellar mass. Compared with the first FASHI absorption catalog presented by \citet{Zhang2025ab}, which was based on 19 absorbers, the larger current sample (72 associated absorbers) reveals a weaker correlation accompanied by substantially larger scatter. This suggests that the relation between \HI\ column density and stellar mass is more complex than previously inferred. Although more massive galaxies tend to exhibit lower line-of-sight \HI\ column densities toward their radio cores, the large scatter indicates considerable diversity in the cold-gas content and gas distribution among individual systems.

No significant correlation is found between $W_{50}$ and either stellar mass or star formation rate, implying that the observed line widths are primarily governed by local gas kinematics rather than by global host-galaxy properties.

\subsection{The Concentration of \HI\ Absorption at Intermediate Inclinations}

 \begin{figure}[htp]
 \centering
 \includegraphics[height=0.35\textwidth, angle=0]{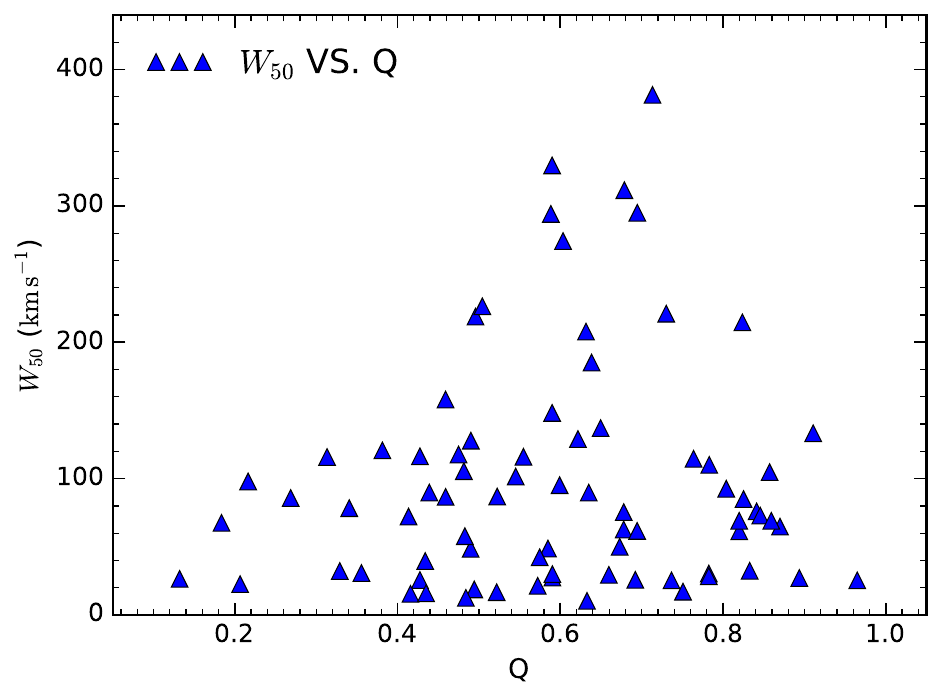}
 \caption{Distribution of the \HI\ line width $W_{50}$ as a function of the axial ratio $Q = b/a$. Most data points cluster around $Q \approx 0.6$--$0.7$, and the sources with larger $W_{50}$ are also predominantly located in this same range of $Q$.}
 \label{Fig:W50-Q}
 \end{figure}
 
Figure~\ref{Fig:W50-Q} shows that the majority of H\,{\sc i} absorbers, including those with the largest line widths ($W_{50}$), are concentrated around $Q \approx 0.6$--$0.7$, corresponding to inclinations of approximately $45^\circ$--$55^\circ$. This trend is likely driven by a combination of geometric projection effects and dust obscuration.

For galaxies viewed close to face-on ($Q \sim 1$), the projected rotational velocity along the line of sight is relatively small, naturally producing narrower absorption profiles. In contrast, highly inclined systems ($Q \sim 0.2$) are viewed through the dense mid-plane of the disk, where increased dust obscuration and higher spin temperatures may reduce the detectability of \HI\ absorption. Galaxies with intermediate inclinations provide a more favourable configuration, in which the projected rotational velocity remains sufficiently large while the line of sight avoids the most heavily obscured regions.

The fact that the broadest absorption systems are also preferentially found within the same inclination range further suggests that many of these broad profiles originate primarily from rotating gas disks. If the dominant contribution instead arose from randomly oriented outflows, a much weaker dependence on inclination would be expected.

\subsection{Velocity Offsets: From Symmetric Distributions to Blueshifted Tails}
\label{sec:vel_offset}

 \begin{figure}[htp]
 \centering
 \includegraphics[width=0.48\textwidth, angle=0]{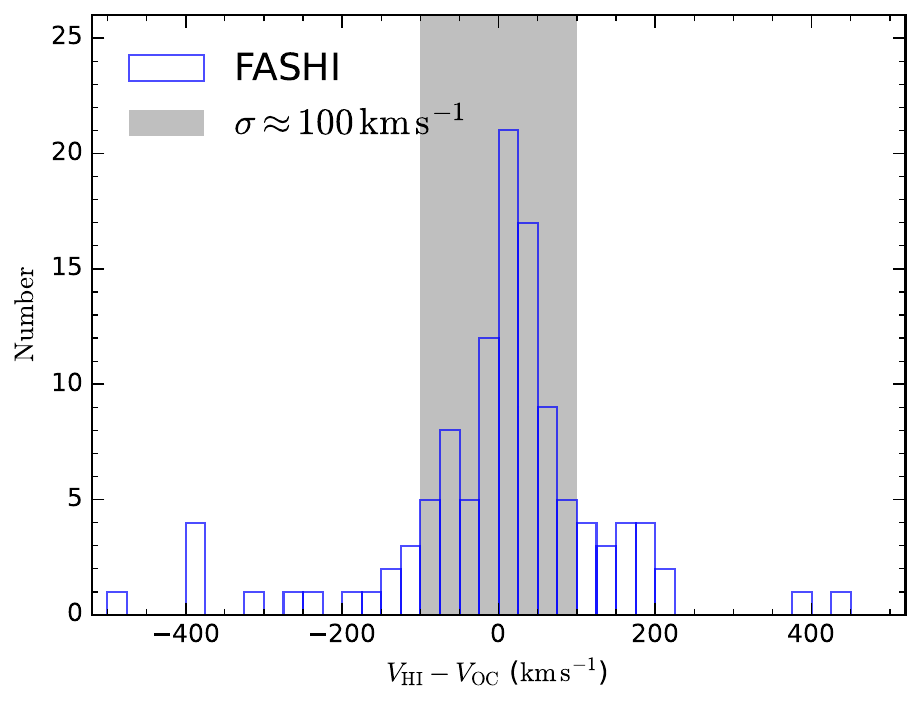}
 \caption{The histogram of velocities for the principal \HI\ absorption components with respect to the systemic velocity estimated from optical counterparts. The derived velocity error is about $\sigma=100\,\kms$ shown with a shaded bar.}
 \label{Fig:hist_vel}
 \end{figure}

The velocity offset between the \HI\ absorption line and the systemic velocity of the host galaxy provides important information about the kinematics of the absorbing gas. Figure~\ref{Fig:hist_vel} presents the velocity-offset distribution for FASHI absorbers with available optical spectroscopic redshifts.

Among the 120 absorbers in the sample, 41 exhibit blueshifted absorption, 21 show redshifted absorption, and 58 have velocity offsets close to zero. The overall distribution is approximately symmetric around zero velocity, suggesting that a large fraction of the absorbers originate from regularly rotating gas structures associated with the host galaxies.

However, a significant asymmetry becomes apparent at large velocity offsets. Six absorbers exhibit blueshifts exceeding $-300$ km\,s$^{-1}$, whereas no absorber shows a similarly large redshifted velocity offset. A comparable excess of high-velocity blueshifted absorbers has previously been reported by \citet{Vermeulen2003} and \citet{Gereb2015}. These systems are commonly interpreted as signatures of AGN-driven neutral gas outflows, which are preferentially detected on the near side of the host galaxy. In contrast, the relatively small number of strongly redshifted narrow absorbers may instead trace infalling clouds or unsettled gas at larger galactocentric radii.

\subsection{Statistical Insights from the Expanded Sample}

Compared with the first FASHI \HI\ absorption catalog presented by \citet{Zhang2025ab}, the current sample of more than 130 absorbers provides a substantially improved statistical view of the low-redshift \HI\ absorption population.

Several trends become clearer in the expanded sample. First, the previously reported anti-correlation between $N_{\rm HI}$ and stellar mass remains present across a larger dynamic range, although the relation becomes weaker and exhibits significantly larger scatter. This indicates a broader diversity in the cold-gas properties of massive galaxies than previously recognized.

Second, the larger sample reveals a distinct excess of high-velocity blueshifted absorbers, including six systems with velocity offsets exceeding $-300$ km\,s$^{-1}$. This strengthens the interpretation that a subset of the absorbers trace AGN-driven neutral gas outflows.

Third, the FASHI blind survey demonstrates that \HI\ absorption can be detected in galaxies with continuum flux densities as low as $\sim2.6$ mJy, extending the known radio-luminosity range of associated absorbers by nearly two orders of magnitude relative to many earlier flux-limited surveys.

\subsection{Limitations of Stacking Caused by Redshift Offsets}
\label{sec:stacking}

 \begin{figure}[htp]
 \centering
 \includegraphics[width=0.47\textwidth, angle=0]{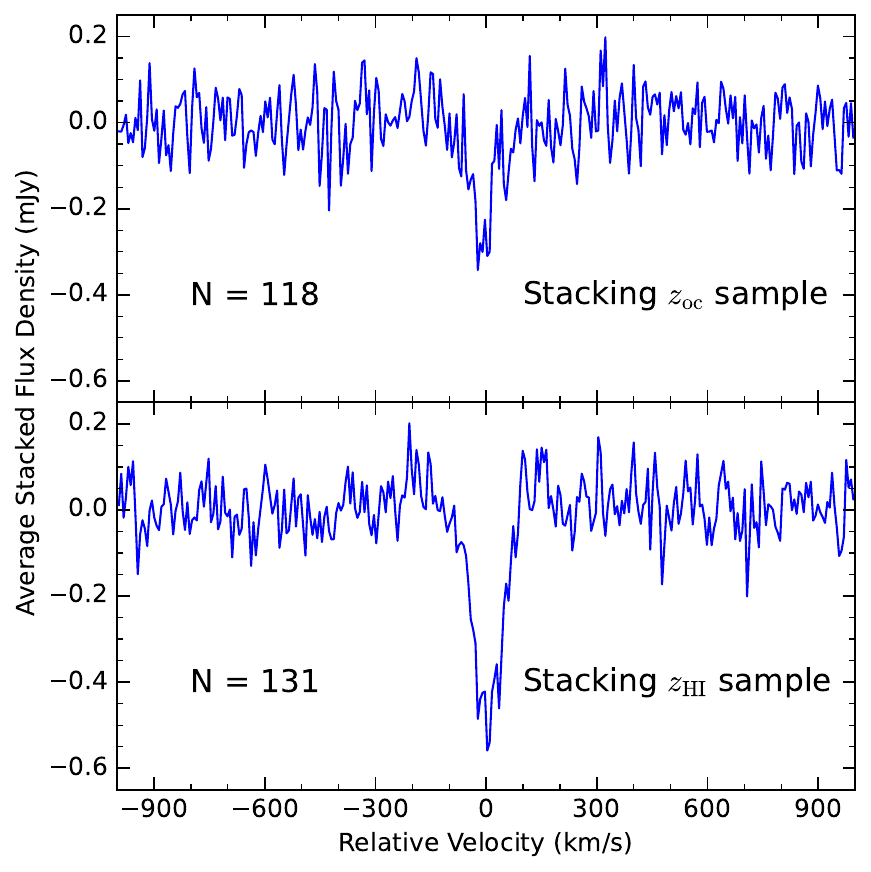}
\caption{Stacked \HI\ spectra of the known absorber sample (Table~\ref{tab:name}). Upper: aligned by optical redshifts. Lower: aligned by \HI\ absorption line redshifts.}
 \label{Fig:stack}
 \end{figure}

As discussed in Section~\ref{sec:vel_offset} and illustrated in Figure~\ref{Fig:hist_vel}, the optical redshifts of host galaxies do not always accurately trace the true velocities of the absorbing \HI\ gas. In some cases, the velocity offsets reach several hundred km\,s$^{-1}$. This effect introduces an important limitation for stacking analyses that rely solely on optical redshifts.

To illustrate this issue, we perform stacking experiments for the known absorbers listed in Table~1 using two different alignment methods (Figure~\ref{Fig:stack}). When the spectra are aligned using the optical redshifts of the host galaxies (upper panel), the resulting stacked profile is broad and shallow, with a peak flux density of only $F_{\rm peak}\sim -0.3$\,mJy. In contrast, when the spectra are aligned using the measured \HI\ absorption centroids (lower panel), the stacked profile becomes substantially narrower and deeper, reaching $F_{\rm peak}\sim -0.5$\,mJy. At the same time, the FWHM decreases from approximately $150$ km\,s$^{-1}$ to $\sim50$ km\,s$^{-1}$.

These results demonstrate that even moderate velocity offsets between optical and \HI\ redshifts can significantly dilute stacked absorption signals. Consequently, for samples in which individual \HI\ absorption lines remain undetected and their intrinsic velocities are unknown, stacking based solely on optical redshifts may fail to recover a representative average absorption profile. This effect should therefore be carefully considered in the design and interpretation of future blind-survey stacking experiments.

\section{Summary}
\label{sec:summary}

We have conducted an untargeted survey of 21 cm \HI\ absorption galaxies using the FASHI DR2 data release, covering approximately 19,500 deg$^2$ at $z \lesssim 0.09$. A total of 132 \HI\ absorbers were identified, including $\sim$60 new discoveries. This constitutes the largest homogeneous sample of low-redshift \HI\ absorbers assembled to date, nearly three times larger than the next largest flux-limited sample. The overall detection rate is 0.007 sources per square degree, corresponding to an occurrence probability of approximately 1/1180 among all detected \HI\ galaxies.

Compared with previous flux-limited surveys, the FASHI absorbers extend to substantially lower radio continuum luminosities, with continuum flux densities as low as $2.6 \pm 0.4$ mJy. Narrow absorption systems ($W_{50}<100$ km\,s$^{-1}$) dominate the sample, accounting for approximately 65.2\% of all absorbers, whereas broad systems ($W_{50}>200$ km\,s$^{-1}$) represent about 13.6\%. Most absorbers are optically thin, with a median optical depth of $\tau_{\rm HI} \approx 0.14$.

The velocity-offset distribution is broadly symmetric around the systemic velocity of the host galaxies, suggesting that many absorbers originate from regularly rotating gas structures. However, six absorbers exhibit blueshifts exceeding $-300$ km\,s$^{-1}$, while no comparably redshifted systems are identified. This asymmetry is consistent with the presence of AGN-driven neutral gas outflows in a subset of the sample. In addition, most absorbers are concentrated around $Q \approx 0.6$--$0.7$ (inclinations of $\sim45^\circ$--$55^\circ$), indicating that intermediate inclinations provide favourable conditions for \HI\ absorption detection.

The host galaxies of the associated absorbers are preferentially massive and actively star-forming systems, with a median stellar mass of $\log M_{\star}=10.89$. An overall anti-correlation is found between $N_{\rm HI}$ and stellar mass. Compared with the first FASHI absorption catalog, the larger current sample reveals a weaker correlation accompanied by substantially larger scatter, indicating increased diversity in the cold-gas properties of massive galaxies.

Finally, we demonstrate that even moderate velocity offsets between optical and \HI\ redshifts can significantly dilute stacked absorption signals. This effect represents an important limitation for stacking analyses based solely on optical redshifts and should be carefully considered in future blind-survey stacking experiments.

\section*{Acknowledgements}
\addcontentsline{toc}{section}{Acknowledgements}

We thank the anonymous referee for their careful reading and constructive comments that helped improve the clarity and quality of this paper. This work is supported by the National SKA Program of China (No.\,2025SKA0150100), the National Key R\&D Program of China (No.\,2025YFE0202300), the National Natural Science Foundation of China (Nos.\,12225303, 12288102, 12595313, 12373011), the science research grants from the China Manned Space Project (No.\,CMS-CSST-2021-A05), the Guizhou Provincial Science and Technology Projects (Nos.\,QKHFQ[2023]003, QKHPTRC-ZDSYS[2023]003, QKHFQ[2024]001, QKHJCMS[2025]015), the CAS Project for Young Scientists in Basic Research (Nos.\,YSBR-063, YSBR-092), and the Office of Science and Technology, Shanghai Municipal Government (Nos.\,24DX1400100, ZJ2023-ZD-001). FAST is a Chinese national mega-science facility, operated by the National Astronomical Observatories of Chinese Academy of Sciences (NAOC).

This research has made use of the SIMBAD database, operated at CDS, Strasbourg, France. This research has made use of the NASA/IPAC Extragalactic Database (NED), which is funded by the National Aeronautics and Space Administration and operated by the California Institute of Technology.

\bibliography{references}{}
\bibliographystyle{aasjournal}

 \appendix
  

 \section{Figures on individual sources}
 
  \begin{figure*}[htp]
 \centering
 \renewcommand{\thefigure}{\arabic{figure}}
 \addtocounter{figure}{-1}
 \includegraphics[height=0.22\textwidth, angle=0]{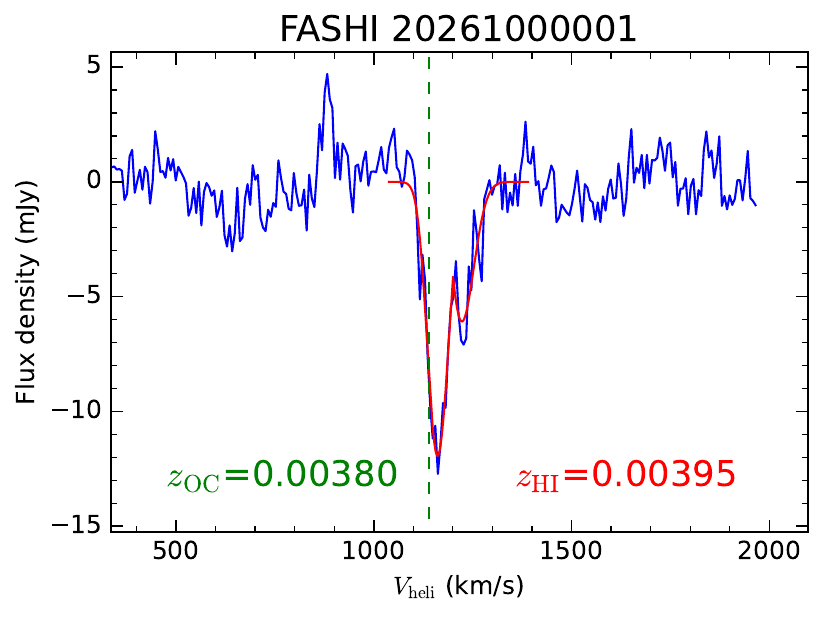}
 \includegraphics[height=0.27\textwidth, angle=0]{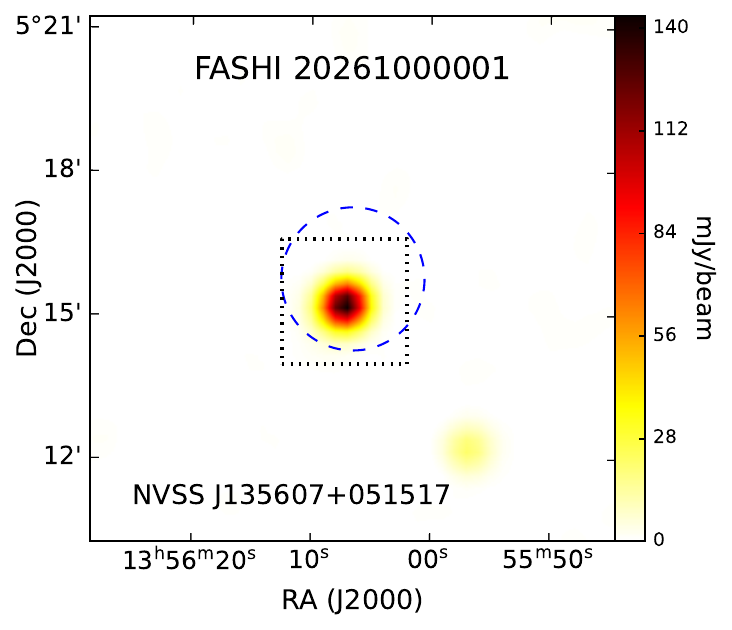}
 \includegraphics[height=0.27\textwidth, angle=0]{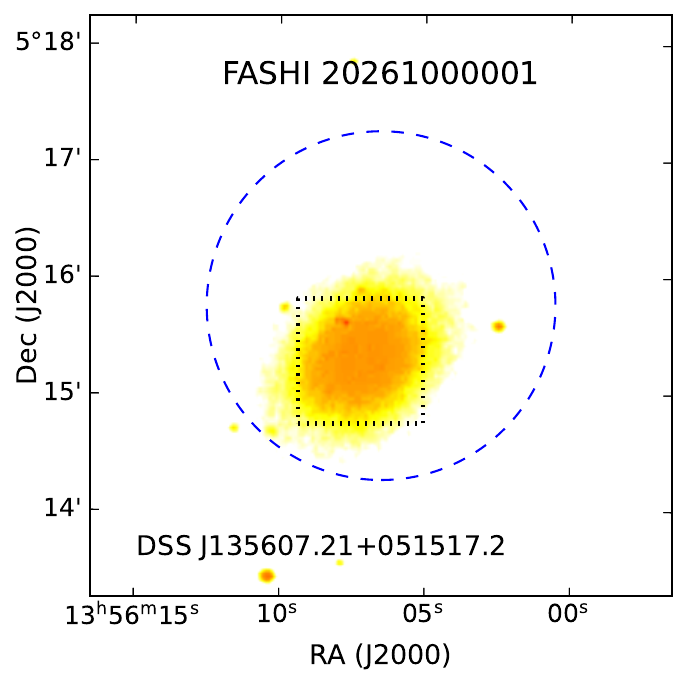}
 \includegraphics[height=0.22\textwidth, angle=0]{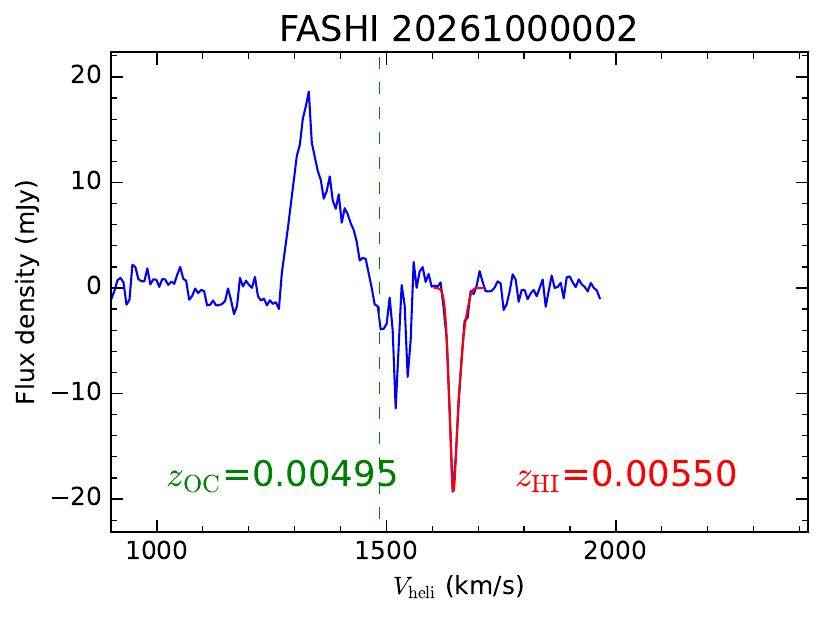}
 \includegraphics[height=0.27\textwidth, angle=0]{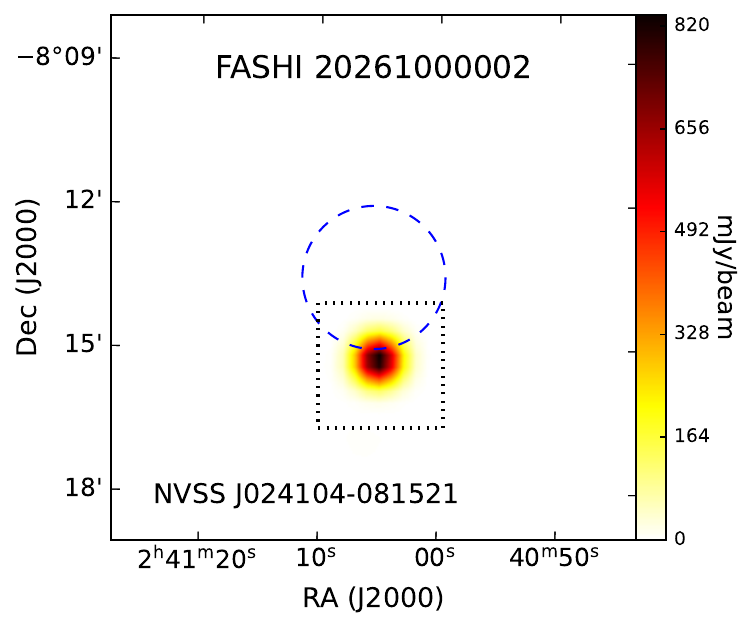}
 \includegraphics[height=0.27\textwidth, angle=0]{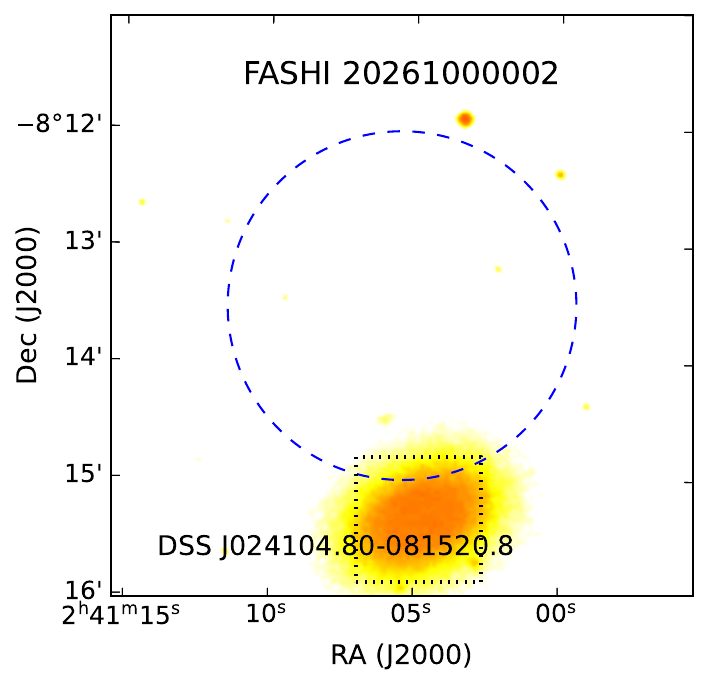}
 \includegraphics[height=0.22\textwidth, angle=0]{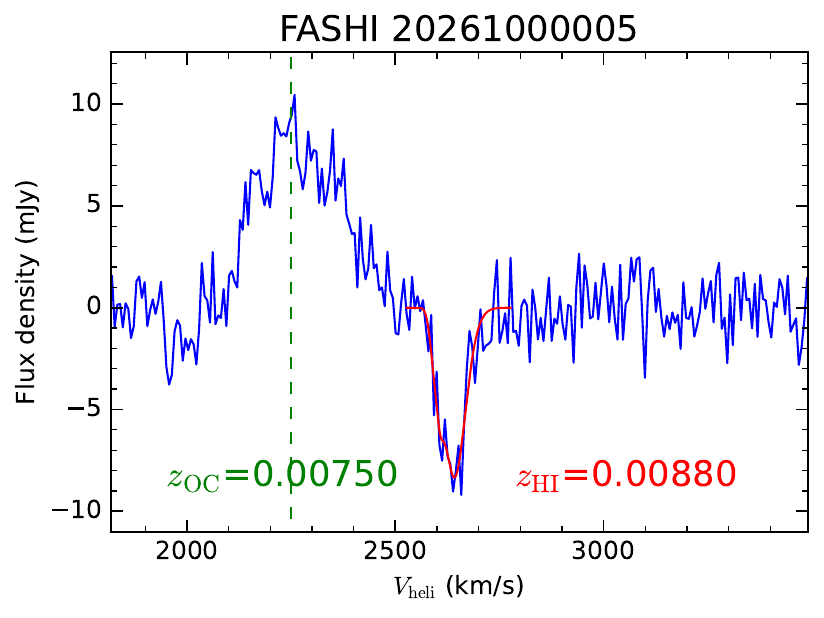}
 \includegraphics[height=0.27\textwidth, angle=0]{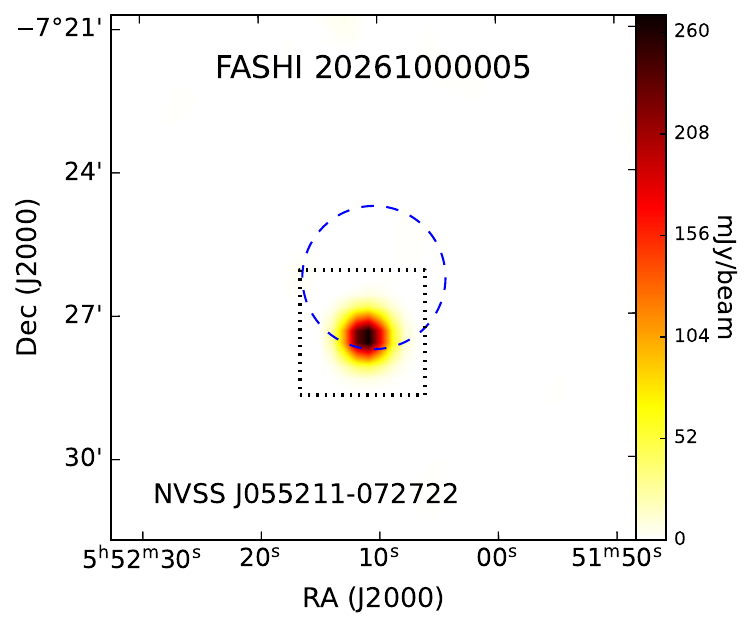}
 \includegraphics[height=0.27\textwidth, angle=0]{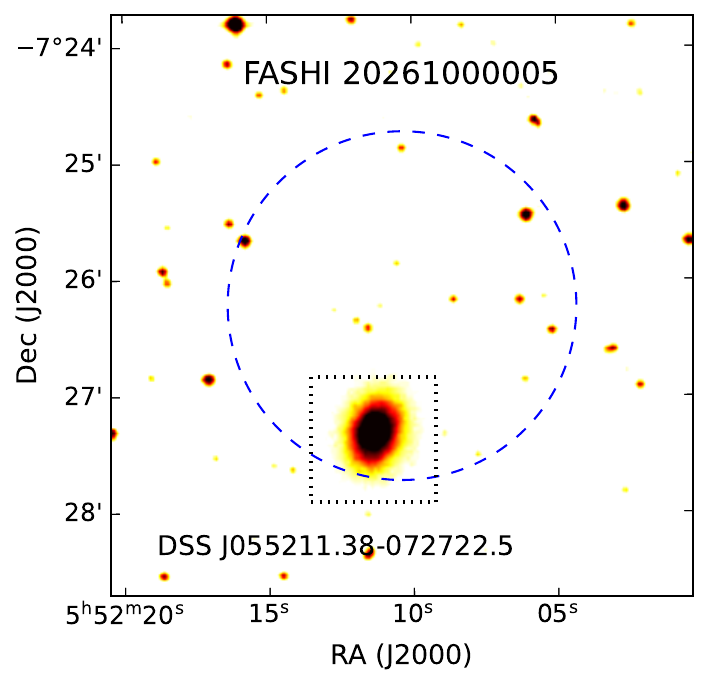}
 \includegraphics[height=0.22\textwidth, angle=0]{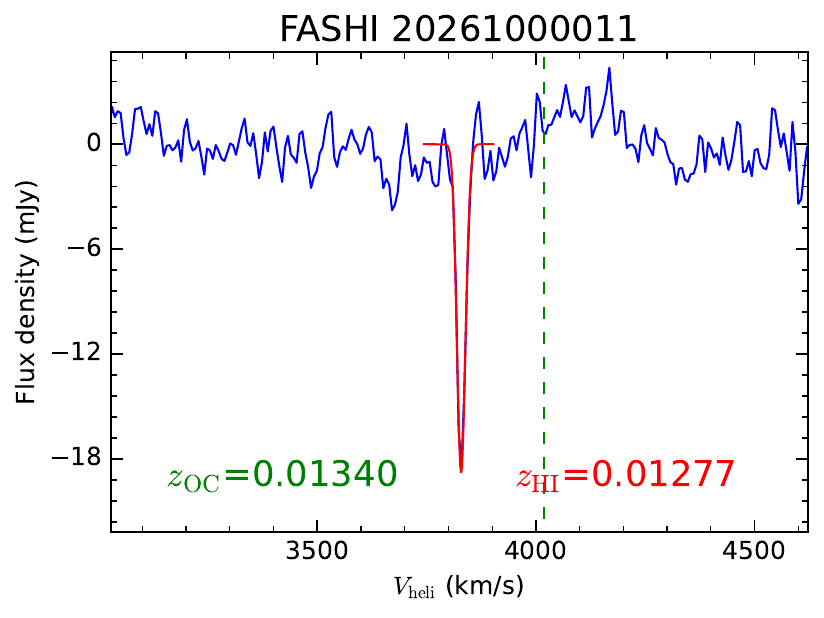}
 \includegraphics[height=0.27\textwidth, angle=0]{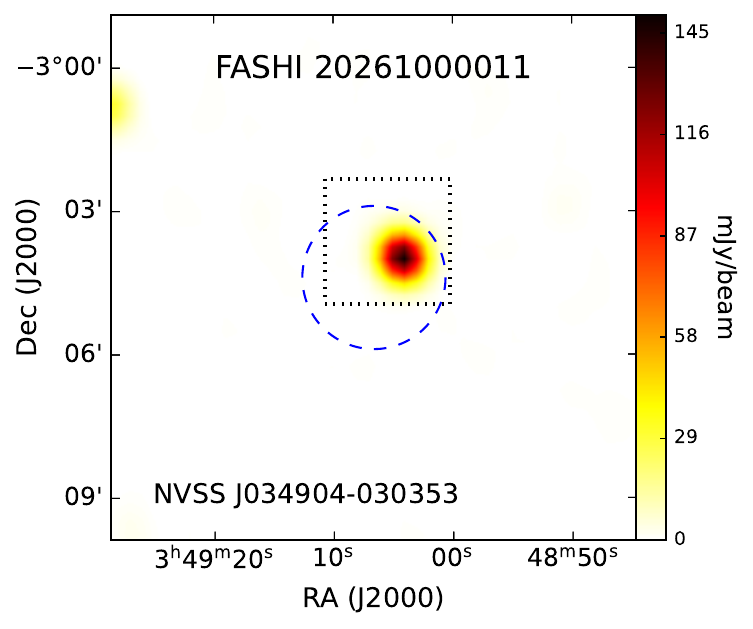}
 \includegraphics[height=0.27\textwidth, angle=0]{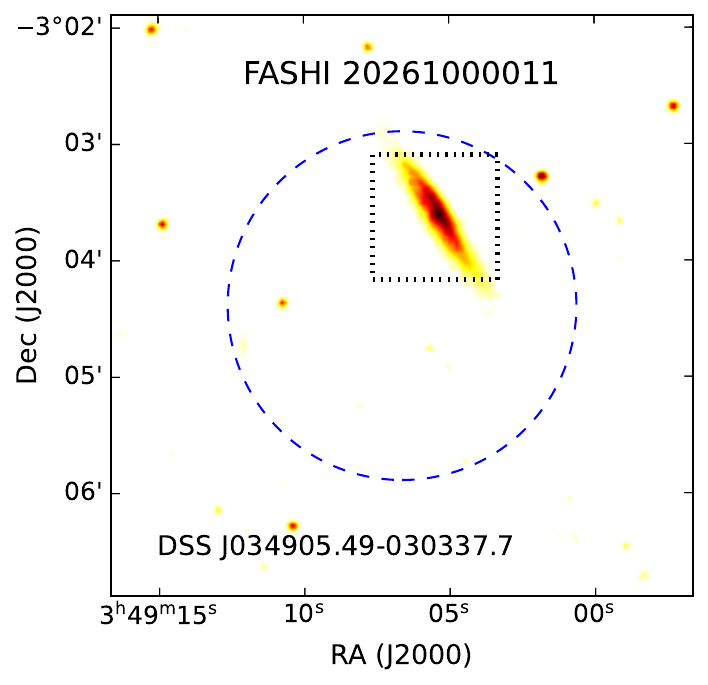}
 \caption{See caption in Figure\,\ref{Fig:FASHI_hi}}
 \label{Fig:FASHI-Continued}
 \end{figure*} 

 \begin{figure*}[htp]
 \centering
 \renewcommand{\thefigure}{\arabic{figure} (Continued)}
 \addtocounter{figure}{-1}
 \includegraphics[height=0.22\textwidth, angle=0]{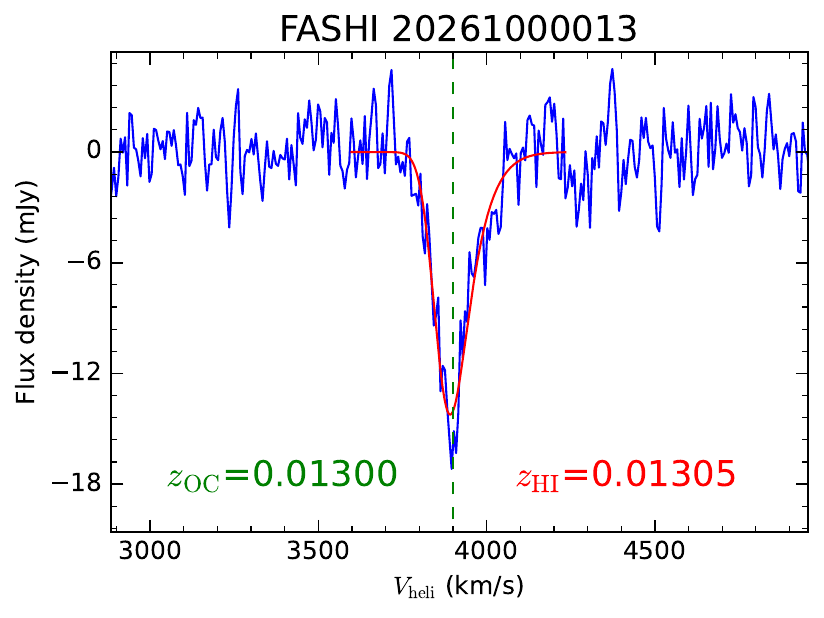}
 \includegraphics[height=0.27\textwidth, angle=0]{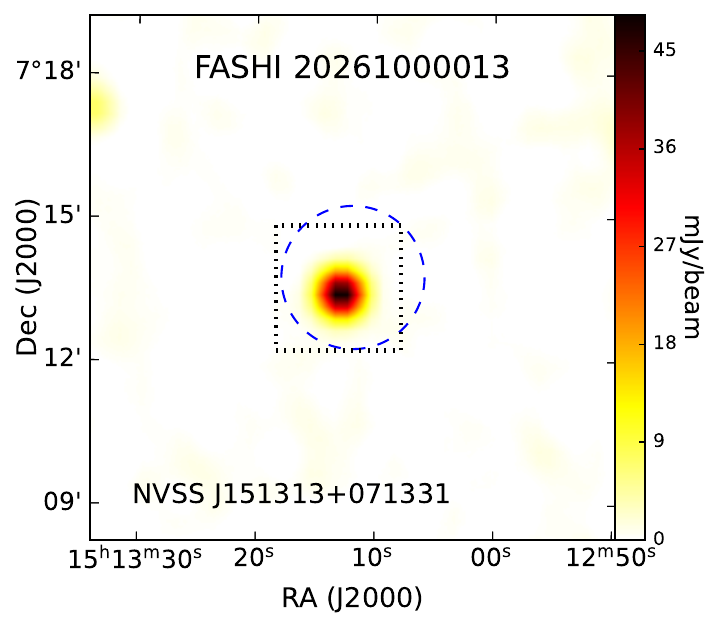}
 \includegraphics[height=0.27\textwidth, angle=0]{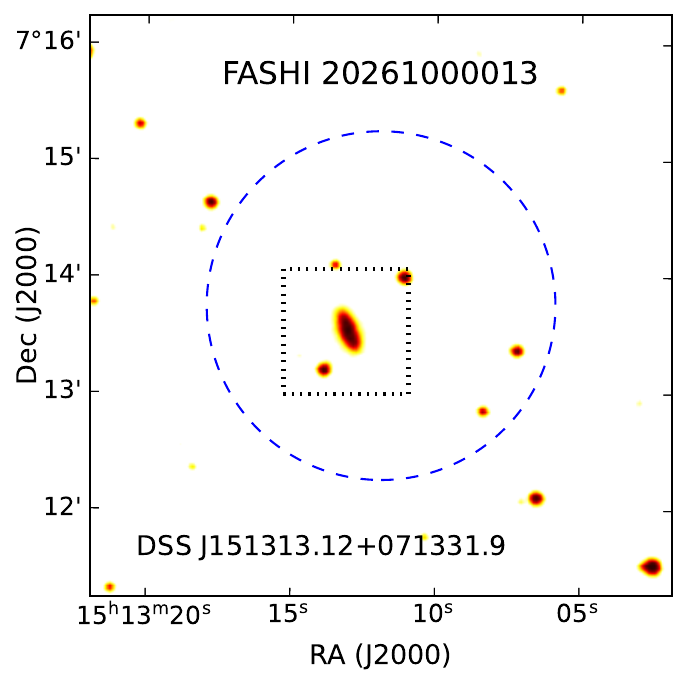}
 \includegraphics[height=0.22\textwidth, angle=0]{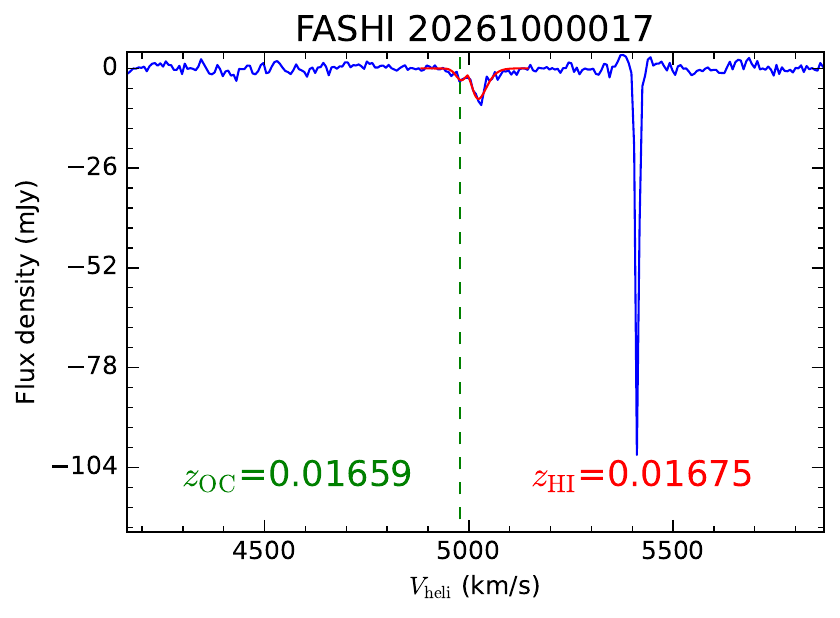}
 \includegraphics[height=0.27\textwidth, angle=0]{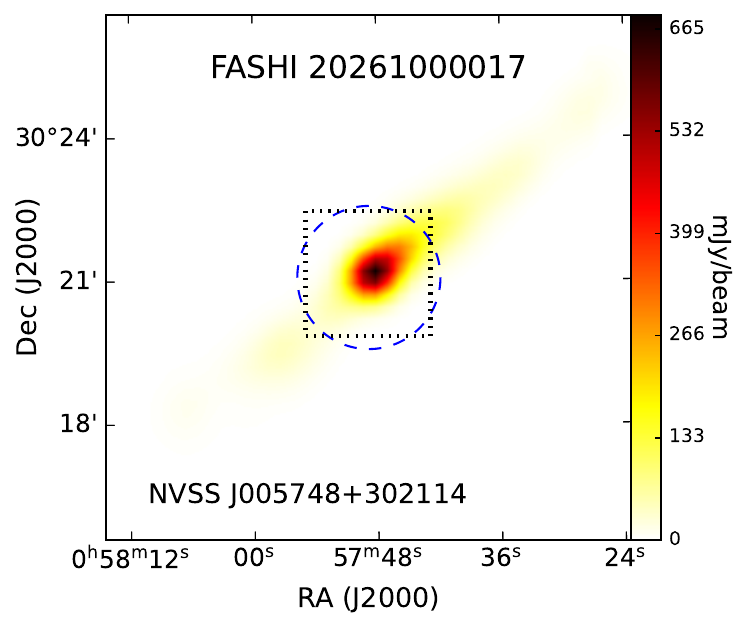}
 \includegraphics[height=0.27\textwidth, angle=0]{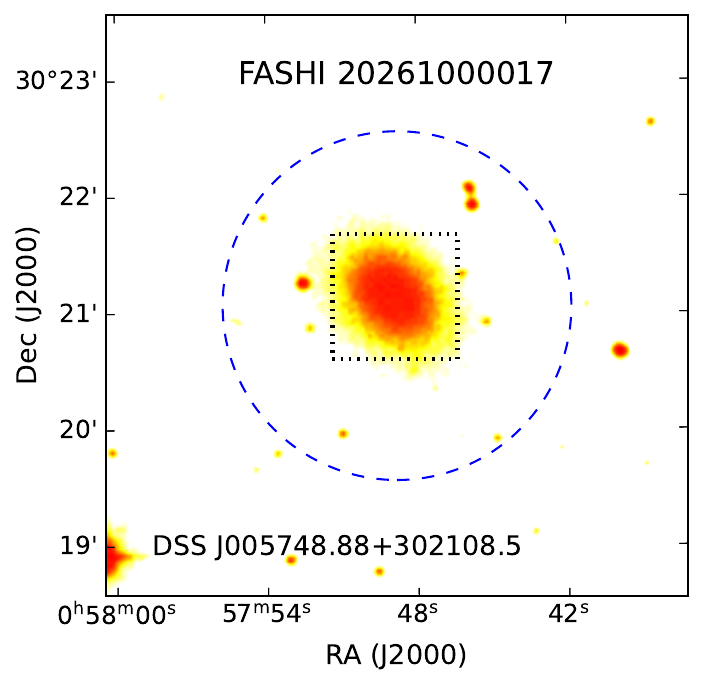}
 \includegraphics[height=0.22\textwidth, angle=0]{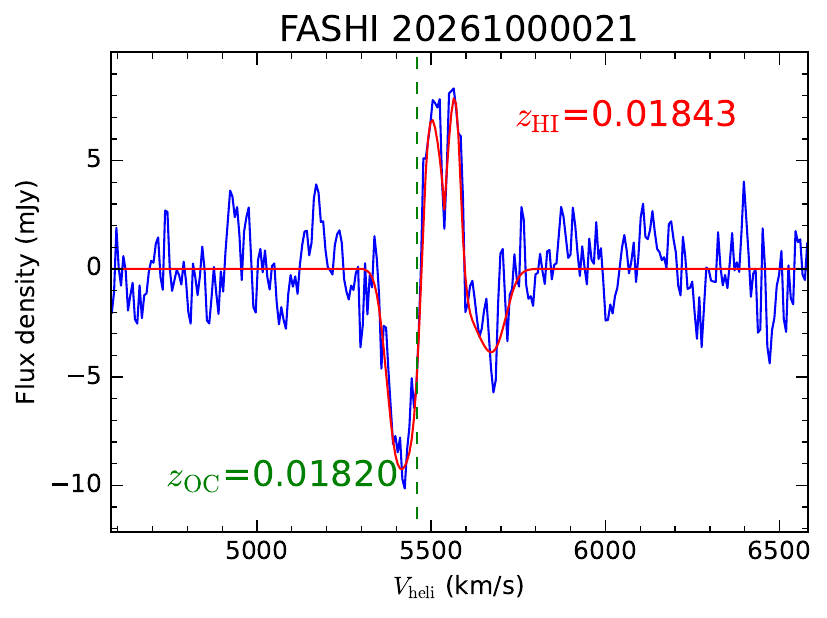}
 \includegraphics[height=0.27\textwidth, angle=0]{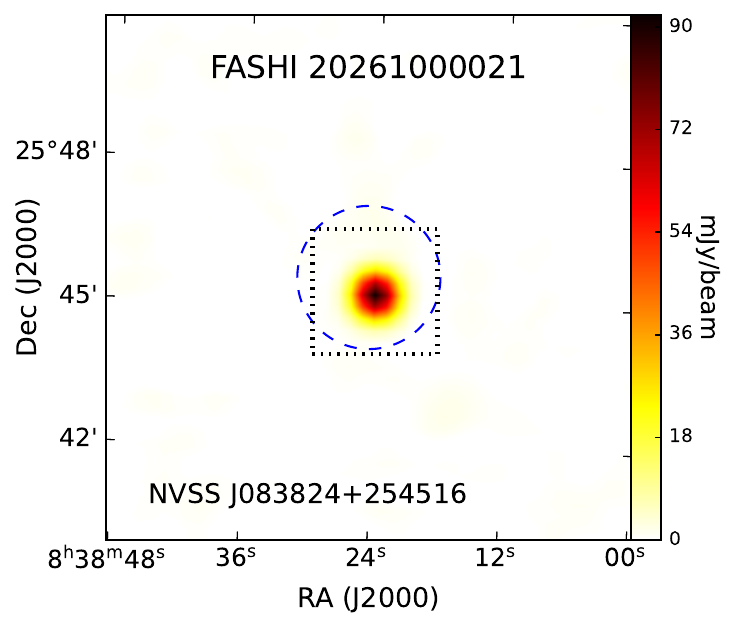}
 \includegraphics[height=0.27\textwidth, angle=0]{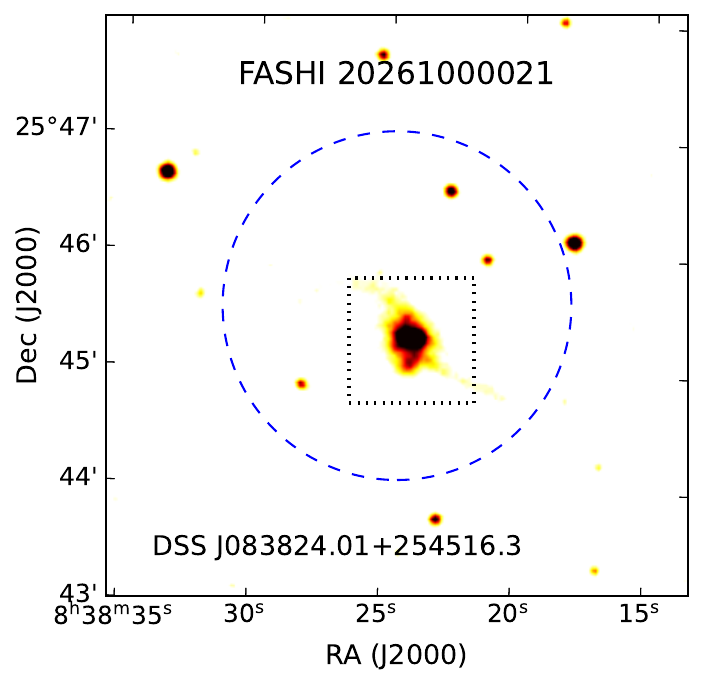}
 \includegraphics[height=0.22\textwidth, angle=0]{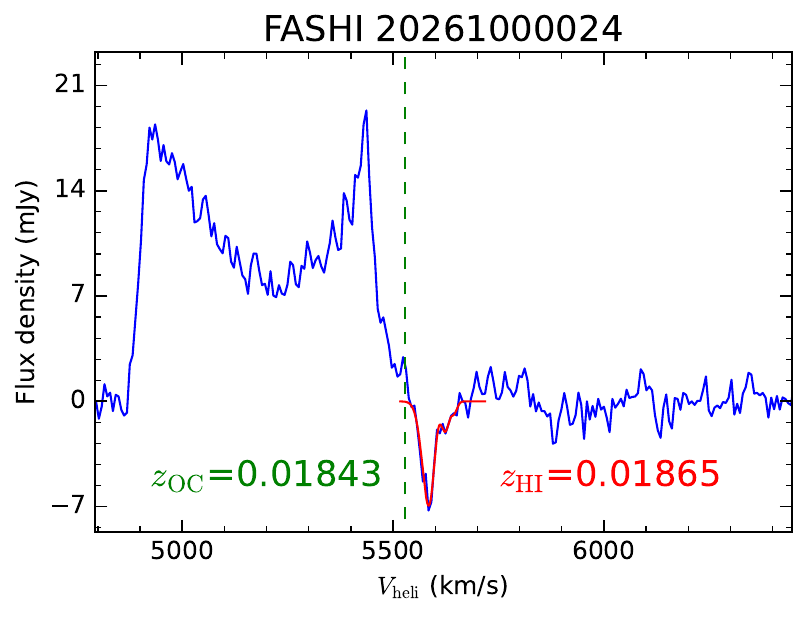}
 \includegraphics[height=0.27\textwidth, angle=0]{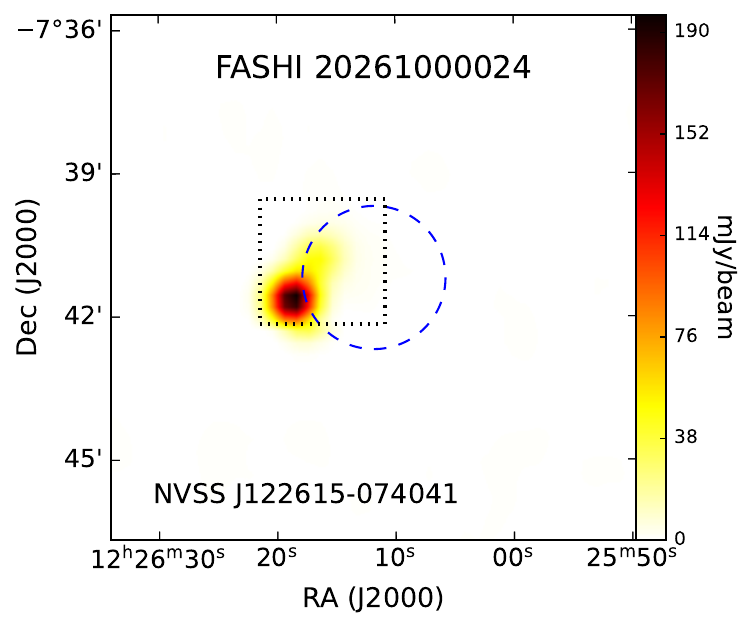}
 \includegraphics[height=0.27\textwidth, angle=0]{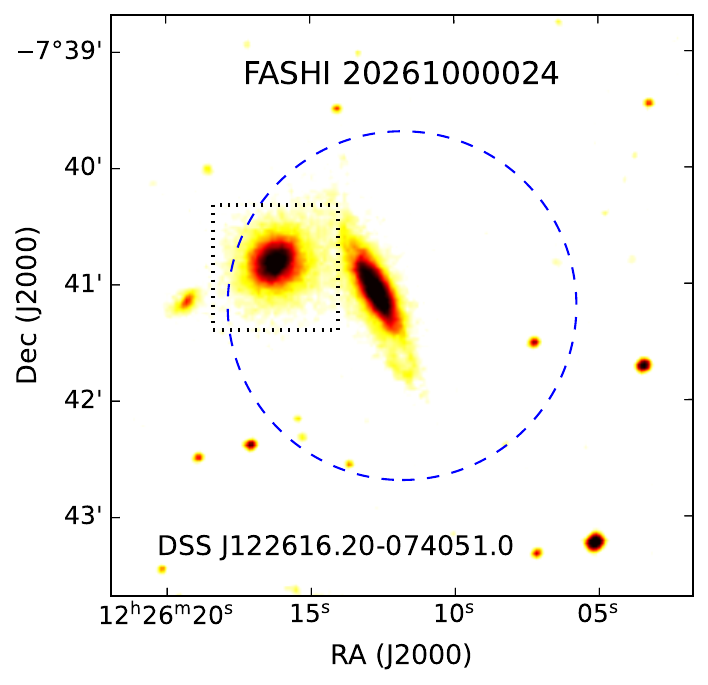}
 \caption{See caption in Figure\,\ref{Fig:FASHI_hi}}
 \end{figure*} 

 \begin{figure*}[htp]
 \centering
 \renewcommand{\thefigure}{\arabic{figure} (Continued)}
 \addtocounter{figure}{-1}
 \includegraphics[height=0.22\textwidth, angle=0]{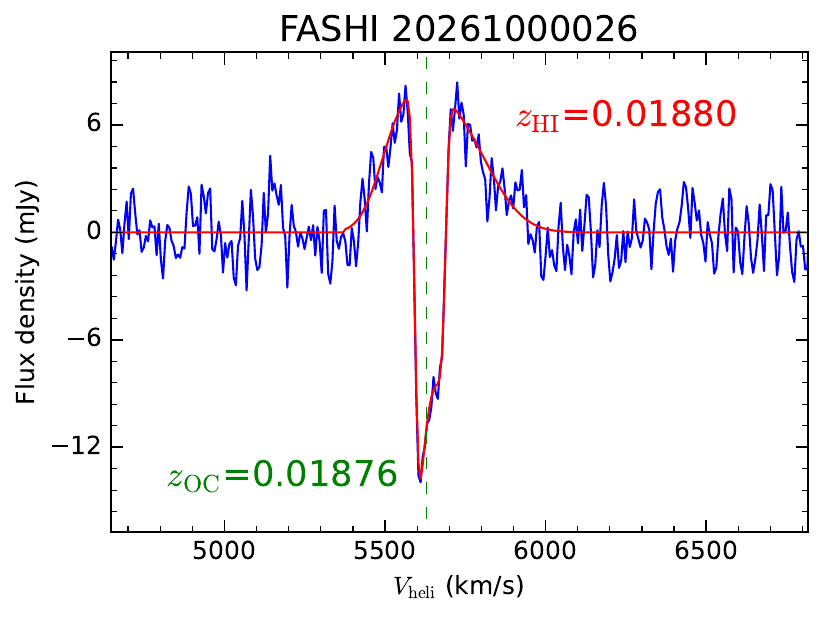}
 \includegraphics[height=0.27\textwidth, angle=0]{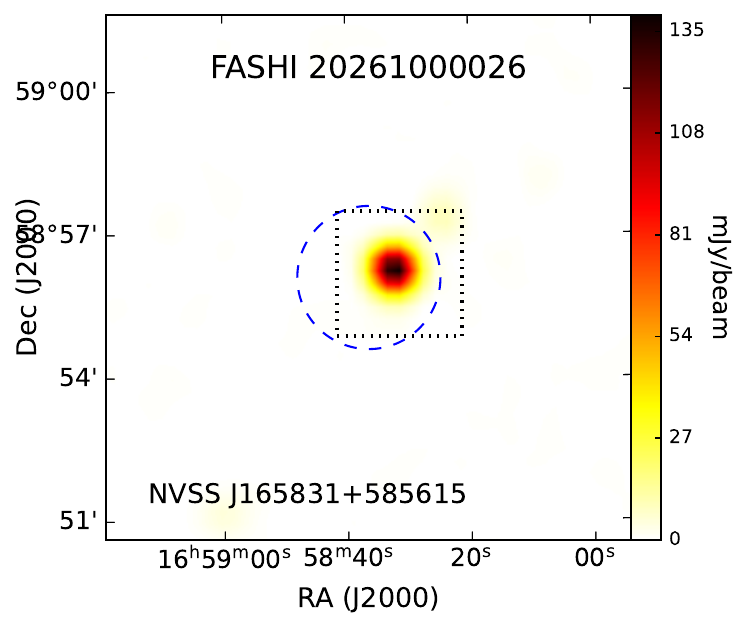}
 \includegraphics[height=0.27\textwidth, angle=0]{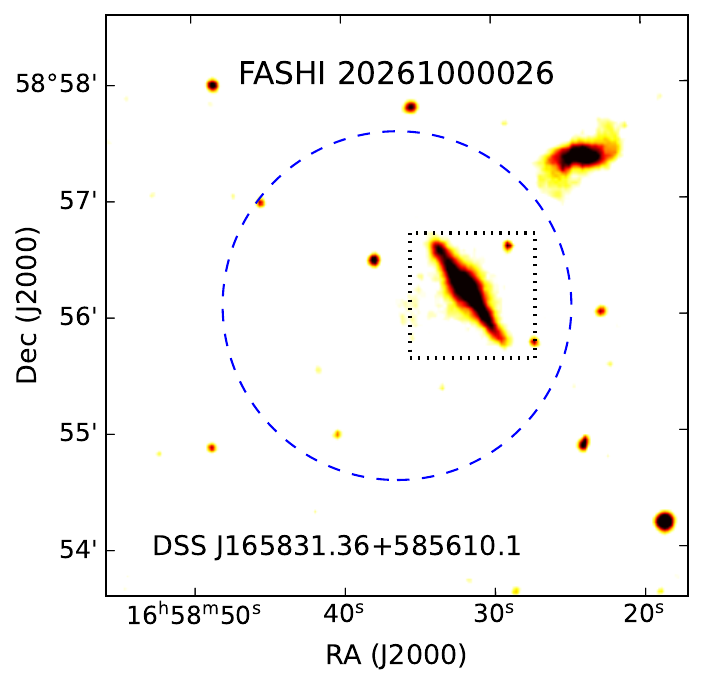}
 \includegraphics[height=0.22\textwidth, angle=0]{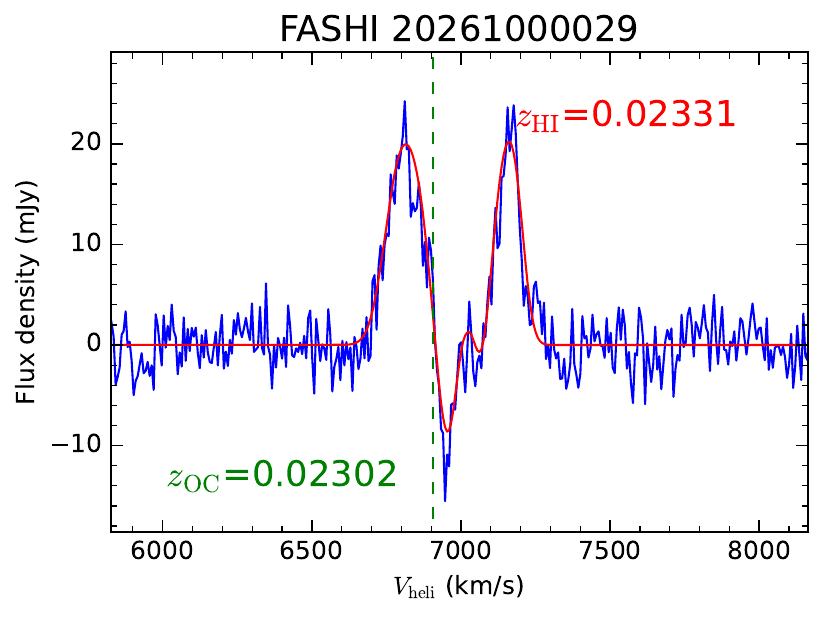}
 \includegraphics[height=0.27\textwidth, angle=0]{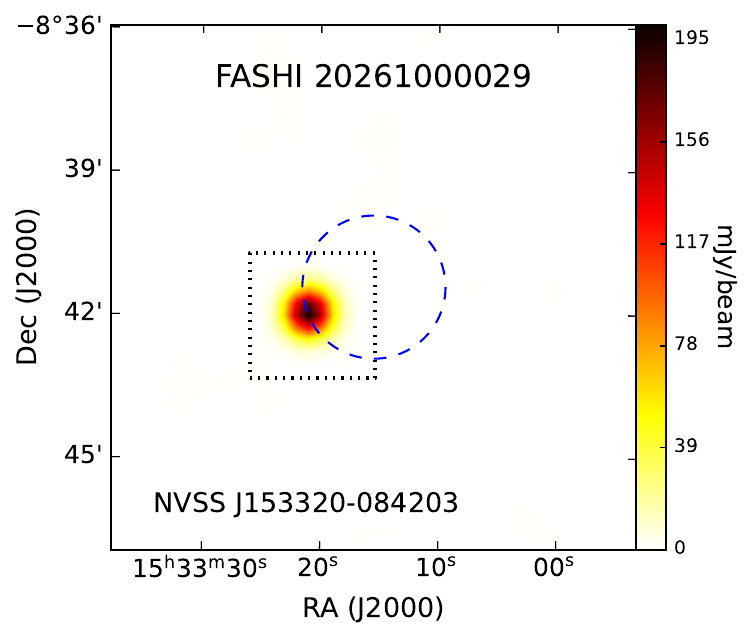}
 \includegraphics[height=0.27\textwidth, angle=0]{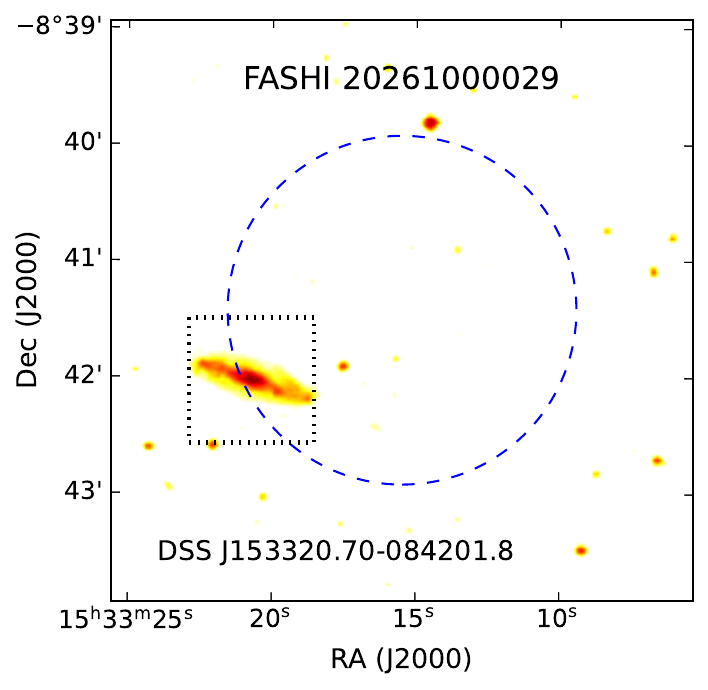}
 \includegraphics[height=0.22\textwidth, angle=0]{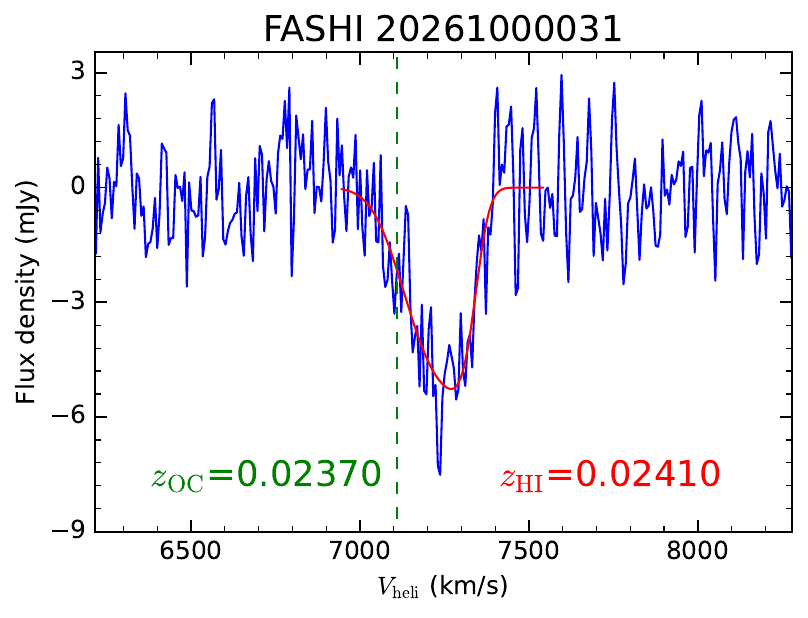}
 \includegraphics[height=0.27\textwidth, angle=0]{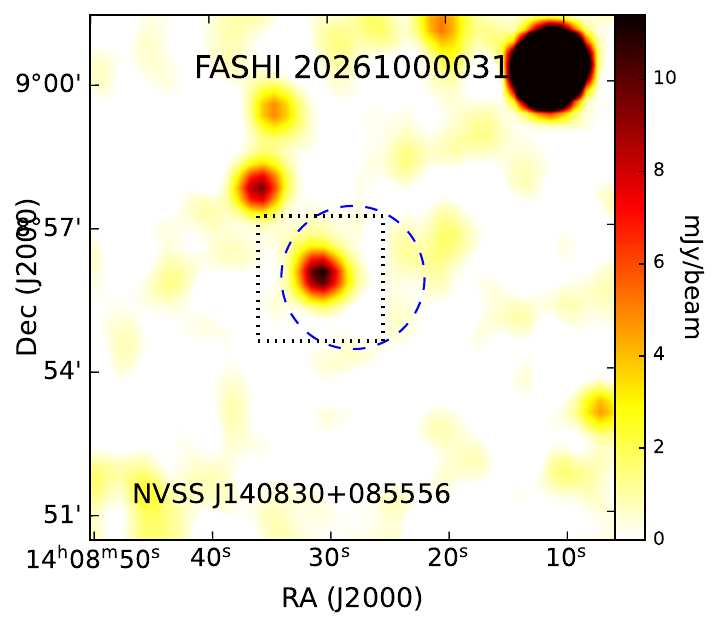}
 \includegraphics[height=0.27\textwidth, angle=0]{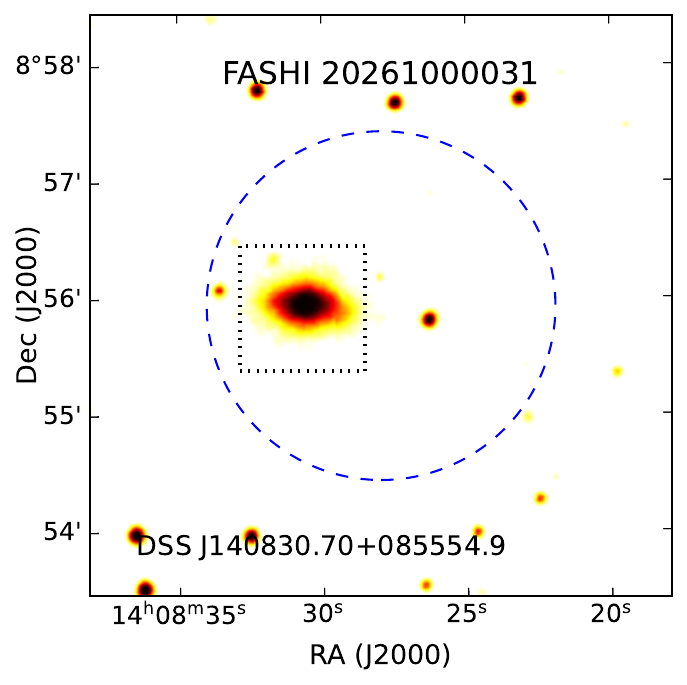}
 \includegraphics[height=0.22\textwidth, angle=0]{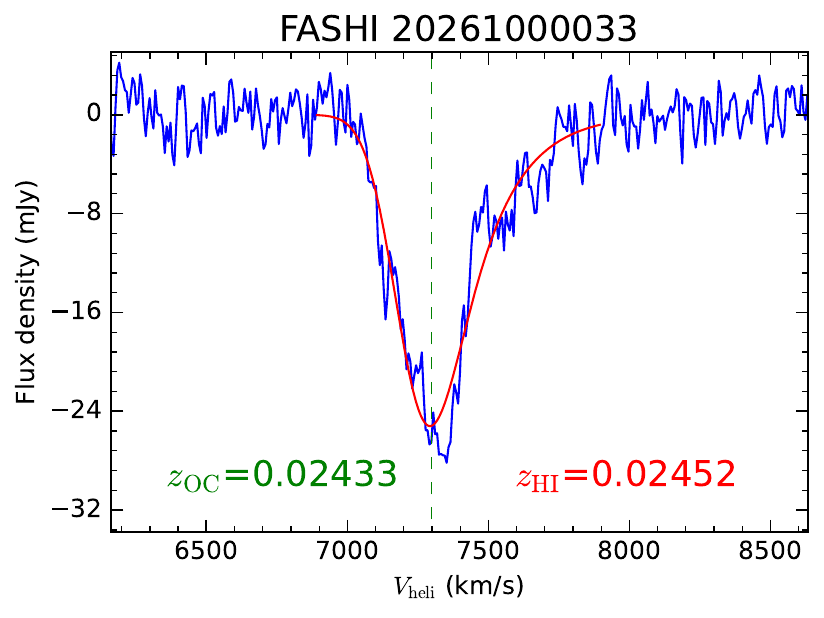}
 \includegraphics[height=0.27\textwidth, angle=0]{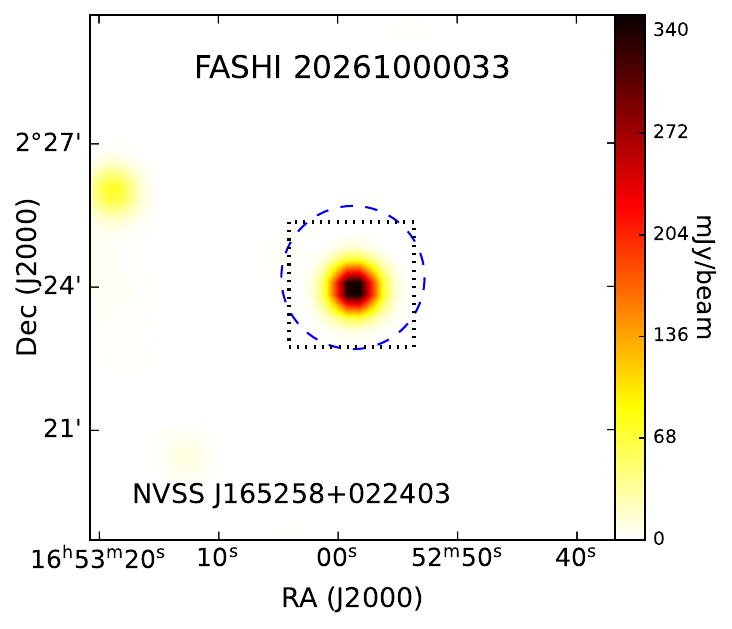}
 \includegraphics[height=0.27\textwidth, angle=0]{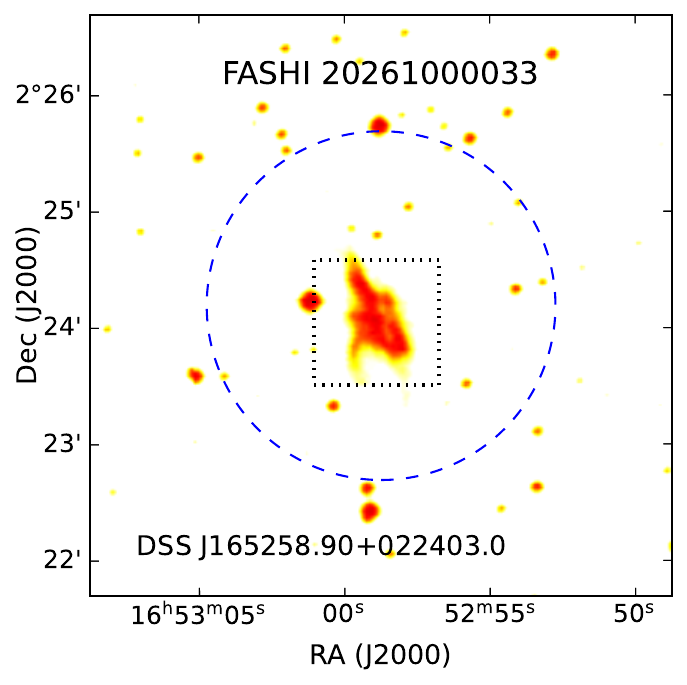}
 \caption{See caption in Figure\,\ref{Fig:FASHI_hi}}
 \end{figure*} 

 \begin{figure*}[htp]
 \centering
 \renewcommand{\thefigure}{\arabic{figure} (Continued)}
 \addtocounter{figure}{-1}
 \includegraphics[height=0.22\textwidth, angle=0]{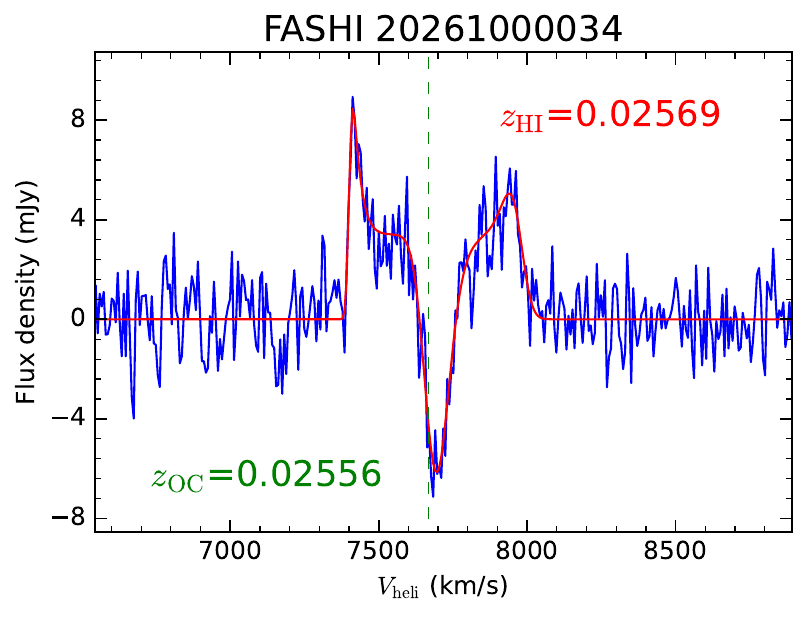}
 \includegraphics[height=0.27\textwidth, angle=0]{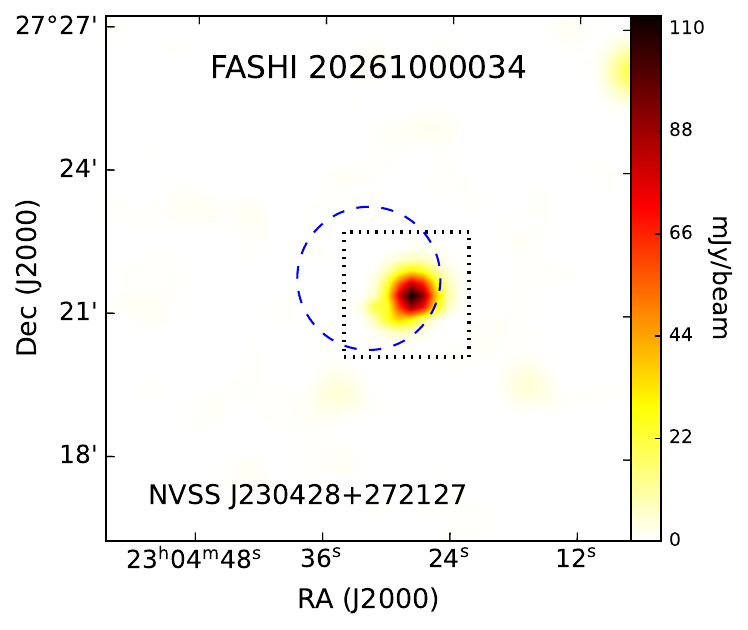}
 \includegraphics[height=0.27\textwidth, angle=0]{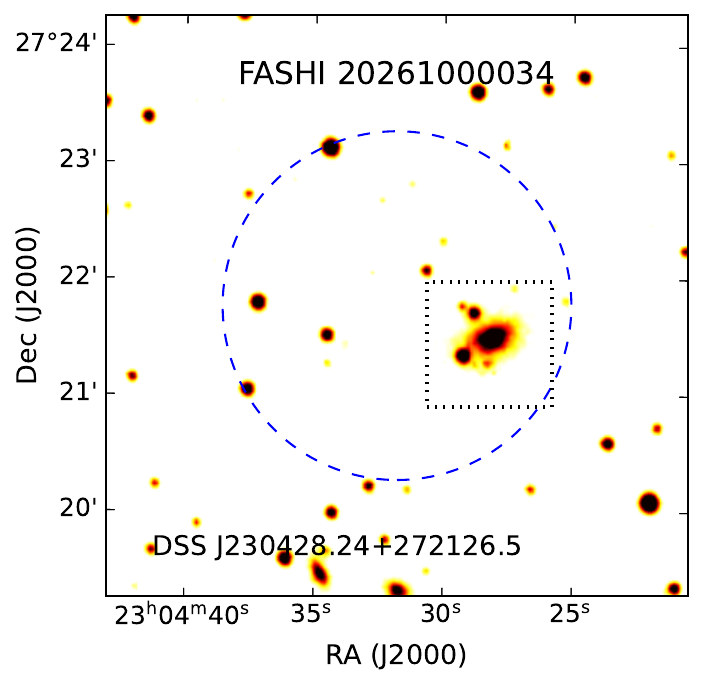}
 \includegraphics[height=0.22\textwidth, angle=0]{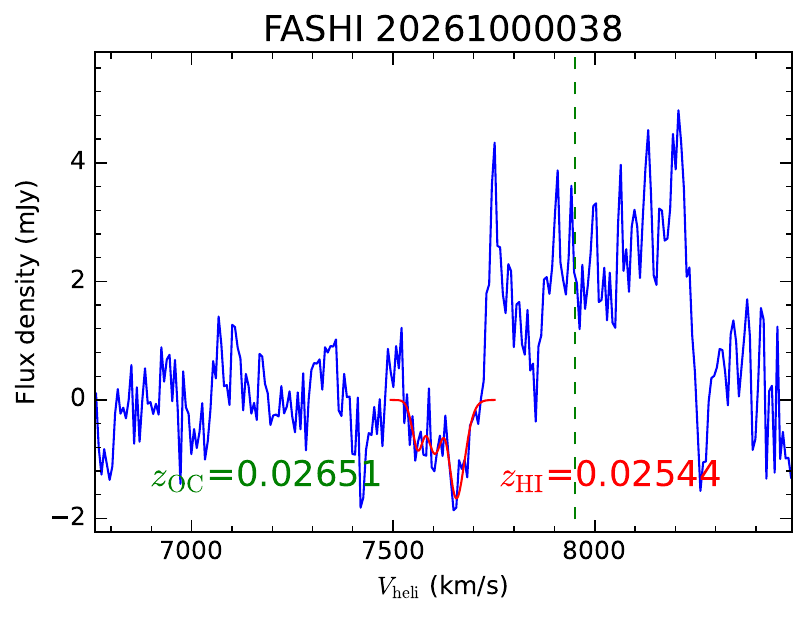}
 \includegraphics[height=0.27\textwidth, angle=0]{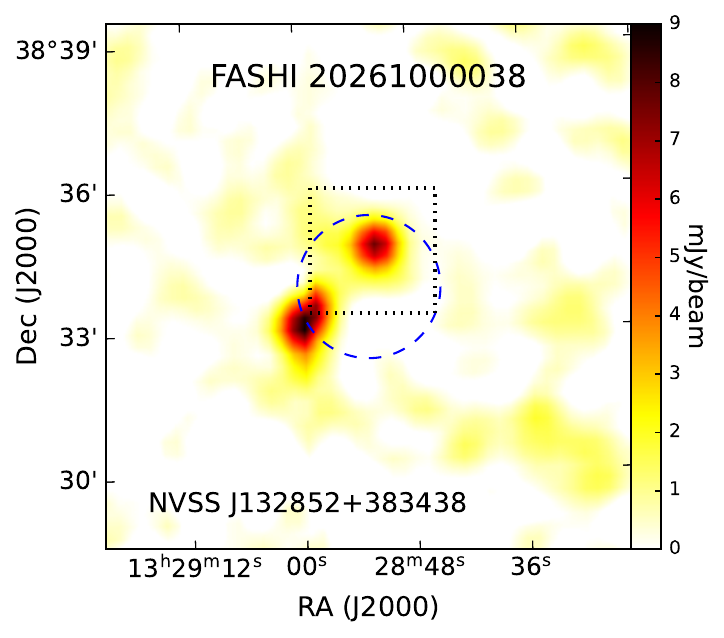}
 \includegraphics[height=0.27\textwidth, angle=0]{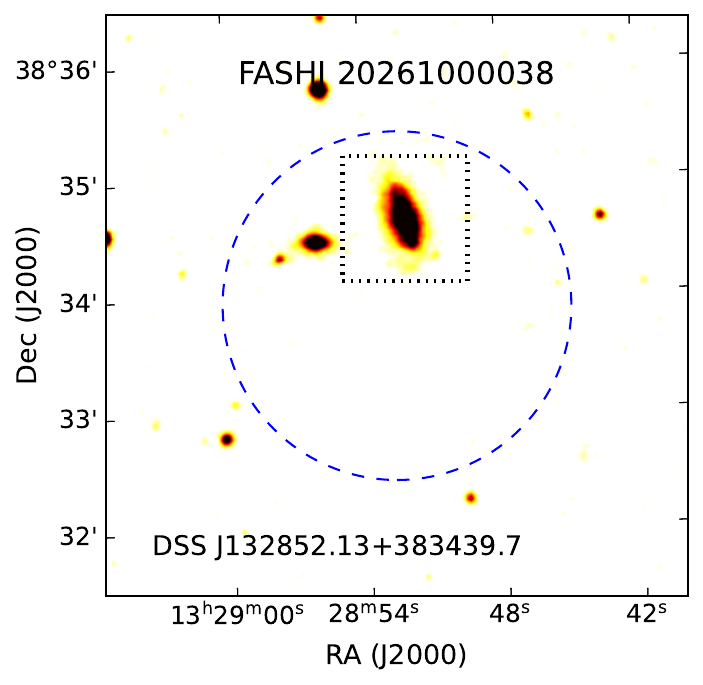}
 \includegraphics[height=0.22\textwidth, angle=0]{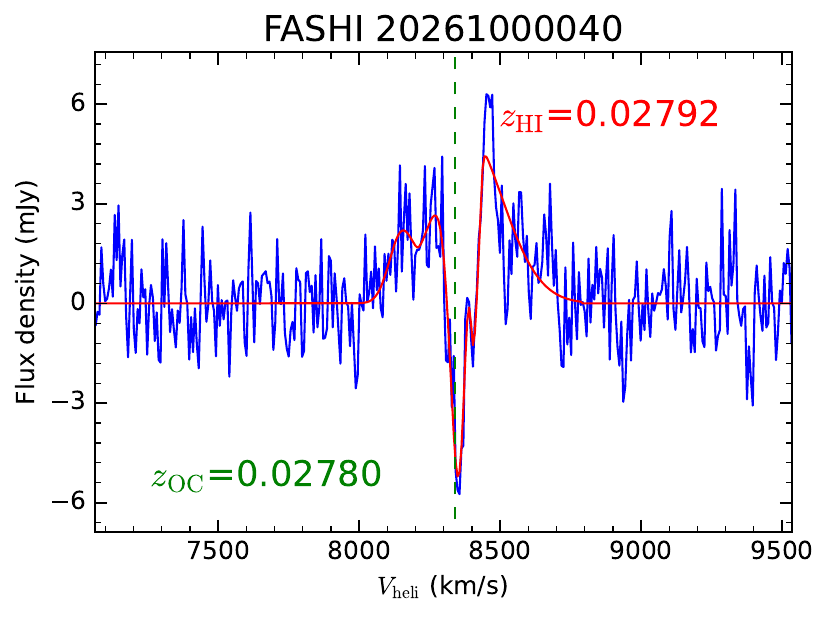}
 \includegraphics[height=0.27\textwidth, angle=0]{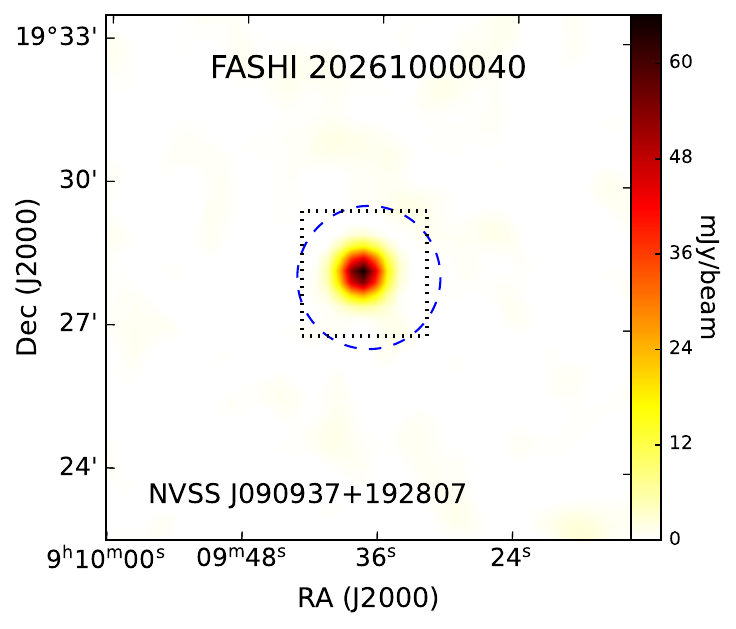}
 \includegraphics[height=0.27\textwidth, angle=0]{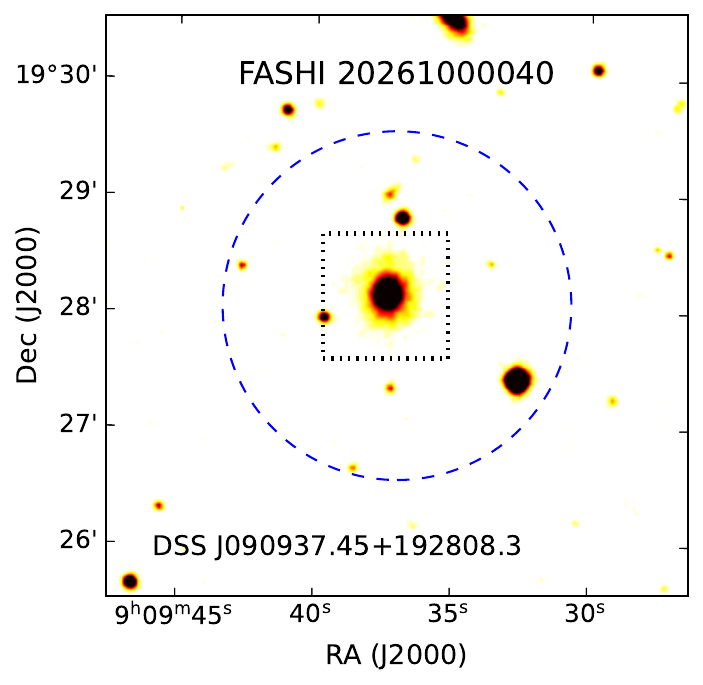}
 \includegraphics[height=0.22\textwidth, angle=0]{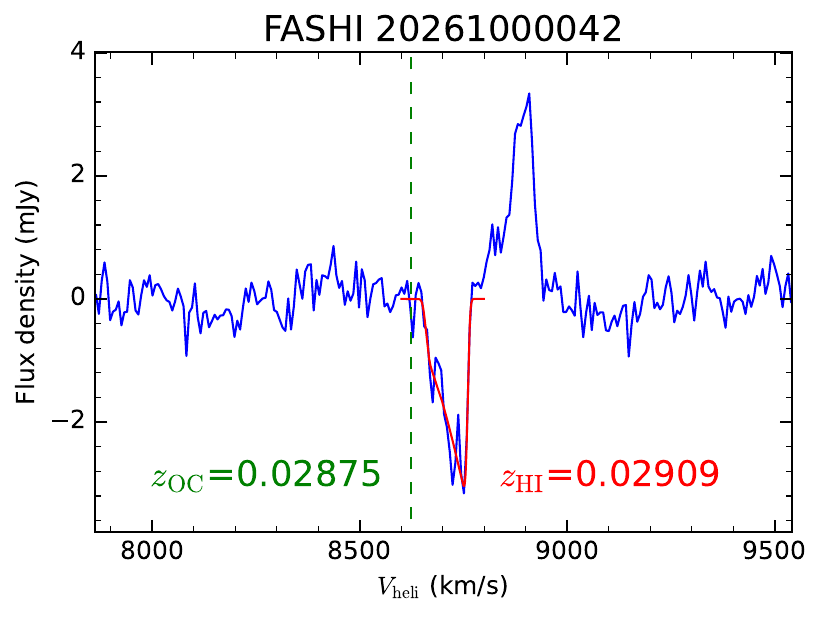}
 \includegraphics[height=0.27\textwidth, angle=0]{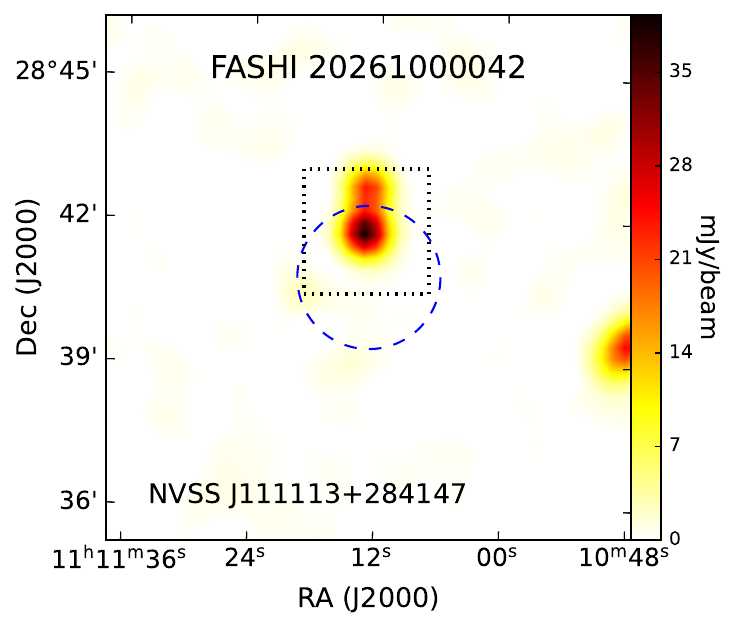}
 \includegraphics[height=0.27\textwidth, angle=0]{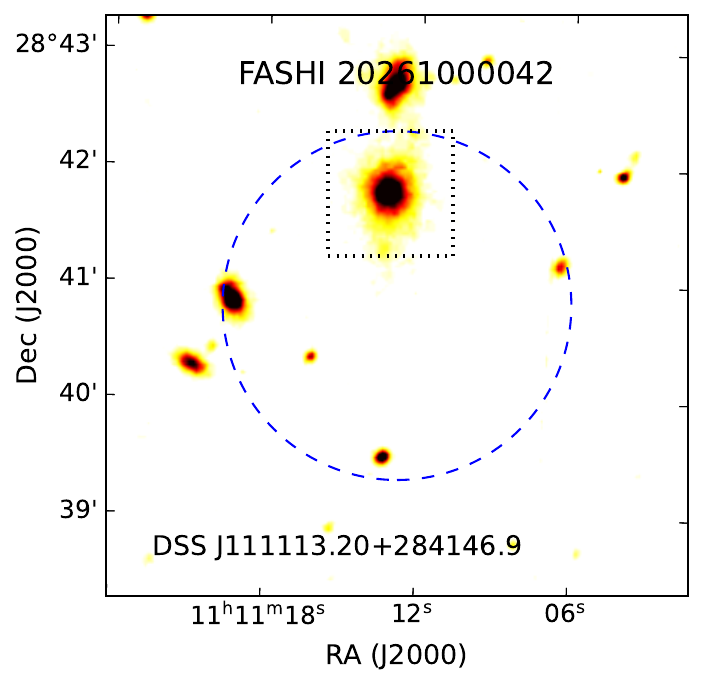}
 \caption{See caption in Figure\,\ref{Fig:FASHI_hi}}
 \end{figure*} 

 \begin{figure*}[htp]
 \centering
 \renewcommand{\thefigure}{\arabic{figure} (Continued)}
 \addtocounter{figure}{-1}
 \includegraphics[height=0.22\textwidth, angle=0]{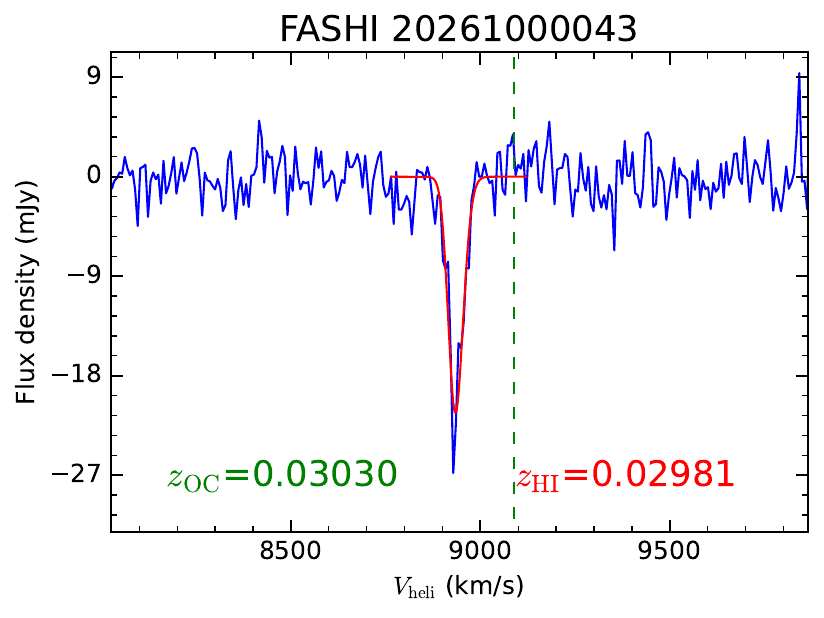}
 \includegraphics[height=0.27\textwidth, angle=0]{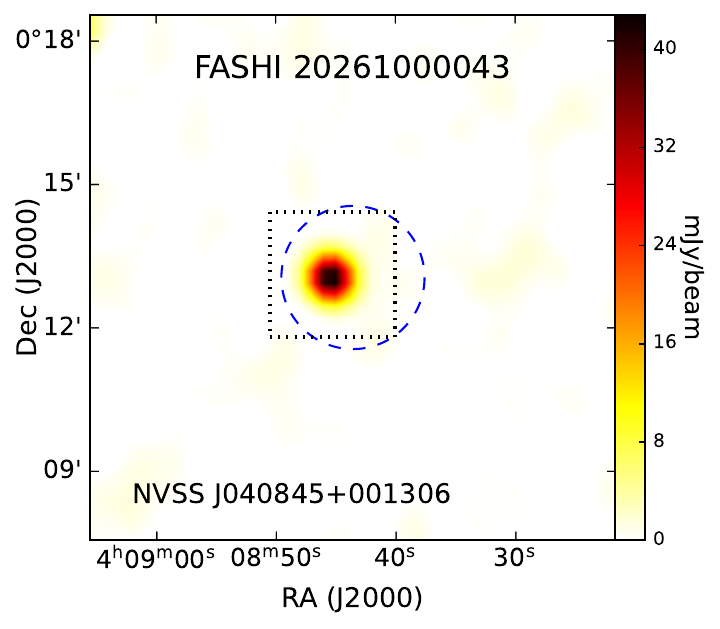}
 \includegraphics[height=0.27\textwidth, angle=0]{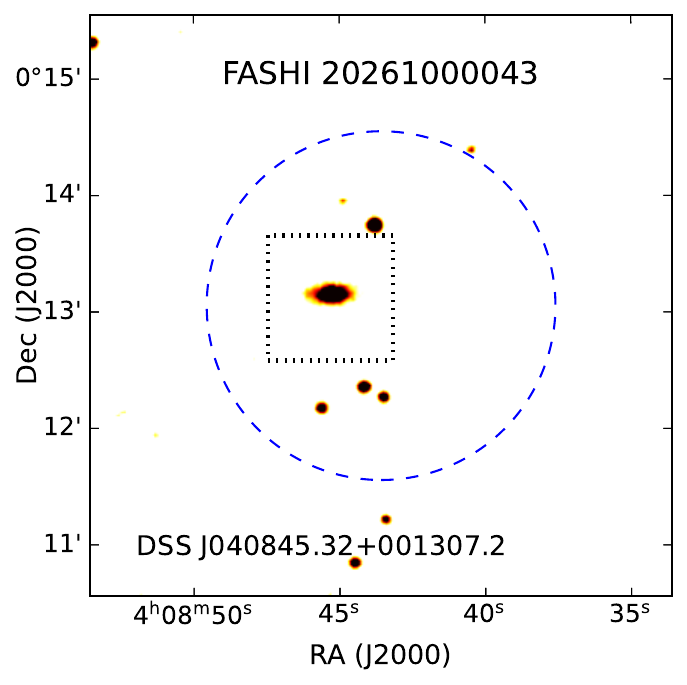}
 \includegraphics[height=0.22\textwidth, angle=0]{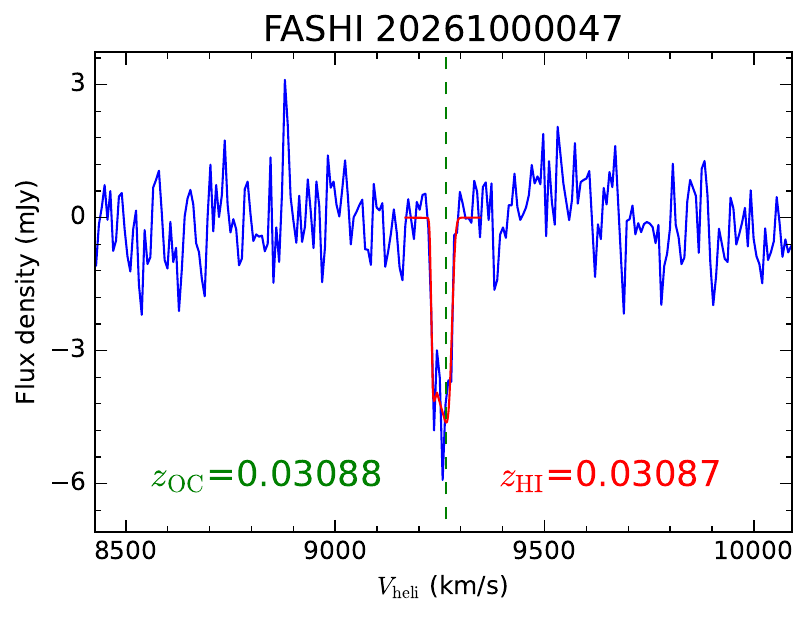}
 \includegraphics[height=0.27\textwidth, angle=0]{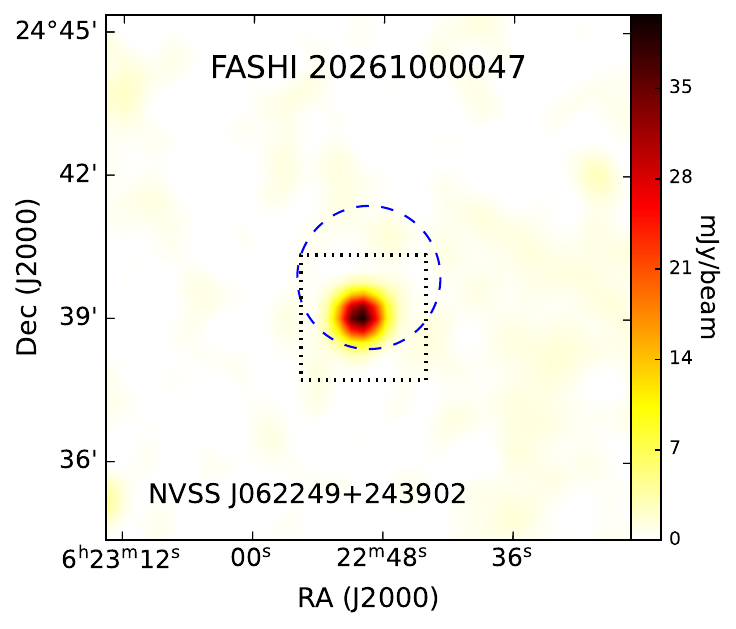}
 \includegraphics[height=0.27\textwidth, angle=0]{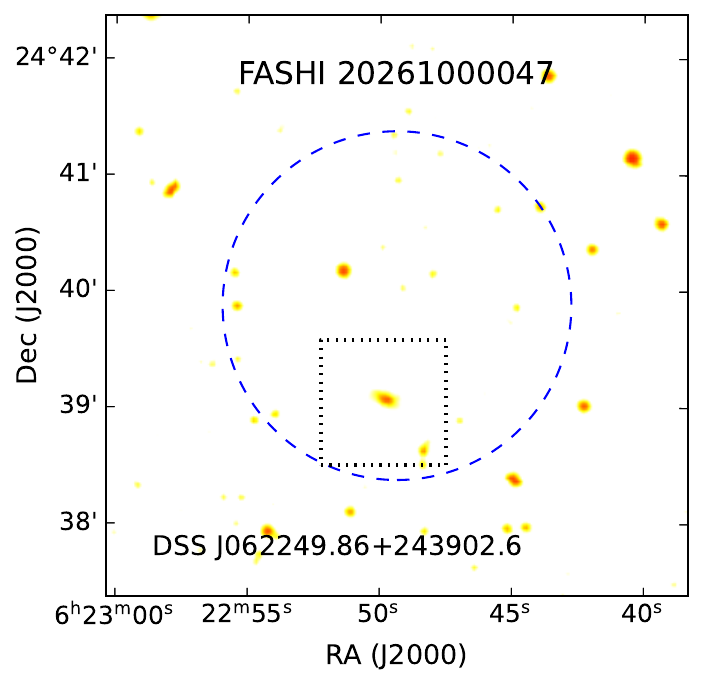}
 \includegraphics[height=0.22\textwidth, angle=0]{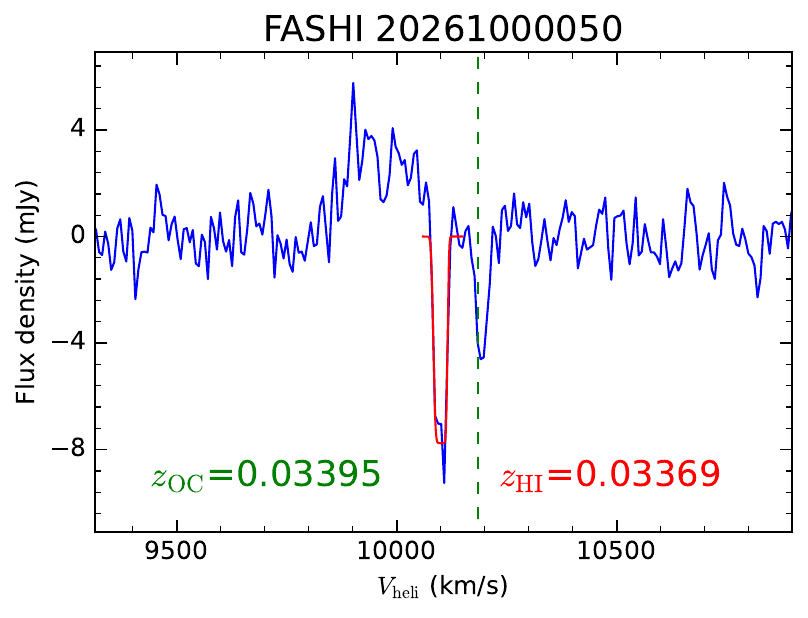}
 \includegraphics[height=0.27\textwidth, angle=0]{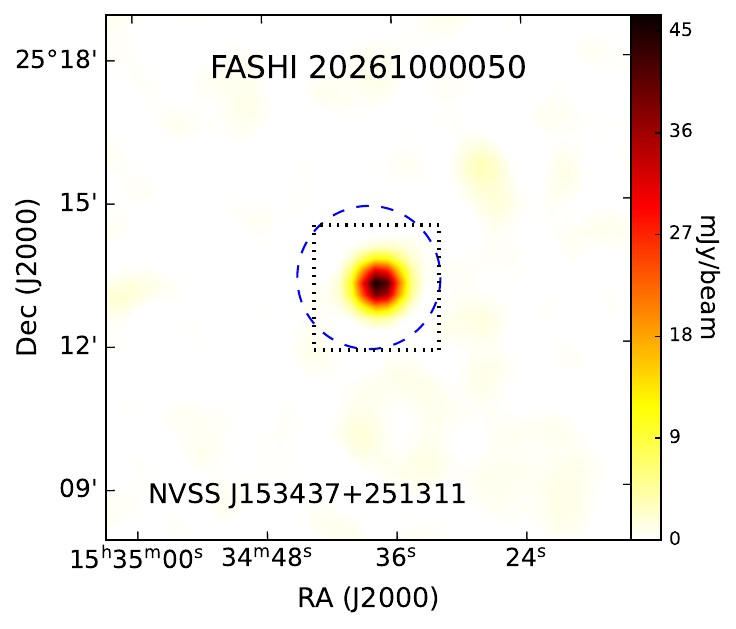}
 \includegraphics[height=0.27\textwidth, angle=0]{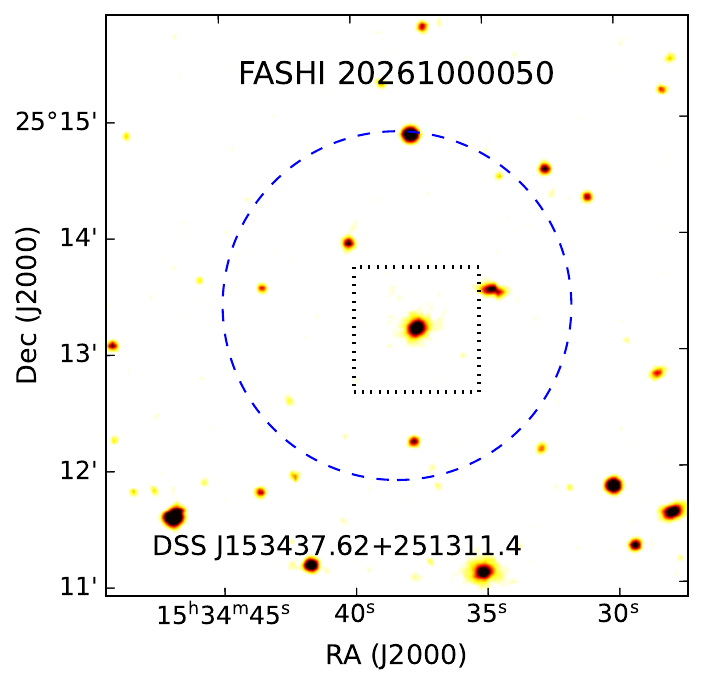}
 \includegraphics[height=0.22\textwidth, angle=0]{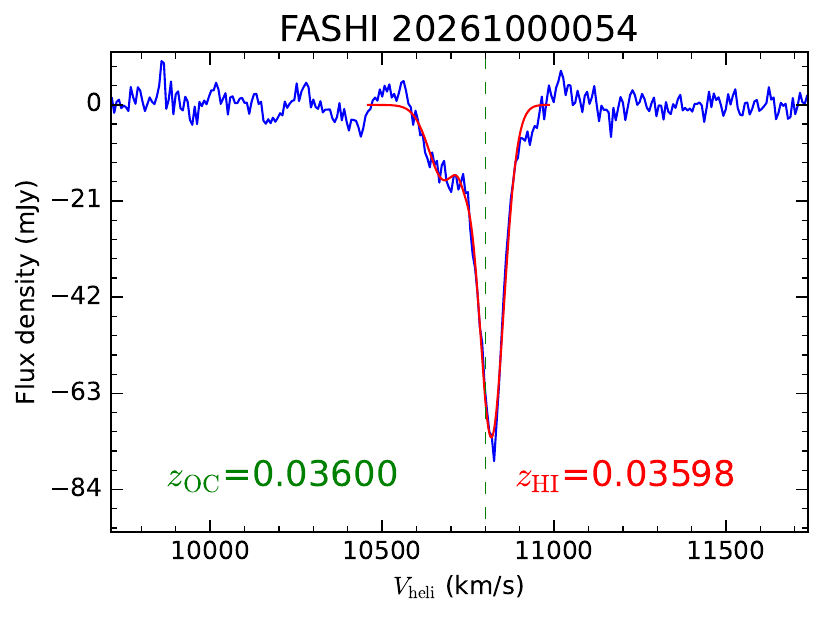}
 \includegraphics[height=0.27\textwidth, angle=0]{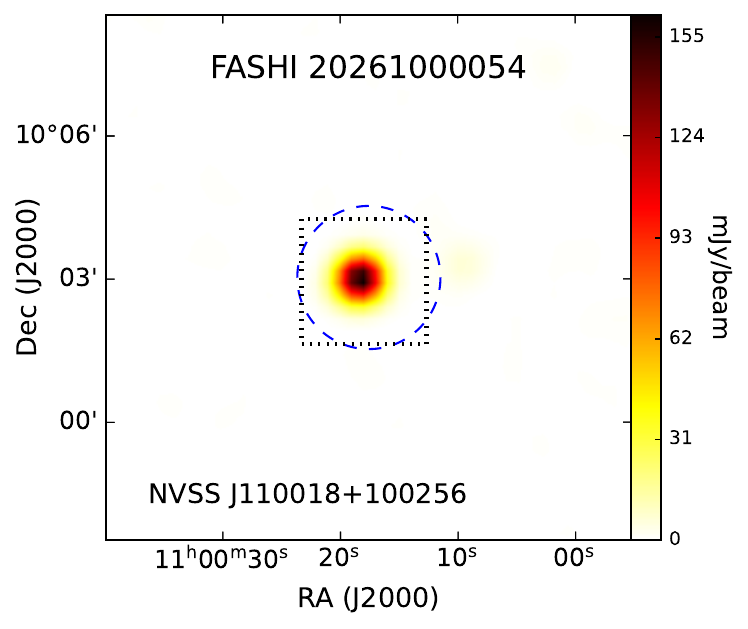}
 \includegraphics[height=0.27\textwidth, angle=0]{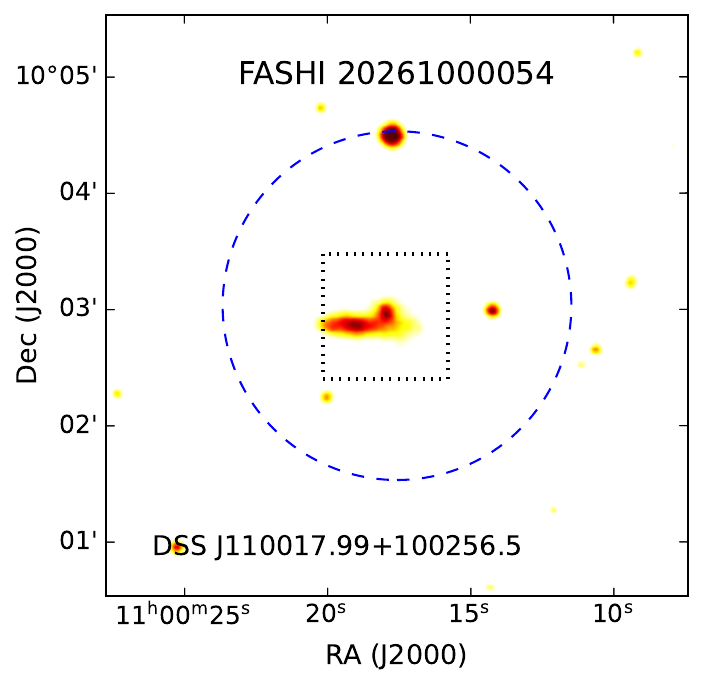}
 \caption{See caption in Figure\,\ref{Fig:FASHI_hi}}
 \end{figure*} 

 \begin{figure*}[htp]
 \centering
 \renewcommand{\thefigure}{\arabic{figure} (Continued)}
 \addtocounter{figure}{-1}
 \includegraphics[height=0.22\textwidth, angle=0]{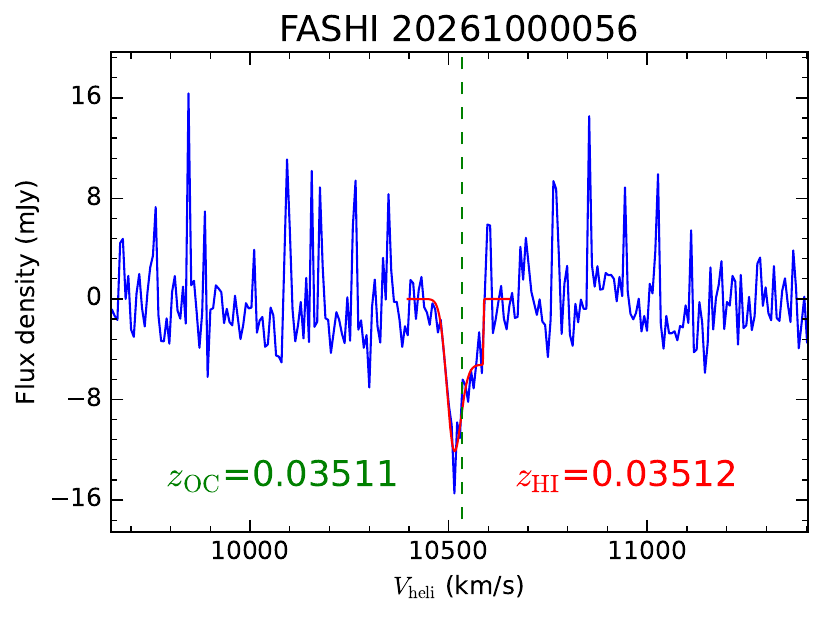}
 \includegraphics[height=0.27\textwidth, angle=0]{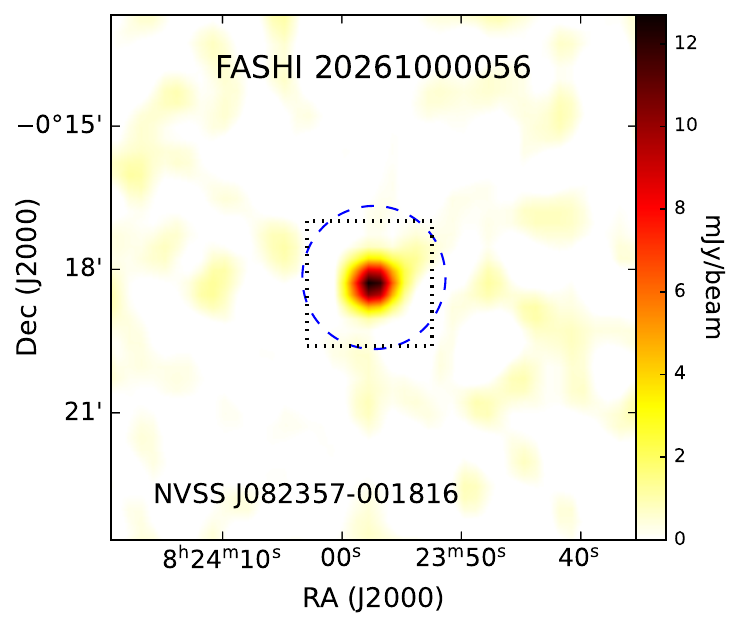}
 \includegraphics[height=0.27\textwidth, angle=0]{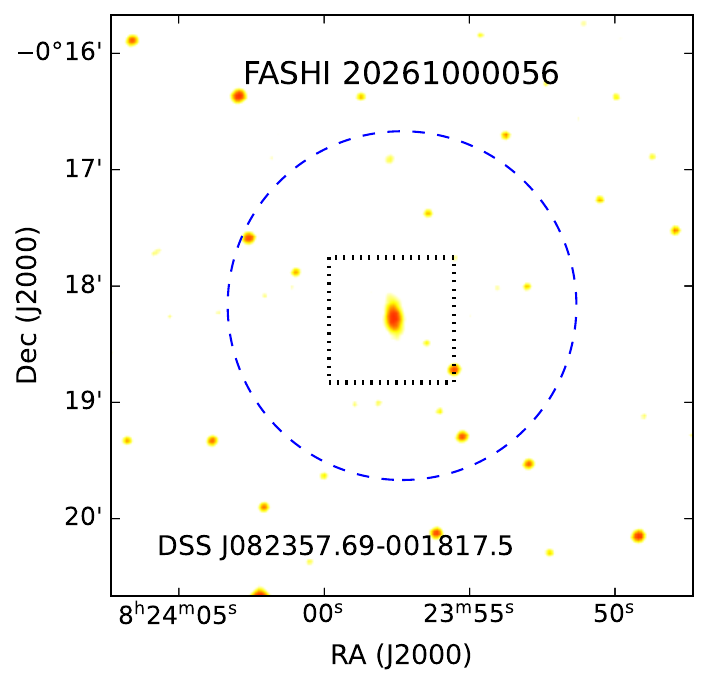}
 \includegraphics[height=0.22\textwidth, angle=0]{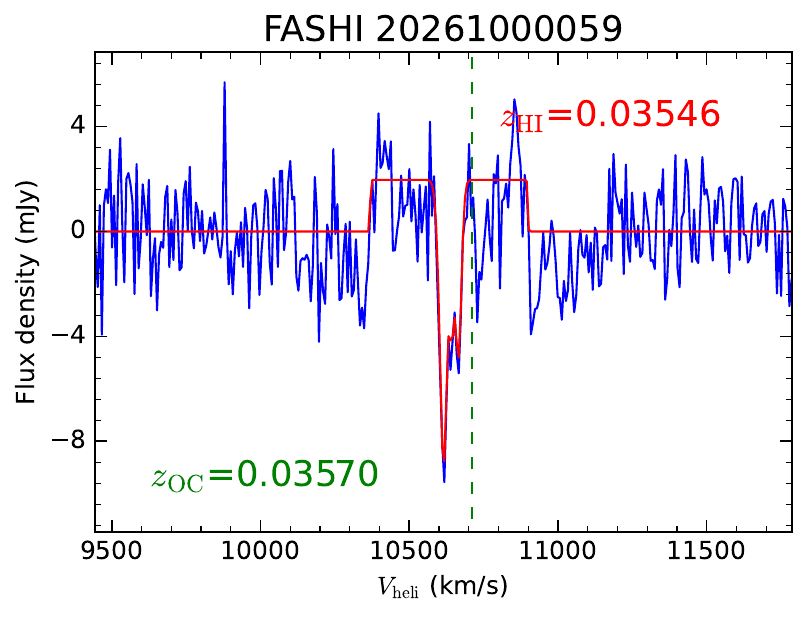}
 \includegraphics[height=0.27\textwidth, angle=0]{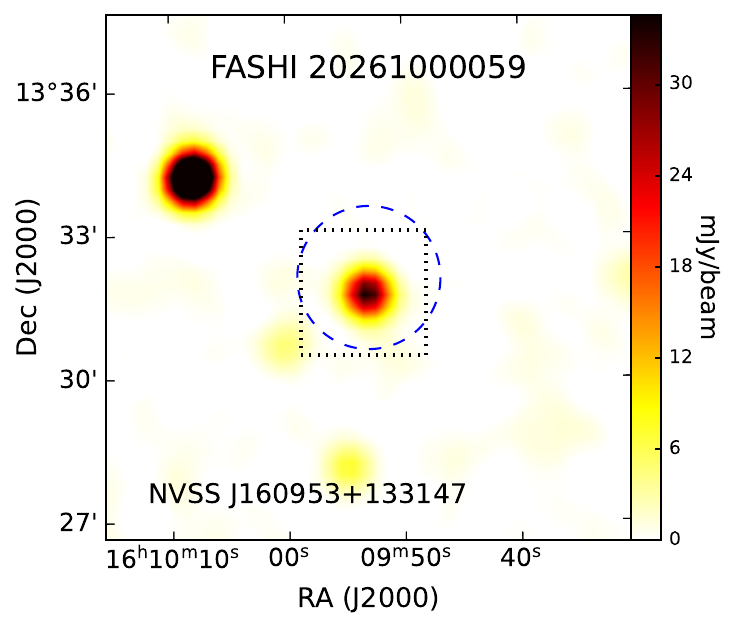}
 \includegraphics[height=0.27\textwidth, angle=0]{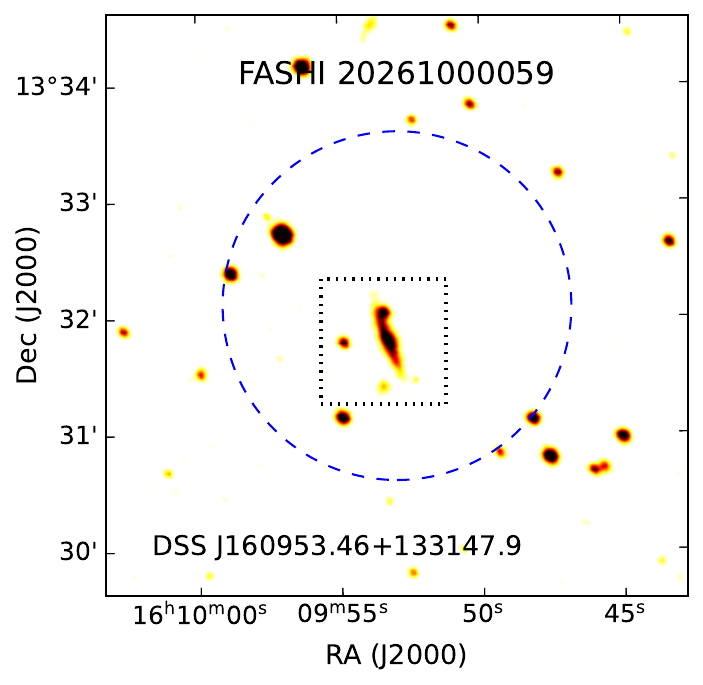}
 \includegraphics[height=0.22\textwidth, angle=0]{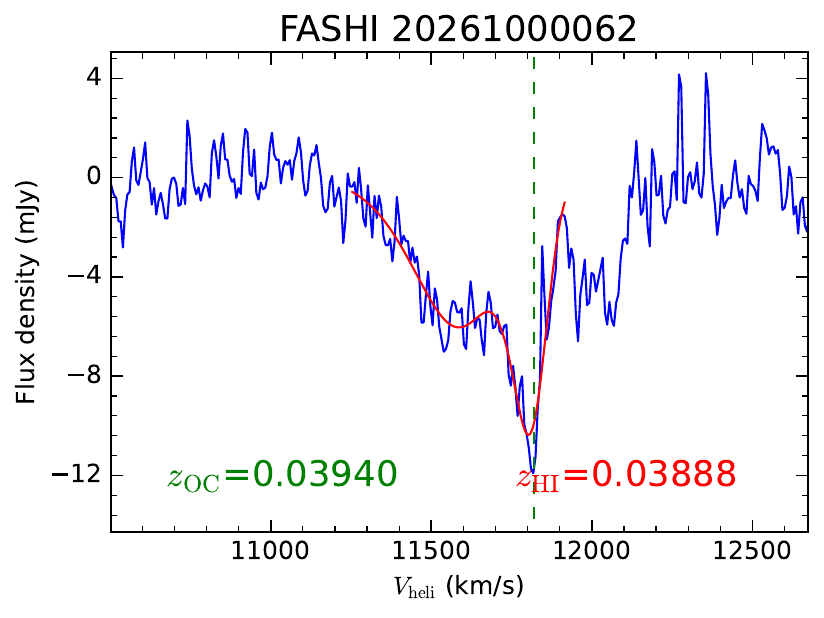}
 \includegraphics[height=0.27\textwidth, angle=0]{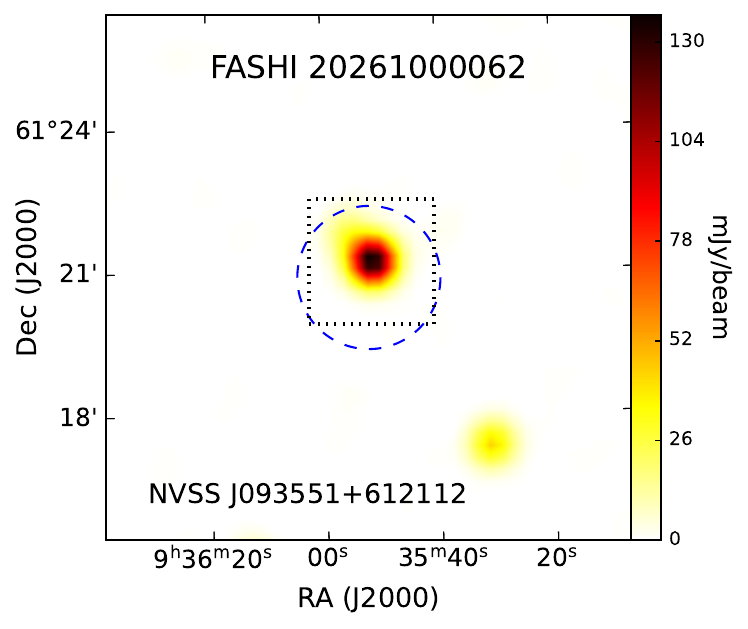}
 \includegraphics[height=0.27\textwidth, angle=0]{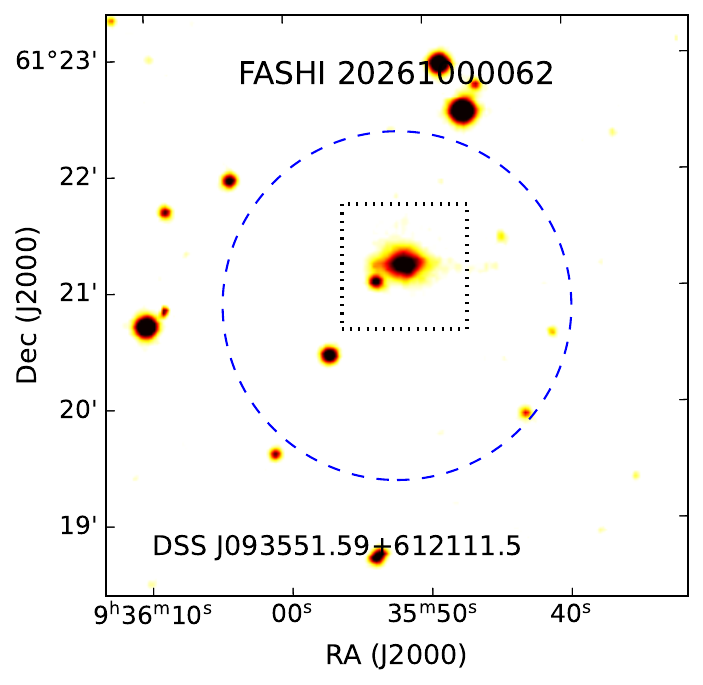}
 \includegraphics[height=0.22\textwidth, angle=0]{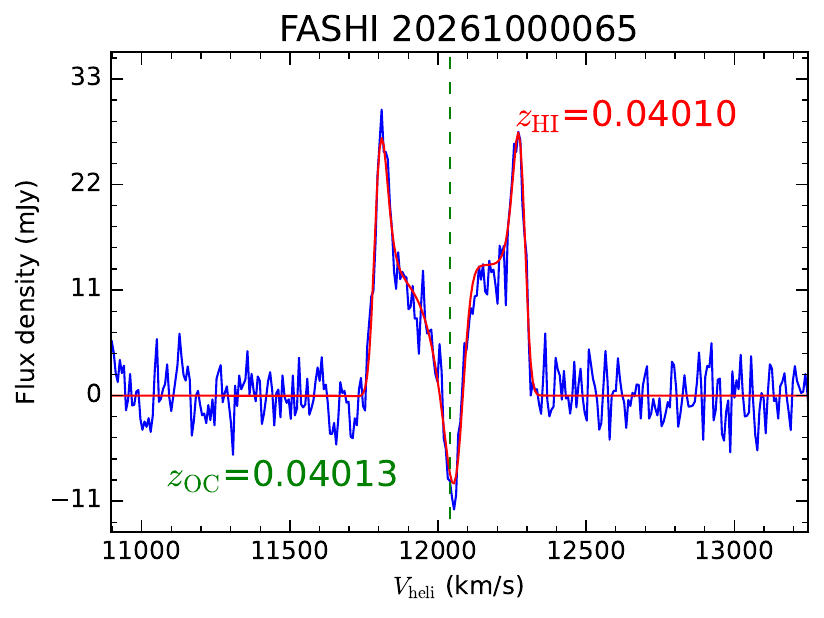}
 \includegraphics[height=0.27\textwidth, angle=0]{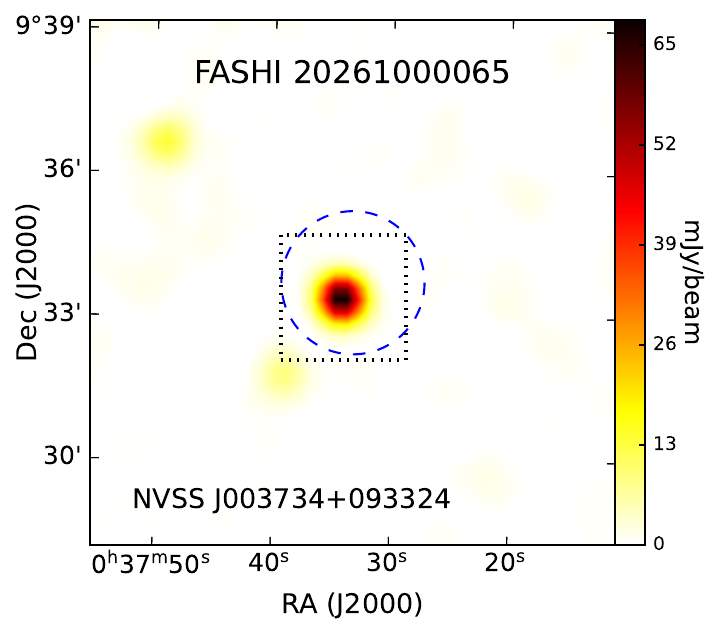}
 \includegraphics[height=0.27\textwidth, angle=0]{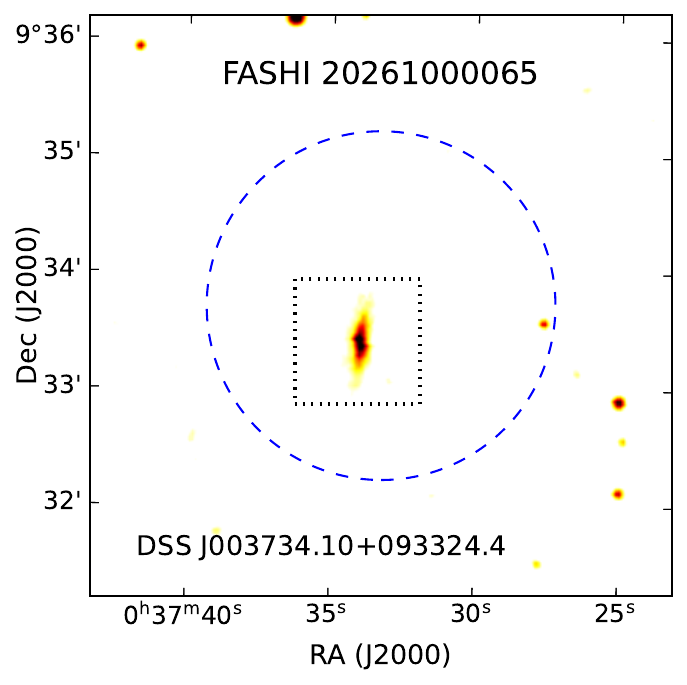}
 \caption{See caption in Figure\,\ref{Fig:FASHI_hi}}
 \end{figure*} 

 \begin{figure*}[htp]
 \centering
 \renewcommand{\thefigure}{\arabic{figure} (Continued)}
 \addtocounter{figure}{-1}
 \includegraphics[height=0.22\textwidth, angle=0]{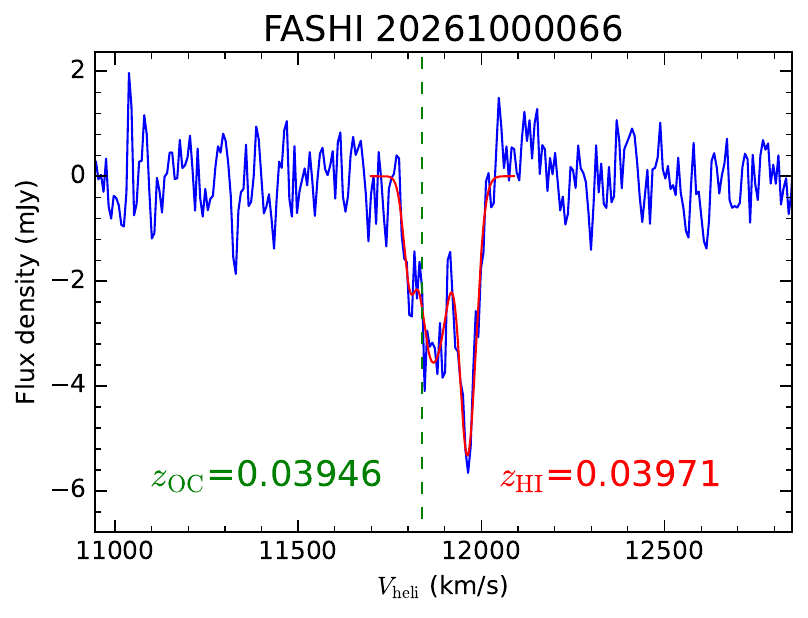}
 \includegraphics[height=0.27\textwidth, angle=0]{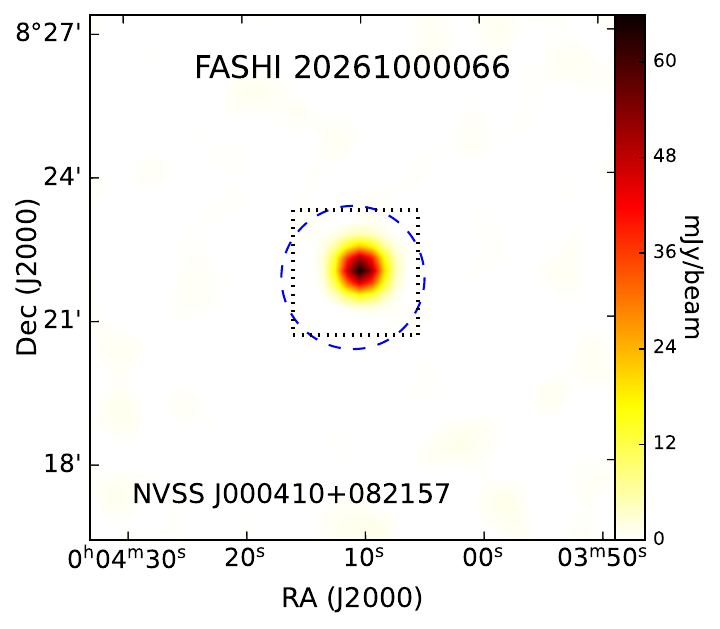}
 \includegraphics[height=0.27\textwidth, angle=0]{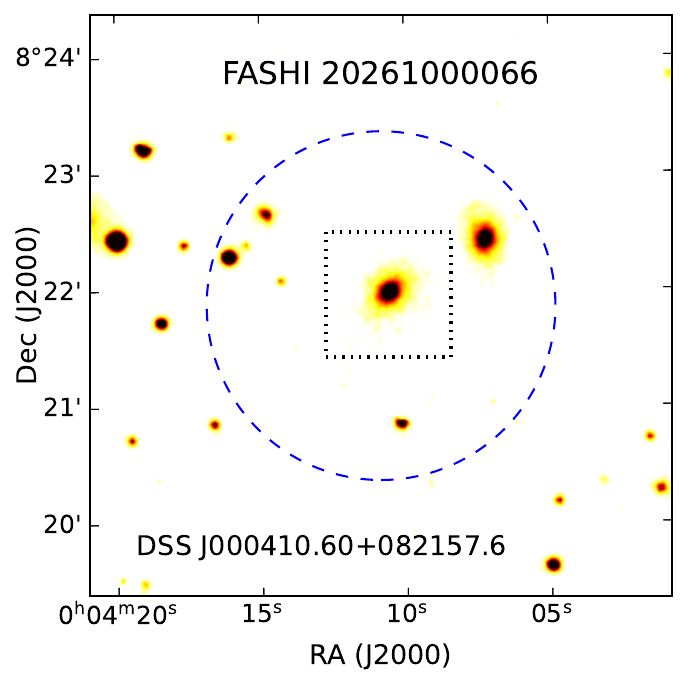}
 \includegraphics[height=0.22\textwidth, angle=0]{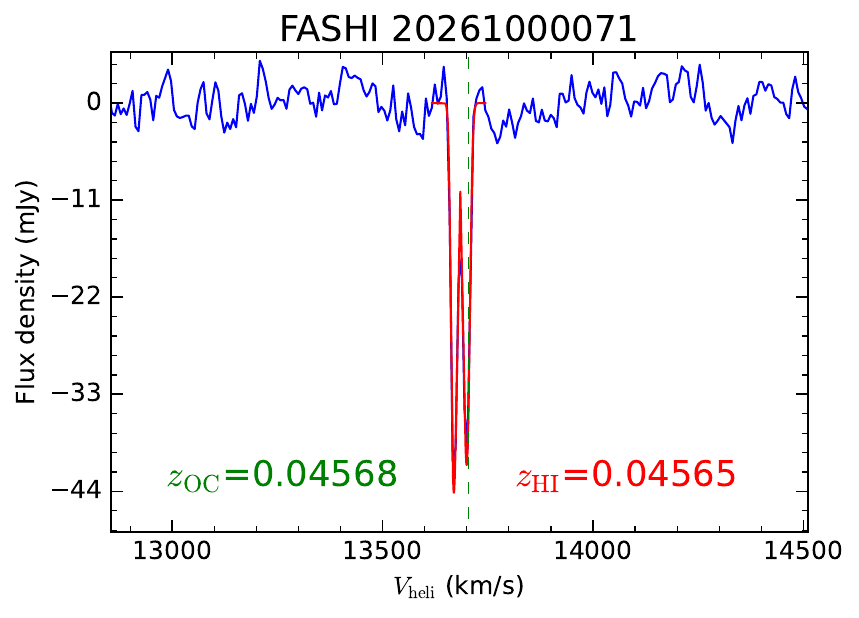}
 \includegraphics[height=0.27\textwidth, angle=0]{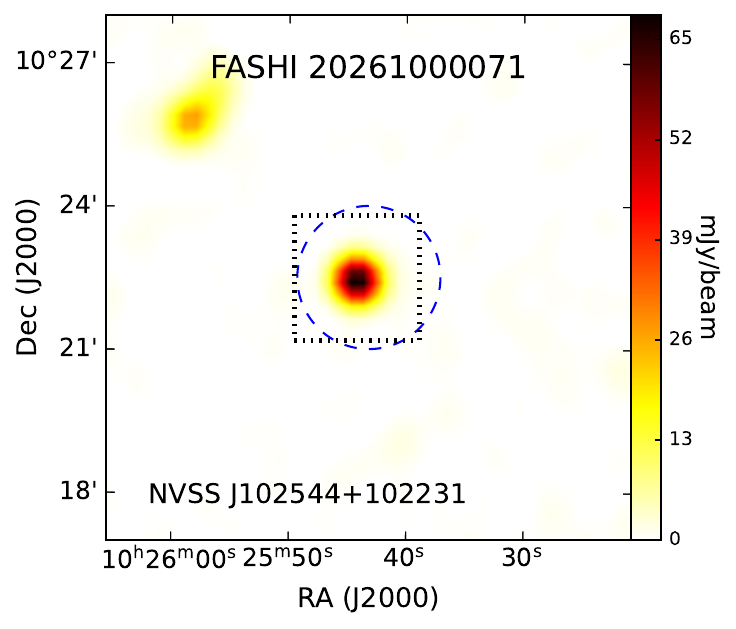}
 \includegraphics[height=0.27\textwidth, angle=0]{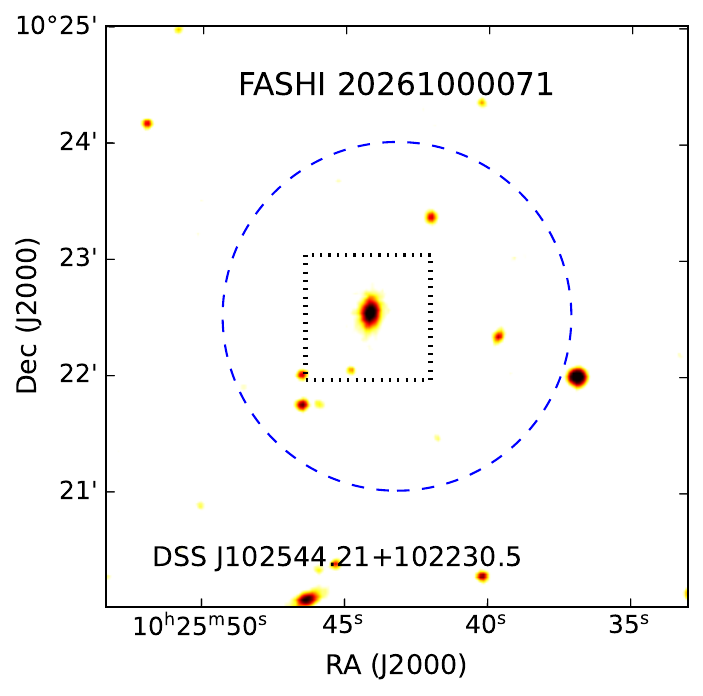}
 \includegraphics[height=0.22\textwidth, angle=0]{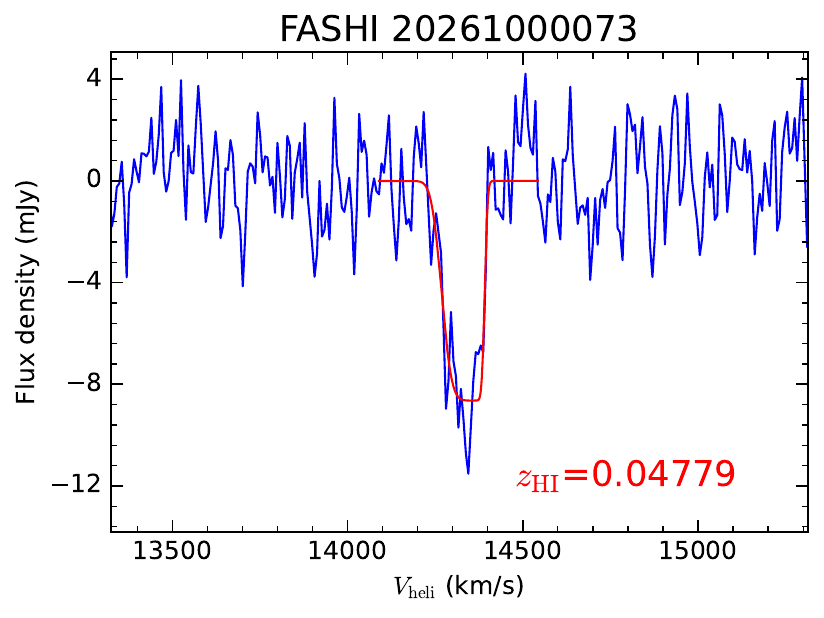}
 \includegraphics[height=0.27\textwidth, angle=0]{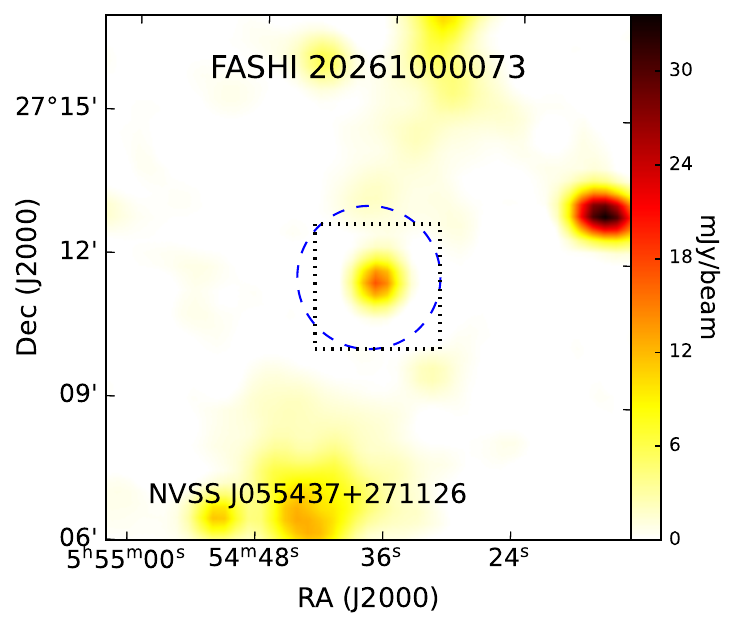}
 \includegraphics[height=0.27\textwidth, angle=0]{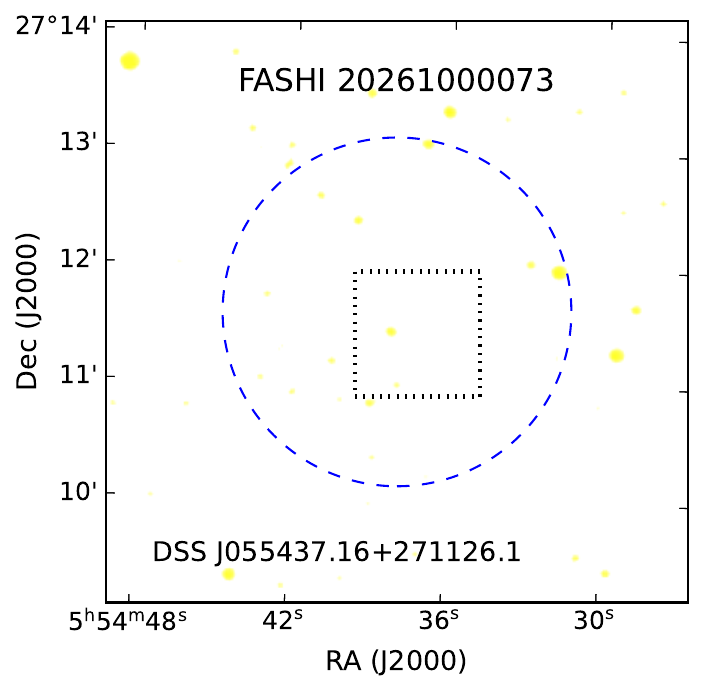}
 \includegraphics[height=0.22\textwidth, angle=0]{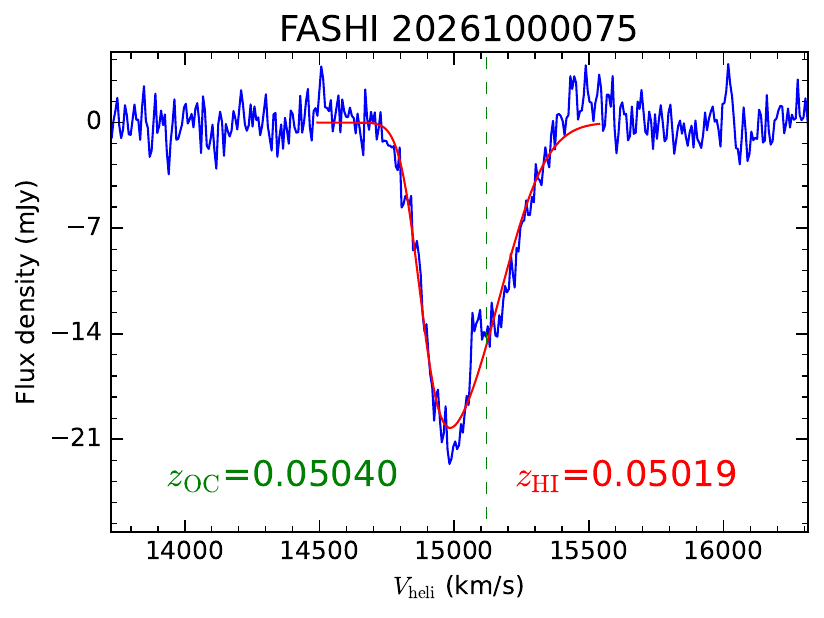}
 \includegraphics[height=0.27\textwidth, angle=0]{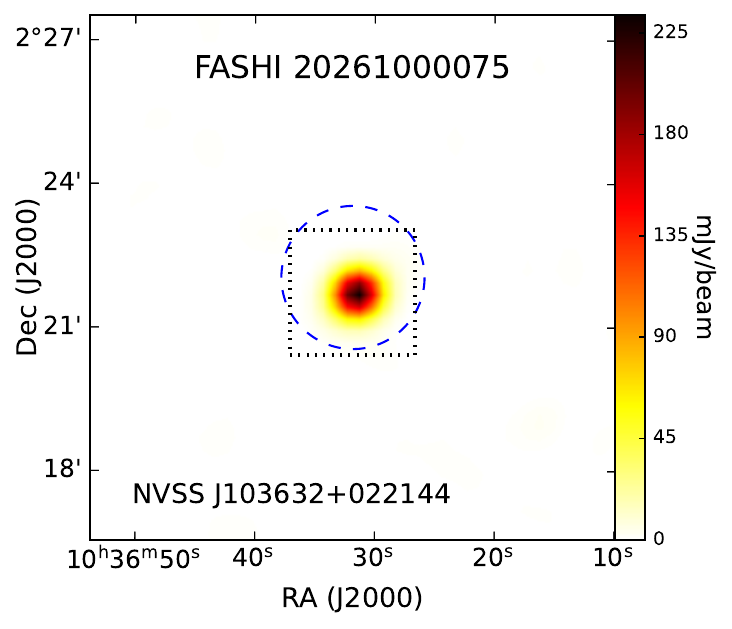}
 \includegraphics[height=0.27\textwidth, angle=0]{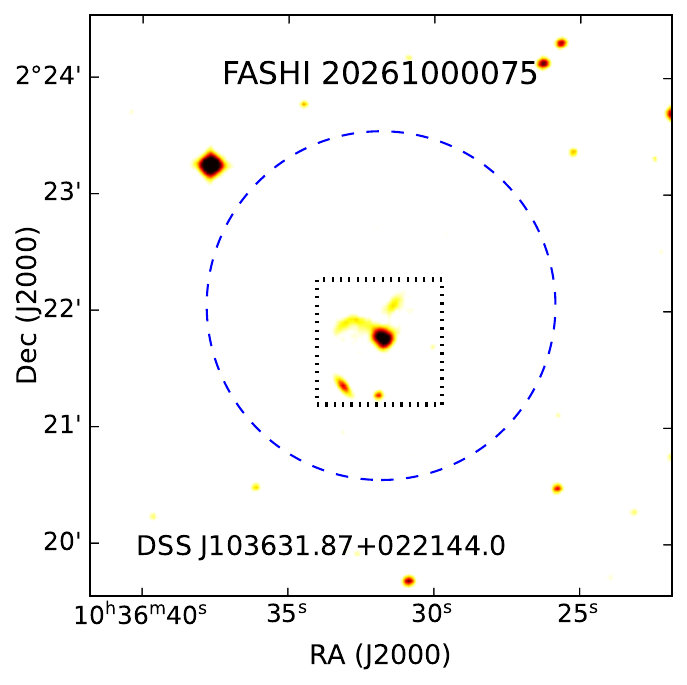}
 \caption{See caption in Figure\,\ref{Fig:FASHI_hi}}
 \end{figure*} 

 \begin{figure*}[htp]
 \centering
 \renewcommand{\thefigure}{\arabic{figure} (Continued)}
 \addtocounter{figure}{-1}
 \includegraphics[height=0.22\textwidth, angle=0]{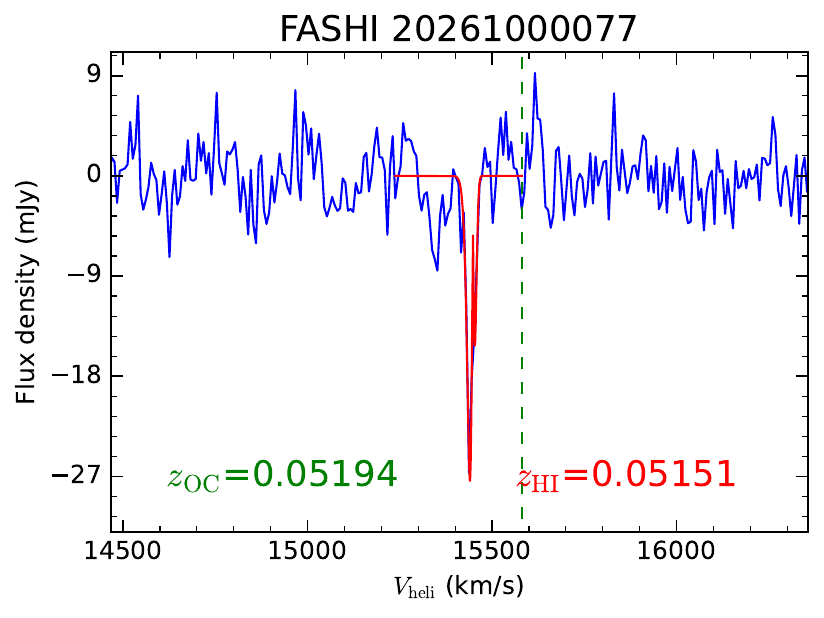}
 \includegraphics[height=0.27\textwidth, angle=0]{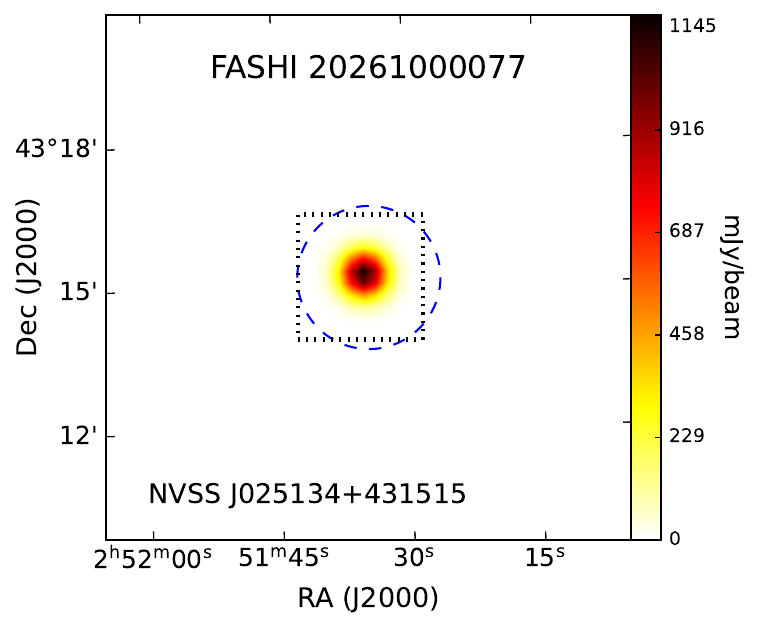}
 \includegraphics[height=0.27\textwidth, angle=0]{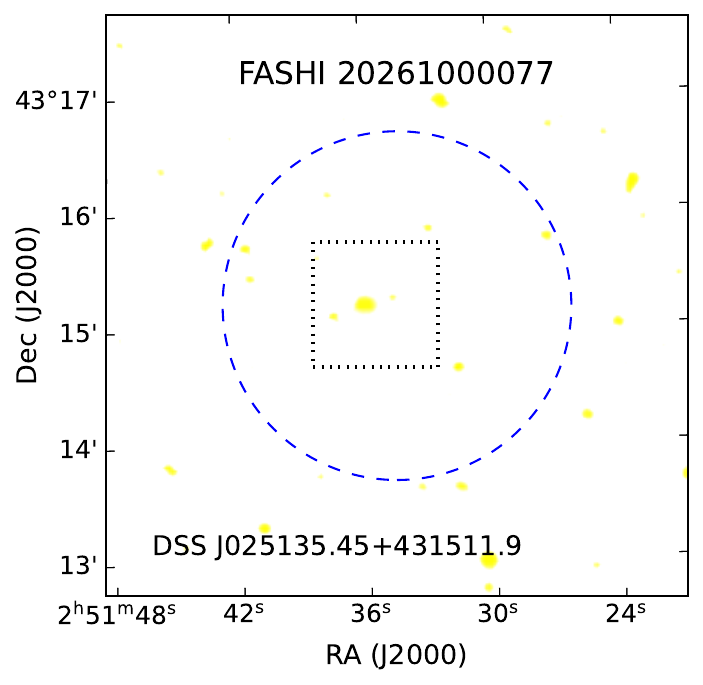}
 \includegraphics[height=0.22\textwidth, angle=0]{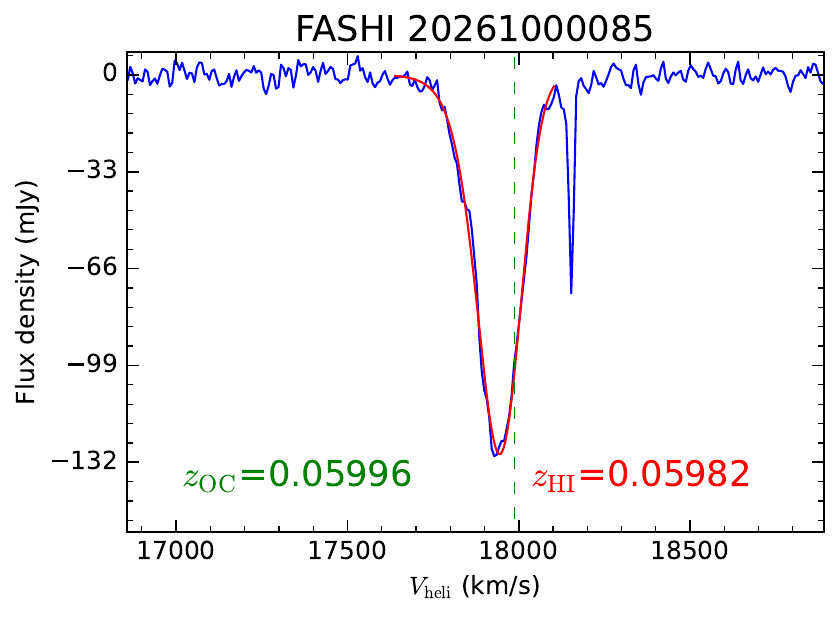}
 \includegraphics[height=0.27\textwidth, angle=0]{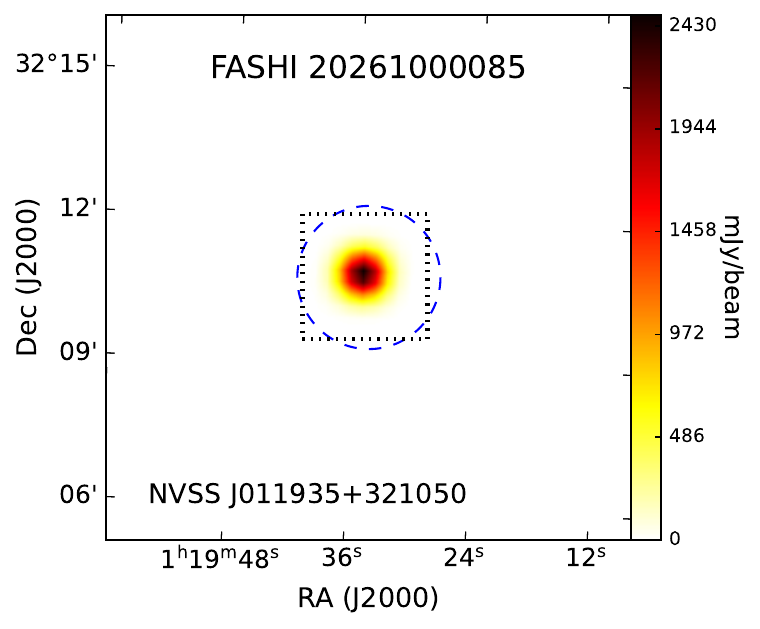}
 \includegraphics[height=0.27\textwidth, angle=0]{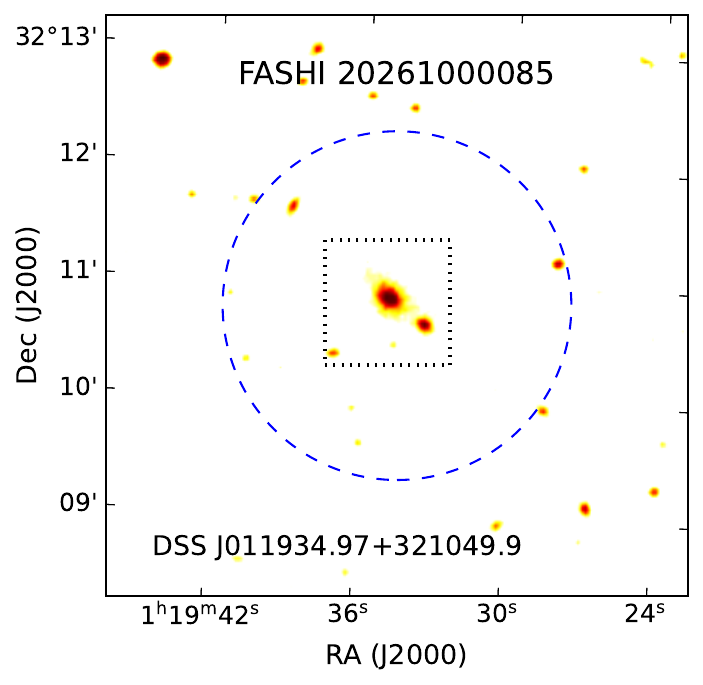}
 \includegraphics[height=0.22\textwidth, angle=0]{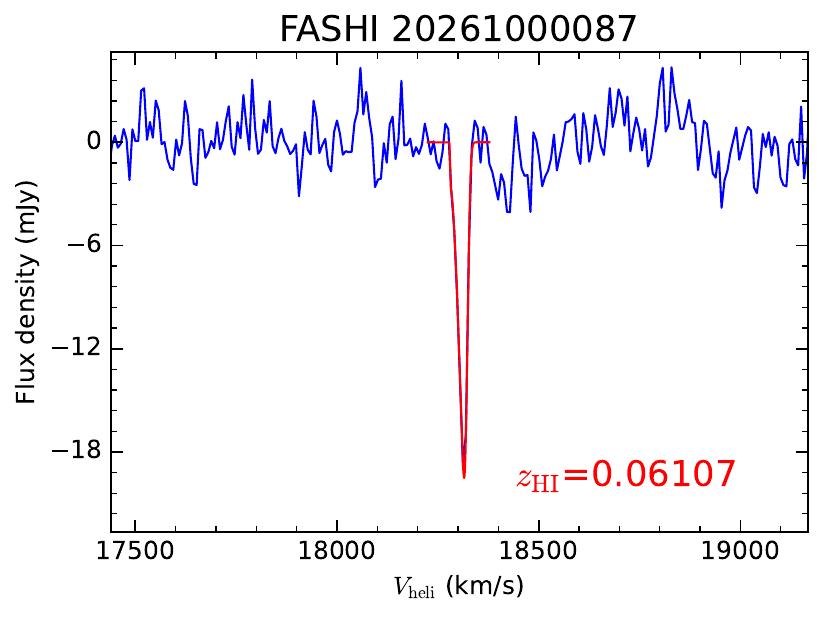}
 \includegraphics[height=0.27\textwidth, angle=0]{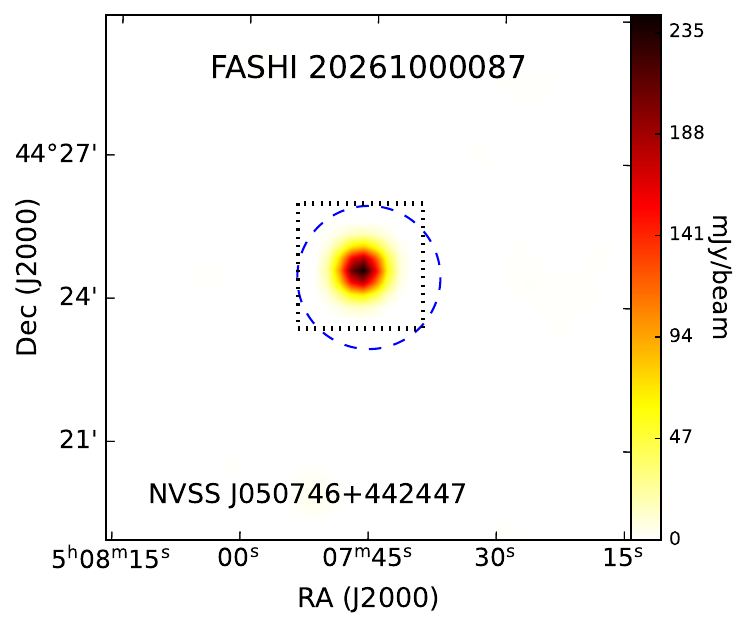}
 \includegraphics[height=0.27\textwidth, angle=0]{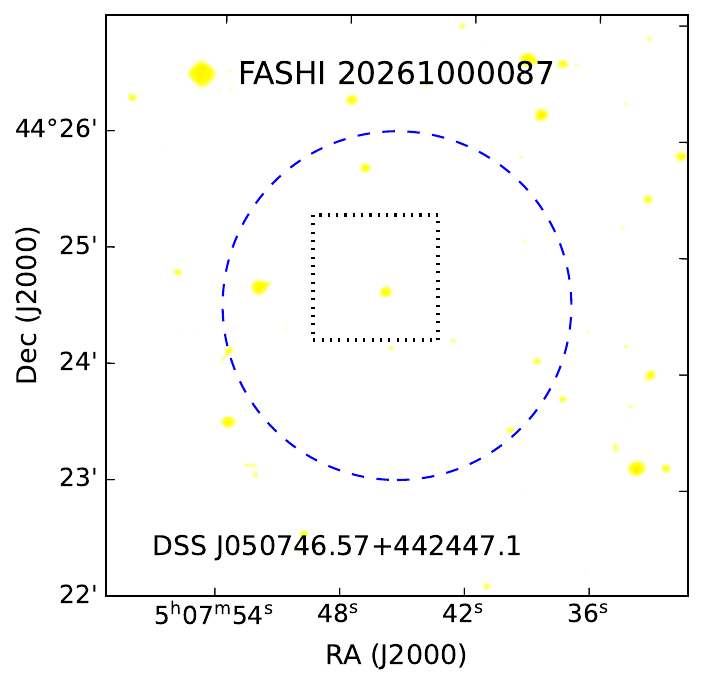}
 \includegraphics[height=0.22\textwidth, angle=0]{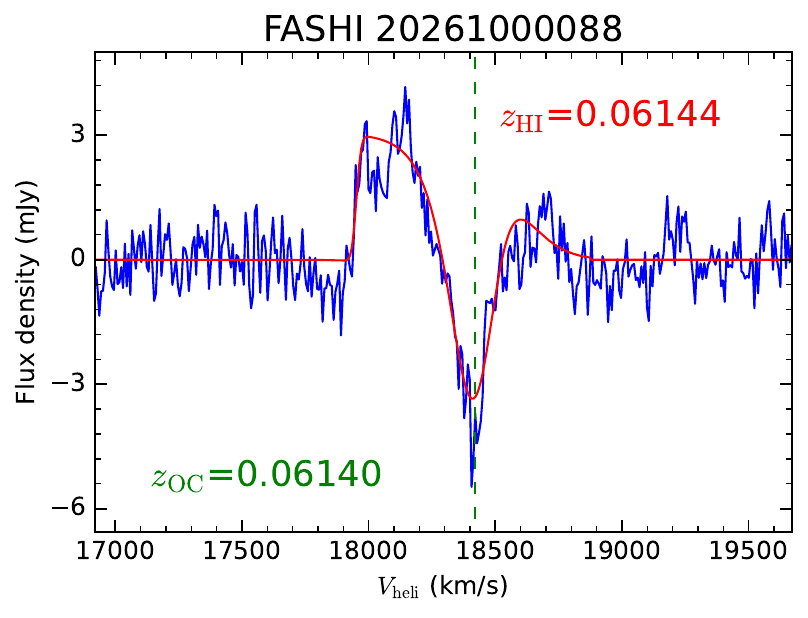}
 \includegraphics[height=0.27\textwidth, angle=0]{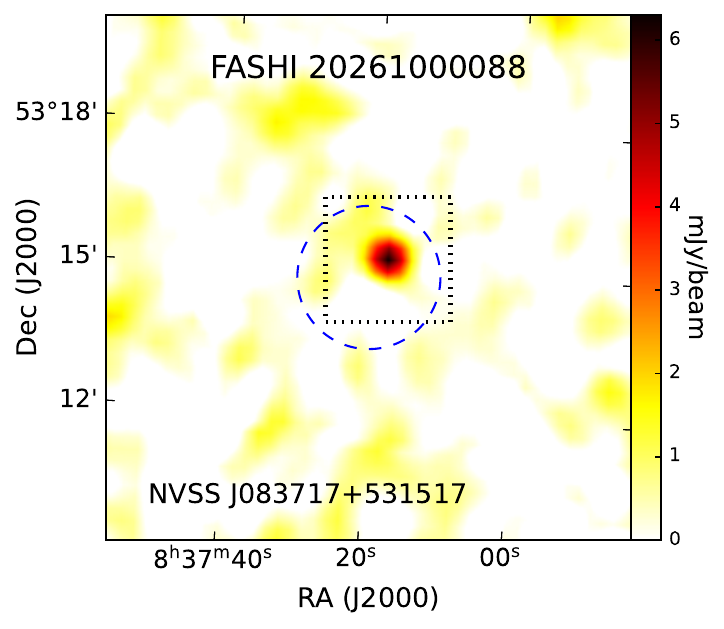}
 \includegraphics[height=0.27\textwidth, angle=0]{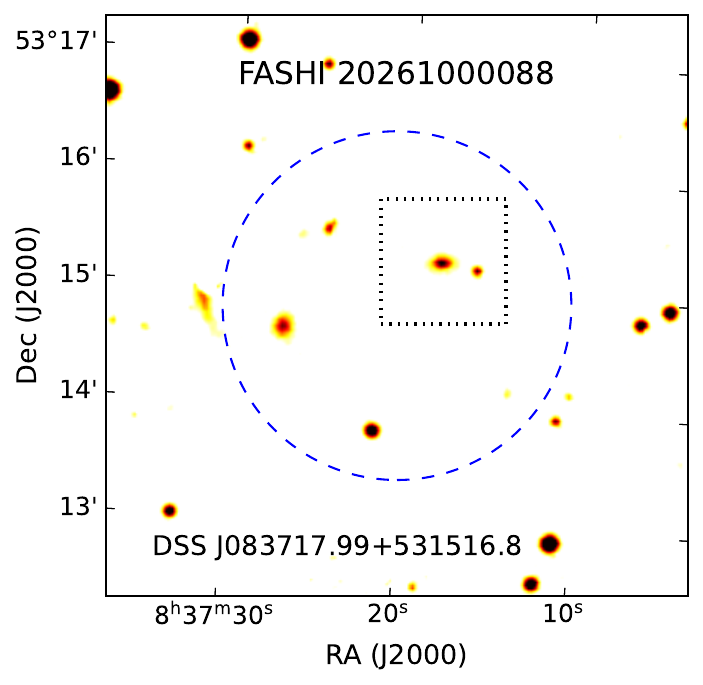}
 \caption{See caption in Figure\,\ref{Fig:FASHI_hi}}
 \end{figure*} 

 \begin{figure*}[htp]
 \centering
 \renewcommand{\thefigure}{\arabic{figure} (Continued)}
 \addtocounter{figure}{-1}
 \includegraphics[height=0.22\textwidth, angle=0]{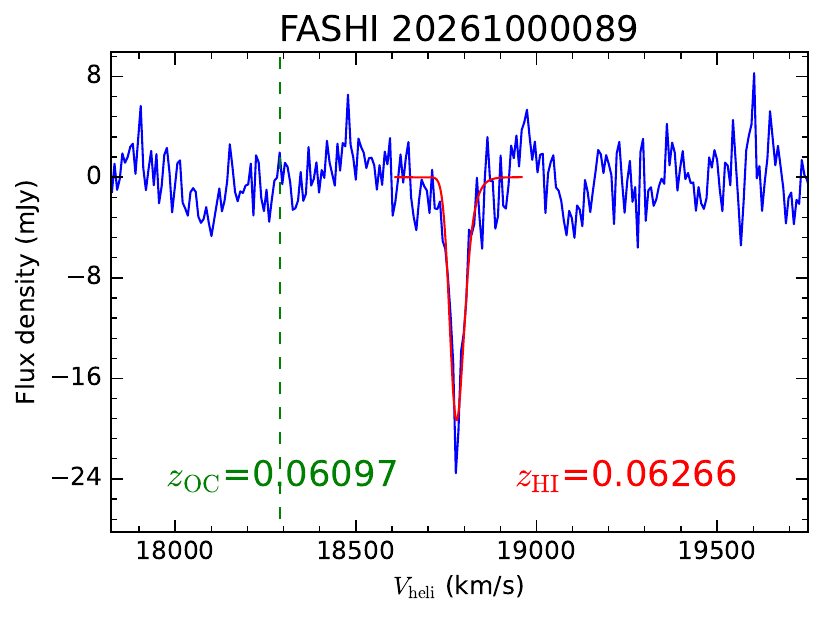}
 \includegraphics[height=0.27\textwidth, angle=0]{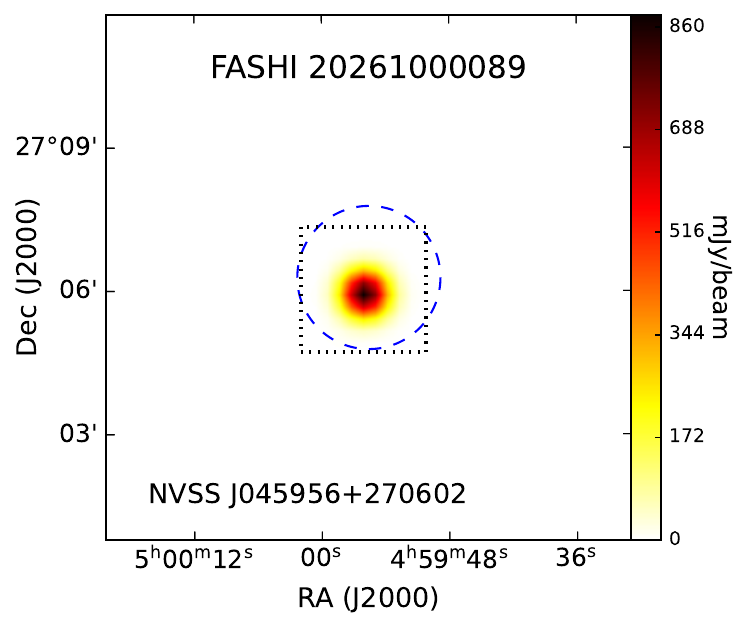}
 \includegraphics[height=0.27\textwidth, angle=0]{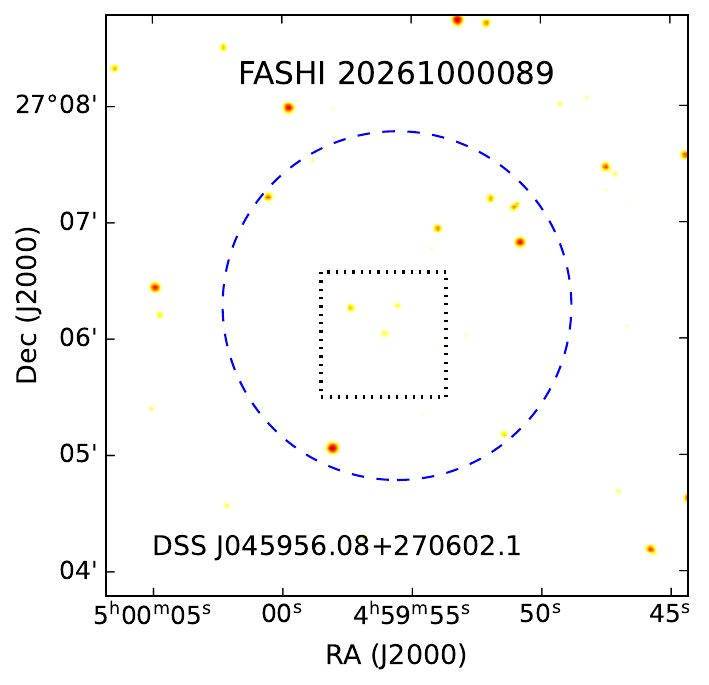}
 \includegraphics[height=0.22\textwidth, angle=0]{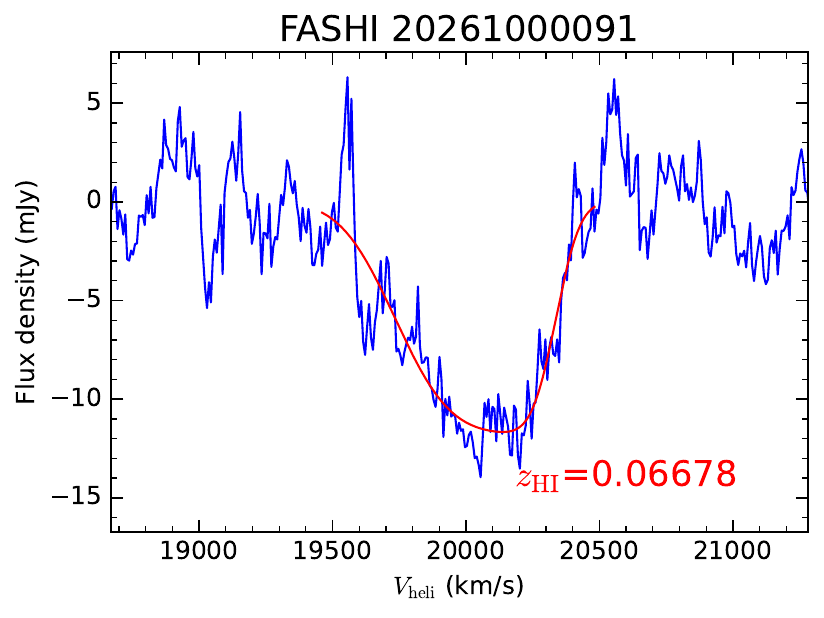}
 \includegraphics[height=0.27\textwidth, angle=0]{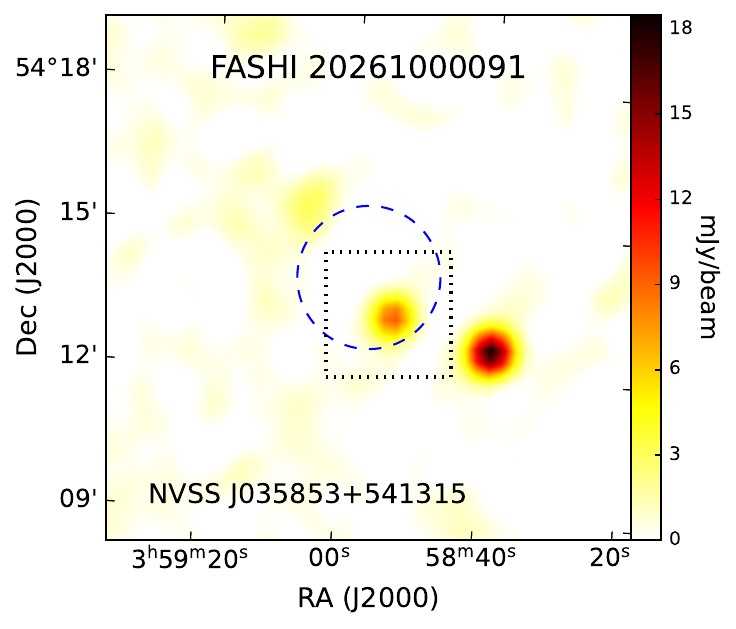}
 \includegraphics[height=0.27\textwidth, angle=0]{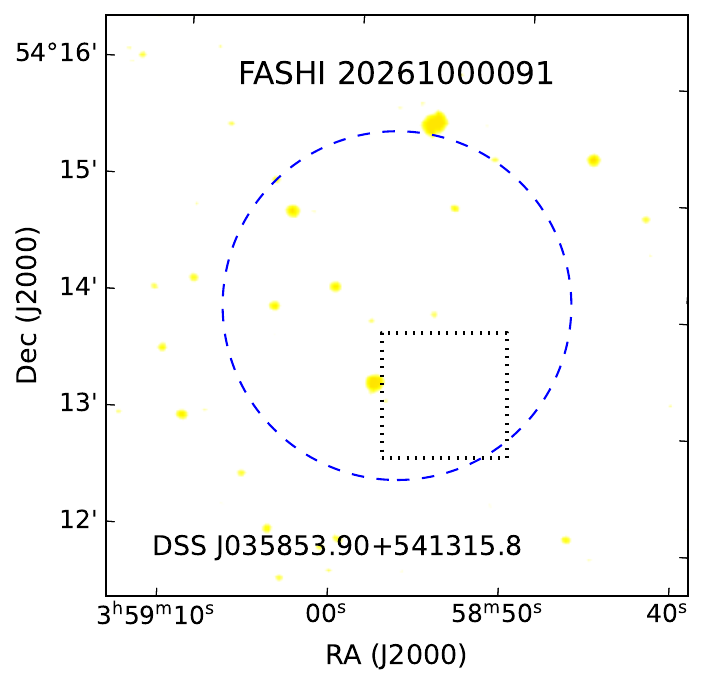}
 \includegraphics[height=0.22\textwidth, angle=0]{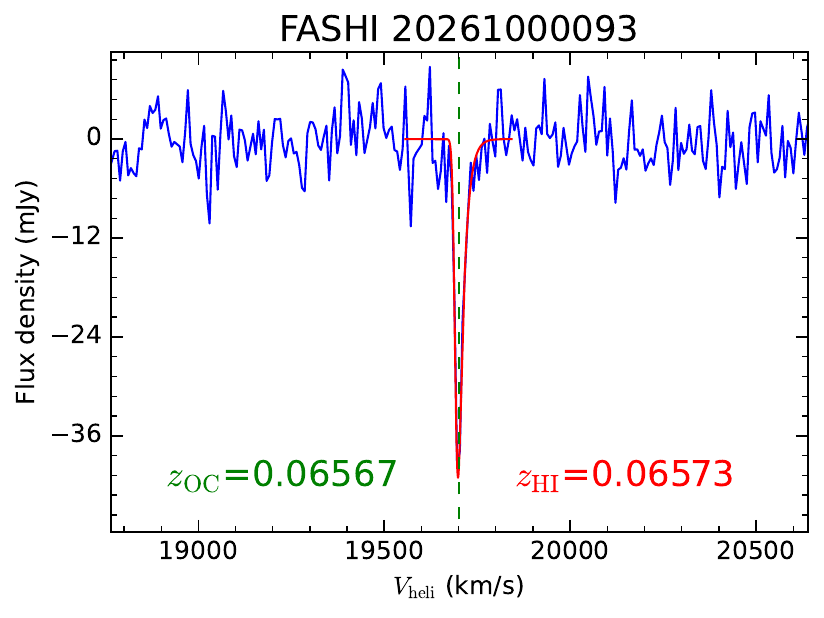}
 \includegraphics[height=0.27\textwidth, angle=0]{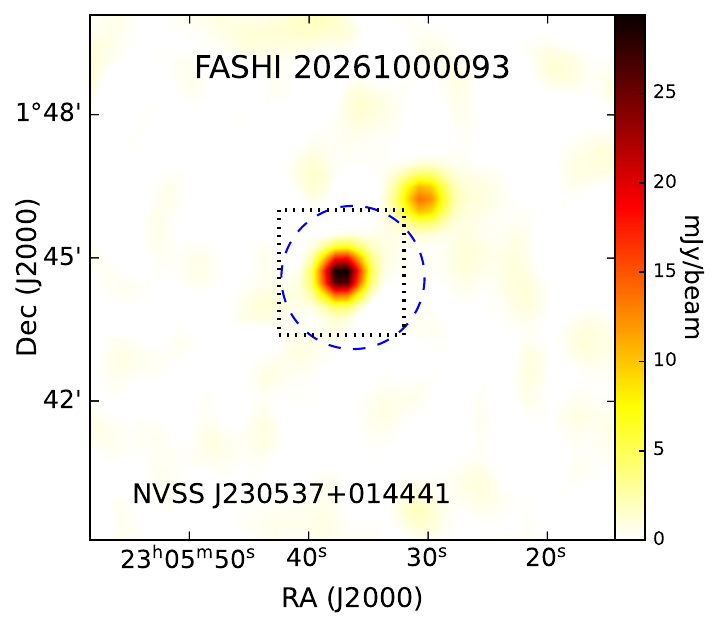}
 \includegraphics[height=0.27\textwidth, angle=0]{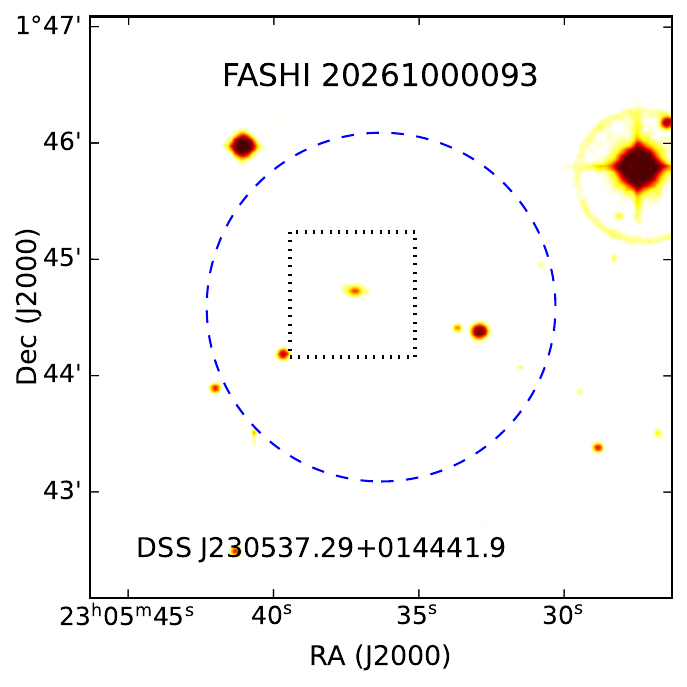}
 \includegraphics[height=0.22\textwidth, angle=0]{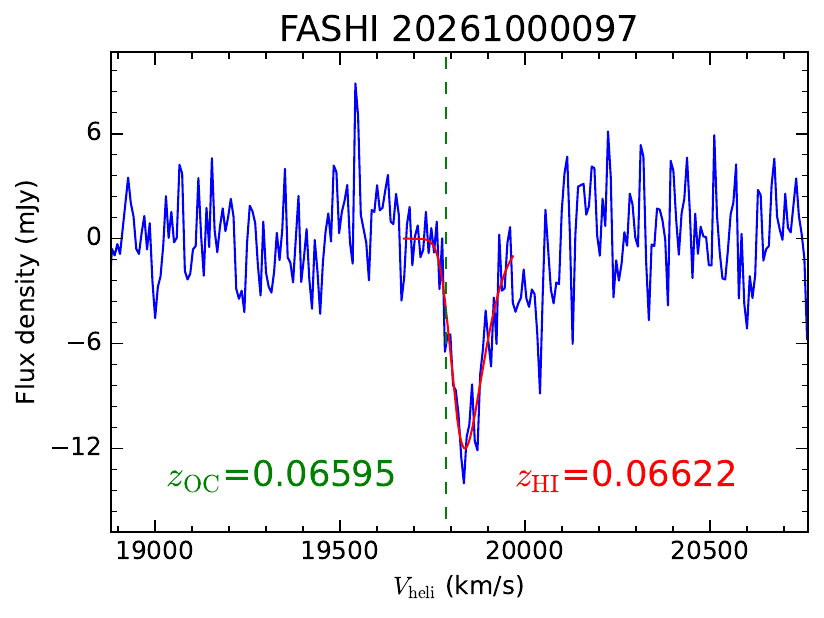}
 \includegraphics[height=0.27\textwidth, angle=0]{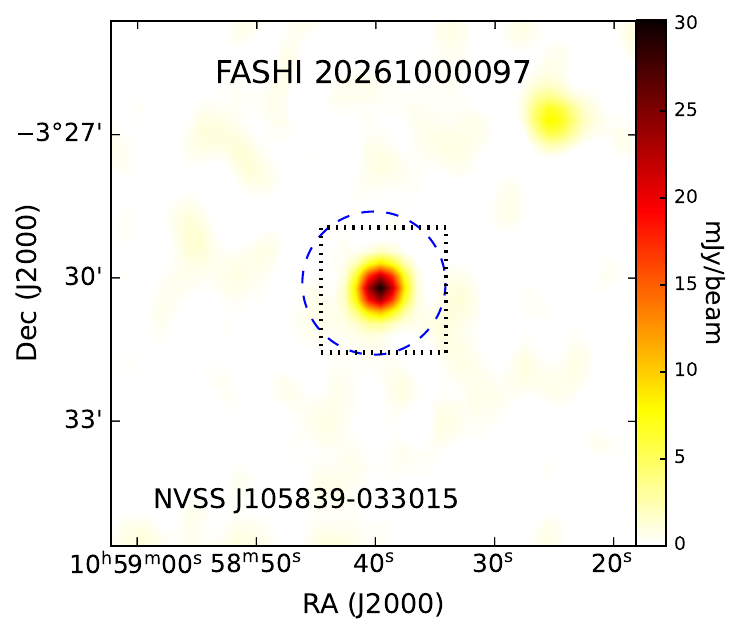}
 \includegraphics[height=0.27\textwidth, angle=0]{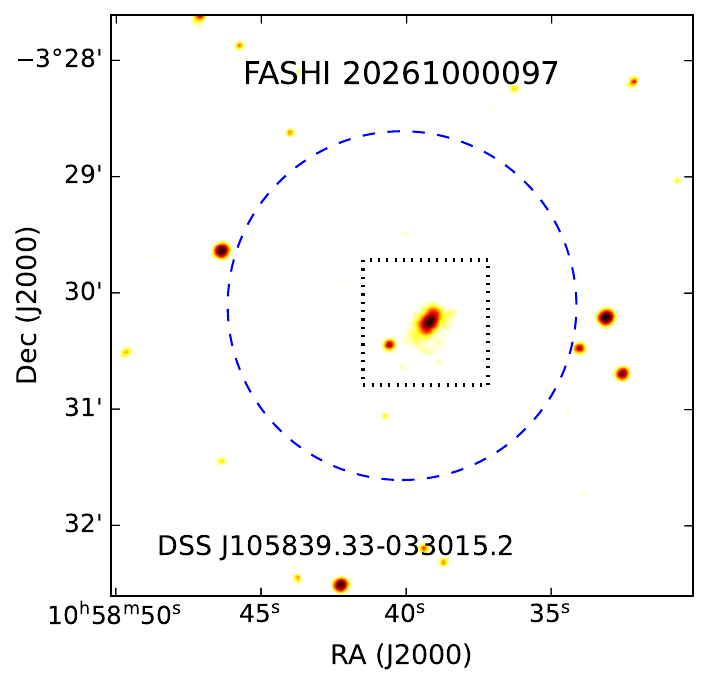}
 \caption{See caption in Figure\,\ref{Fig:FASHI_hi}}
 \end{figure*} 

 \begin{figure*}[htp]
 \centering
 \renewcommand{\thefigure}{\arabic{figure} (Continued)}
 \addtocounter{figure}{-1}
 \includegraphics[height=0.22\textwidth, angle=0]{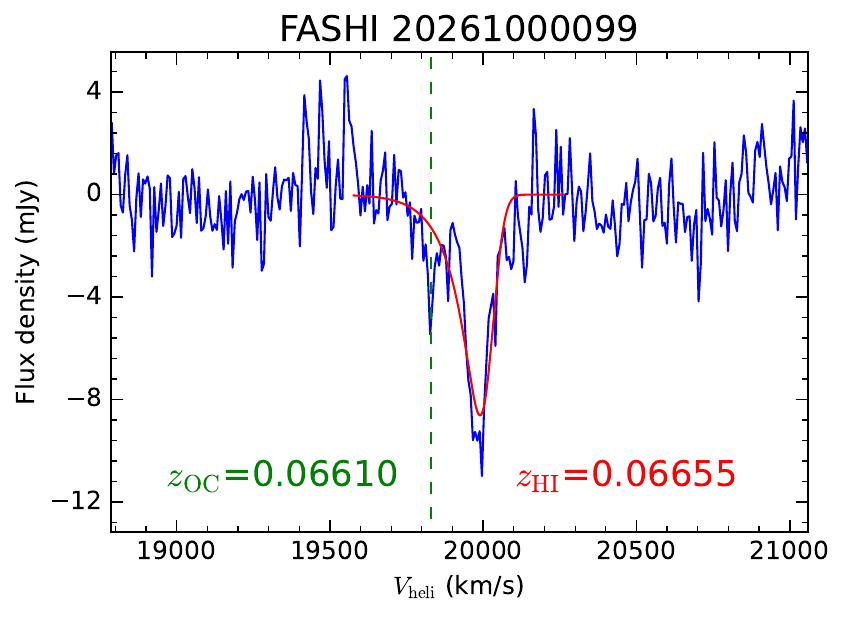}
 \includegraphics[height=0.27\textwidth, angle=0]{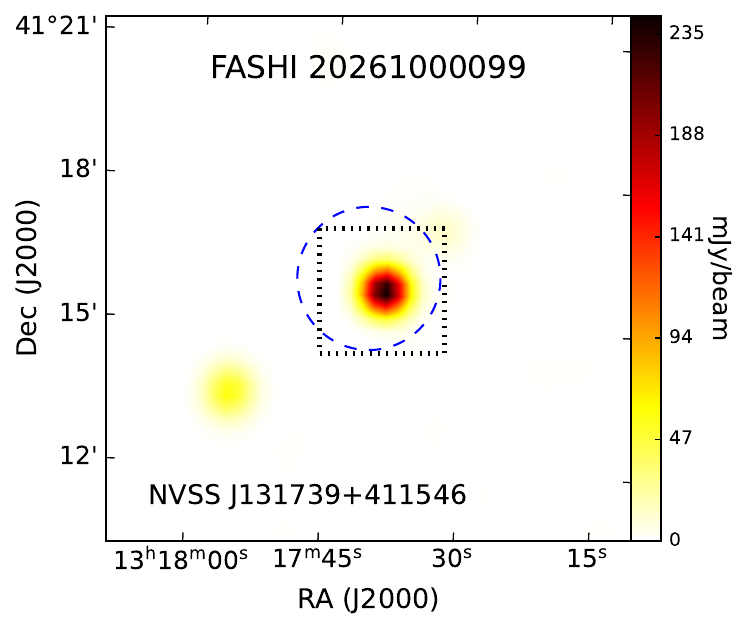}
 \includegraphics[height=0.27\textwidth, angle=0]{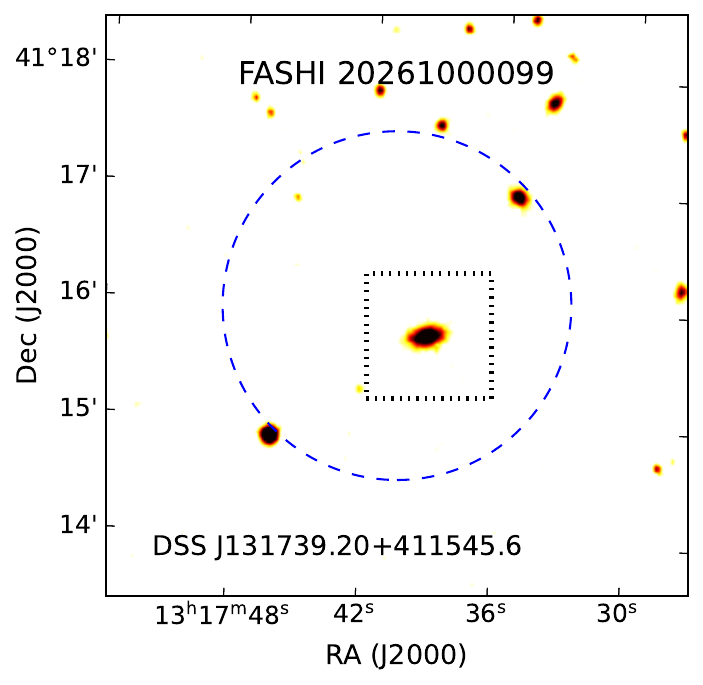}
 \includegraphics[height=0.22\textwidth, angle=0]{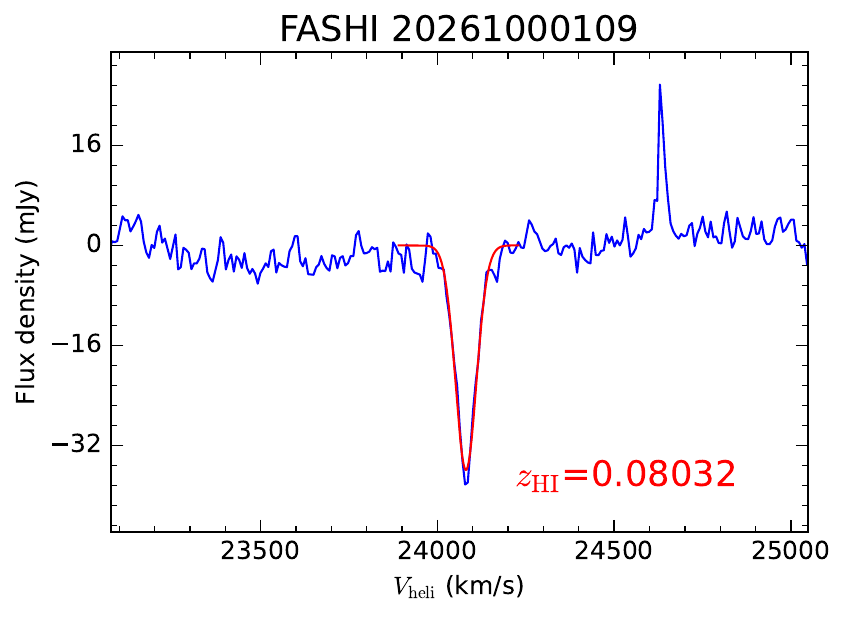}
 \includegraphics[height=0.27\textwidth, angle=0]{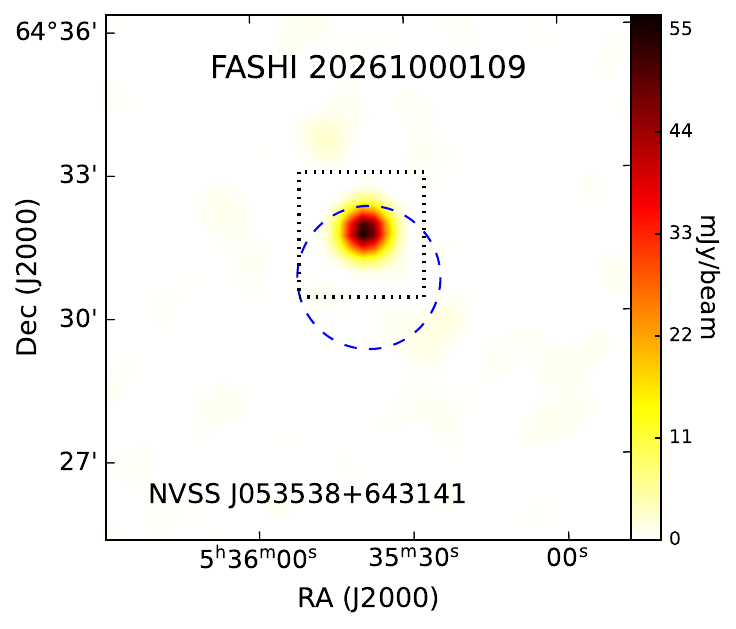}
 \includegraphics[height=0.27\textwidth, angle=0]{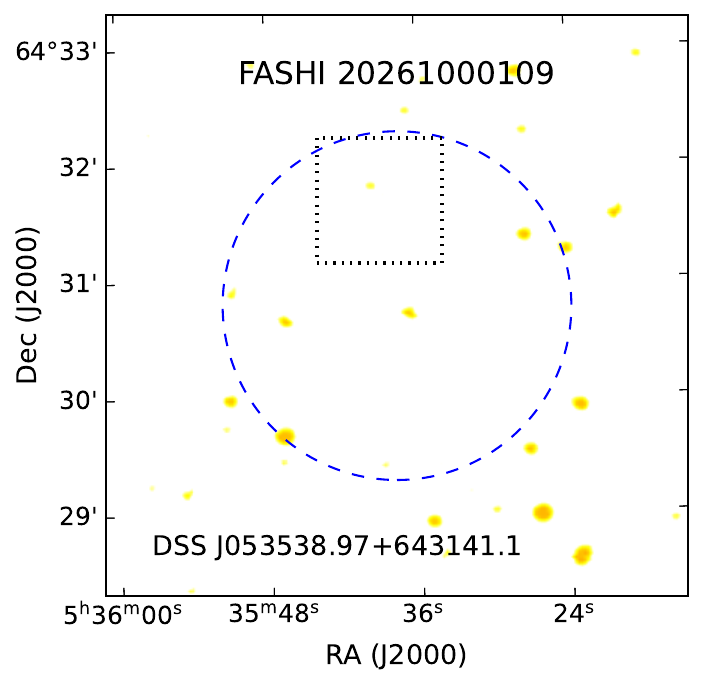}
 \includegraphics[height=0.22\textwidth, angle=0]{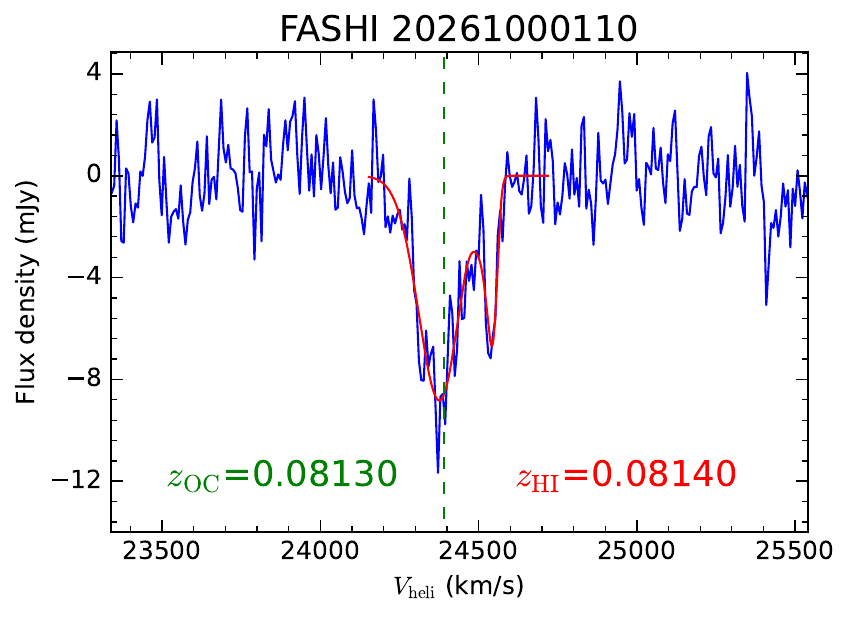}
 \includegraphics[height=0.27\textwidth, angle=0]{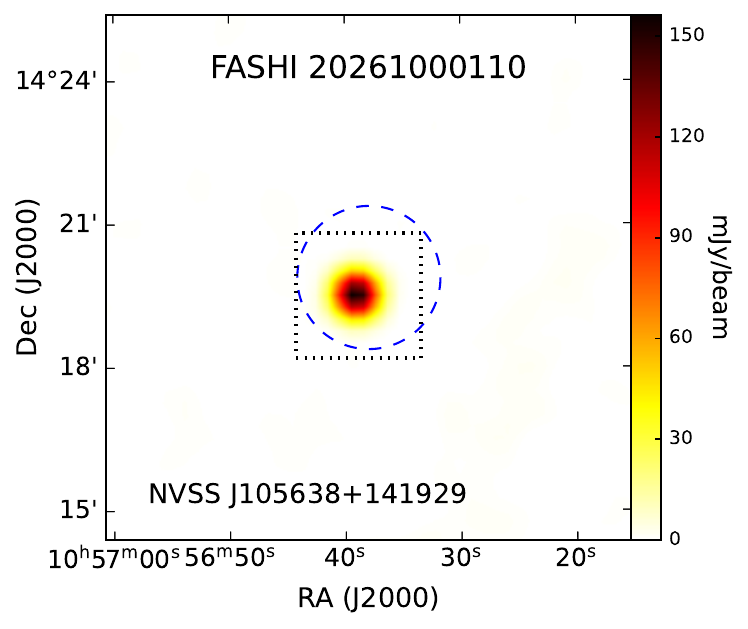}
 \includegraphics[height=0.27\textwidth, angle=0]{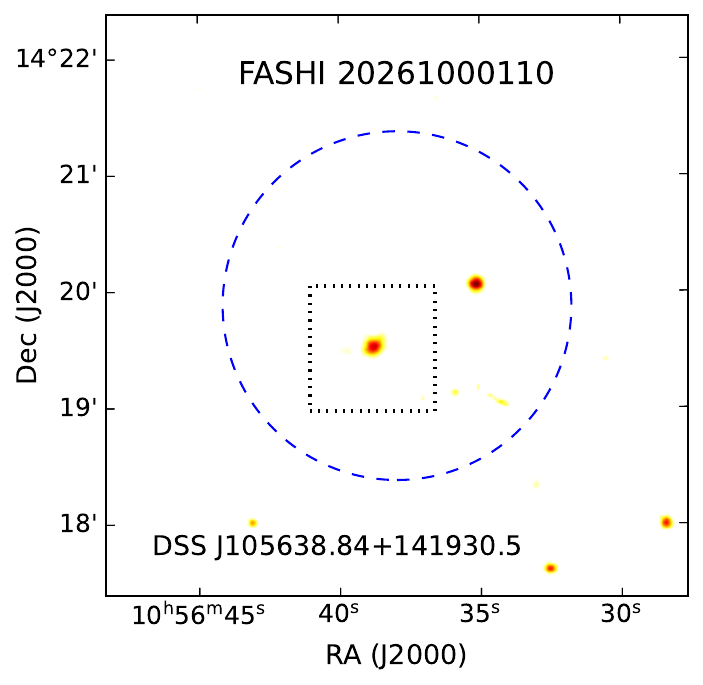}
 \includegraphics[height=0.22\textwidth, angle=0]{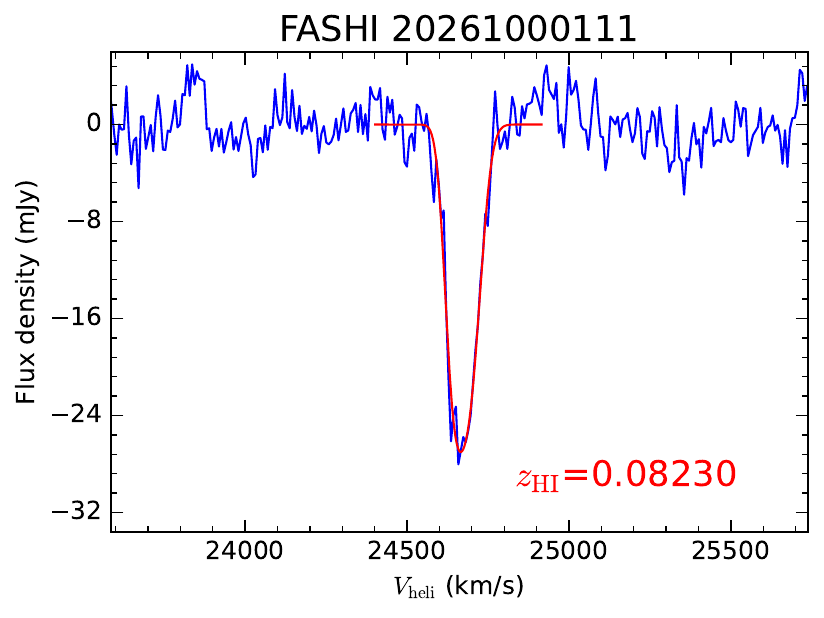}
 \includegraphics[height=0.27\textwidth, angle=0]{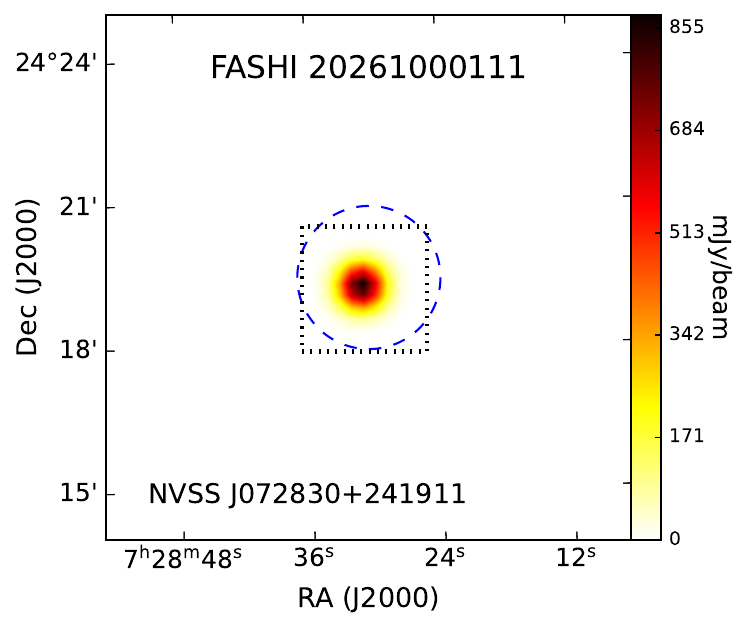}
 \includegraphics[height=0.27\textwidth, angle=0]{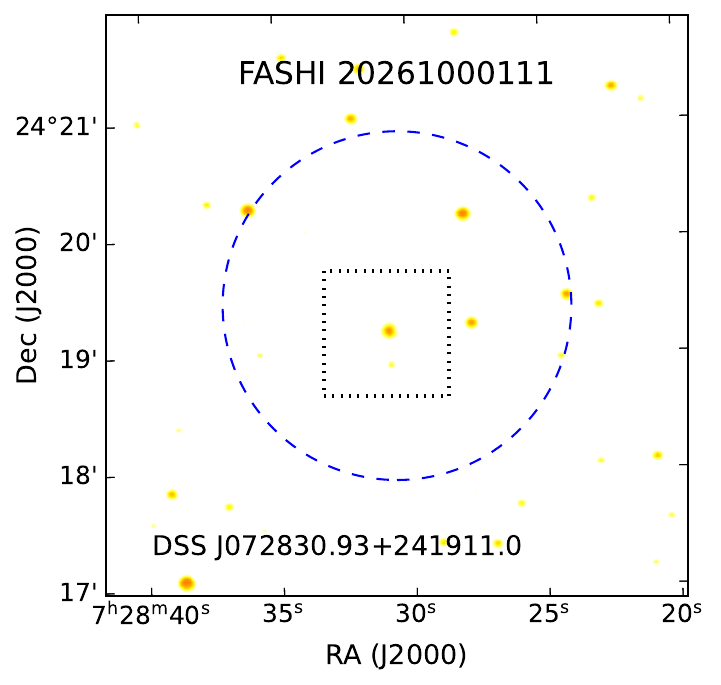}
 \caption{See caption in Figure\,\ref{Fig:FASHI_hi}}
 \end{figure*} 

 \begin{figure*}[htp]
 \centering
 \renewcommand{\thefigure}{\arabic{figure} (Continued)}
 \addtocounter{figure}{-1}
 \includegraphics[height=0.22\textwidth, angle=0]{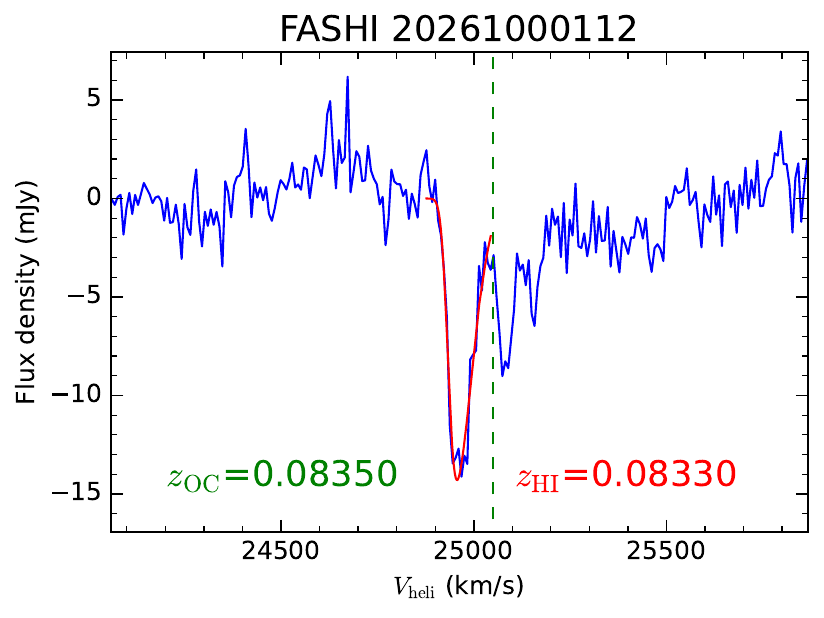}
 \includegraphics[height=0.27\textwidth, angle=0]{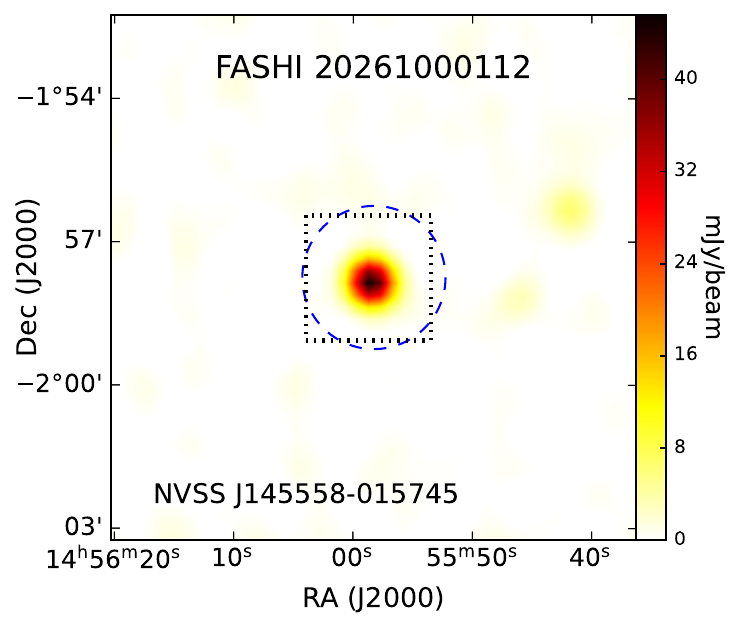}
 \includegraphics[height=0.27\textwidth, angle=0]{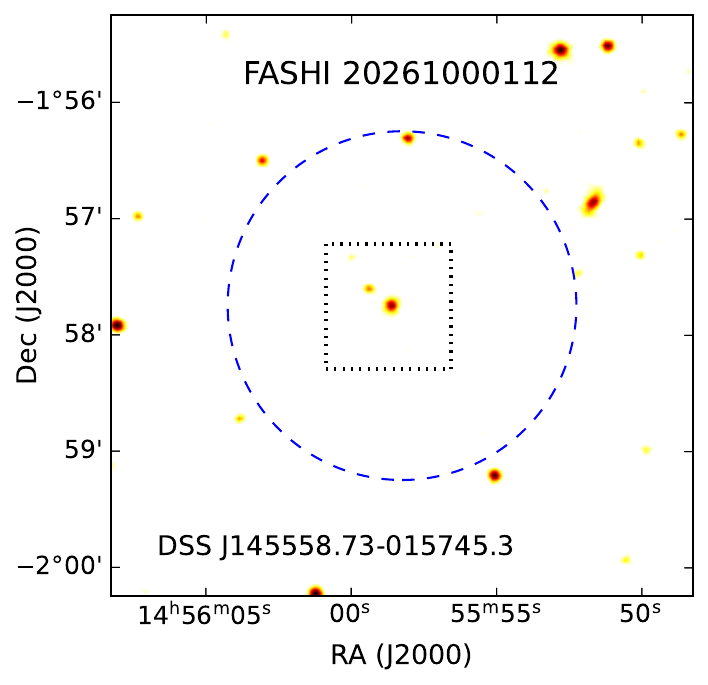}
 \includegraphics[height=0.22\textwidth, angle=0]{./figures/fashi_20261000113_spec.pdf}
 \includegraphics[height=0.27\textwidth, angle=0]{./figures/fashi_20261000113_nvss.pdf}
 \includegraphics[height=0.27\textwidth, angle=0]{./figures/fashi_20261000113_dss.pdf}
 \includegraphics[height=0.22\textwidth, angle=0]{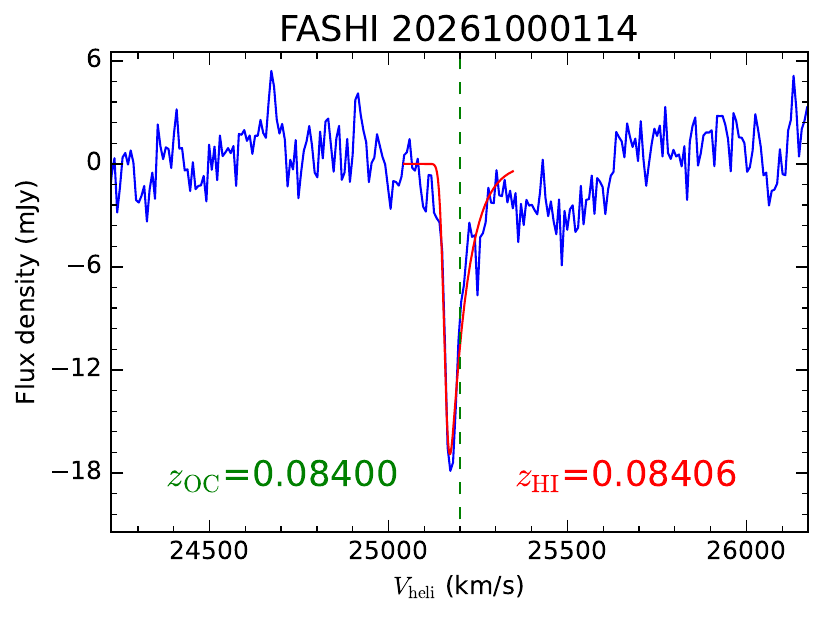}
 \includegraphics[height=0.27\textwidth, angle=0]{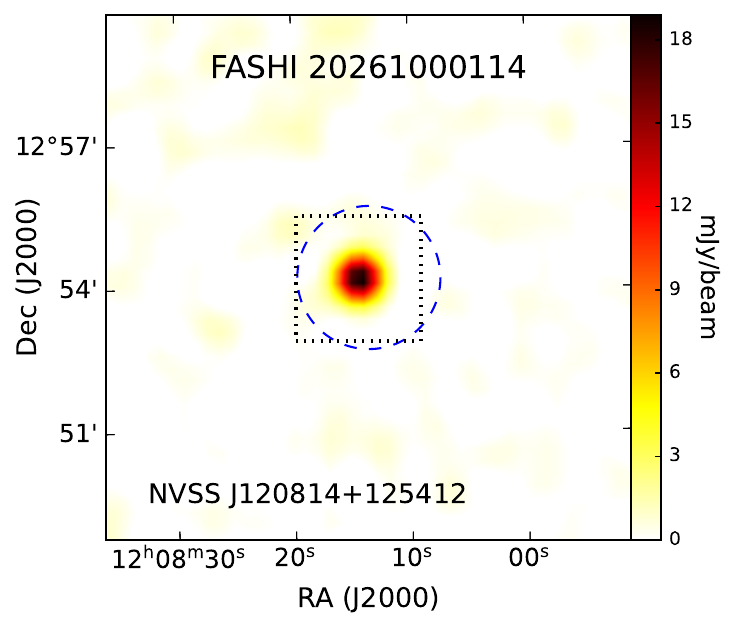}
 \includegraphics[height=0.27\textwidth, angle=0]{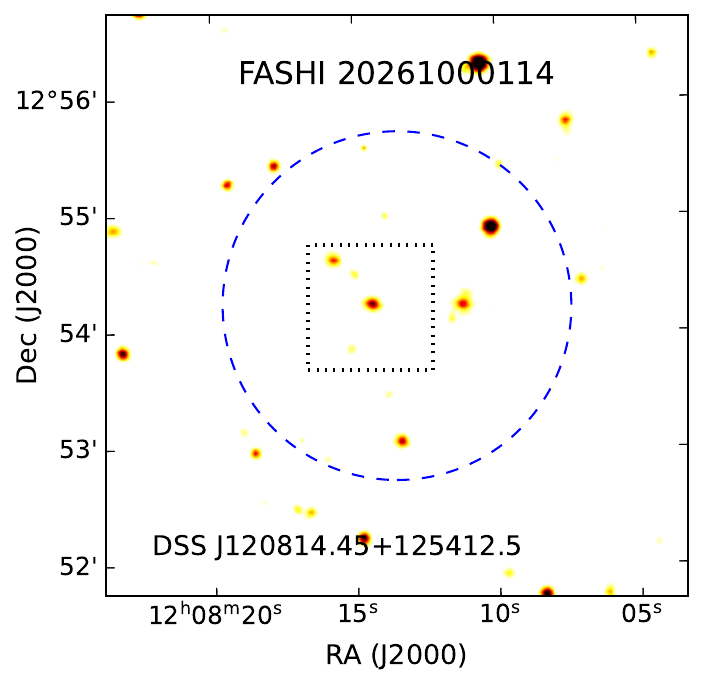}
 \includegraphics[height=0.22\textwidth, angle=0]{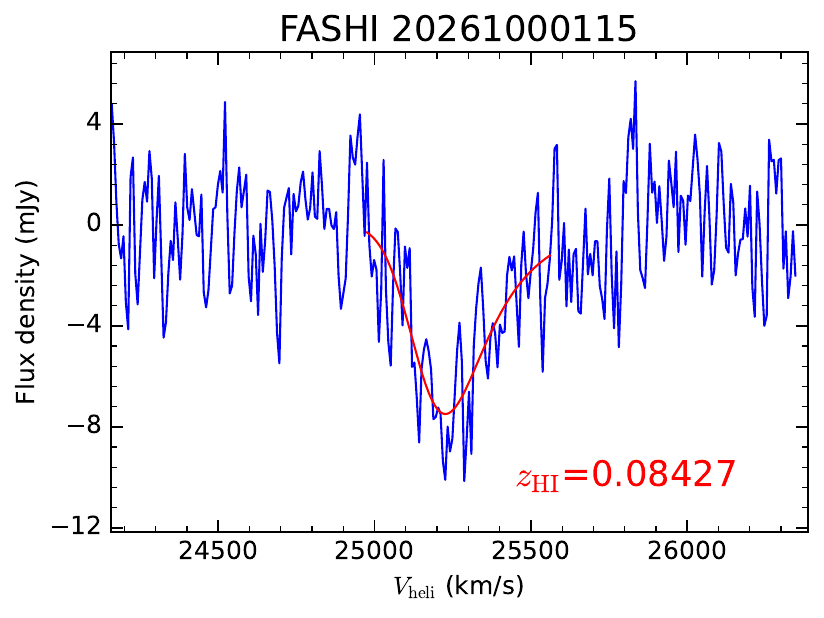}
 \includegraphics[height=0.27\textwidth, angle=0]{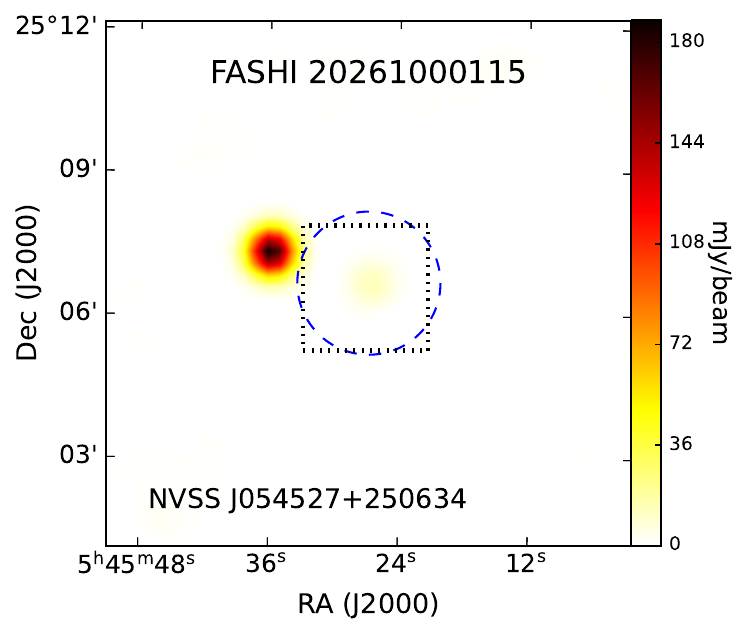}
 \includegraphics[height=0.27\textwidth, angle=0]{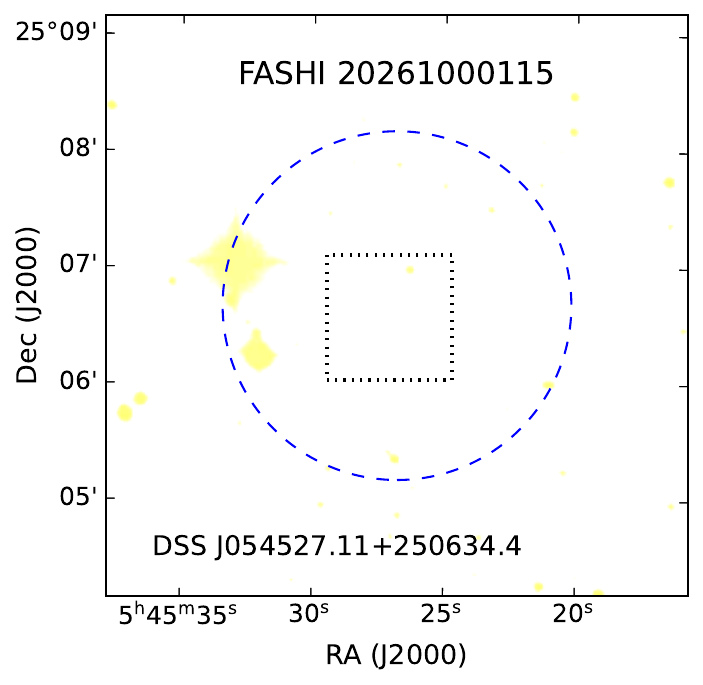}
 \caption{See caption in Figure\,\ref{Fig:FASHI_hi}}
 \end{figure*} 

 \begin{figure*}[htp]
 \centering
 \renewcommand{\thefigure}{\arabic{figure} (Continued)}
 \addtocounter{figure}{-1}
 \includegraphics[height=0.22\textwidth, angle=0]{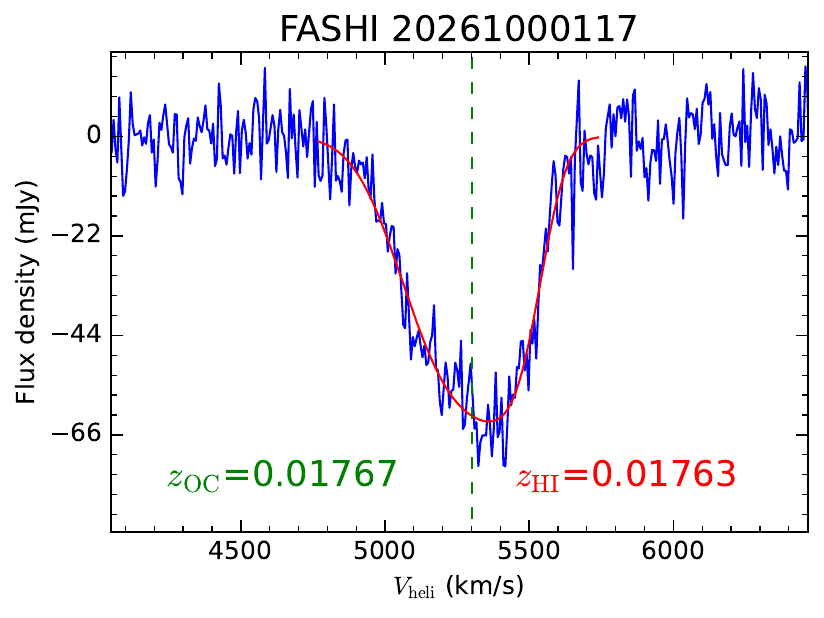}
 \includegraphics[height=0.27\textwidth, angle=0]{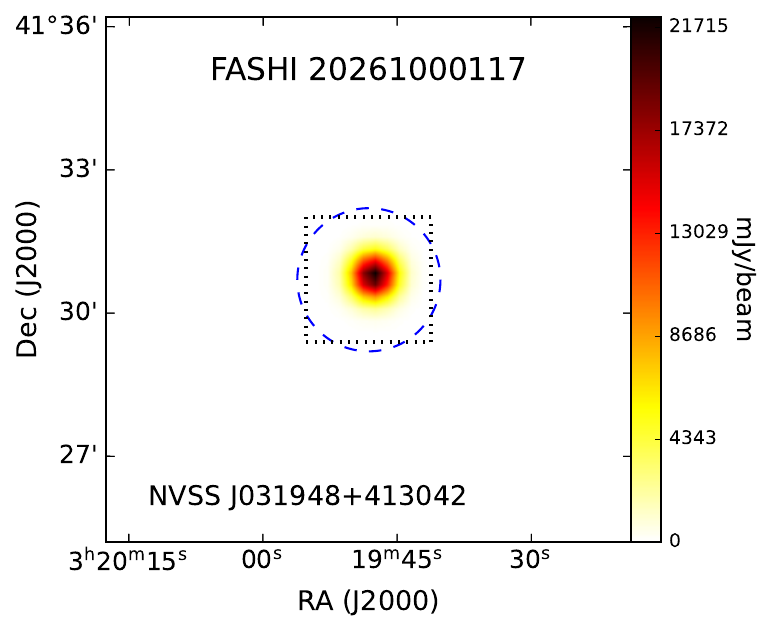}
 \includegraphics[height=0.27\textwidth, angle=0]{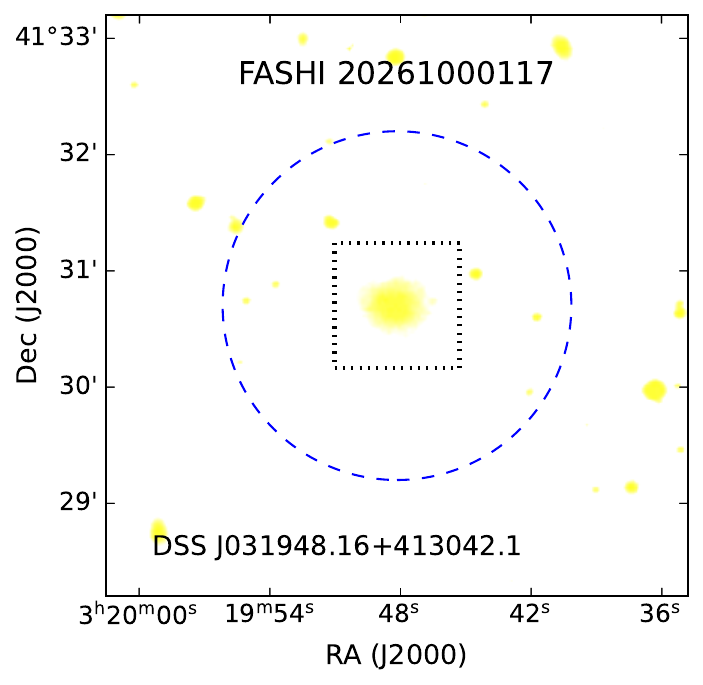}
 \includegraphics[height=0.22\textwidth, angle=0]{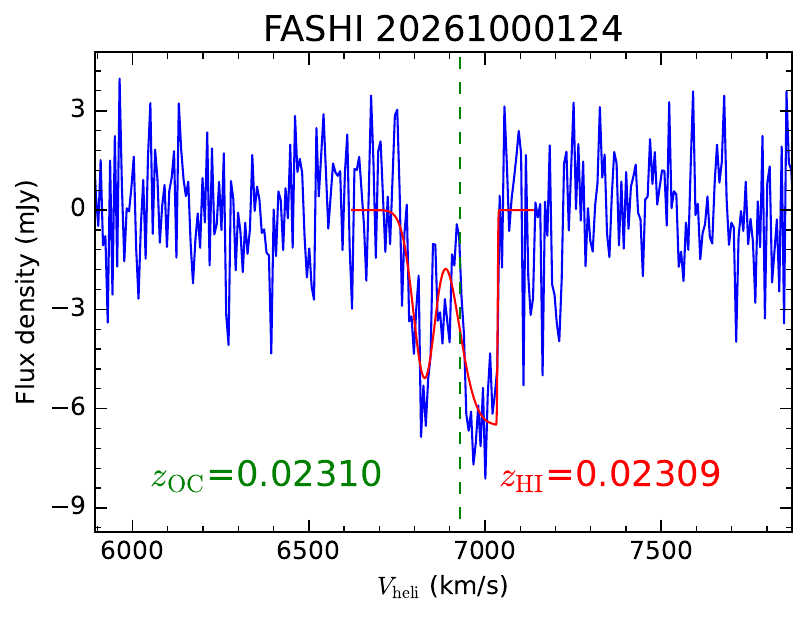}
 \includegraphics[height=0.27\textwidth, angle=0]{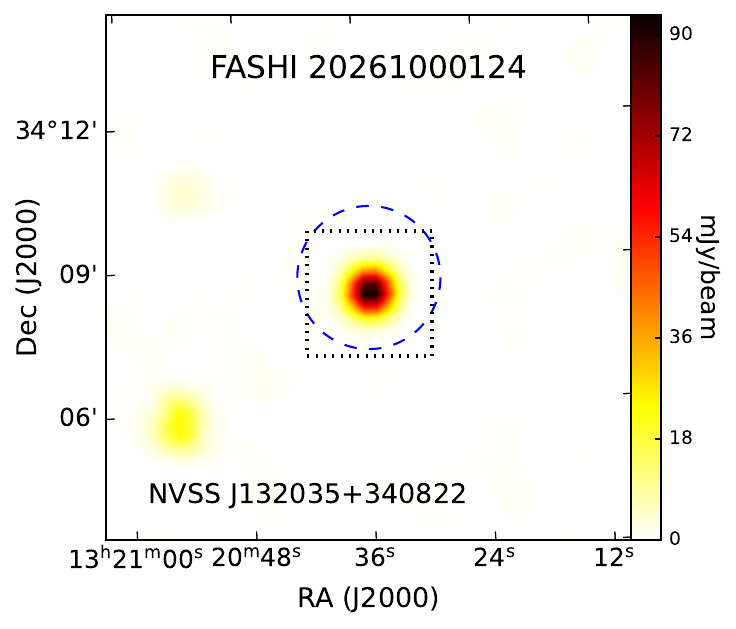}
 \includegraphics[height=0.27\textwidth, angle=0]{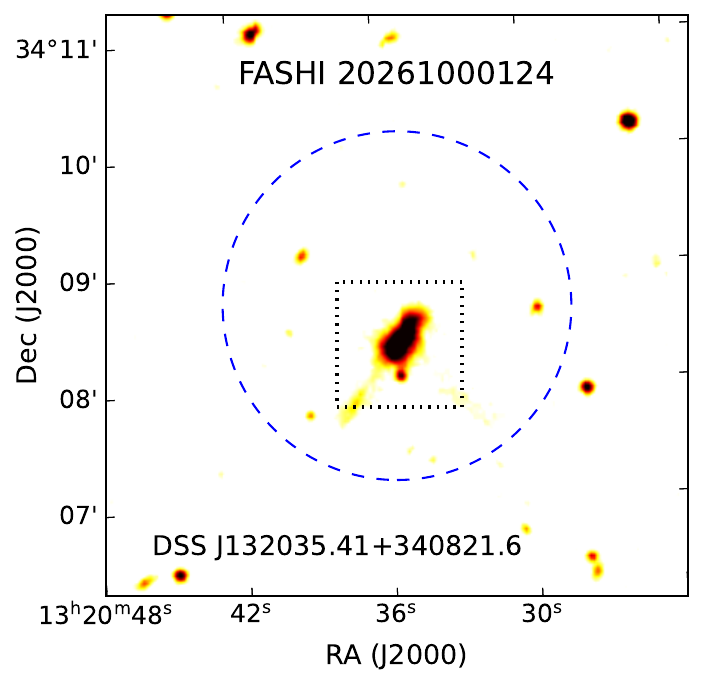}
 \includegraphics[height=0.22\textwidth, angle=0]{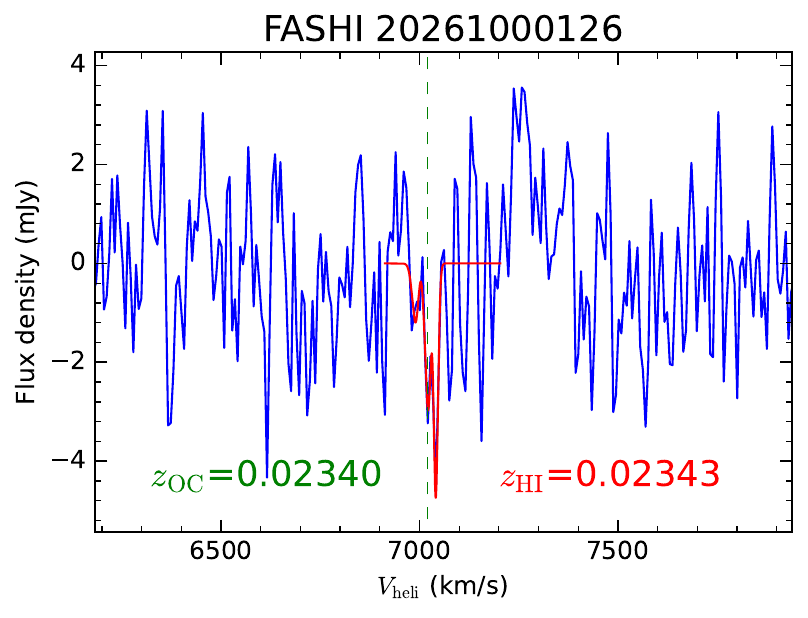}
 \includegraphics[height=0.27\textwidth, angle=0]{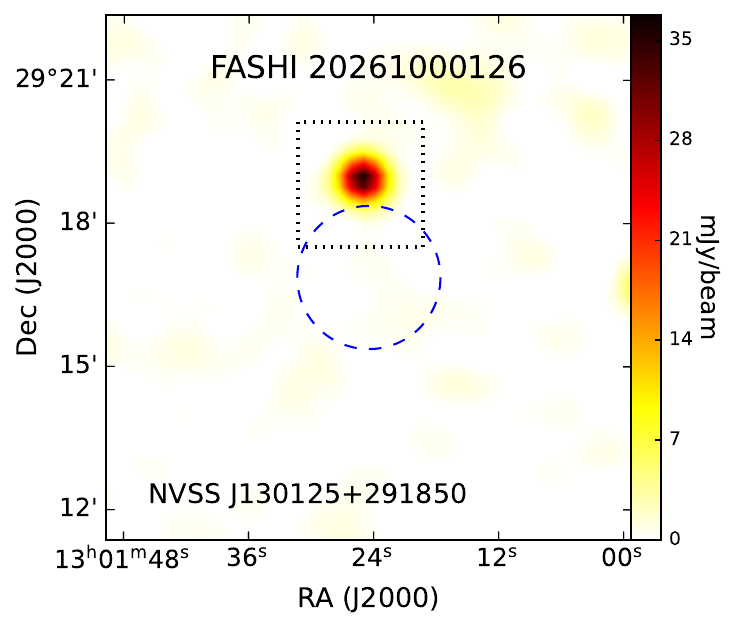}
 \includegraphics[height=0.27\textwidth, angle=0]{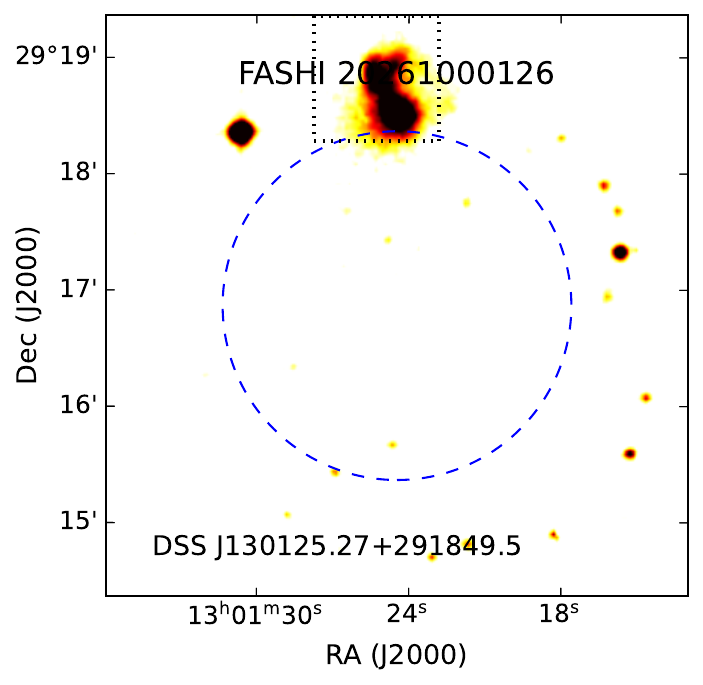}
 \includegraphics[height=0.22\textwidth, angle=0]{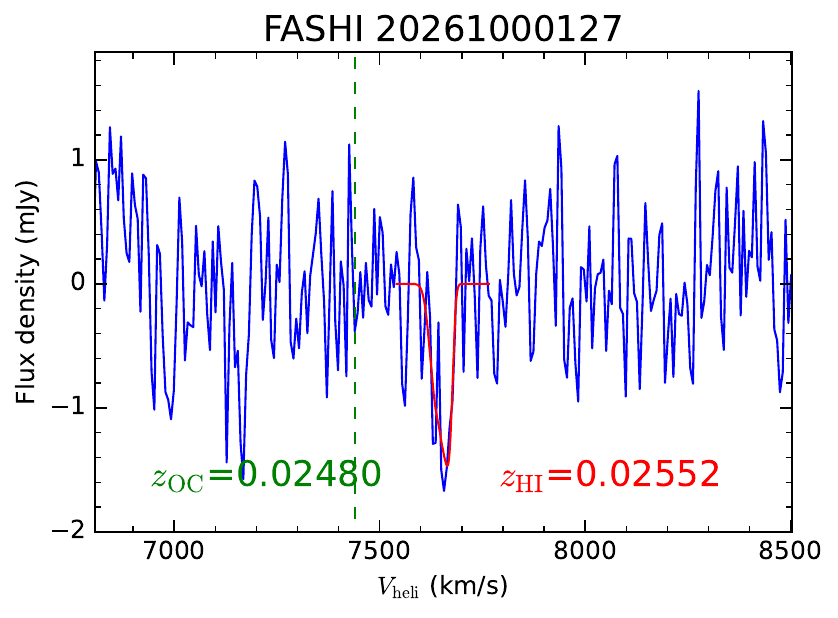}
 \includegraphics[height=0.27\textwidth, angle=0]{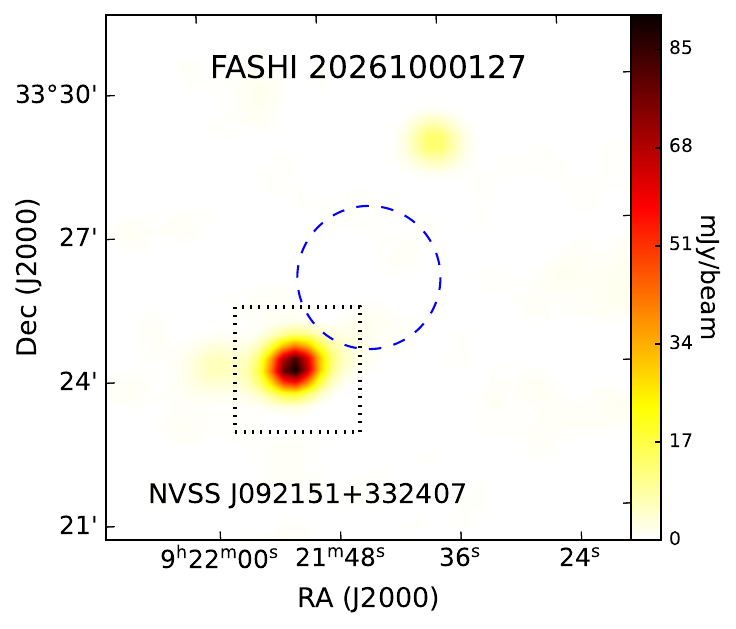}
 \includegraphics[height=0.27\textwidth, angle=0]{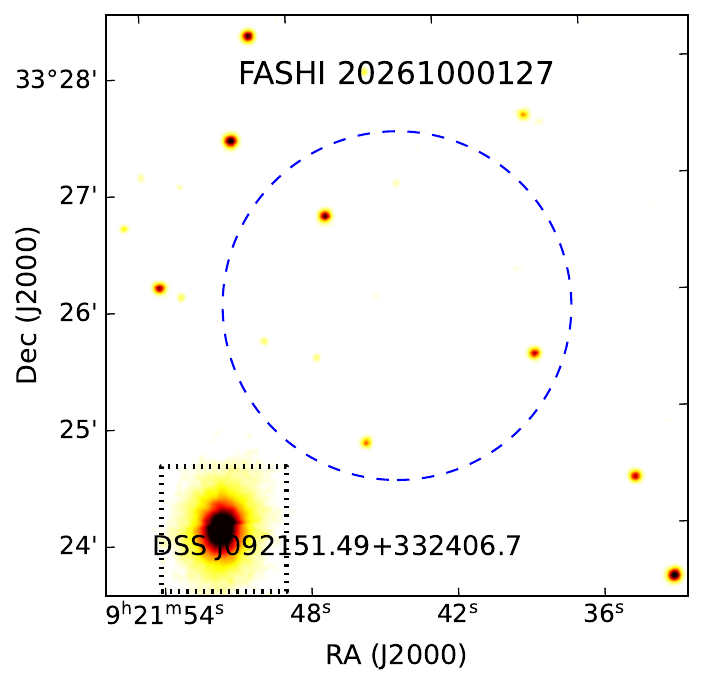}
 \caption{See caption in Figure\,\ref{Fig:FASHI_hi}}
 \end{figure*} 

 \begin{figure*}[htp]
 \centering
 \renewcommand{\thefigure}{\arabic{figure} (Continued)}
 \addtocounter{figure}{-1}
 \includegraphics[height=0.22\textwidth, angle=0]{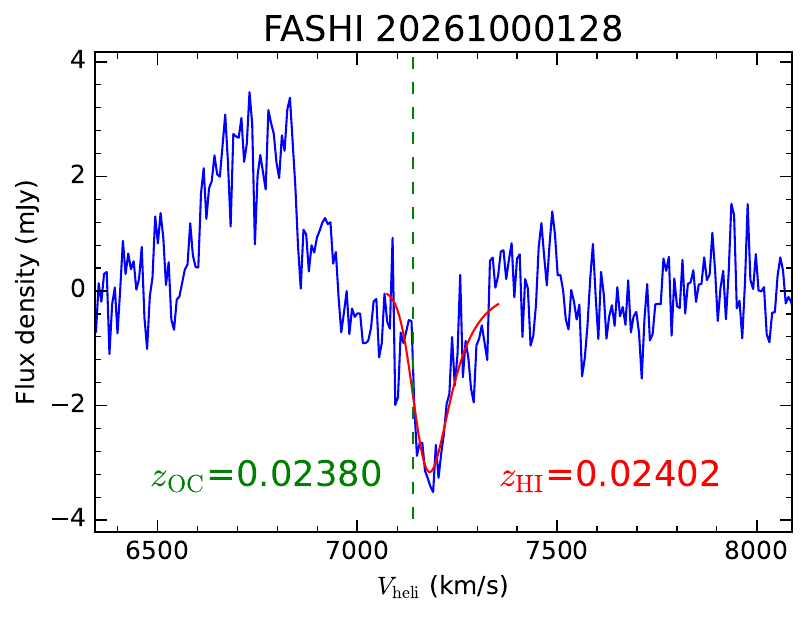}
 \includegraphics[height=0.27\textwidth, angle=0]{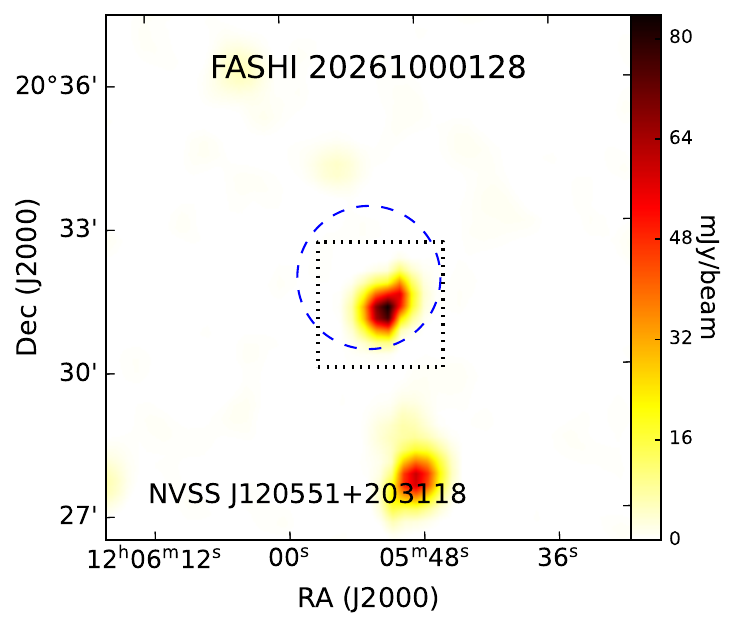}
 \includegraphics[height=0.27\textwidth, angle=0]{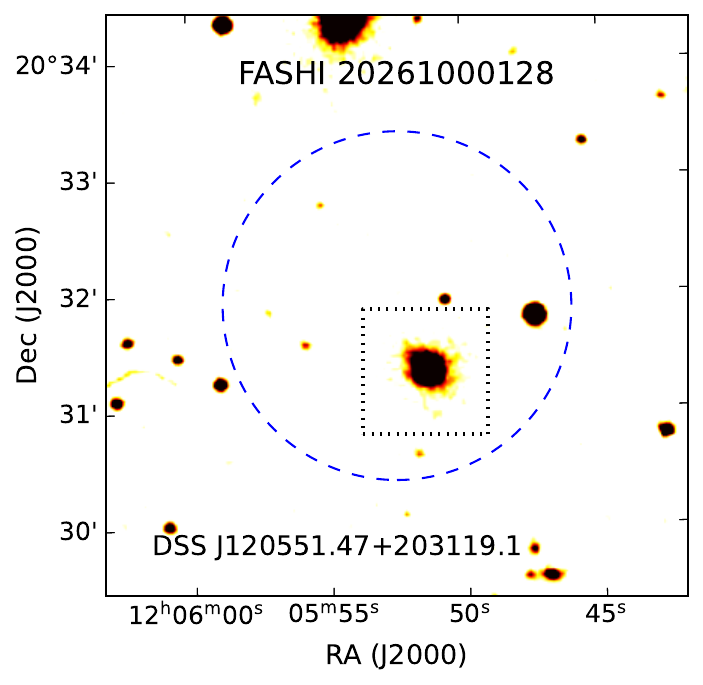}
 \includegraphics[height=0.22\textwidth, angle=0]{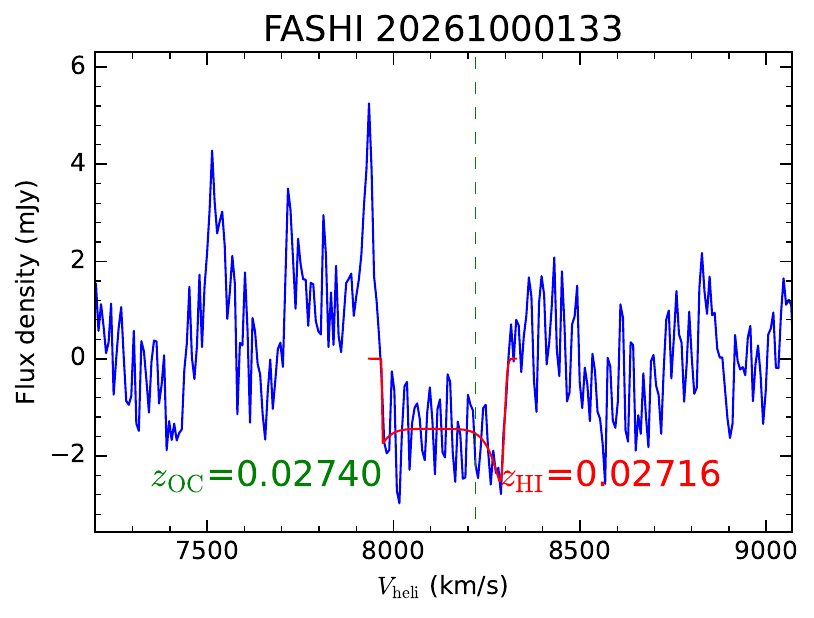}
 \includegraphics[height=0.27\textwidth, angle=0]{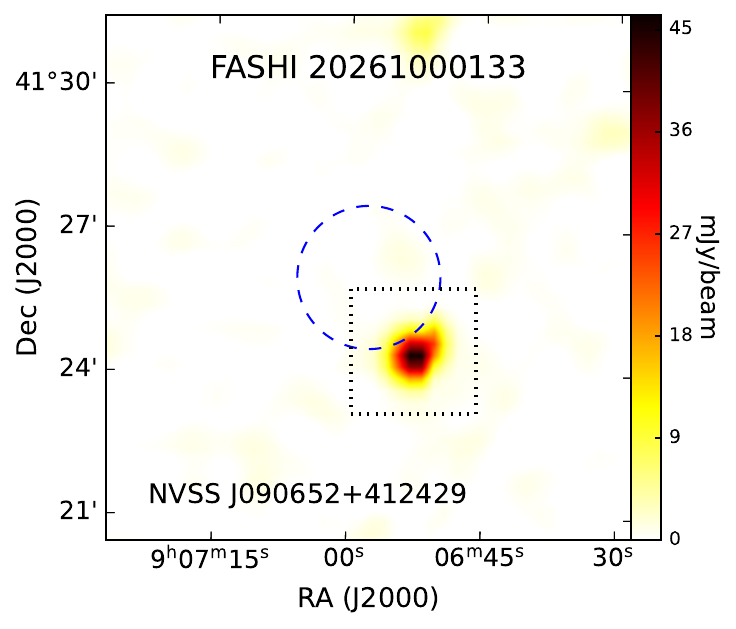}
 \includegraphics[height=0.27\textwidth, angle=0]{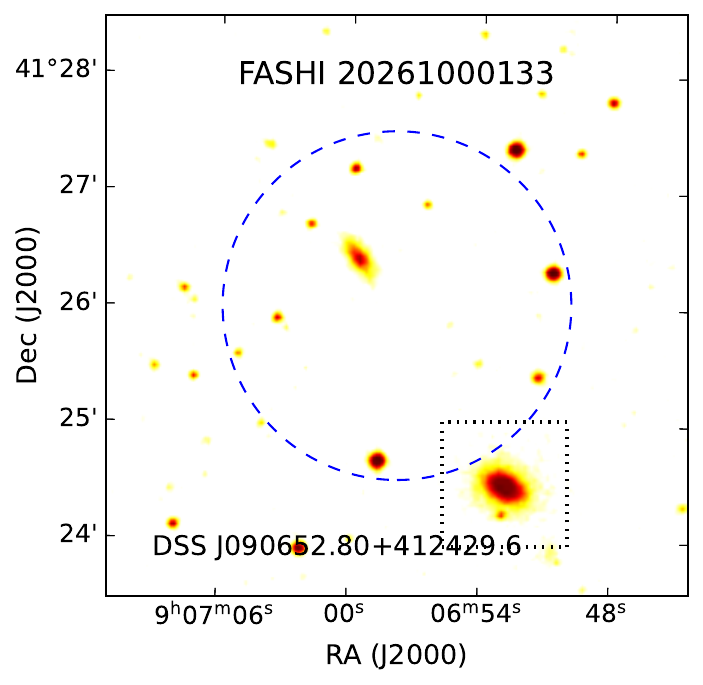}
 \includegraphics[height=0.22\textwidth, angle=0]{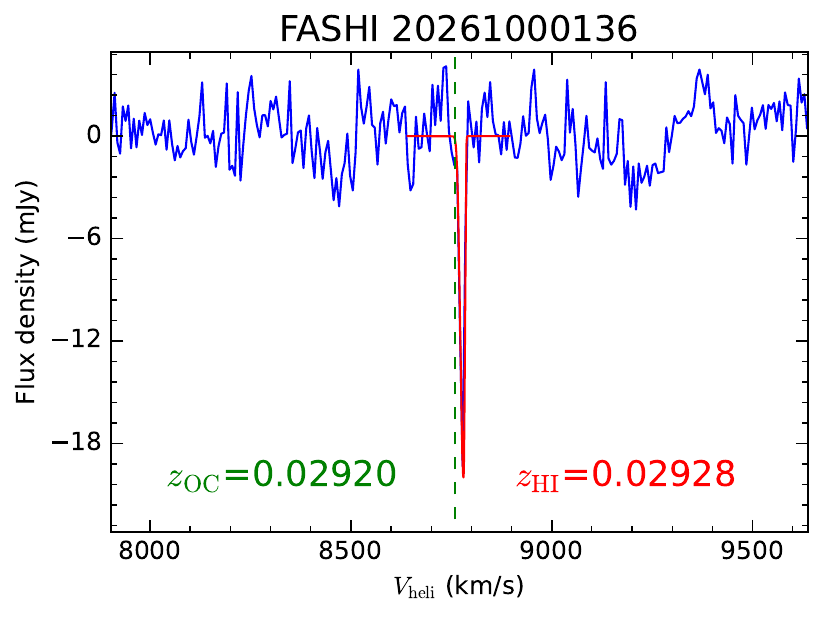}
 \includegraphics[height=0.27\textwidth, angle=0]{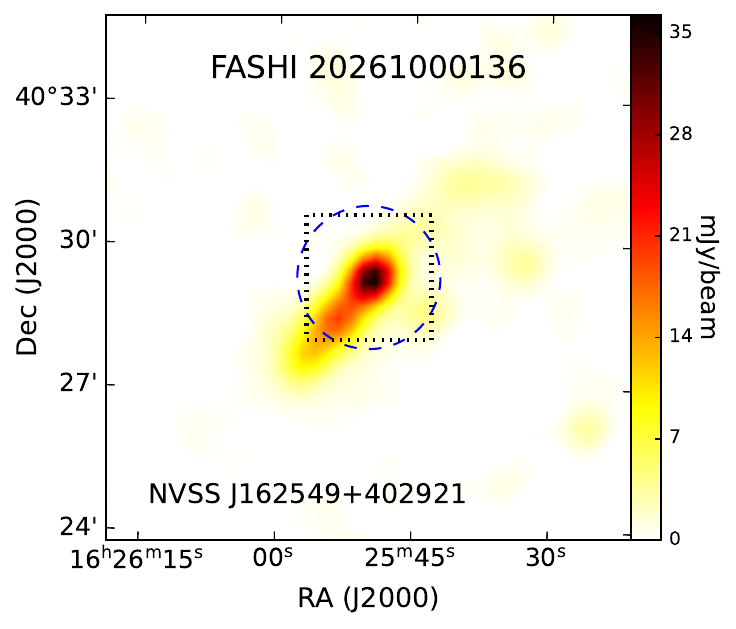}
 \includegraphics[height=0.27\textwidth, angle=0]{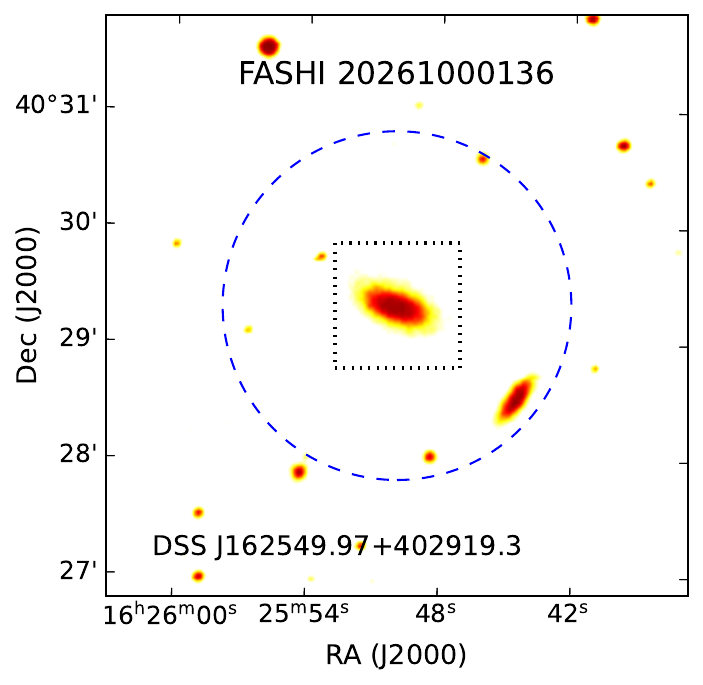}
 \includegraphics[height=0.22\textwidth, angle=0]{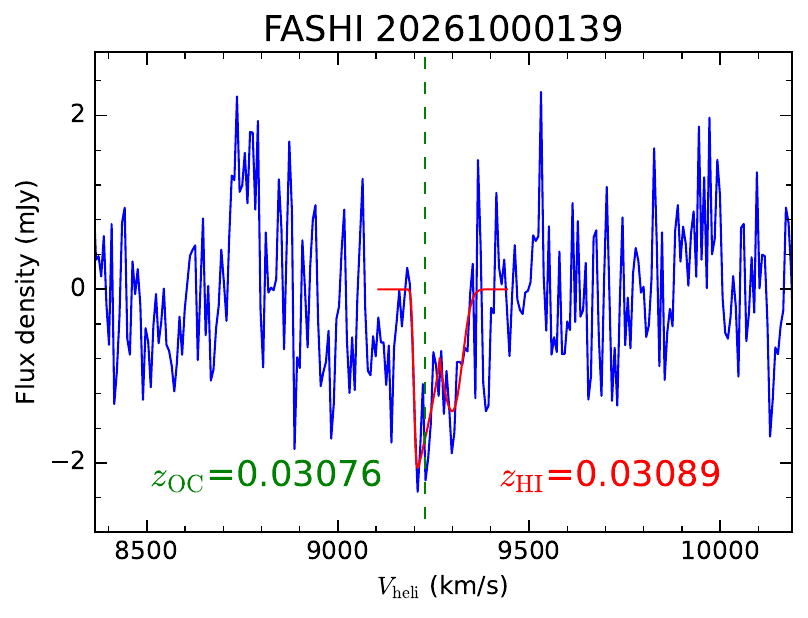}
 \includegraphics[height=0.27\textwidth, angle=0]{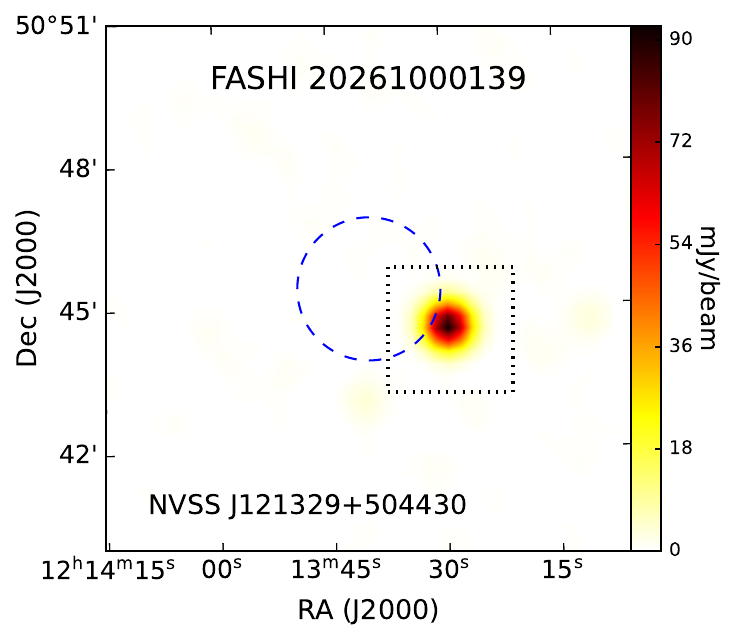}
 \includegraphics[height=0.27\textwidth, angle=0]{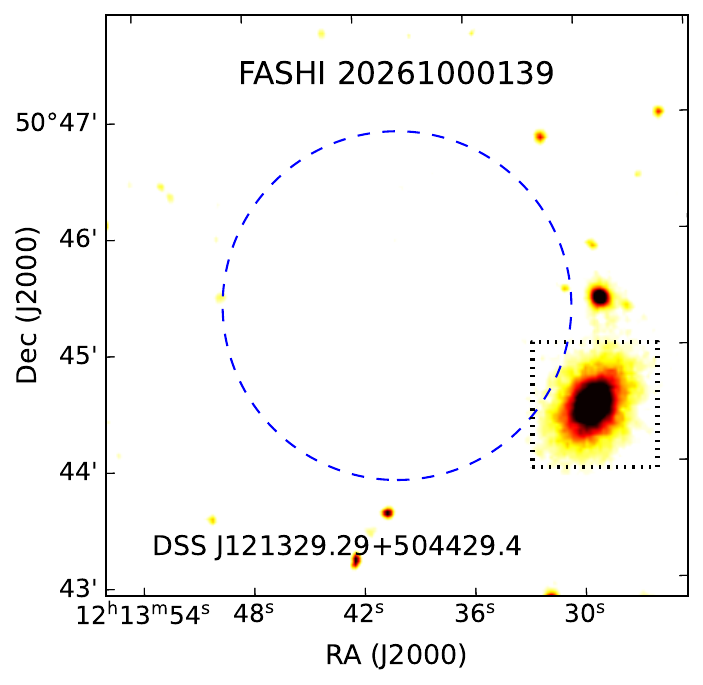}
 \caption{See caption in Figure\,\ref{Fig:FASHI_hi}}
 \end{figure*} 

 \begin{figure*}[htp]
 \centering
 \renewcommand{\thefigure}{\arabic{figure} (Continued)}
 \addtocounter{figure}{-1}
 \includegraphics[height=0.22\textwidth, angle=0]{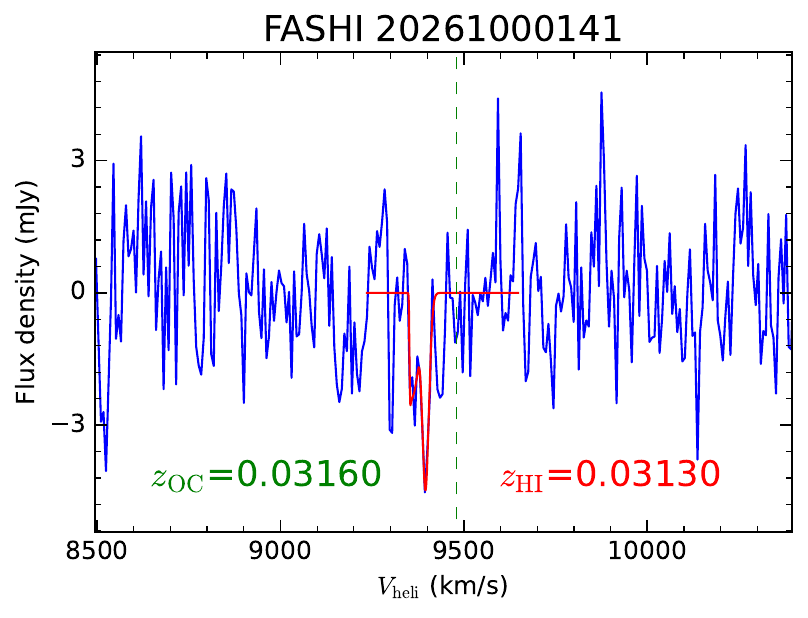}
 \includegraphics[height=0.27\textwidth, angle=0]{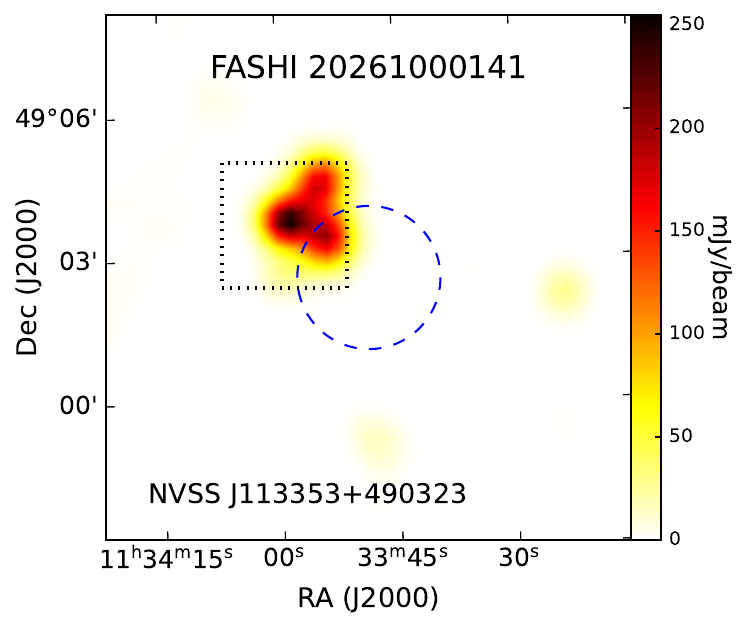}
 \includegraphics[height=0.27\textwidth, angle=0]{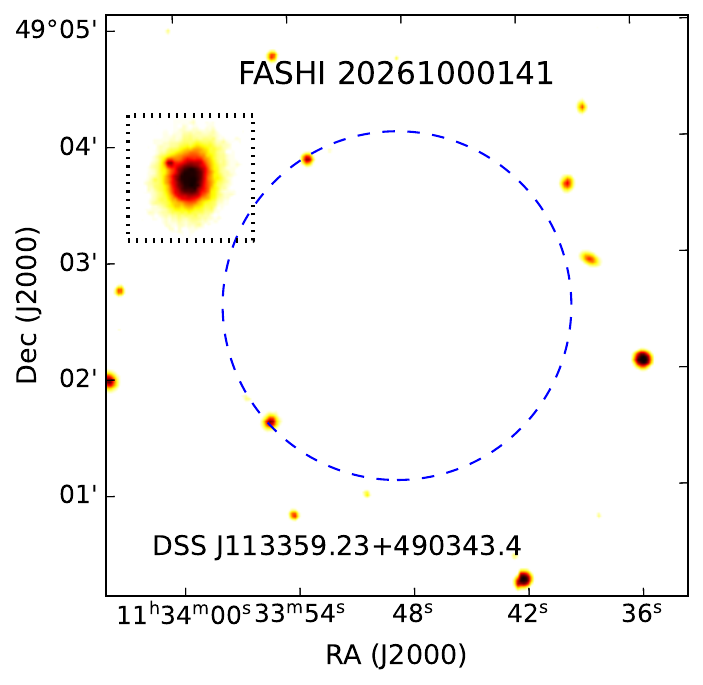}
 \includegraphics[height=0.22\textwidth, angle=0]{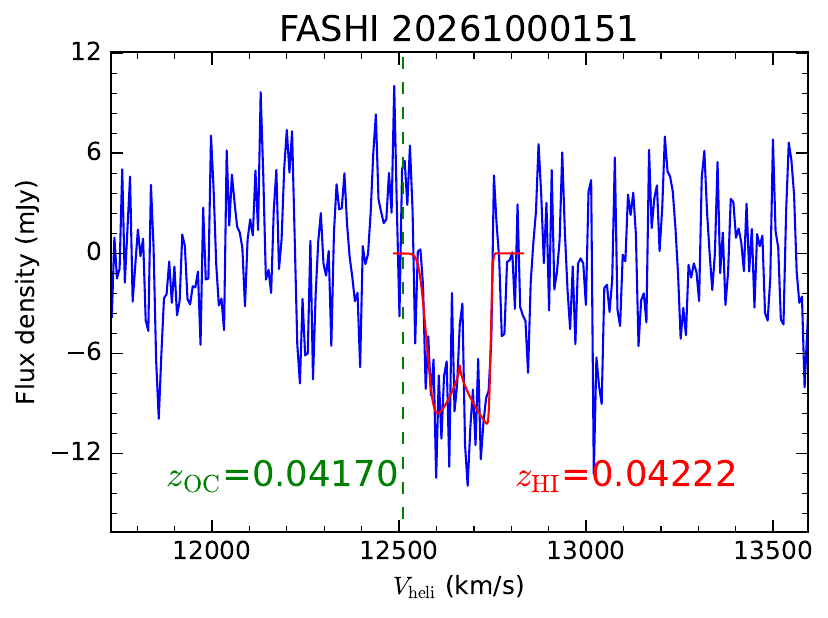}
 \includegraphics[height=0.27\textwidth, angle=0]{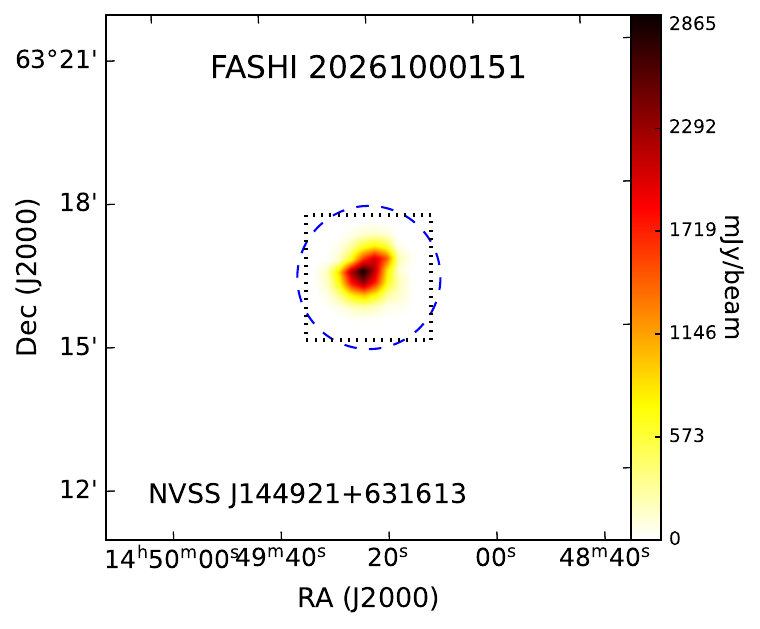}
 \includegraphics[height=0.27\textwidth, angle=0]{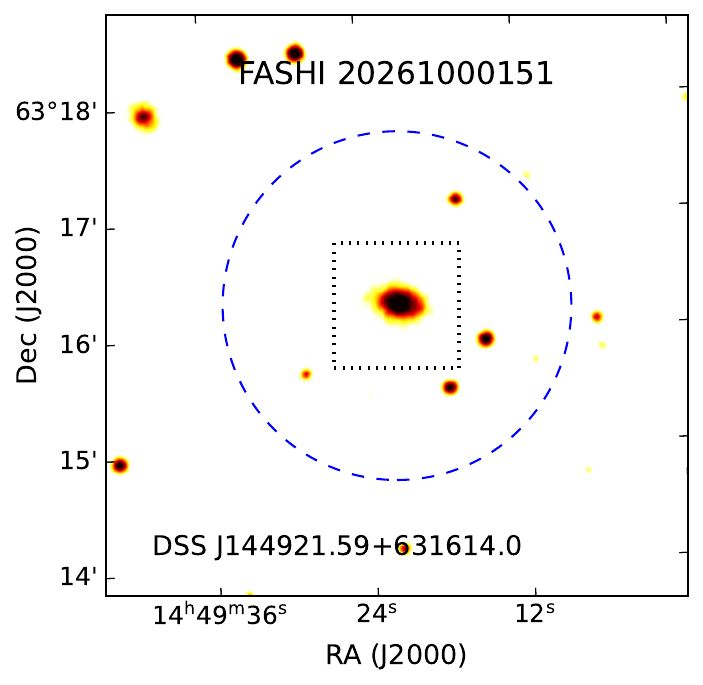}
 \includegraphics[height=0.22\textwidth, angle=0]{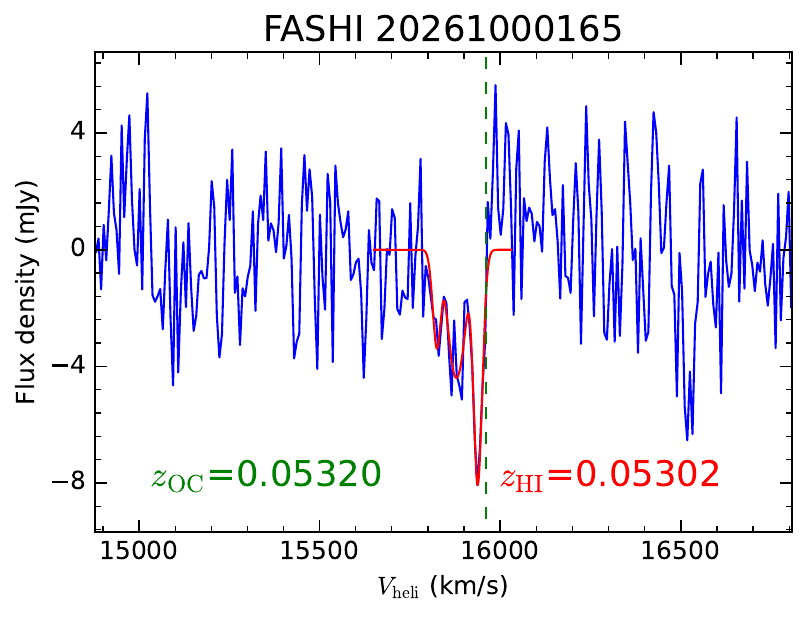}
 \includegraphics[height=0.27\textwidth, angle=0]{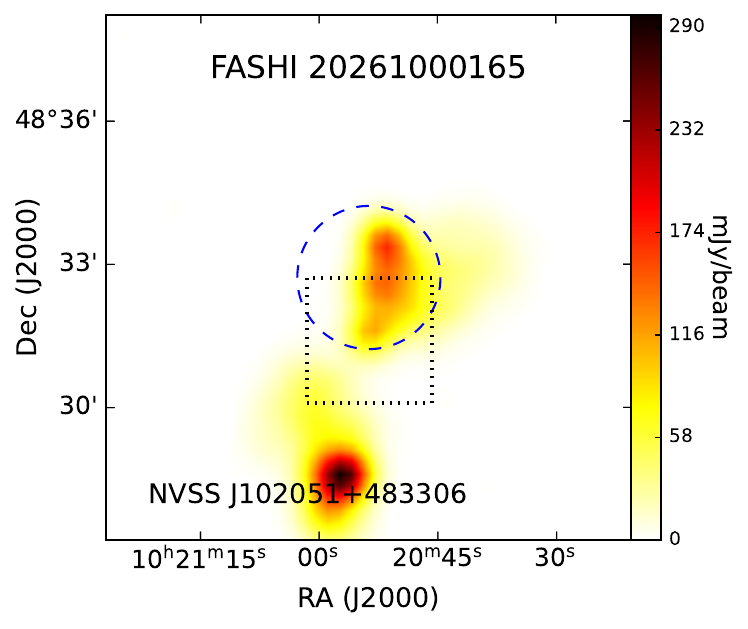}
 \includegraphics[height=0.27\textwidth, angle=0]{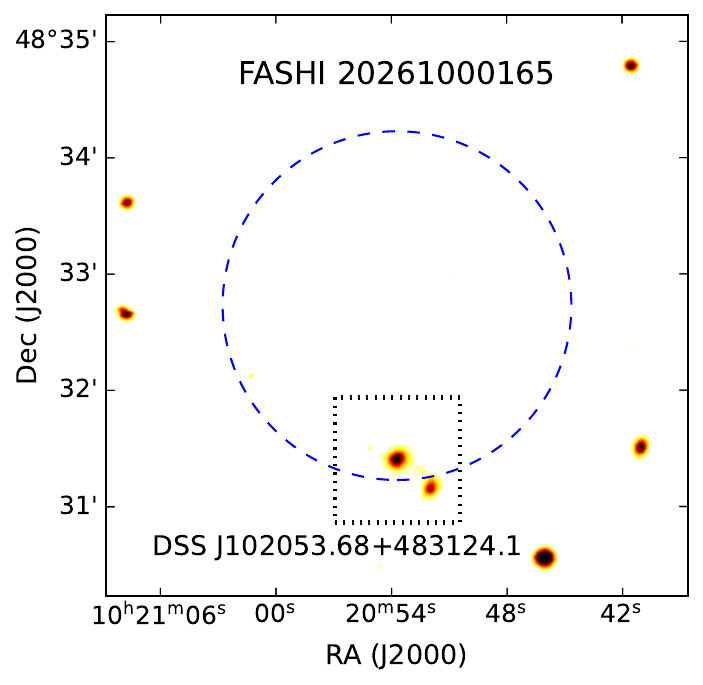}
 \includegraphics[height=0.22\textwidth, angle=0]{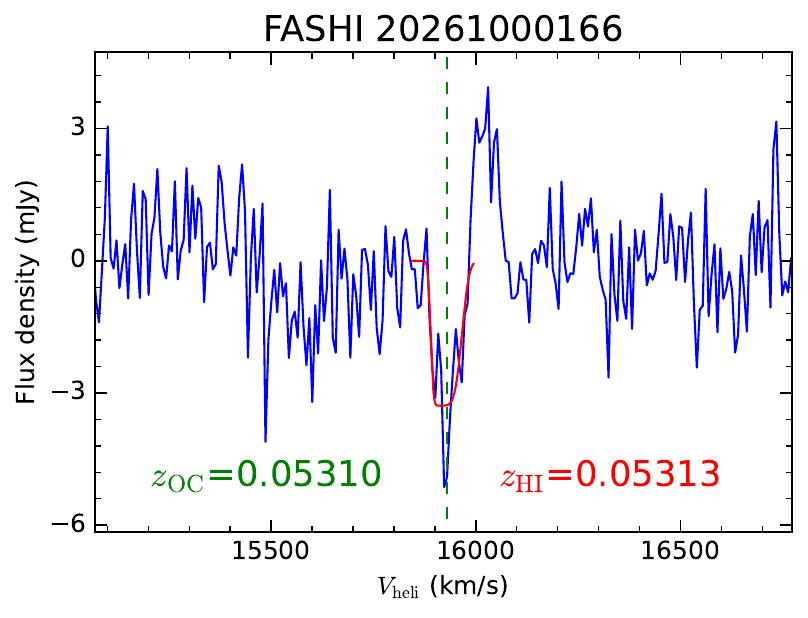}
 \includegraphics[height=0.27\textwidth, angle=0]{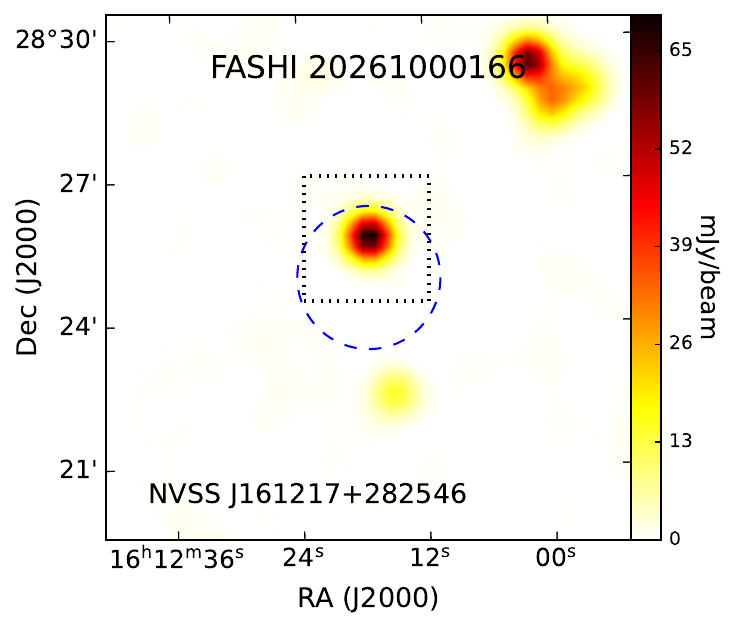}
 \includegraphics[height=0.27\textwidth, angle=0]{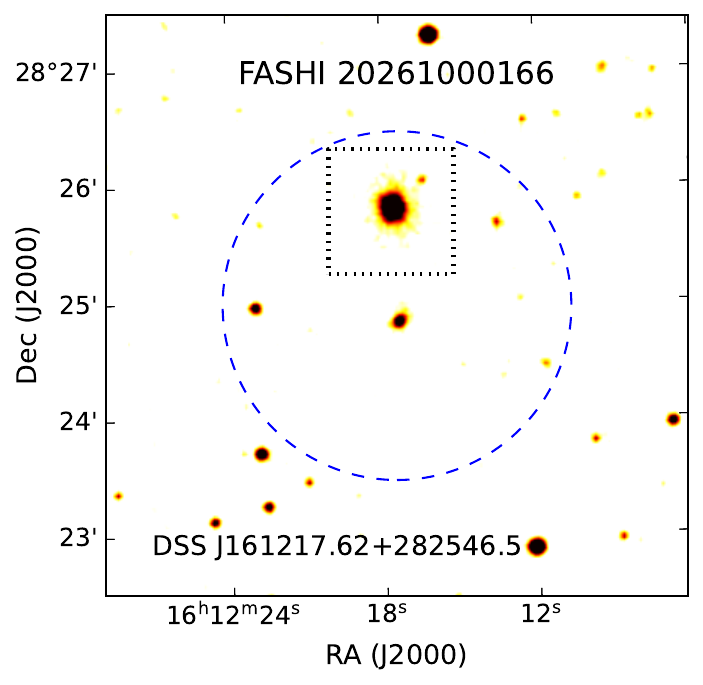}
 \caption{See caption in Figure\,\ref{Fig:FASHI_hi}}
 \end{figure*} 

 \begin{figure*}[htp]
 \centering
 \renewcommand{\thefigure}{\arabic{figure} (Continued)}
 \addtocounter{figure}{-1}
 \includegraphics[height=0.22\textwidth, angle=0]{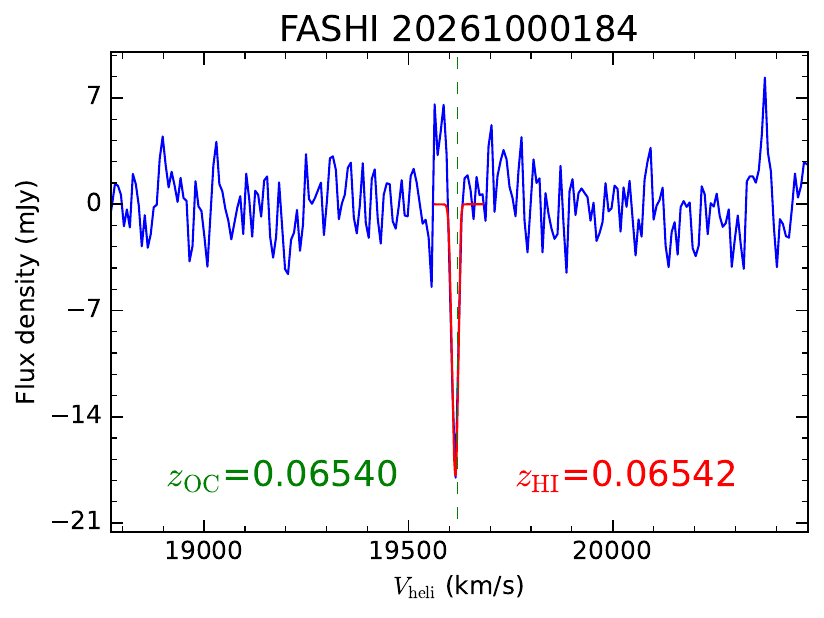}
 \includegraphics[height=0.27\textwidth, angle=0]{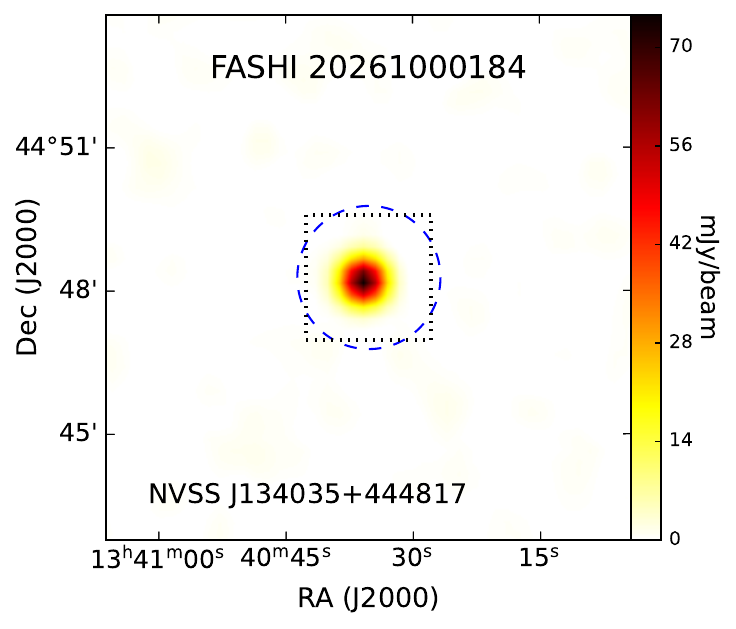}
 \includegraphics[height=0.27\textwidth, angle=0]{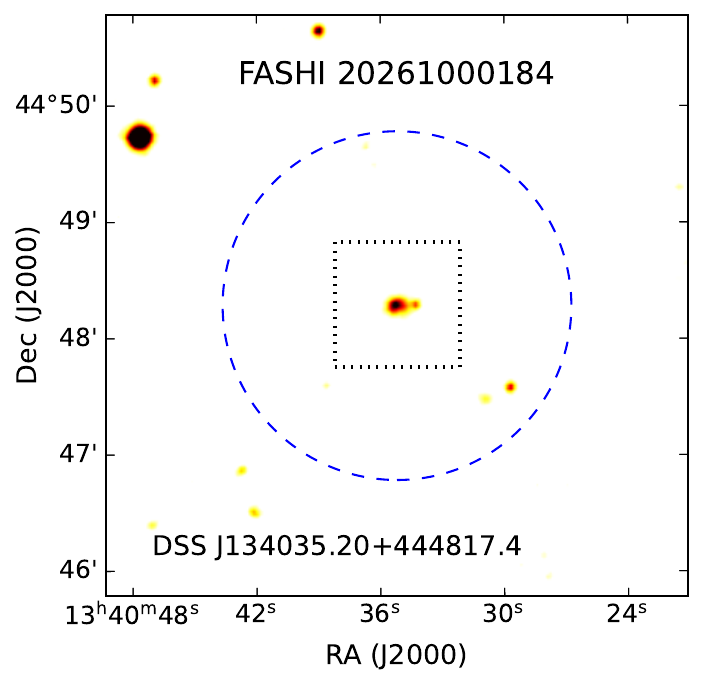}
 \includegraphics[height=0.22\textwidth, angle=0]{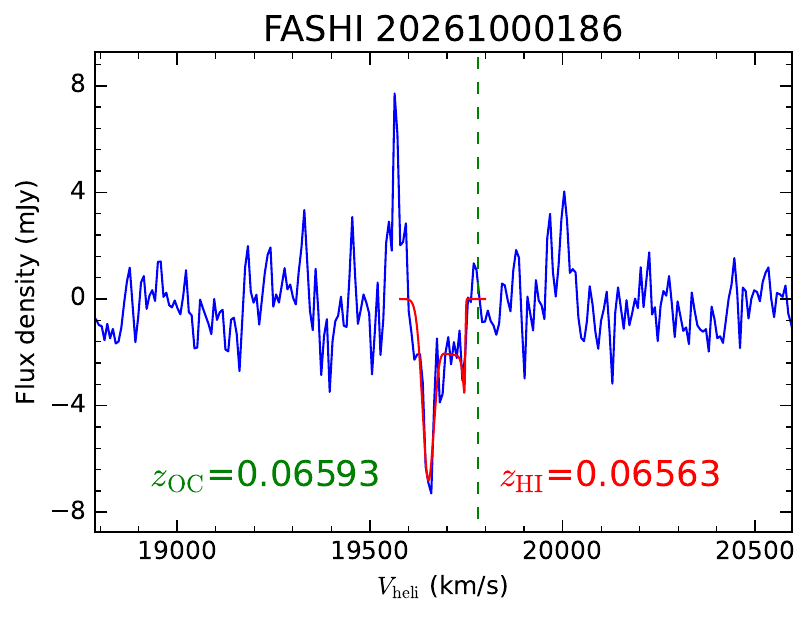}
 \includegraphics[height=0.27\textwidth, angle=0]{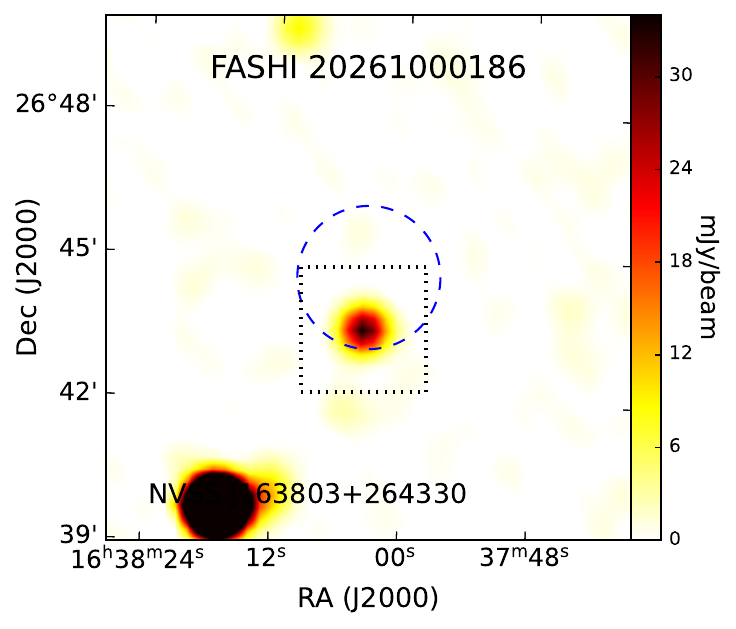}
 \includegraphics[height=0.27\textwidth, angle=0]{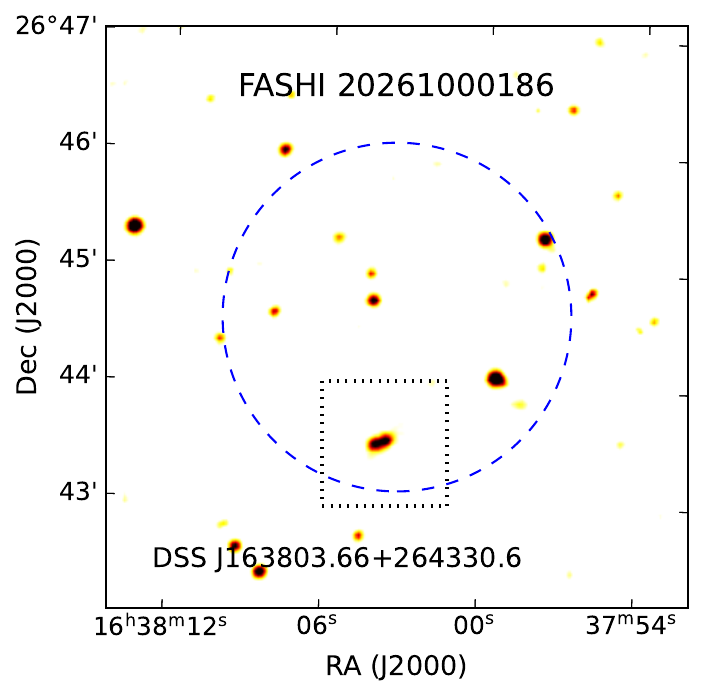}
 \includegraphics[height=0.22\textwidth, angle=0]{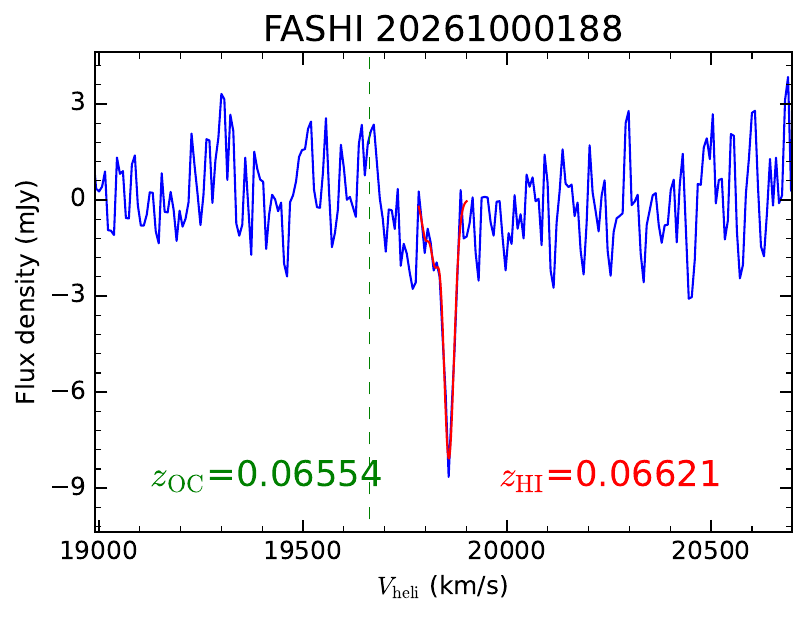}
 \includegraphics[height=0.27\textwidth, angle=0]{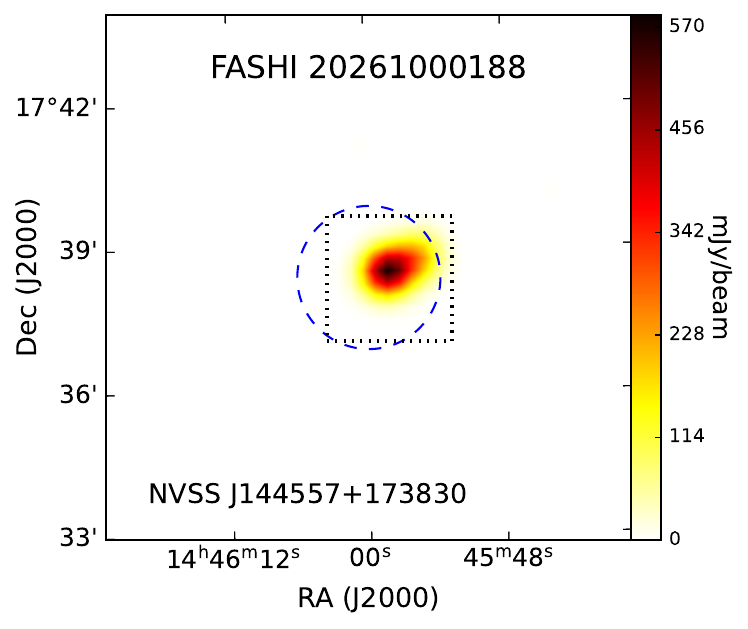}
 \includegraphics[height=0.27\textwidth, angle=0]{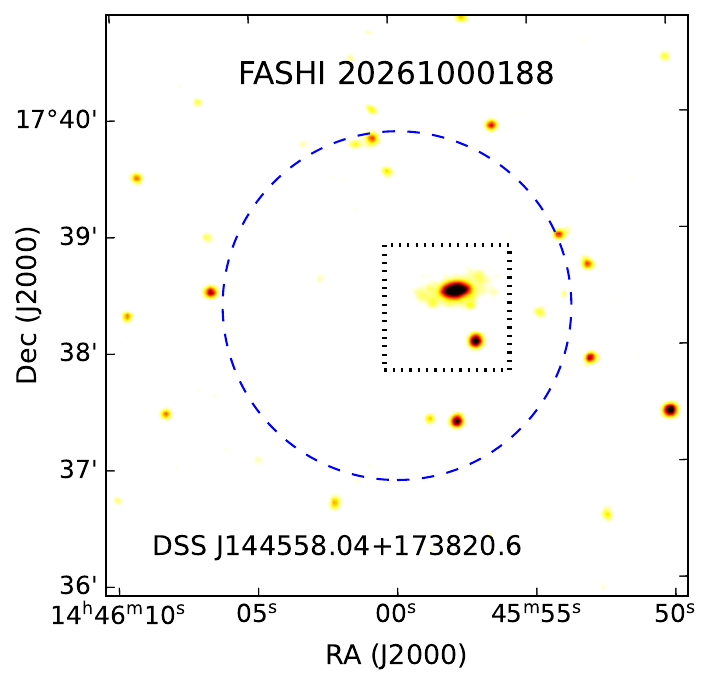}
 \includegraphics[height=0.22\textwidth, angle=0]{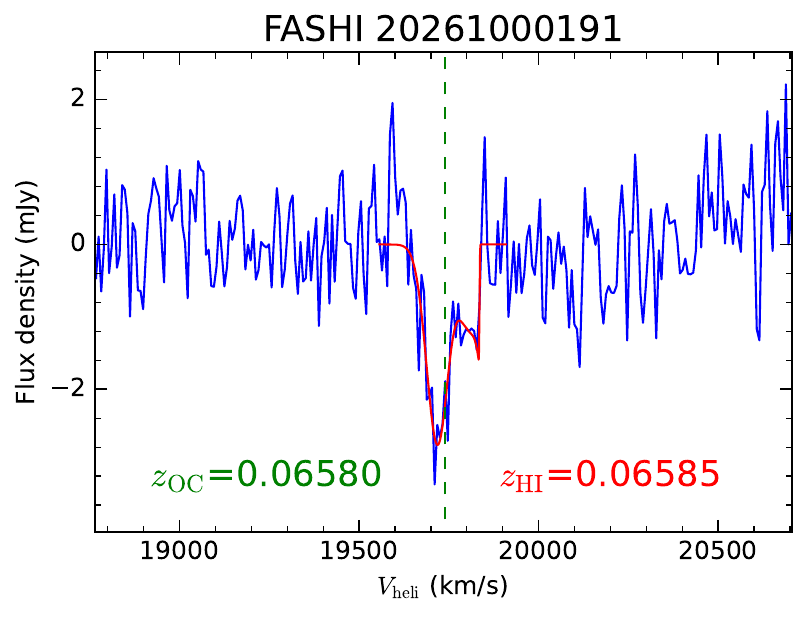}
 \includegraphics[height=0.27\textwidth, angle=0]{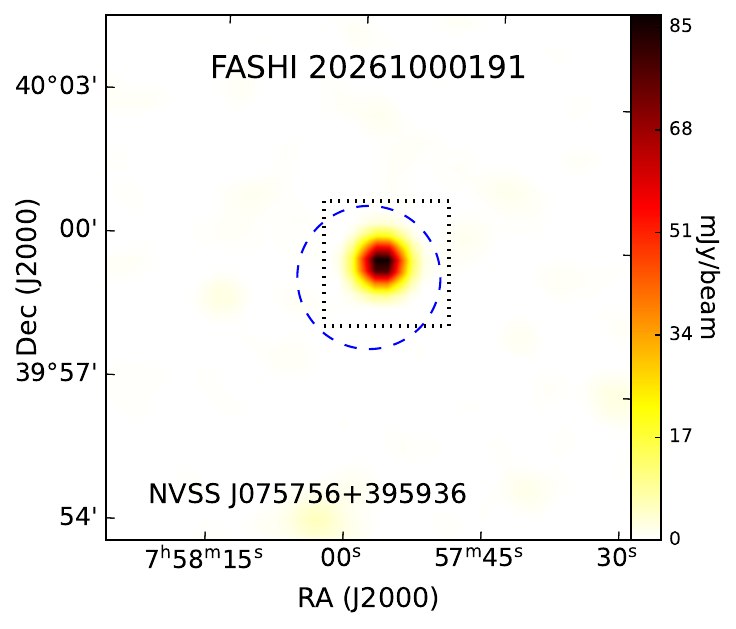}
 \includegraphics[height=0.27\textwidth, angle=0]{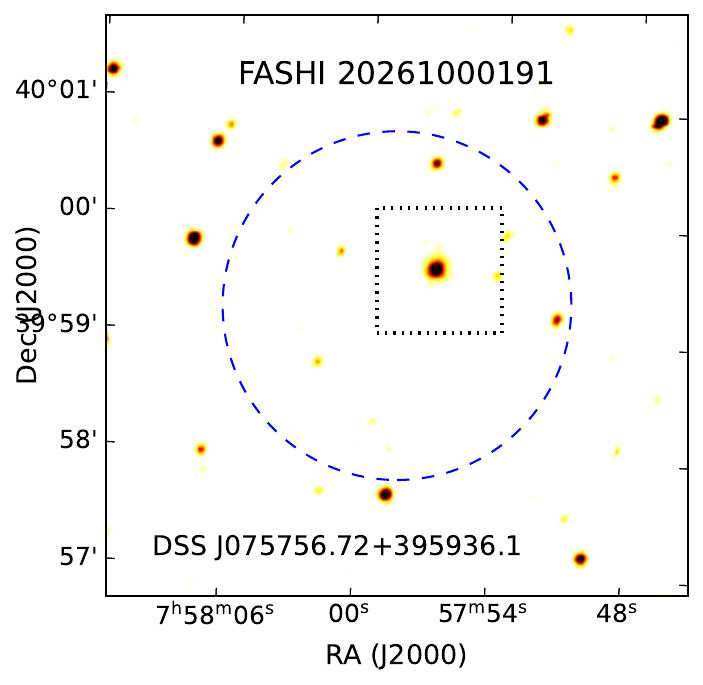}
 \caption{See caption in Figure\,\ref{Fig:FASHI_hi}}
 \end{figure*} 

 \begin{figure*}[htp]
 \centering
 \renewcommand{\thefigure}{\arabic{figure} (Continued)}
 \addtocounter{figure}{-1}
 \includegraphics[height=0.22\textwidth, angle=0]{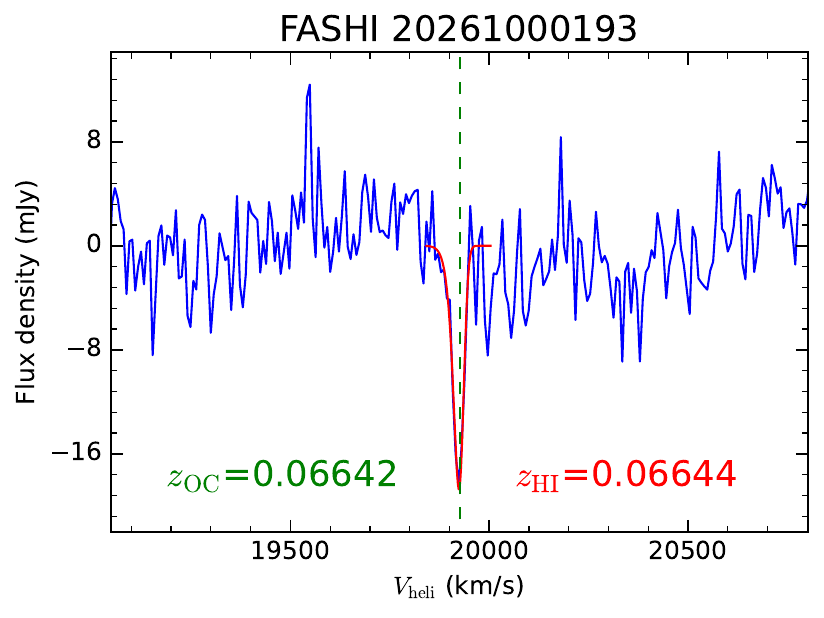}
 \includegraphics[height=0.27\textwidth, angle=0]{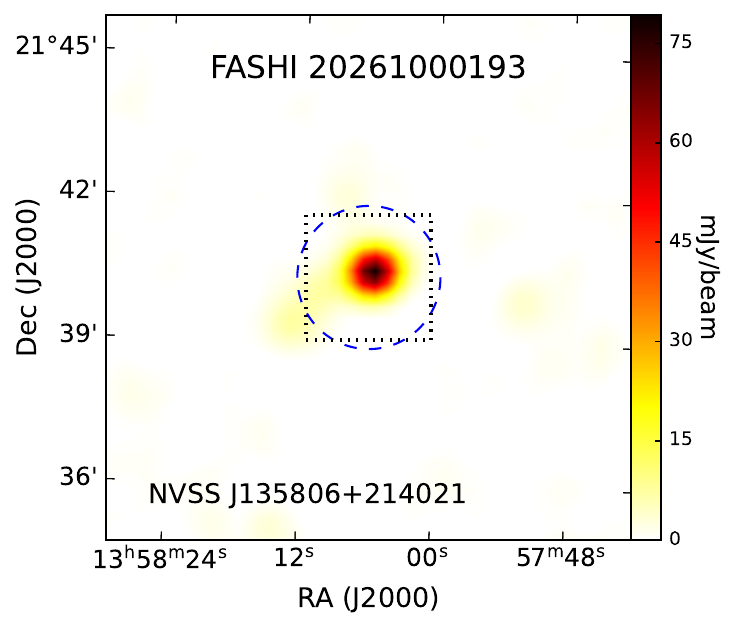}
 \includegraphics[height=0.27\textwidth, angle=0]{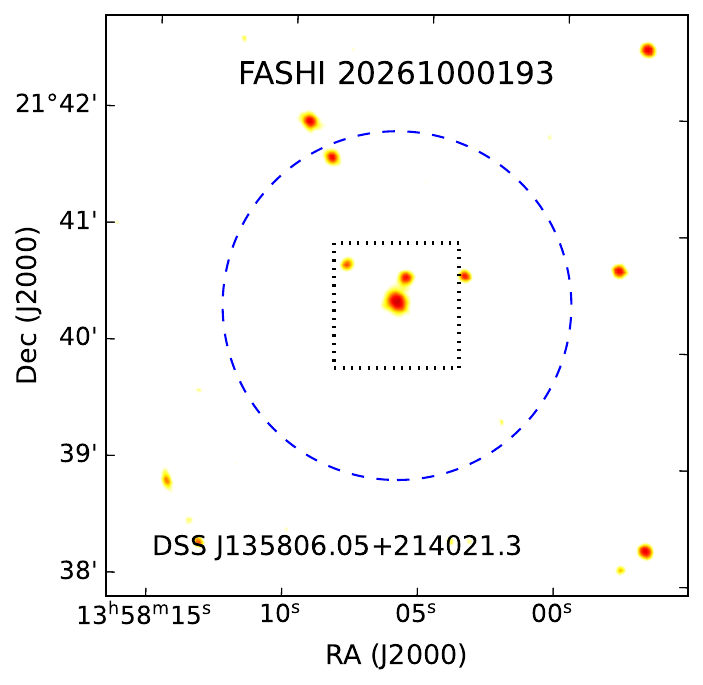}
 \includegraphics[height=0.22\textwidth, angle=0]{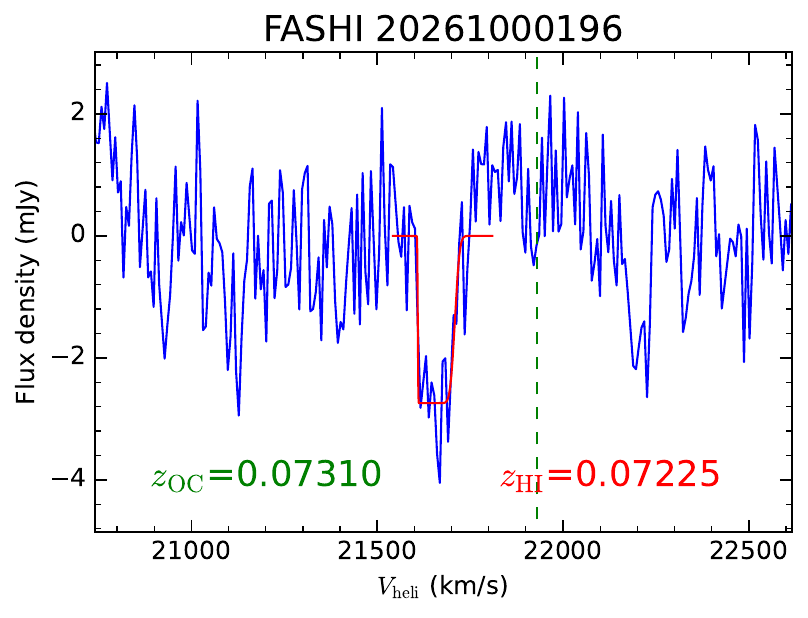}
 \includegraphics[height=0.27\textwidth, angle=0]{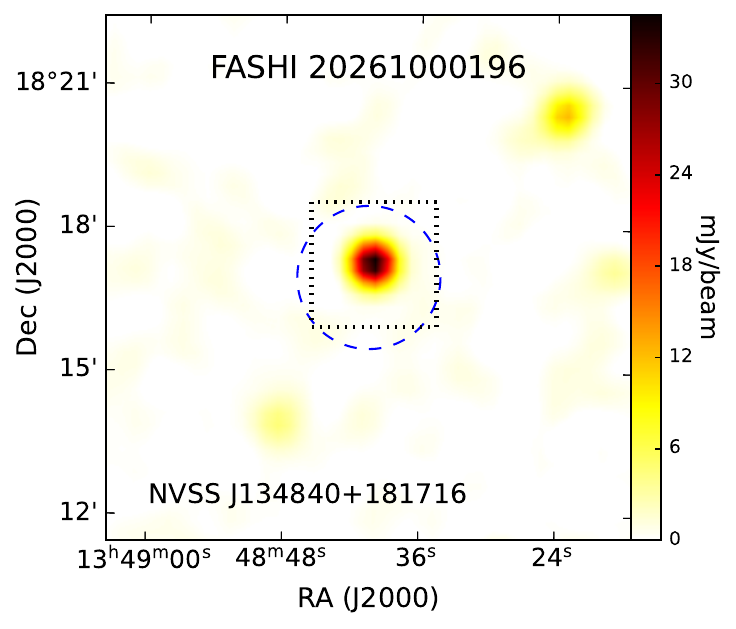}
 \includegraphics[height=0.27\textwidth, angle=0]{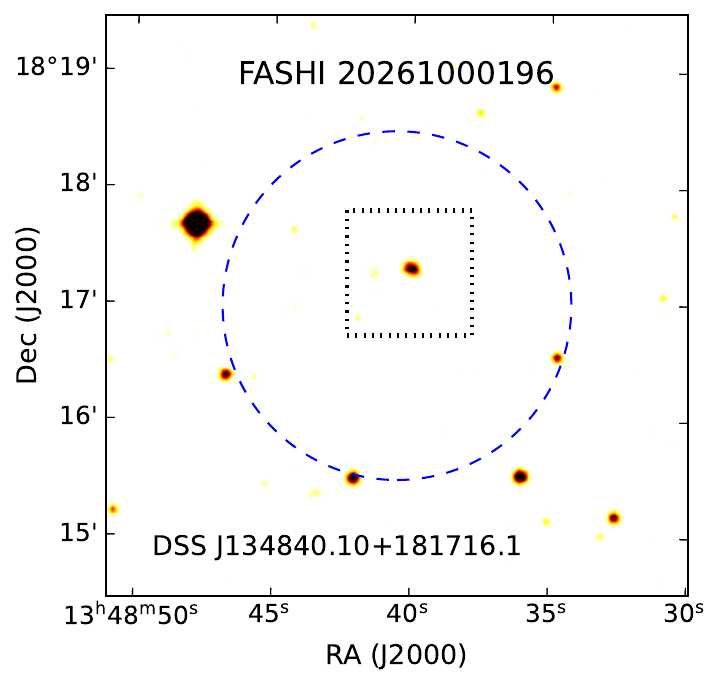}
 \includegraphics[height=0.22\textwidth, angle=0]{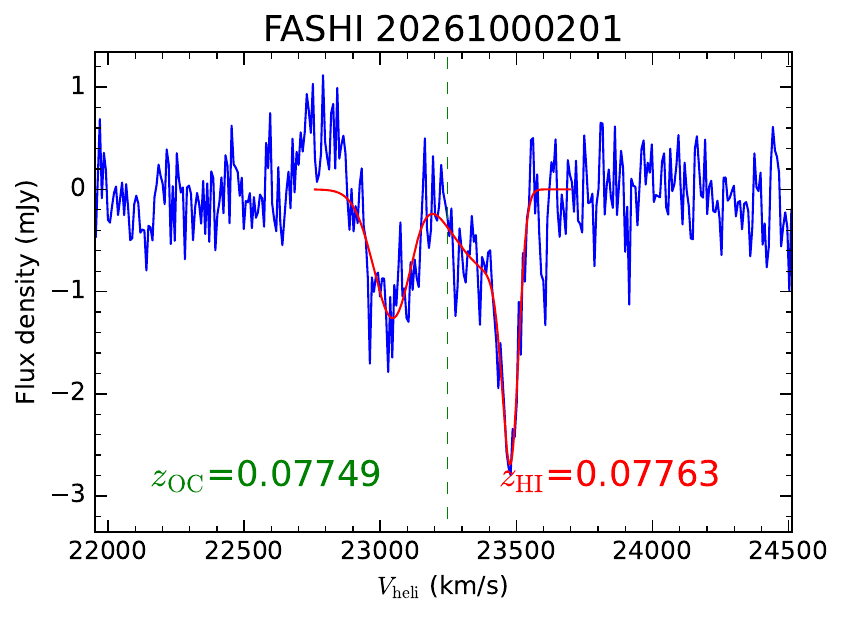}
 \includegraphics[height=0.27\textwidth, angle=0]{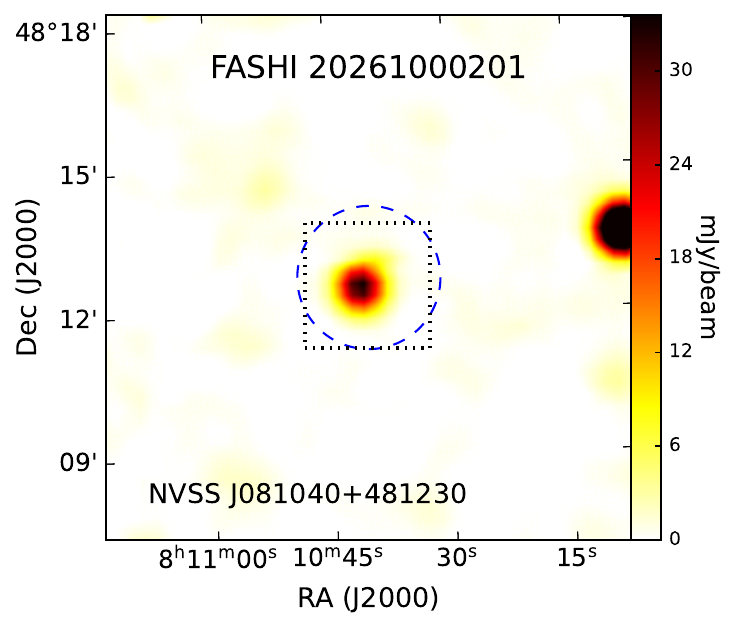}
 \includegraphics[height=0.27\textwidth, angle=0]{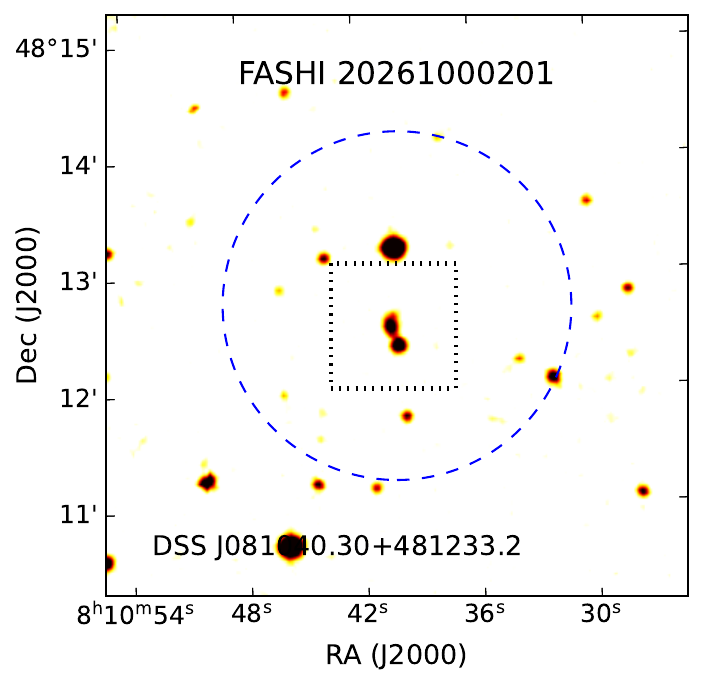}
 \includegraphics[height=0.22\textwidth, angle=0]{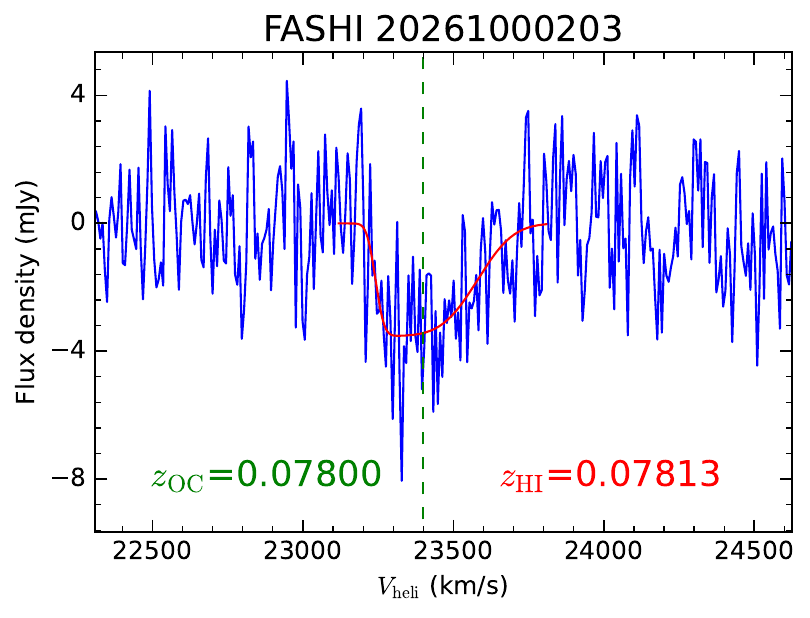}
 \includegraphics[height=0.27\textwidth, angle=0]{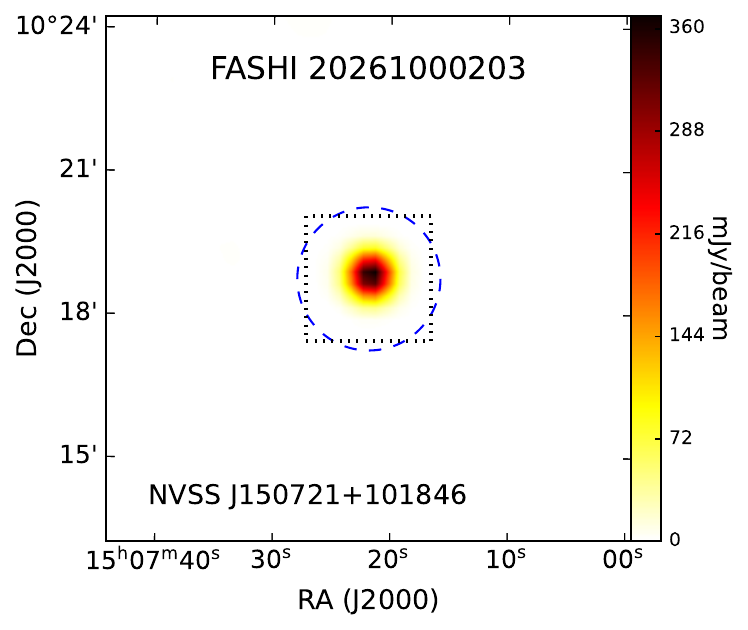}
 \includegraphics[height=0.27\textwidth, angle=0]{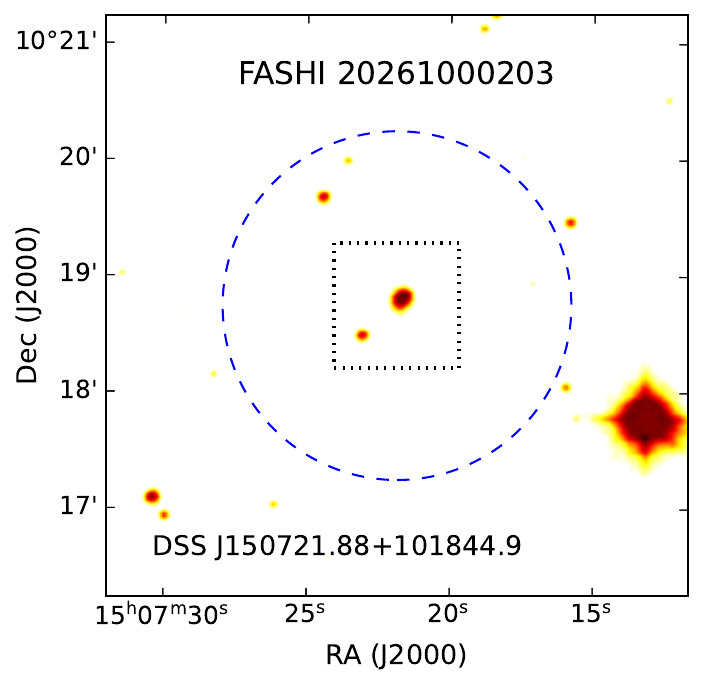}
 \caption{See caption in Figure\,\ref{Fig:FASHI_hi}}
 \end{figure*} 

 \begin{figure*}[htp]
 \centering
 \renewcommand{\thefigure}{\arabic{figure} (Continued)}
 \addtocounter{figure}{-1}
 \includegraphics[height=0.22\textwidth, angle=0]{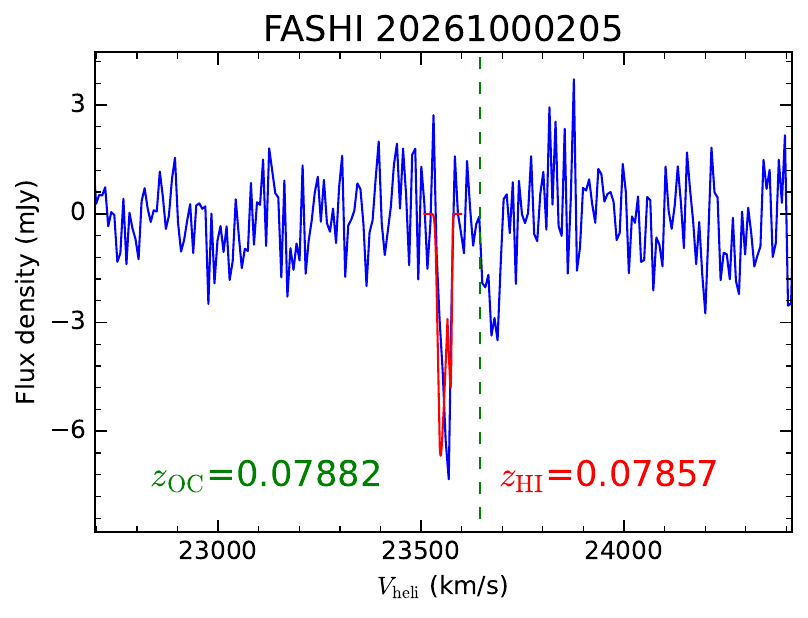}
 \includegraphics[height=0.27\textwidth, angle=0]{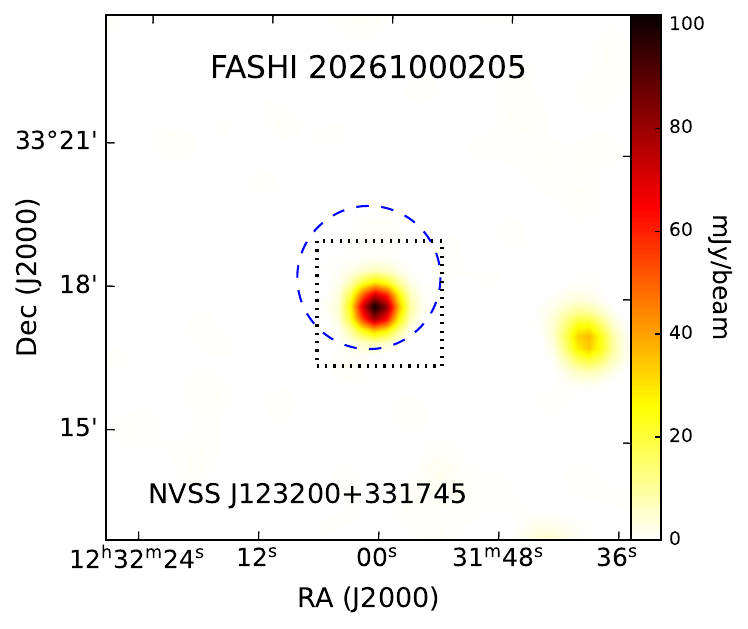}
 \includegraphics[height=0.27\textwidth, angle=0]{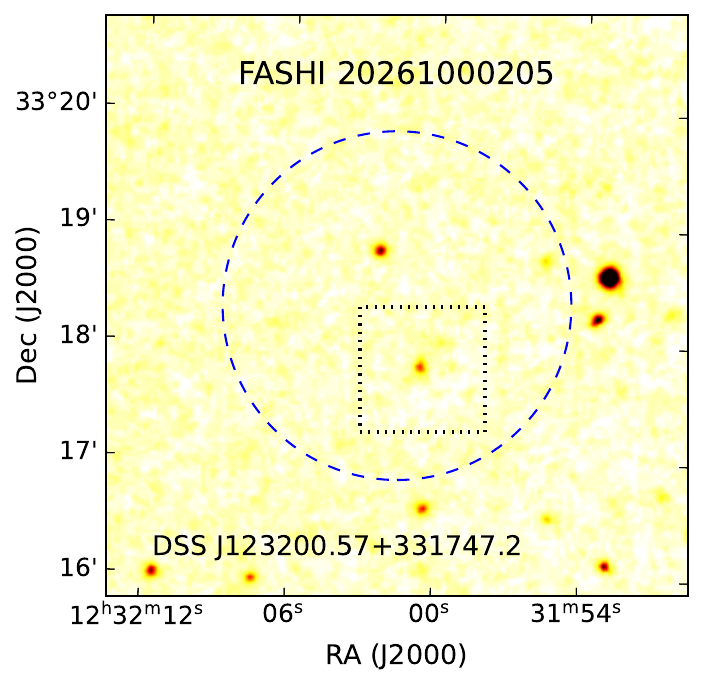}
 \includegraphics[height=0.22\textwidth, angle=0]{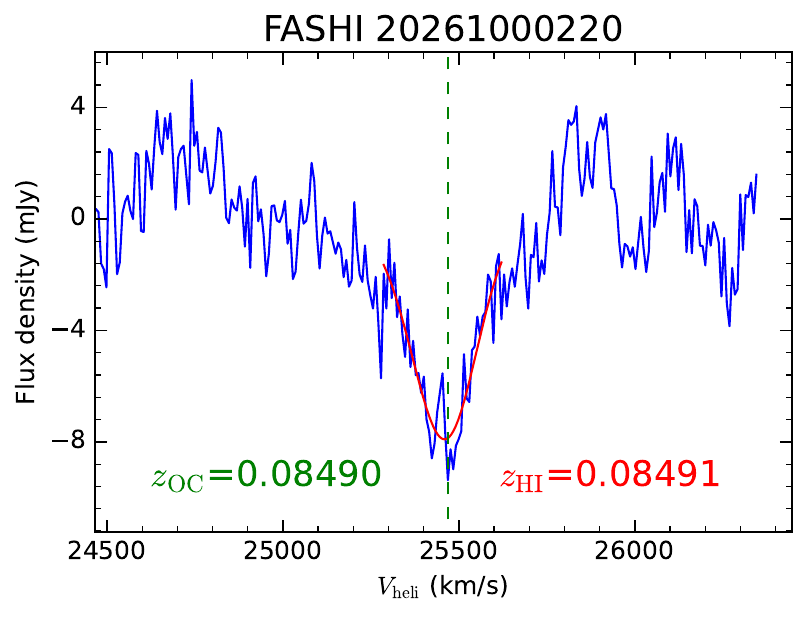}
 \includegraphics[height=0.27\textwidth, angle=0]{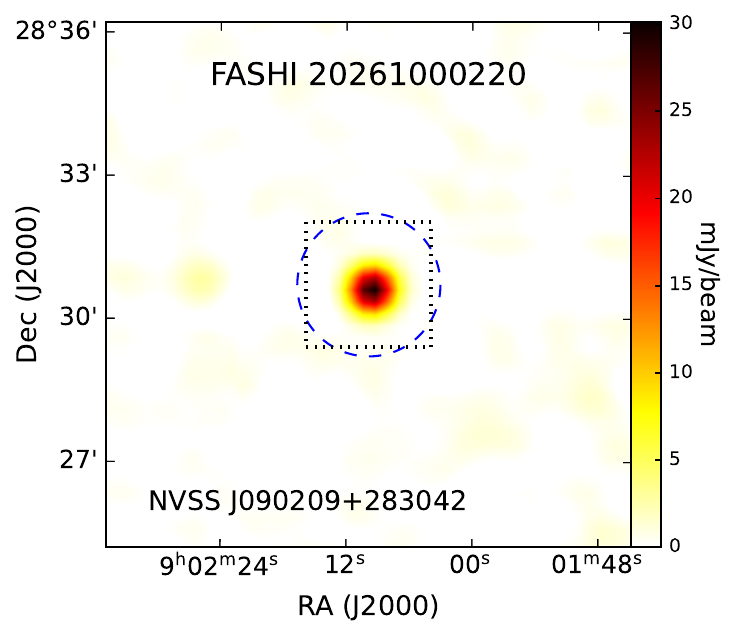}
 \includegraphics[height=0.27\textwidth, angle=0]{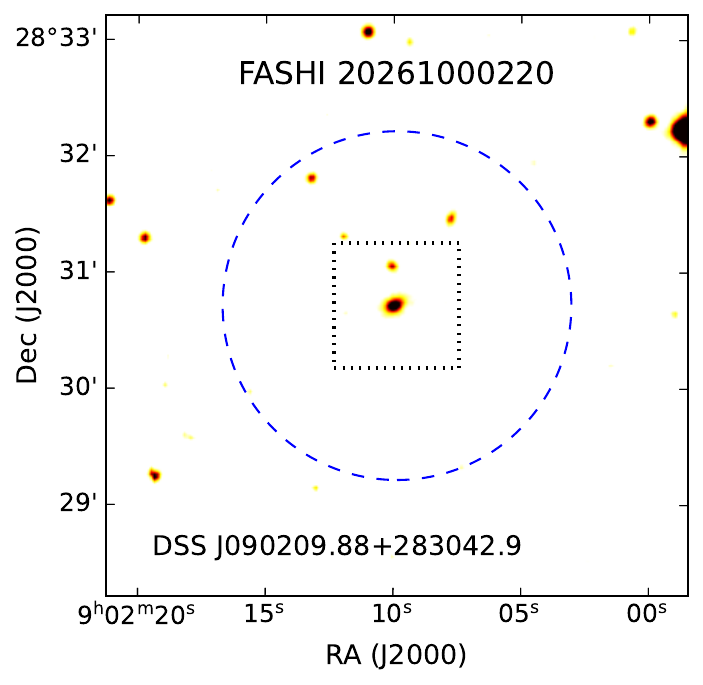}
 \includegraphics[height=0.22\textwidth, angle=0]{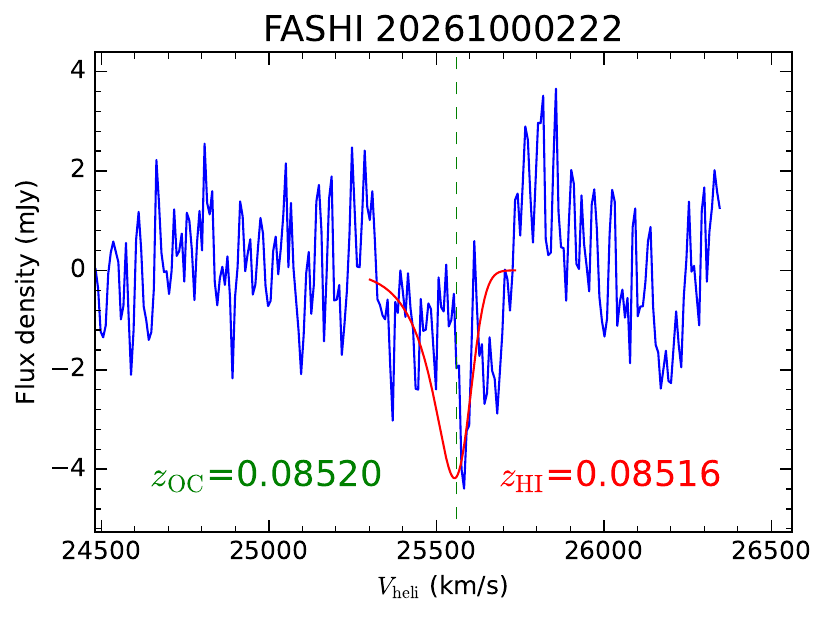}
 \includegraphics[height=0.27\textwidth, angle=0]{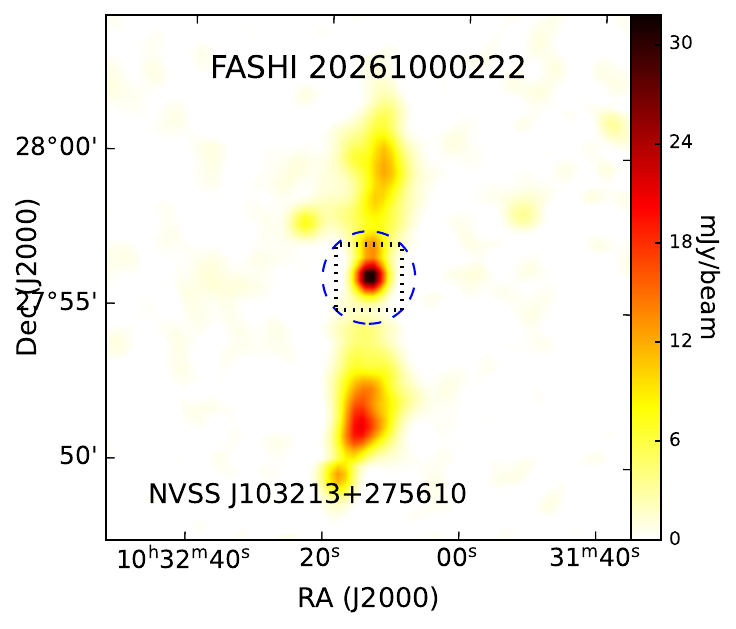}
 \includegraphics[height=0.27\textwidth, angle=0]{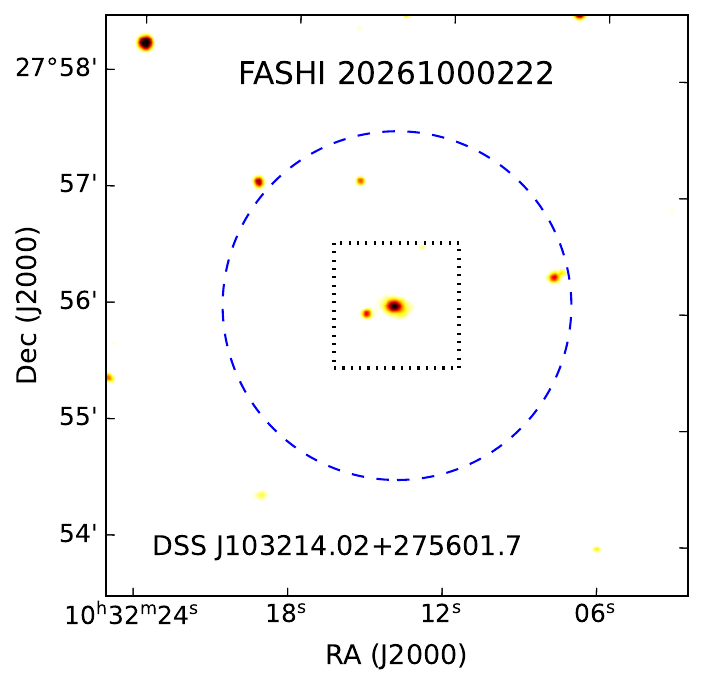}
 \includegraphics[height=0.22\textwidth, angle=0]{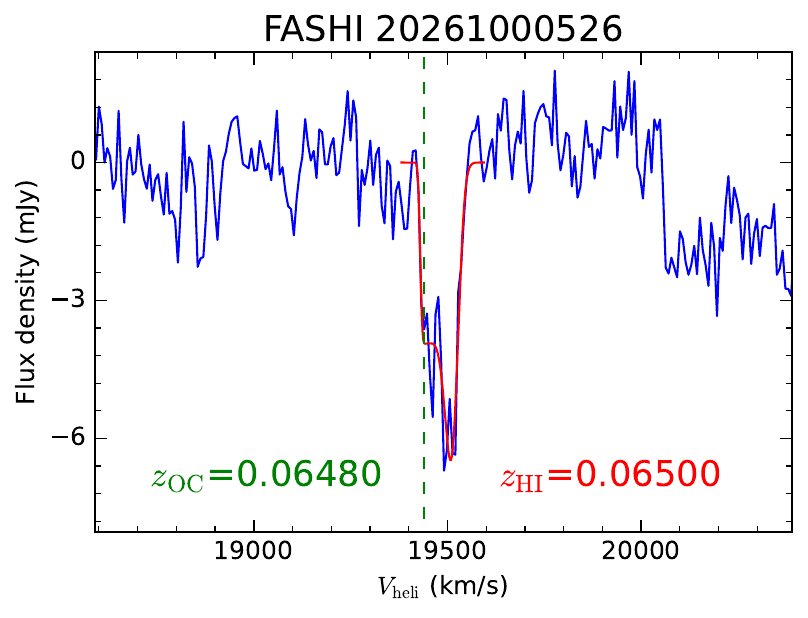}
 \includegraphics[height=0.27\textwidth, angle=0]{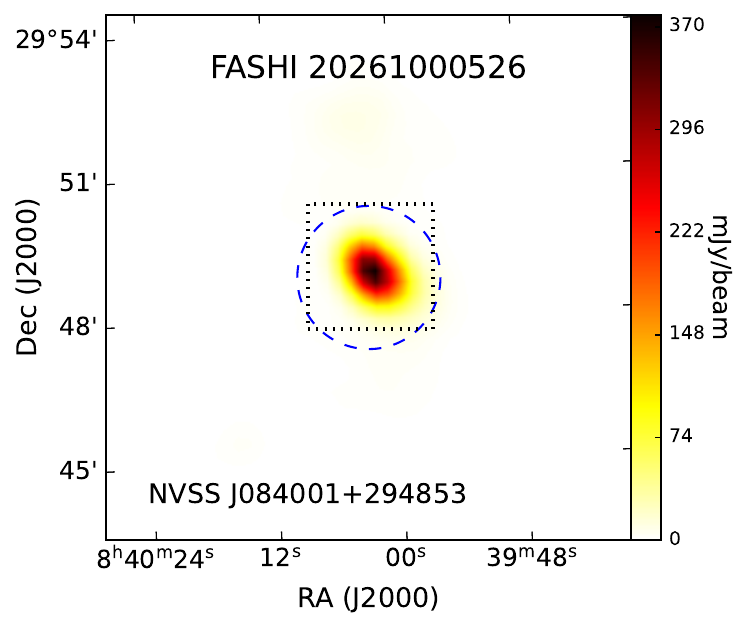}
 \includegraphics[height=0.27\textwidth, angle=0]{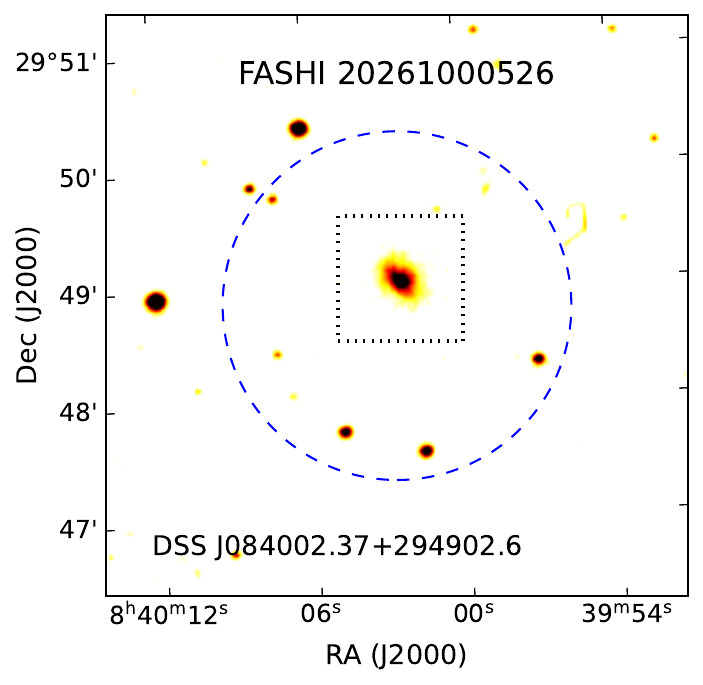}
 \caption{See caption in Figure\,\ref{Fig:FASHI_hi}}
 \end{figure*} 

 \begin{figure*}[htp]
 \centering
 \renewcommand{\thefigure}{\arabic{figure} (Continued)}
 \addtocounter{figure}{-1}
 \includegraphics[height=0.22\textwidth, angle=0]{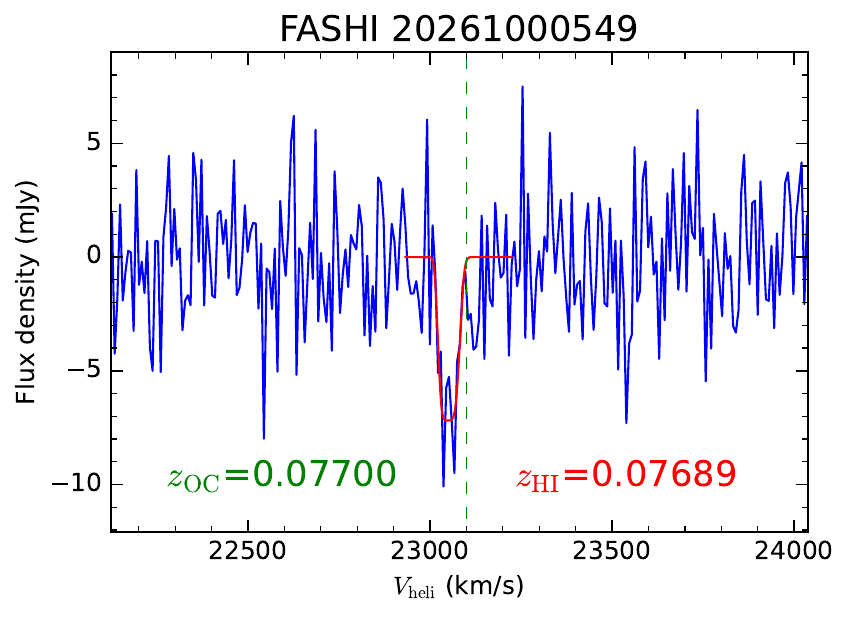}
 \includegraphics[height=0.27\textwidth, angle=0]{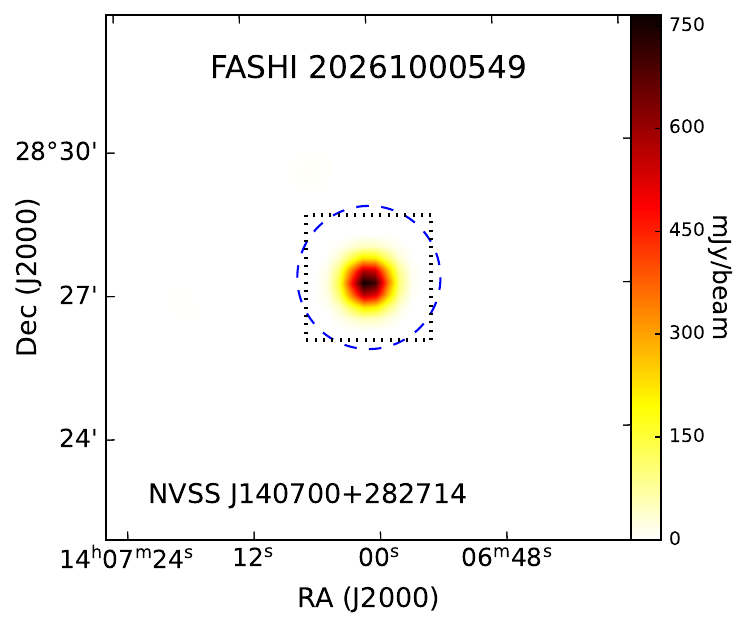}
 \includegraphics[height=0.27\textwidth, angle=0]{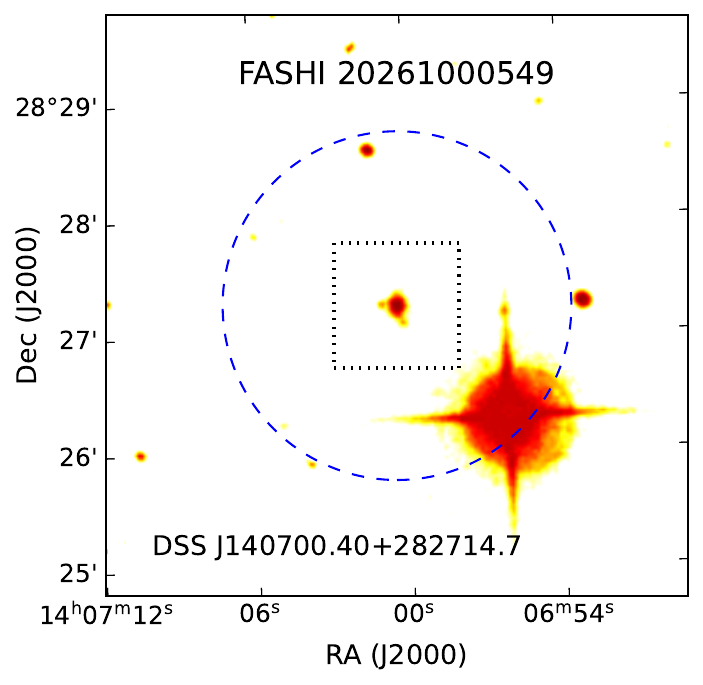}
 \includegraphics[height=0.22\textwidth, angle=0]{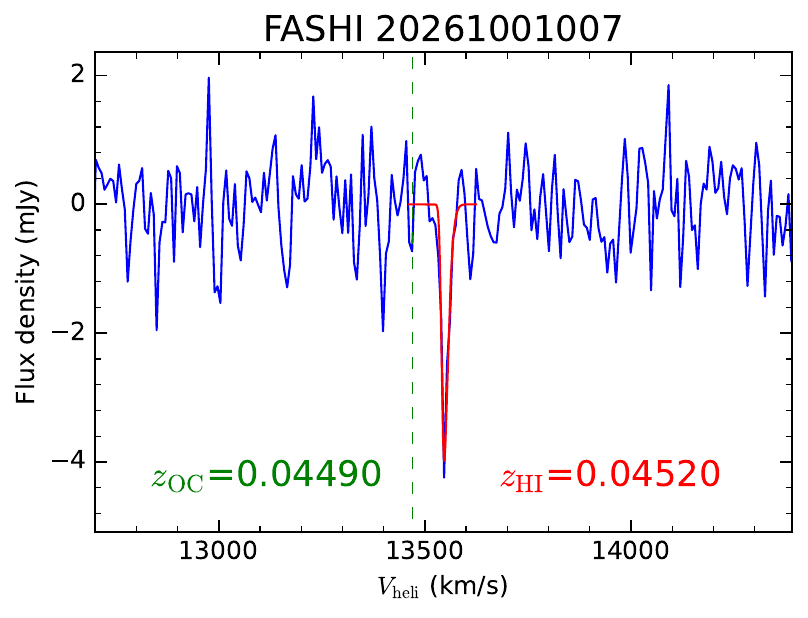}
 \includegraphics[height=0.27\textwidth, angle=0]{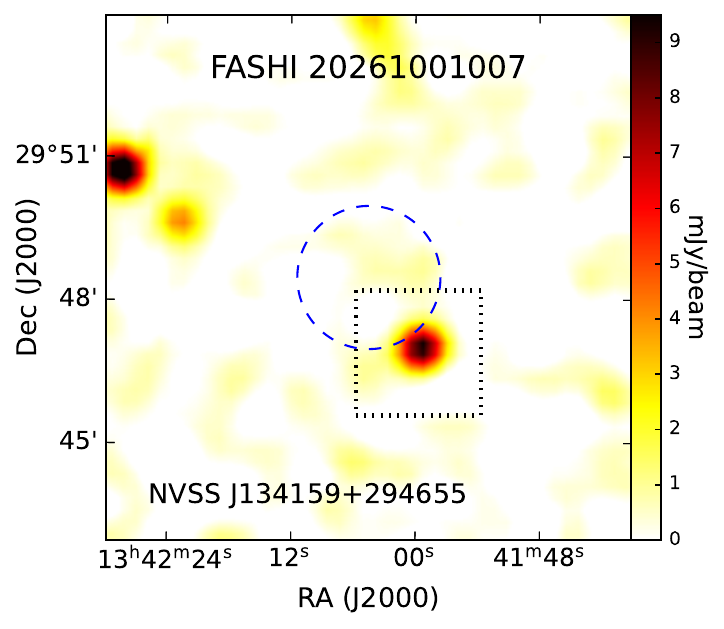}
 \includegraphics[height=0.27\textwidth, angle=0]{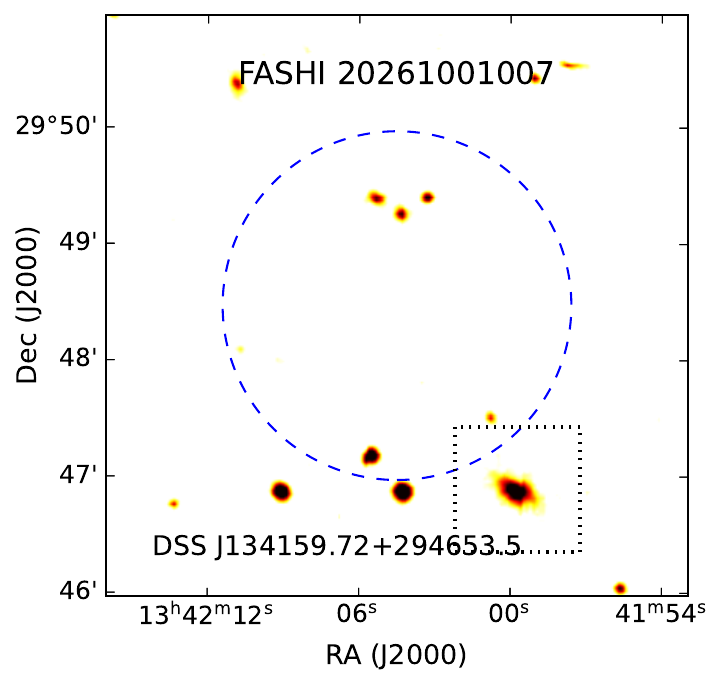}
 \includegraphics[height=0.22\textwidth, angle=0]{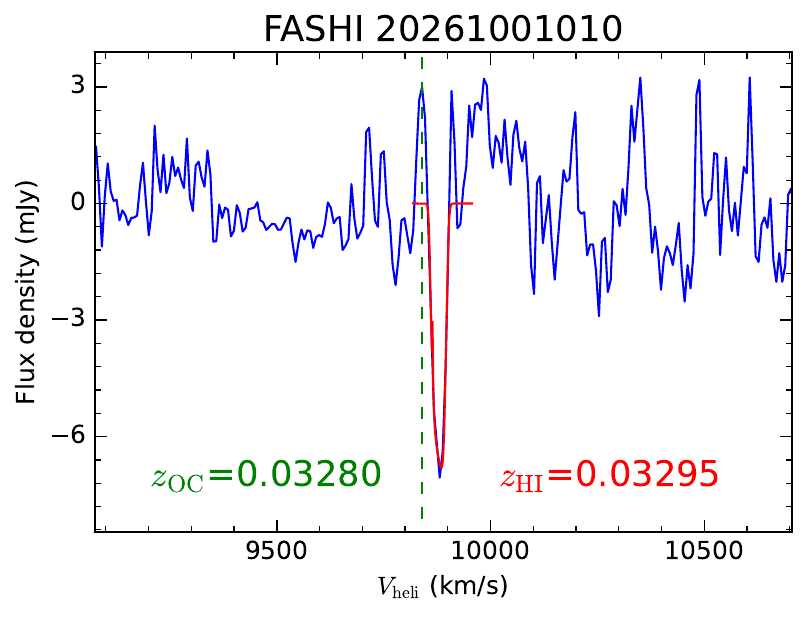}
 \includegraphics[height=0.27\textwidth, angle=0]{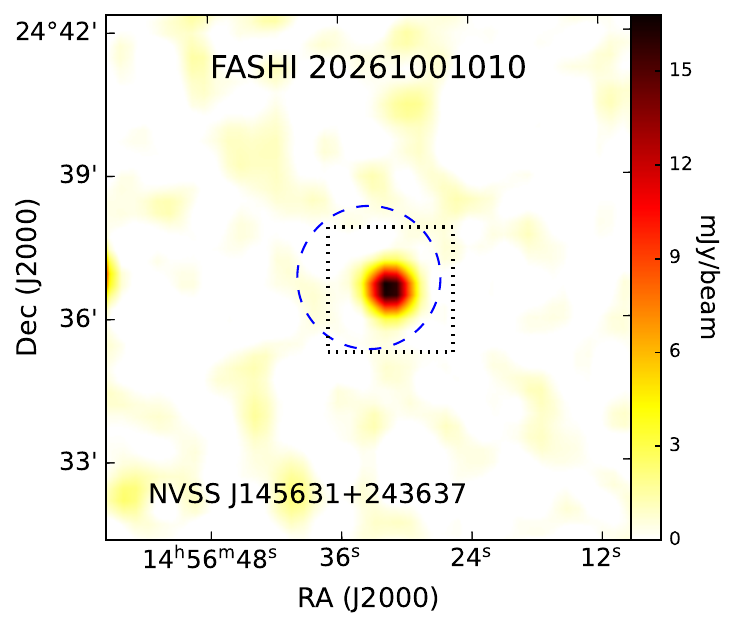}
 \includegraphics[height=0.27\textwidth, angle=0]{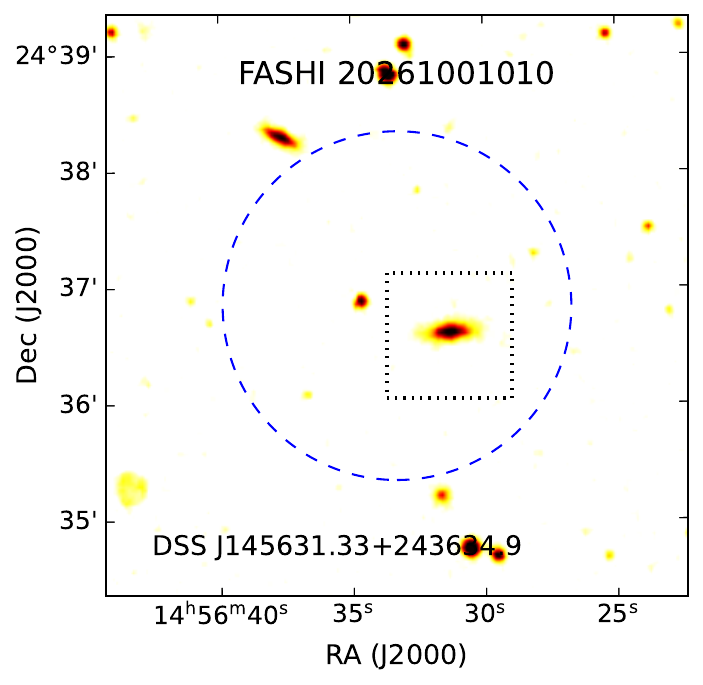}
 \includegraphics[height=0.22\textwidth, angle=0]{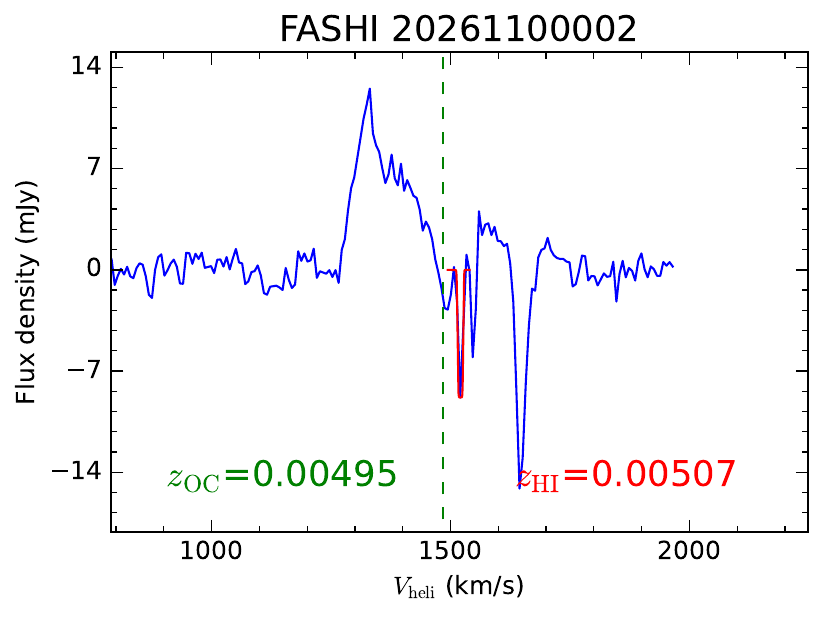}
 \includegraphics[height=0.27\textwidth, angle=0]{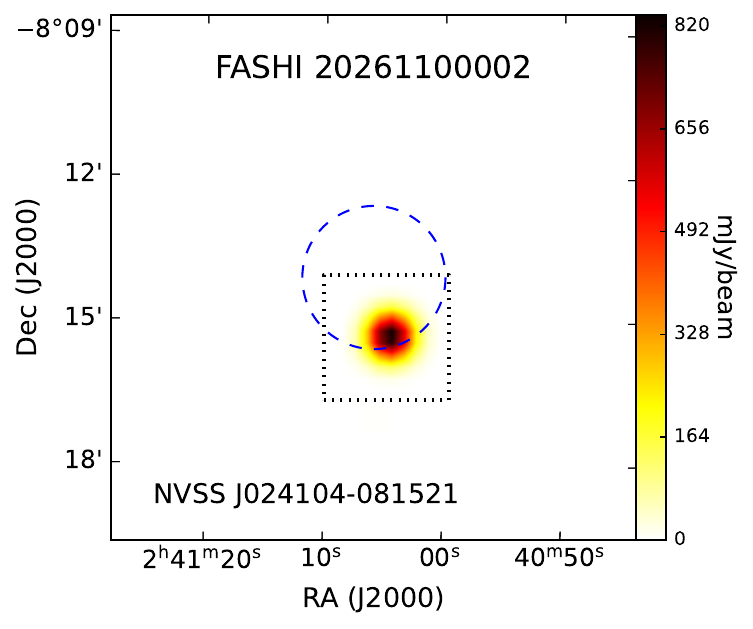}
 \includegraphics[height=0.27\textwidth, angle=0]{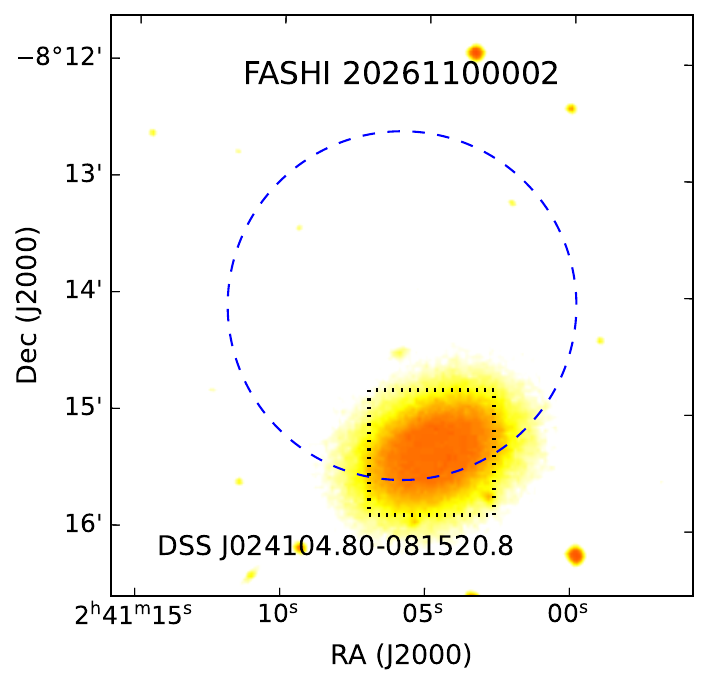}
 \caption{See caption in Figure\,\ref{Fig:FASHI_hi}}
 \end{figure*} 

 \begin{figure*}[htp]
 \centering
 \renewcommand{\thefigure}{\arabic{figure} (Continued)}
 \addtocounter{figure}{-1}
 \includegraphics[height=0.22\textwidth, angle=0]{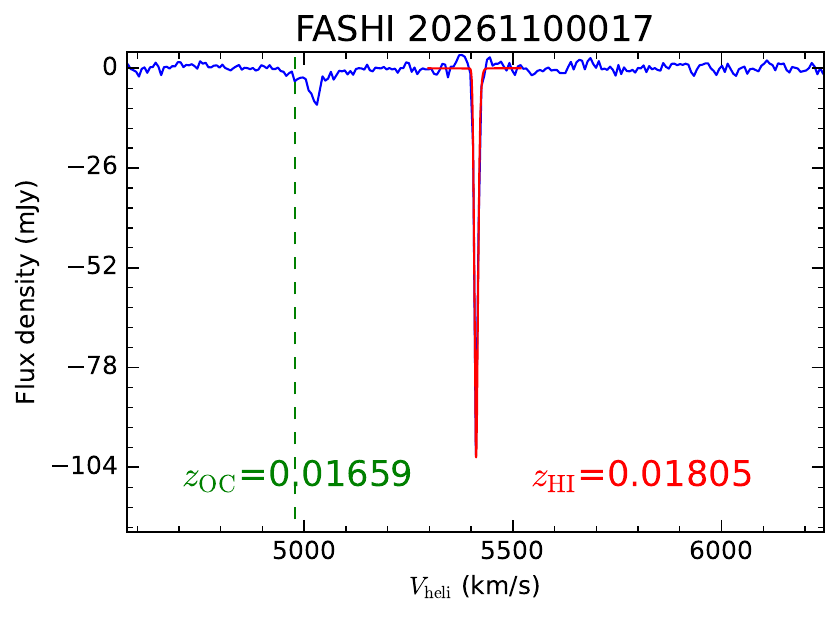}
 \includegraphics[height=0.27\textwidth, angle=0]{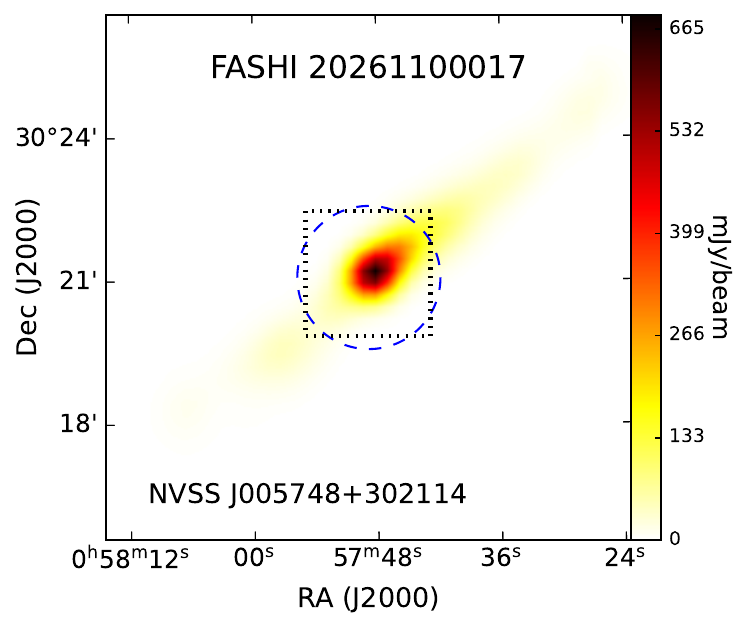}
 \includegraphics[height=0.27\textwidth, angle=0]{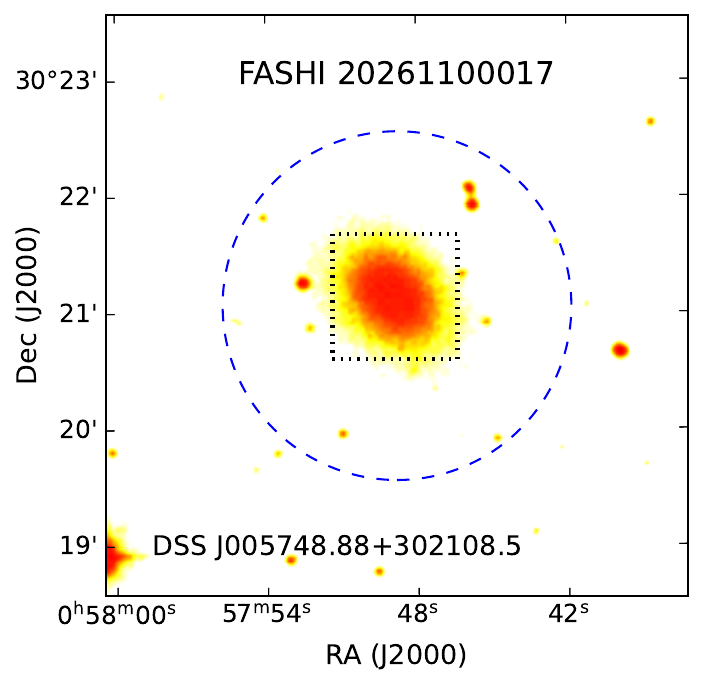}
 \includegraphics[height=0.22\textwidth, angle=0]{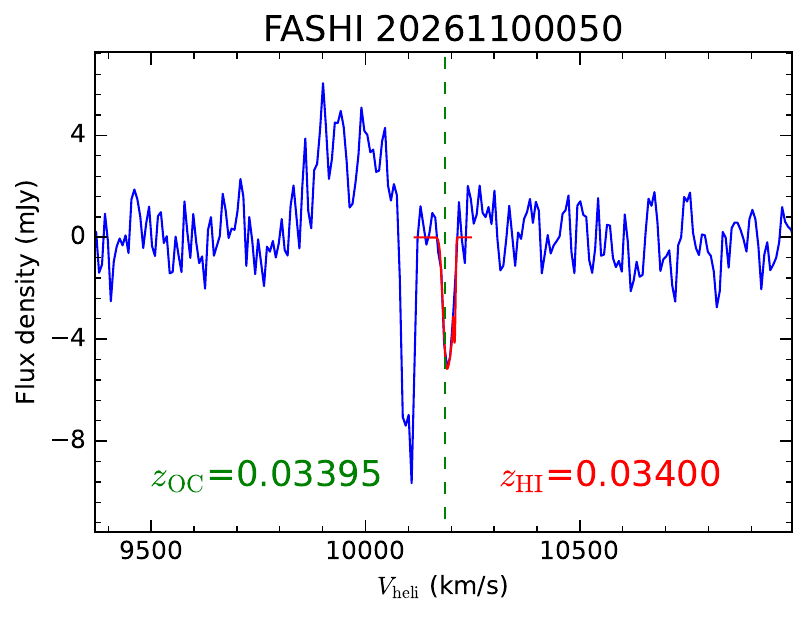}
 \includegraphics[height=0.27\textwidth, angle=0]{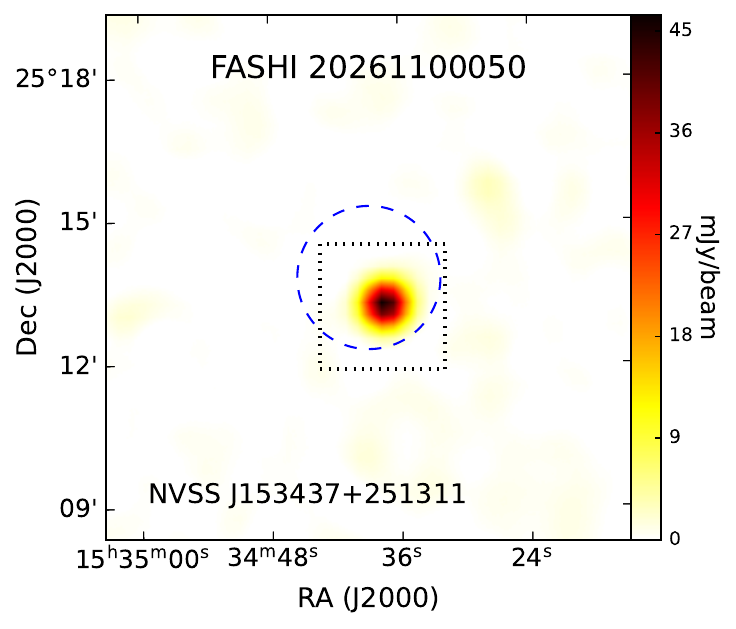}
 \includegraphics[height=0.27\textwidth, angle=0]{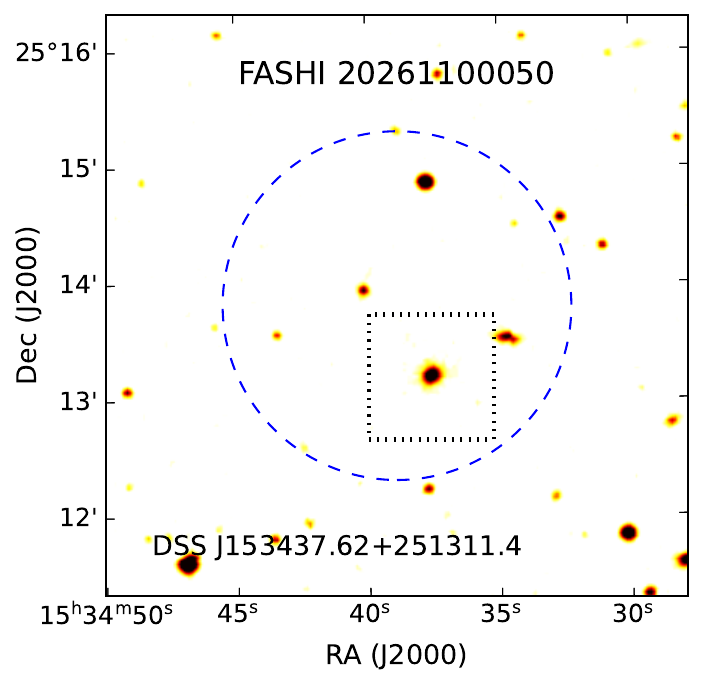}
 \includegraphics[height=0.22\textwidth, angle=0]{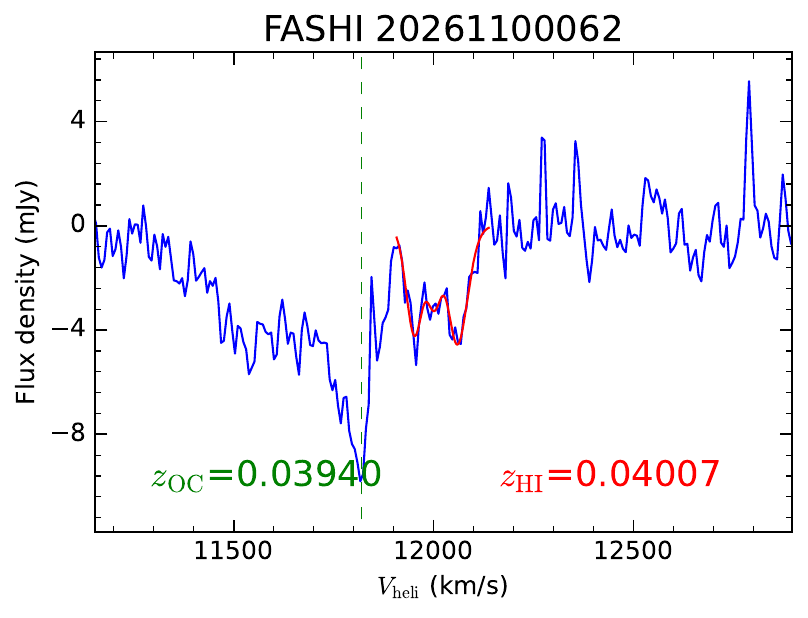}
 \includegraphics[height=0.27\textwidth, angle=0]{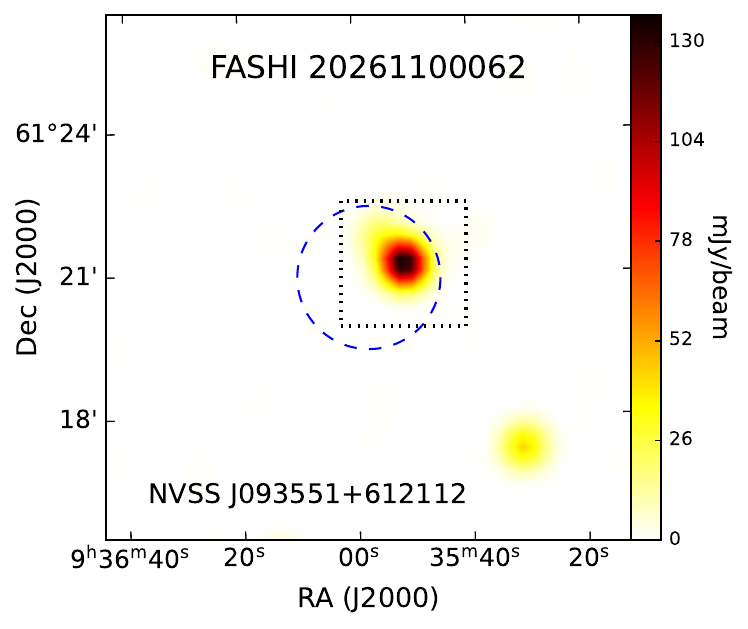}
 \includegraphics[height=0.27\textwidth, angle=0]{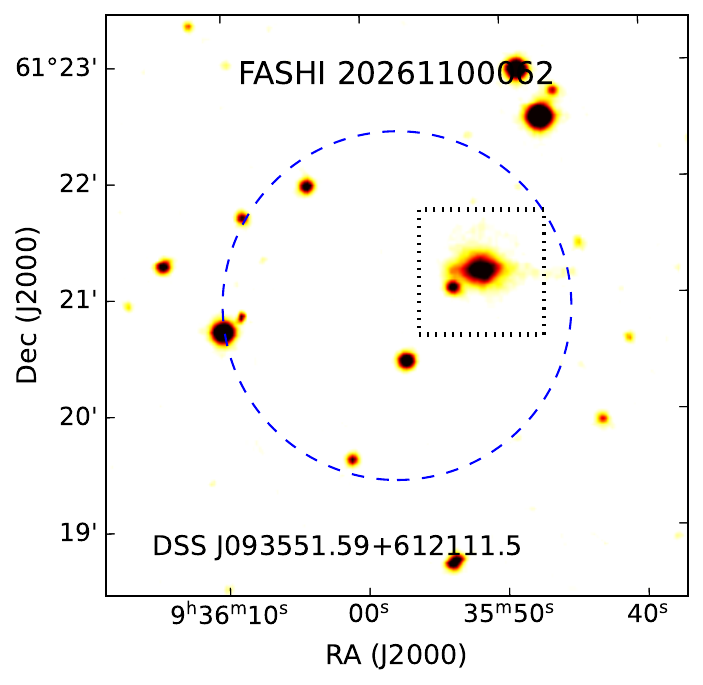}
 \includegraphics[height=0.22\textwidth, angle=0]{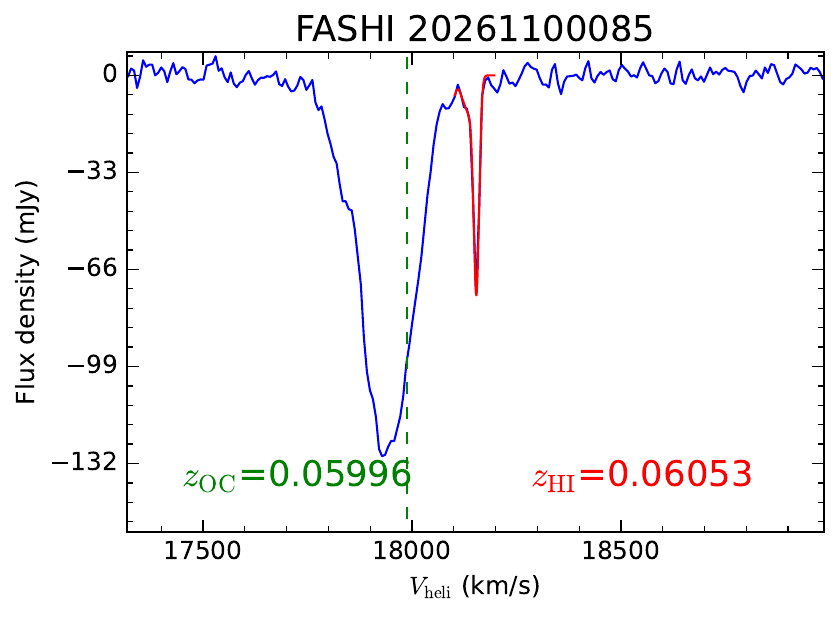}
 \includegraphics[height=0.27\textwidth, angle=0]{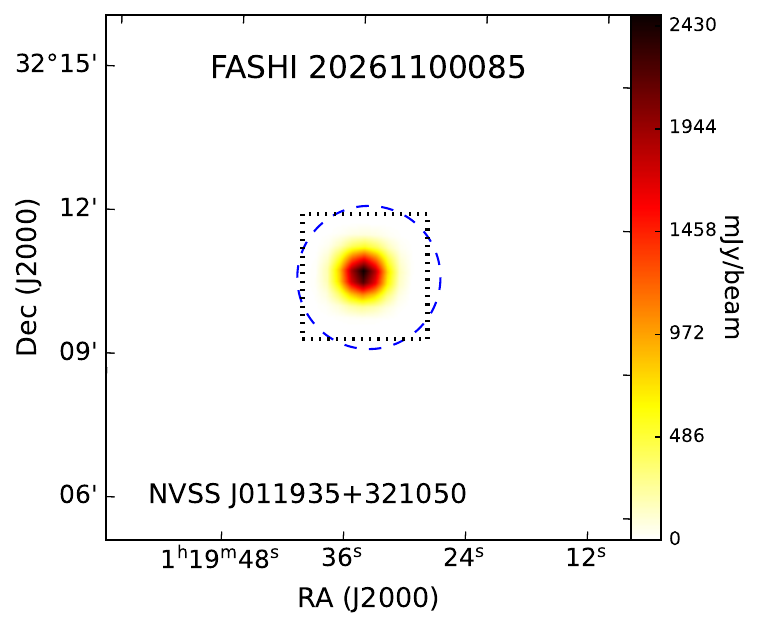}
 \includegraphics[height=0.27\textwidth, angle=0]{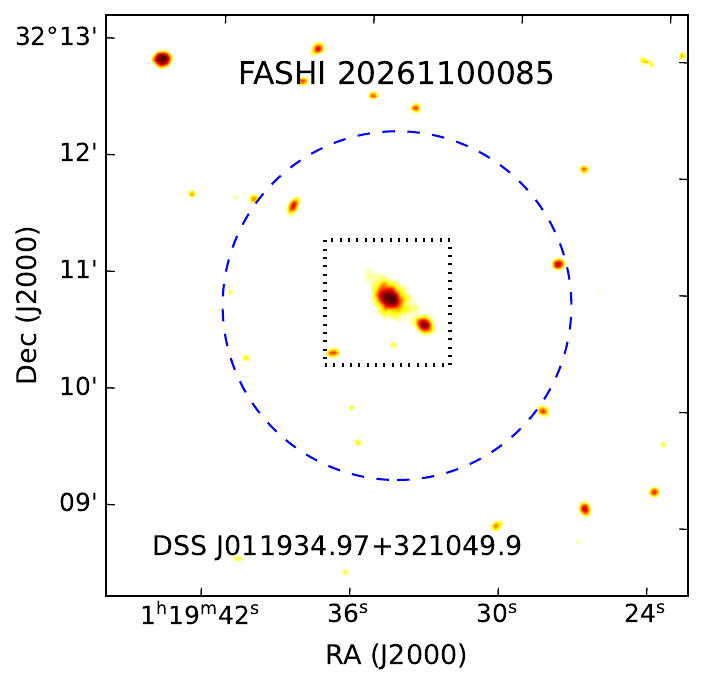}
 \caption{See caption in Figure\,\ref{Fig:FASHI_hi}}
 \end{figure*} 

 \begin{figure*}[htp]
 \centering
 \renewcommand{\thefigure}{\arabic{figure} (Continued)}
 \addtocounter{figure}{-1}
 \includegraphics[height=0.22\textwidth, angle=0]{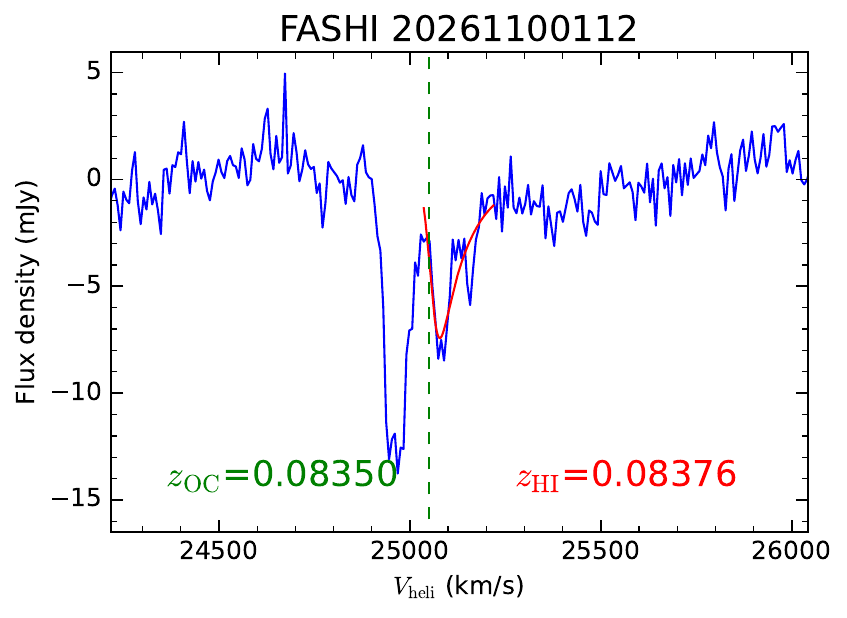}
 \includegraphics[height=0.27\textwidth, angle=0]{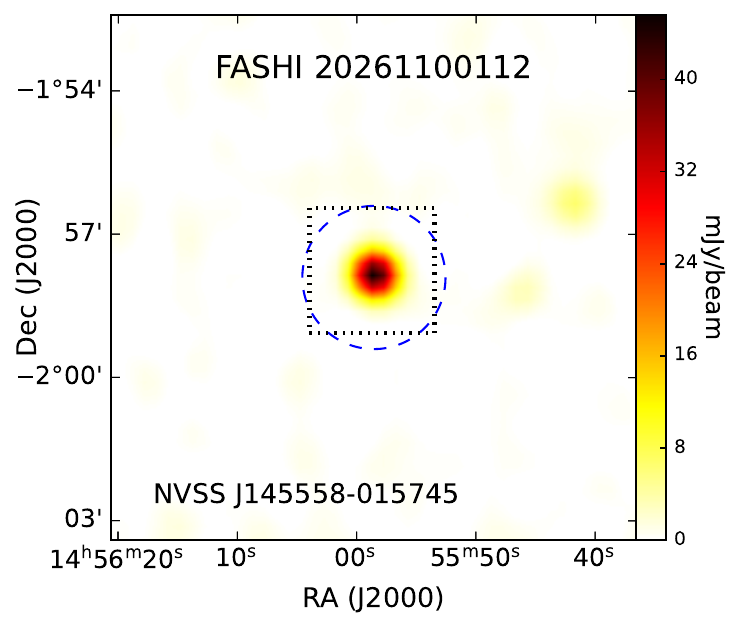}
 \includegraphics[height=0.27\textwidth, angle=0]{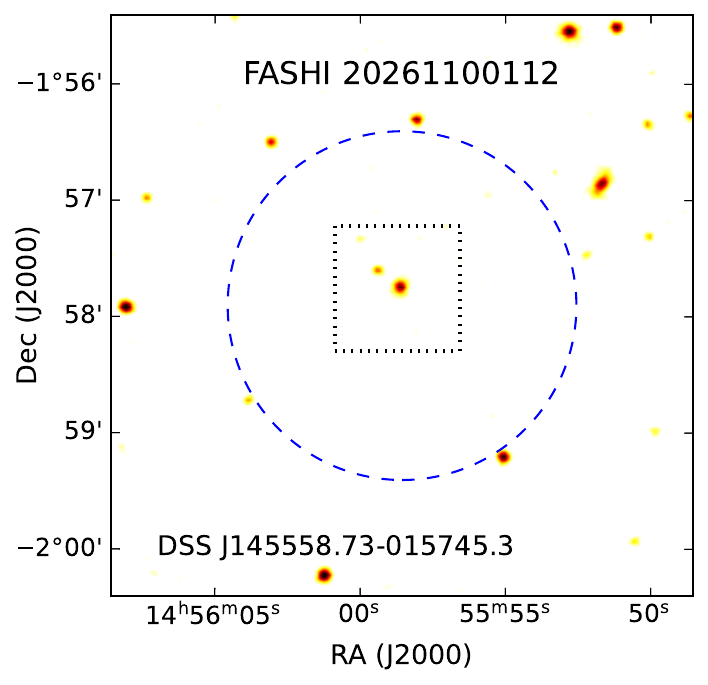}
 \includegraphics[height=0.22\textwidth, angle=0]{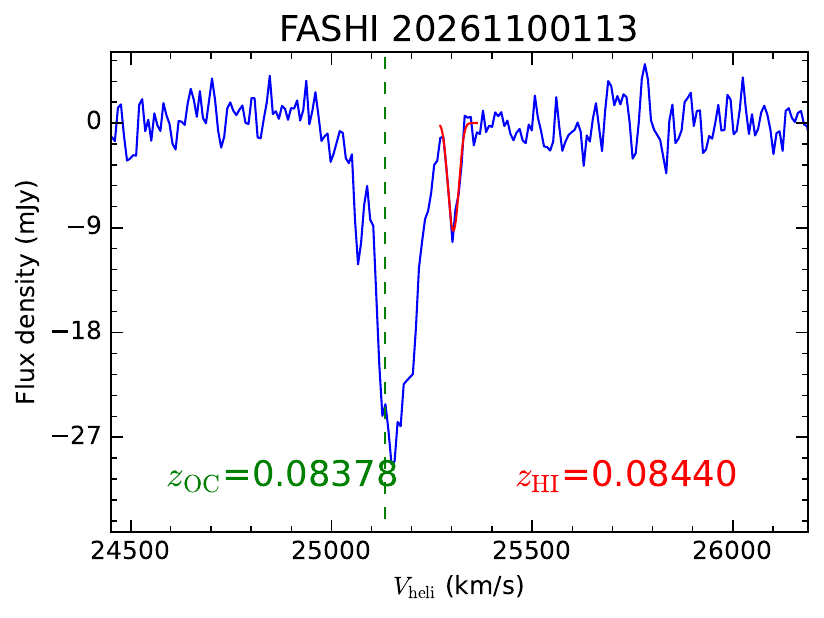}
 \includegraphics[height=0.27\textwidth, angle=0]{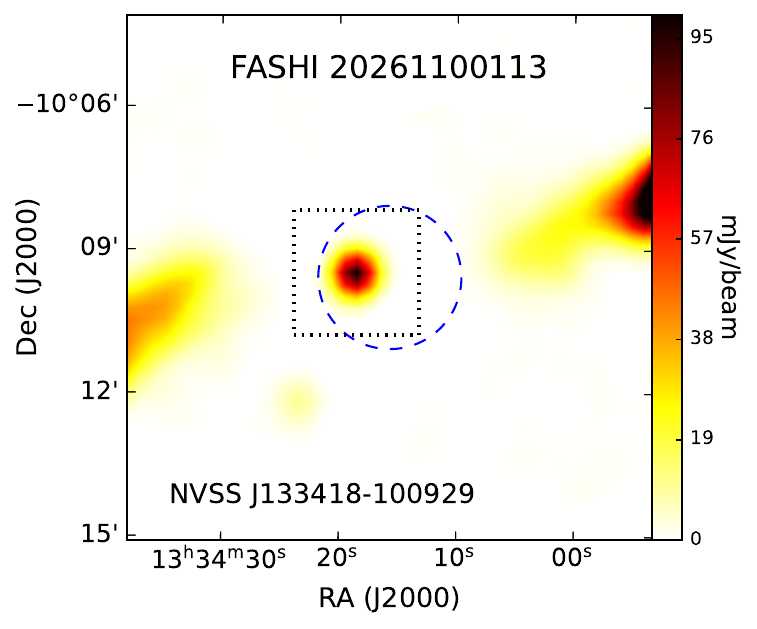}
 \includegraphics[height=0.27\textwidth, angle=0]{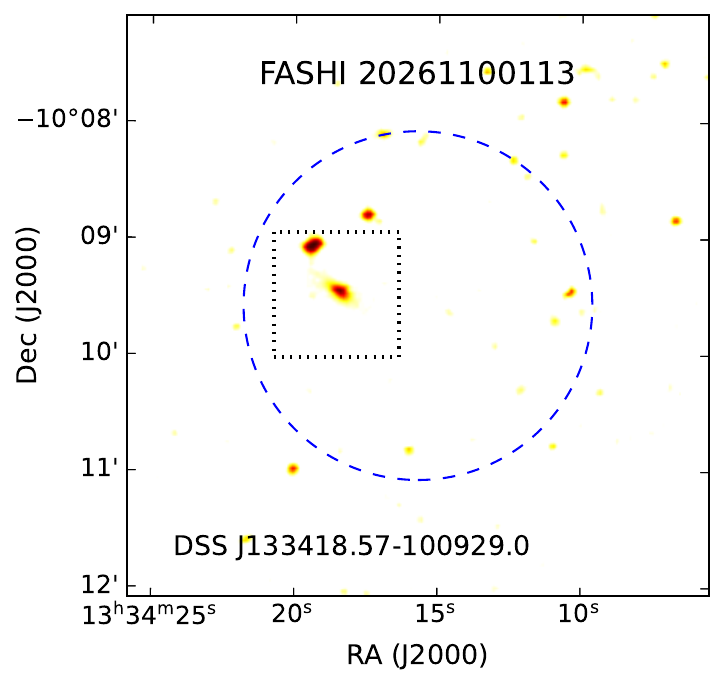}
 \includegraphics[height=0.22\textwidth, angle=0]{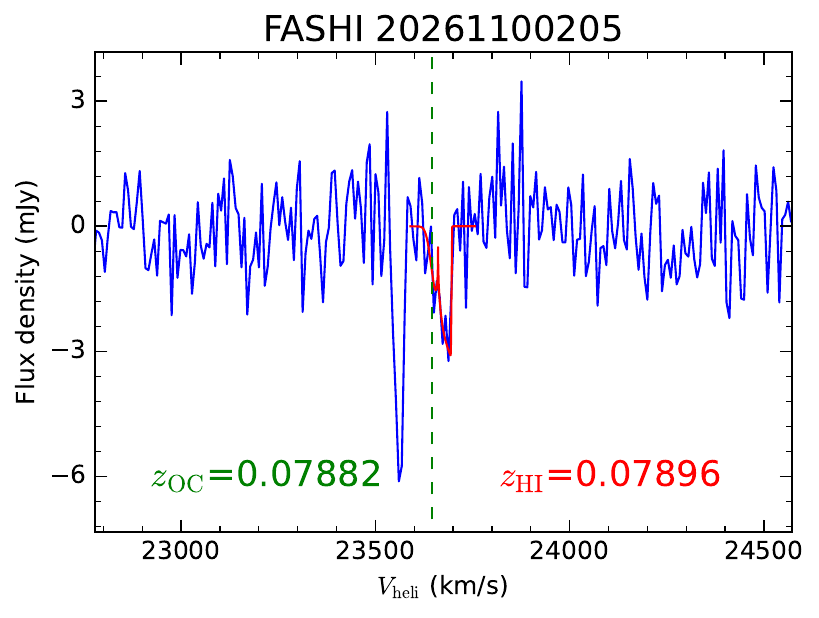}
 \includegraphics[height=0.27\textwidth, angle=0]{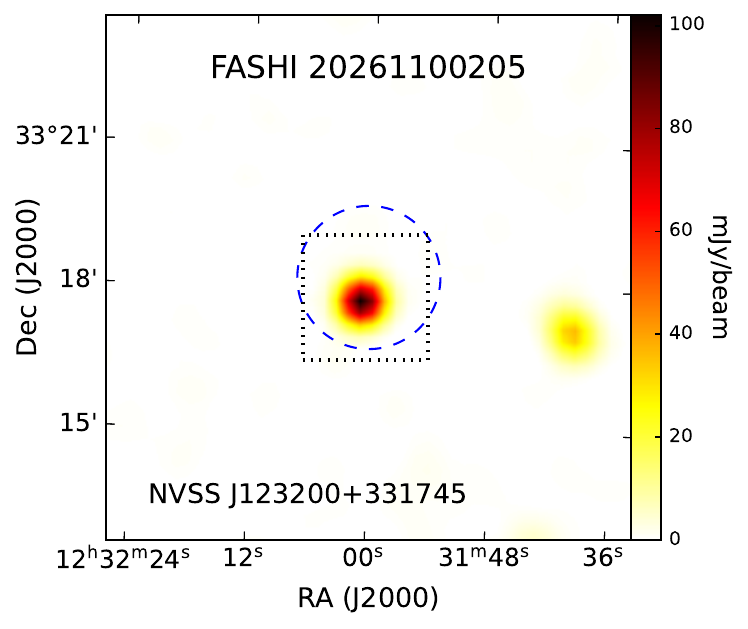}
 \includegraphics[height=0.27\textwidth, angle=0]{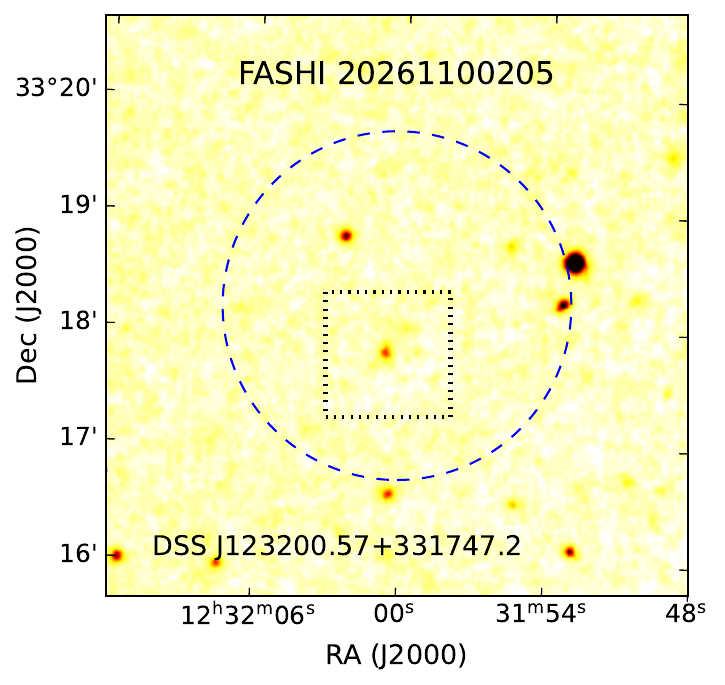}
 \includegraphics[height=0.22\textwidth, angle=0]{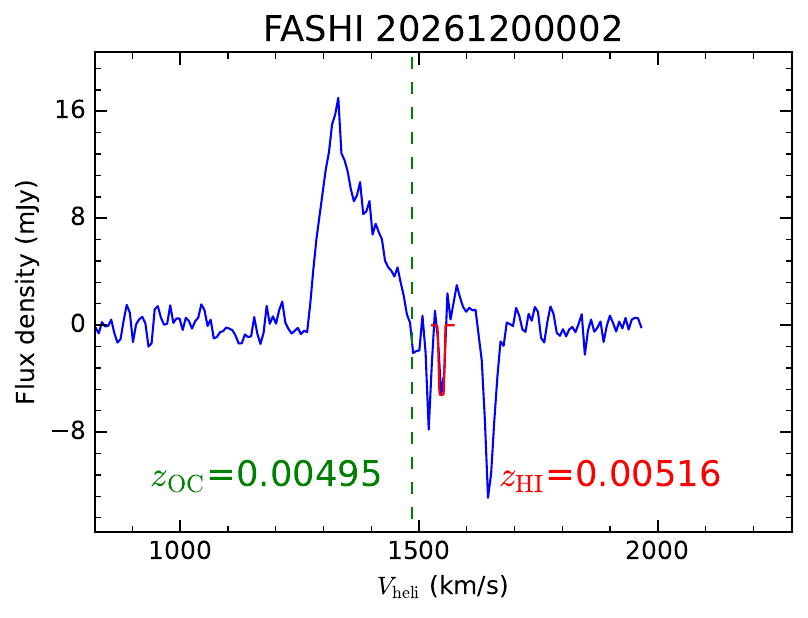}
 \includegraphics[height=0.27\textwidth, angle=0]{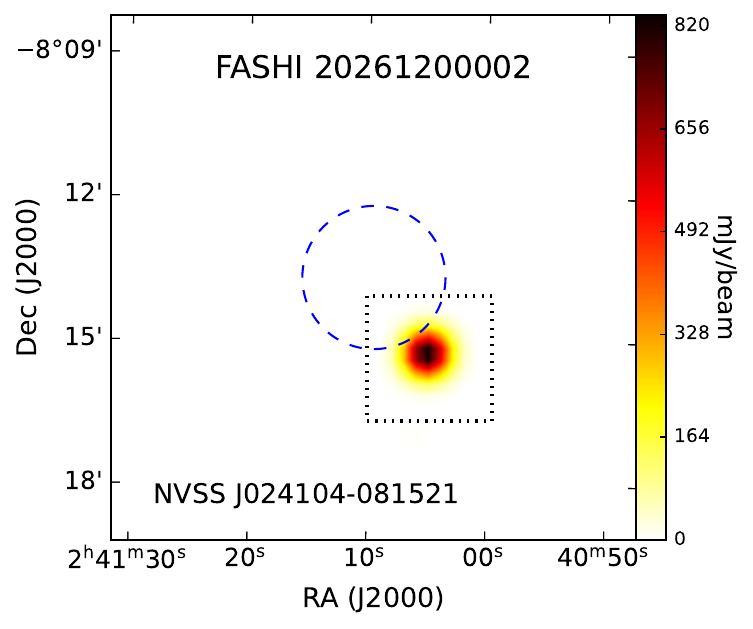}
 \includegraphics[height=0.27\textwidth, angle=0]{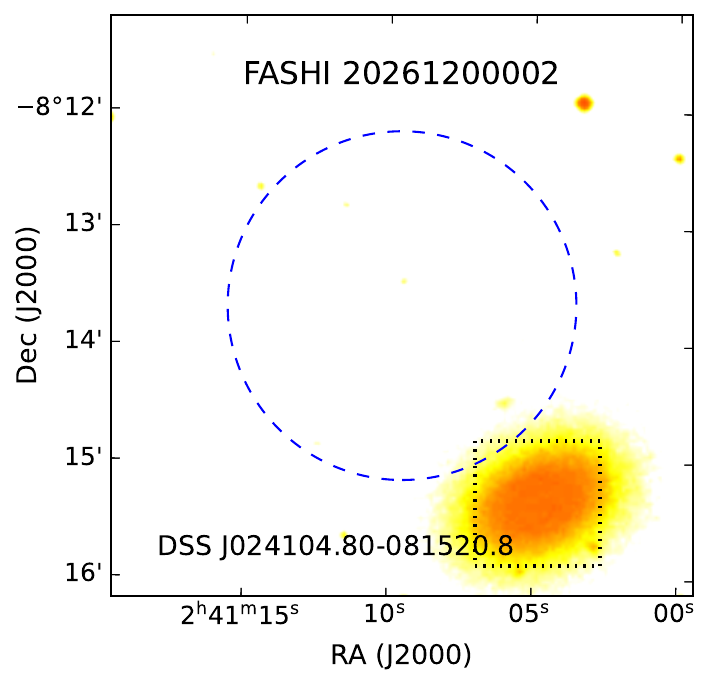}
 \caption{See caption in Figure\,\ref{Fig:FASHI_hi}}
 \end{figure*} 

 \begin{figure*}[htp]
 \centering
 \renewcommand{\thefigure}{\arabic{figure} (Continued)}
 \addtocounter{figure}{-1}
 \includegraphics[height=0.22\textwidth, angle=0]{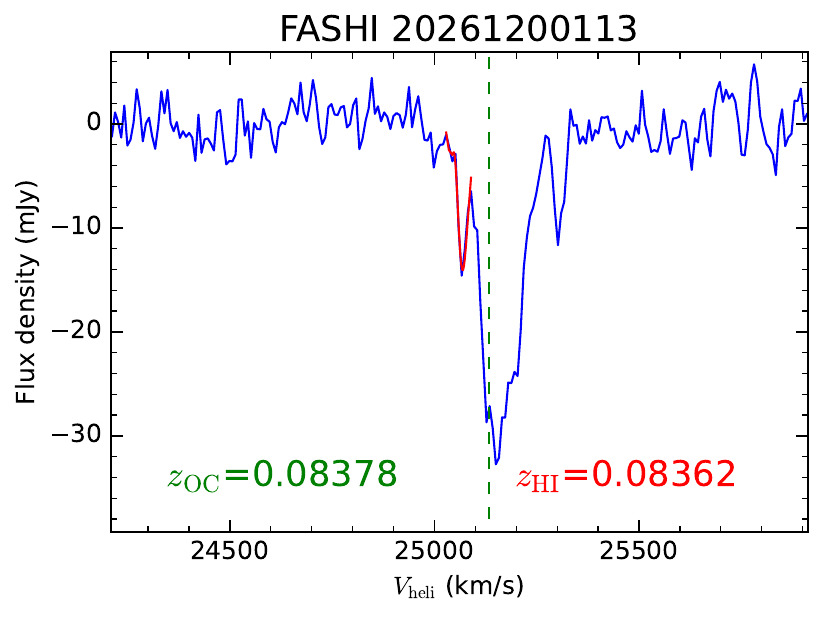}
 \includegraphics[height=0.27\textwidth, angle=0]{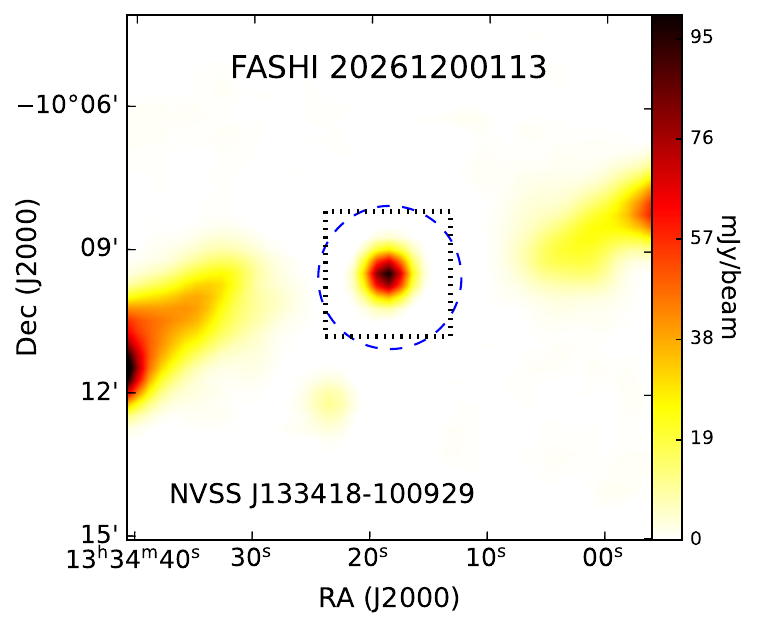}
 \includegraphics[height=0.27\textwidth, angle=0]{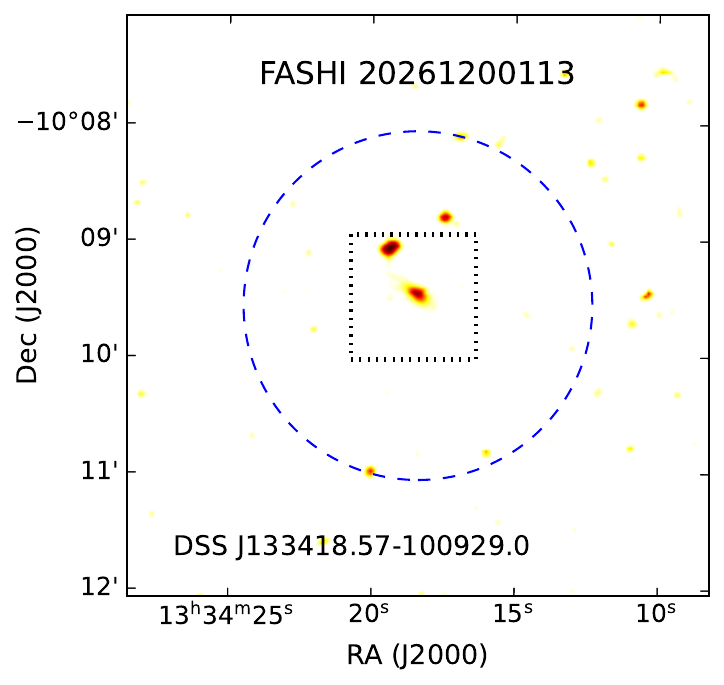}
 \caption{See caption in Figure\,\ref{Fig:FASHI_hi}}
 \end{figure*}

 \clearpage

\end{document}